\documentclass[aps,prx,superscriptaddress,reprint,notitlepage]{revtex4-2}
\usepackage[utf8]{inputenc}

\usepackage[hidelinks]{hyperref}

\usepackage{color}
\hypersetup{colorlinks=true}
\usepackage{graphicx}
\usepackage{bm}
\usepackage{comment}
\usepackage{amsmath}
\usepackage{amssymb}
\usepackage{xspace}
\usepackage{tabularx}
\usepackage[capitalise]{cleveref}
\usepackage{txfonts}
\graphicspath{{./figures/}}
\usepackage{physics}
\usepackage{siunitx}
\usepackage{booktabs}
\usepackage{chemformula}
\usepackage{algorithmic}
\usepackage{algorithm}
\newcommand{\subfigref}[2]{Fig.~\hyperref[#1]{\ref*{#1}#2}}

\renewcommand{\vec}[1]{{\boldsymbol #1}}

\begin{document}
	
	\title{
    Spontaneous Magnon Decays from Nonrelativistic Time-Reversal Symmetry Breaking in Altermagnets
    }
	
	\author{Rintaro Eto}
    \affiliation{Department of Applied Physics, Waseda University, Okubo, Shinjuku-ku, Tokyo 169-8555, Japan}
    \affiliation{Institut für Physik, Johannes Gutenberg Universität Mainz, D-55099 Mainz, Germany}

    \author{Matthias Gohlke}
    \affiliation{Theory of Quantum Matter Unit, Okinawa Institute of Science and Technology Graduate University, Onna-son, Okinawa 904-0495, Japan}
    \affiliation{Max Planck Institute for the Physics of Complex Systems, N\"{o}thnitzer Str. 38, 01187 Dresden, Germany}

    \author{Jairo Sinova}
    \affiliation{Institut für Physik, Johannes Gutenberg Universität Mainz, D-55099 Mainz, Germany}

    \author{Masahito Mochizuki}
    \affiliation{Department of Applied Physics, Waseda University, Okubo, Shinjuku-ku, Tokyo 169-8555, Japan}

    \author{Alexander L. Chernyshev}
    \affiliation{Department of Physics and Astronomy, University of California, Irvine, California 92697, USA}

	\author{Alexander Mook}
	\affiliation{Institut für Physik, Johannes Gutenberg Universität Mainz, D-55099 Mainz, Germany}
	
	\date{\today}
	%

\begin{abstract}
Quasiparticles are central to condensed matter physics, but their stability can be undermined by quantum many-body interactions. Magnons, quasiparticles in quantum magnets, are particularly intriguing because their properties are governed by both real and spin space. While crystal symmetries may be low, spin interactions often remain approximately isotropic, limiting spontaneous magnon decay. Textbook wisdom holds that collinear Heisenberg magnets follow a dichotomy: ferromagnets host stable magnons, while antiferromagnetic magnons may decay depending on dispersion curvature. Up to now, relativistic spin-orbit coupling and noncollinear order that connect spin space to real space, were shown to introduce more complex magnon instability mechanisms. Here, we show that even in nonrelativistic isotropic collinear systems, this conventional dichotomy is disrupted in altermagnets. Altermagnets, a newly identified class of collinear magnets, exhibit compensated spin order with nonrelativistic time-reversal symmetry breaking and even-parity band splitting. Using kinematic analysis, nonlinear spin-wave theory, and quantum simulations, we reveal that even weak band splitting opens a decay phase space, driving quasiparticle breakdown. Additionally, d-wave altermagnets form a rare ``island of stability'' at the Brillouin zone center. Our findings establish a quasiparticle stability trichotomy in collinear Heisenberg magnets and position altermagnets as a promising platform for unconventional spin dynamics.
\end{abstract}

\maketitle





\section*{Introduction}

The concept of a quasiparticle---an emergent, long-lived excitation that weakly interacts with its environment---is foundational to condensed matter physics, shaping our understanding of diverse systems \cite{Venema2016, Pines2018}. A central question is when and how quasiparticles break down at zero temperature due to quantum fluctuations \cite{ZhitomirskyChernyshev2013, Verresen2019}. In bosonic systems, many-body interactions can trigger spontaneous quasiparticle decay, as seen in the instability of Bogoliubov excitations \cite{pitaevskii1959properties, Maris1977} and anharmonic phonon decay in solids \cite{Maradudin1962}. In quantum magnets, magnons---collective spin-wave excitations---experience nonlinear interactions, leading to decay in frustrated and low-dimensional systems \cite{ZhitomirskyChernyshev2013}. Understanding these decay mechanisms is crucial for determining whether quasiparticles remain the appropriate low-energy degrees of freedom or give way to more complex excitations. Experimentally, such decay manifests as spectral broadening and weight redistribution in neutron scattering and optical spectroscopy, offering insight into a material’s quantum many-body physics.

In many magnetically ordered insulators, spin space remains approximately isotropic and their leading-order physics is described by the paradigmatic Heisenberg model. This is the case when relativistic effects are negligible, such that spin space and real space are not connected. Spontaneous magnetic long-range order breaks continuous spin-rotation symmetry, leading to gapless Nambu-Goldstone modes and universal low-energy behavior. In collinear magnets---the simplest form of magnetic order---a dichotomy of magnon stability emerges: ferromagnetic magnons are inherently stable, as their ground state and excitations are exact eigenstates free of quantum fluctuations, whereas antiferromagnetic magnons experience fluctuations that \textit{can} induce spontaneous decay \cite{Harris1971, ZhitomirskyChernyshev2013}. 
This dichotomy collapses when spin space becomes connected to real space by the relativistic spin-orbit coupling, noncollinear order, or dipolar interactions, making magnon decay more intricate \cite{chernyshevzhitomirsky2009, Mourigal2010, Stephanovich2011, ZhitomirskyChernyshev2013, ChernyshevMaksimov2016, Winter2017, McClarty2018, McClarty2019, mook2021, Gohlke2023PRL, Hong2017}.

Altermagnets---a newly identified class of collinear magnets \cite{Smejkal2022altermagnets, Smejkal2022emergent}---challenge this dichotomy even within the nonrelativistic limit. Like antiferromagnets, they feature compensated magnetic order and quantum fluctuations, yet they also break time-reversal symmetry and exhibit nonrelativistic band splitting reminiscent of ferromagnets. Their unconventional beyond-s-wave even-parity spin splitting (d-, g-, or i-wave) of magnons \cite{Naka2019, SmejkalMagnons23, Maier2023, McClarty2025AMPolarizedNeutrons} raises a fundamental question: Can altermagnets support stable magnons, and how does their distinctive spin splitting interact with quantum many-body fluctuations to influence magnon stability? This question has gained urgency following quantum simulations of spin-$1/2$ altermagnets \cite{GarciaGaitan2025, YLiu2025MagnonAltermagnet}, which suggest pronounced nonclassical effects and high-energy magnon instability.

\begin{figure*}
    \centering
    \includegraphics[width=\textwidth]{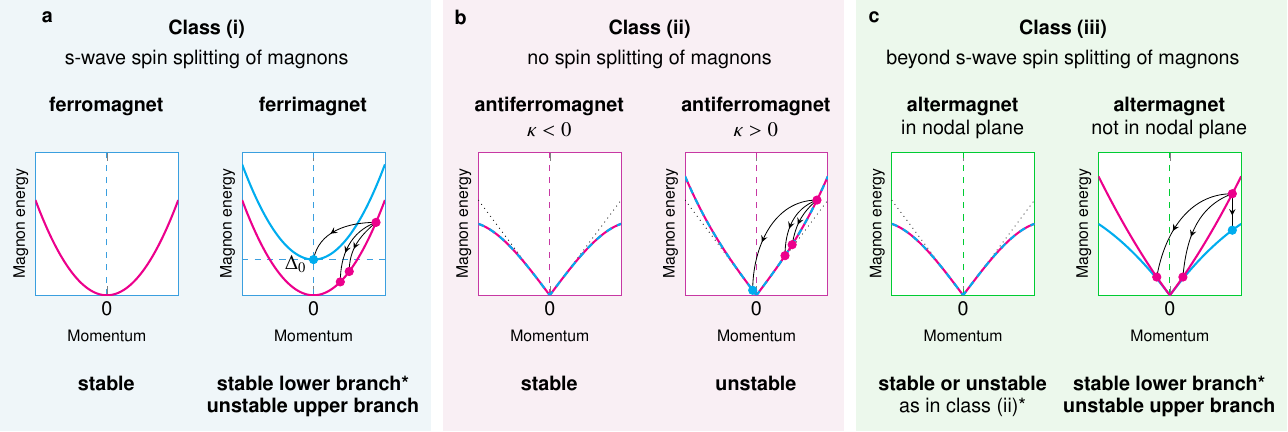}   
    \caption{\textbf{Stability of low-energy magnons in the three classes of \textit{collinear} magnets in the nonrelativistic limit of an isotropic Heisenberg model.} Due to spin conservation, magnons carry a spin quantum number $\pm 1$, indicated by color, and potential decay processes have to conserve spin. \textbf{(a)} Class-(i) magnets: Magnets with s-wave spin splitting, such as ferromagnets and ferrimagnets, exhibit inherently stable magnons at low energies. In ferromagnets, there are no quantum fluctuations and there is only one (acoustic) magnon branch, which can be considered the limit of infinite s-wave splitting. In contrast, ferrimagnets fluctuate. They support two oppositely polarized sublattices with unequal moments, and their spin splitting, i.e., the gap $\Delta_0$, is finite. It prohibits magnon decay at energies below $\Delta_0$, ensuring their stability. The indicated decay process of the magenta magnon is only possible for high-energy magnons above $\Delta_0$, such that one decay product can carry opposite spin (cyan). Thus, the lower magnon branch is stable below the gap $\Delta_0$. \textbf{(b)} Class-(ii) magnets: Antiferromagnets fluctuate and have spin-degenerate magnons. Stability depends on the curvature of the magnon spectrum: negatively curved spectra ($\kappa<0$) lead to stable magnons, while positively curved spectra ($\kappa>0$) result in instability. \textbf{(c)} Class-(iii) magnets: Altermagnets are characterized by unconventional time-reversal symmetry breaking and beyond-s-wave even-parity spin splitting such as d/g/i-wave. In nodal planes, i.e., directions without spin splitting, the stability mirrors that of class-(ii) magnets, where the curvature of the spectrum dictates whether magnons can decay, but is additionally modified by the altermagnetic splitting, as indicated by the asterisk. Along directions with finite spin splitting the upper magnon branch is universally unstable due to the indicated decay process. The lower magnon branch exhibits rich stability physics that depends on the symmetry of the splitting as indicated by the asterisk: In d-wave magnets, it is either completely stable for $\kappa<0$ or exhibits a d-wave-shaped island of stability in the vicinity of the origin for $\kappa>0$. In contrast, there is no such island of stability for general g-wave and i-wave altermagnets.}
    \label{fig:central-message}
\end{figure*}

Here, we combine kinematic analysis, many-body perturbation theory within nonlinear spin-wave theory, and nonperturbative quantum simulations to demonstrate that magnons in altermagnets are generally unstable, with exceptions to be discussed below. Even minimal even-parity spin splitting generically opens a finite phase space for decay, rendering the high-energy magnon branch universally unstable. Strikingly, the symmetry of the spin splitting is crucial for the stability of the lower-energy magnon branch: while d-wave spin splitting can protect lower-energy magnons within a d-wave-shaped ``island of stability'' around the Brillouin zone origin, g-wave and i-wave splittings cannot. We identify quasi-two-dimensional d-wave altermagnets with negligible spin-orbit interaction as ideal platforms for exploring altermagnetic quantum spin dynamics, and predict experimentally resolvable magnon damping in organic altermagnets. These findings establish a quasiparticle stability \textit{trichotomy} of nonrelativistic collinear magnets, with the three principal cases shown in \textbf{Fig.~\ref{fig:central-message}}, highlighting how altermagnets depart from established paradigms in quantum magnetism and spontaneous quasiparticle decay.

\section*{Results}

\subsection*{\textbf{Trichotomy of magnon stability in collinear magnets --- kinematics}}
Nonrelativistic collinear magnets can be categorized into three distinct classes using spin space group arguments, which involve the partial-wave decomposition of the spin splitting in their electronic or magnonic bands \cite{Smejkal2022altermagnets, Smejkal2022emergent,SmejkalMagnons23}, see \textbf{Fig.~\ref{fig:central-message}}.
\textit{(i) s-wave spin splitting:} Magnets in this class, such as ferromagnets and ferrimagnets, exhibit a finite net magnetic moment in real space.
\textit{(ii) No spin splitting:} Characterized by compensated order in real space, these magnets exhibit oppositely ordered sublattices connected by inversion or translation symmetry and are called antiferromagnets. 
\textit{(iii) Even-parity spin splitting beyond s-wave (d-, g-, or i-wave):} These magnets---now recognized as altermagnets \cite{Smejkal2022altermagnets}---display compensated order in real space, but their oppositely ordered sublattices are \textit{not} related by inversion or translation symmetry but instead, e.g., by rotational symmetry. 

For insulators with localized electronic degrees of freedom, all three classes of magnets adhere to the nonrelativistic SO$(3)$-symmetric Heisenberg model ($\hbar = 1$),
\begin{align}
    H = \frac{1}{2} \sum_{\vec{r},\vec{r}'} J_{\vec{r},\vec{r}'} \vec{S}_{\vec{r}} \cdot \vec{S}_{\vec{r}'}, 
    \label{eq:HeisenbergModel}
\end{align}
where $\vec{S}_\vec{r}$ is a local spin operator of length $S$ at site $\vec{r}$, and $J_{\vec{r},\vec{r}'}$ the exchange constant between spins at $\vec{r}$ and $\vec{r}'$. In the collinear ground state, spontaneous symmetry breaking reduces SO$(3)$ to SO$(2)$, with the ordered moment chosen along the $z$-axis. Goldstone modes emerge, and the total spin component $S^z = \sum_\vec{r} S_\vec{r}^z$ remains a good quantum number. Single spin-flip excitations (magnons) thus carry spin $\pm 1$ (also denoted by $\uparrow$ and $\downarrow$) relative to the ground state, and follow dispersions $\epsilon_\pm(\vec{k})$, where $\vec{k}$ is the crystal momentum, and ``$\pm$'' indicates spin, not necessarily energetic order. Optical magnon branches are neglected as they do not impact low-energy physics.

\begin{figure*}
    \centering
    \includegraphics[width=0.8\textwidth]{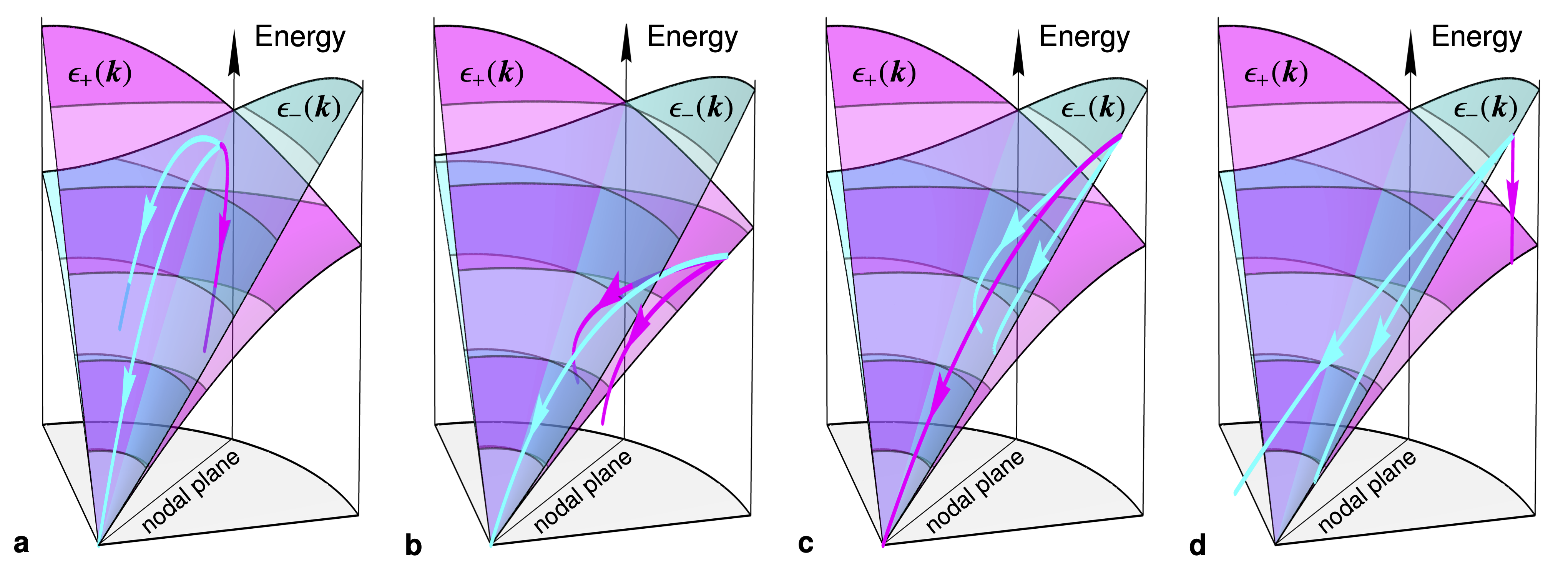}   
    \caption{\textbf{Kinematics of magnon decays in class-(iii) magnets: altermagnets.} Low-energy dispersion of spin-$\uparrow$ and spin-$\downarrow$ magnons with energy $\epsilon_+(\vec{k})$ and $\epsilon_-(\vec{k})$, respectively. Along a nodal plane, the magnon dispersion is degenerate. (\textbf{a}) Decay channel of nodal-plane magnons. They decay into a magnon close to the origin and two magnons of opposite spin in the lower branch to the left and right of the nodal line, respectively. (\textbf{b}) Decay channel of the lower-energy magnon. It decays into a magnon close to the origin and two magnons with the same spin to its left and right, respectively. (\textbf{c}) First decay channel of the higher-energy magnon. It decays into a magnon close to the origin and two magnons with the same spin to its left and right, respectively. (\textbf{d}) Second decay channel of the higher-energy magnon. It decays into two magnons of the same spin close to the origin and a magnon in the lower-energy branch.}
    \label{fig:decay3D}
\end{figure*}

Although spin is conserved, magnon number is not—except in ferromagnets---allowing for quantum many-body interactions that change magnon count while preserving total spin \cite{Harris1971, 
ZhitomirskyChernyshev2013}. Among these, three-magnon processes are forbidden by spin conservation, making four-magnon interactions the lowest-order relevant scattering mechanisms \cite{Harris1971}. These fall into three categories: (1) Two-in-two-out, which require thermally excited magnons and are inactive at zero temperature; (2) None-in-four-out (and \textit{vice versa}), which violate energy conservation and thus do not affect stability; (3) One-in-three-out (and \textit{vice versa}), the only processes relevant at zero temperature. For instance, an initial spin-$\uparrow$ magnon can decay into two spin-$\uparrow$ magnons and one spin-$\downarrow$ magnon, conserving total spin. The feasibility of such a three-magnon decay depends on energy and momentum conservation:
\begin{align}
    \epsilon_\pm(\vec{k}) = \epsilon_\pm(\vec{q}) + \epsilon_\pm(\vec{p}) + \epsilon_\mp(\vec{k}-\vec{q}-\vec{p}).
    \label{eq:energy conservation}
\end{align}
Thus, magnon stability in collinear nonrelativistic magnets at absolute zero reduces to determining whether Eq.~\eqref{eq:energy conservation} has solutions. If it does, decays occur, provided the scattering matrix element is nonzero.

\textbf{Class-(i) magnets:} Ferromagnets inherently support stable magnons because their ground state is an exact eigenstate. Ferrimagnets, however, fluctuate because their sublattice-polarized ground state is not an eigenstate. Having two antiparallel but uncompensated sublattice magnetizations, they feature two magnon branches: a gapless branch, $\epsilon_-(\vec{k}) = A |\vec{k}|^2$, and a gapped branch, $\epsilon_+(\vec{k}) = A |\vec{k}|^2 + \Delta_0$, where $A>0$ is the spin-wave stiffness and $\Delta_0>0$ the gap, see \textbf{Fig.~\ref{fig:central-message}a}. Here, s-wave spin splitting manifests as an isotropic energy difference: $\epsilon_+(\vec{k}) - \epsilon_-(\vec{k}) = \Delta_0$. The positive curvature of the gapless branch implies that any initial magnon could theoretically decay into three magnons at smaller momenta while conserving energy and momentum. However, Eq.~\eqref{eq:energy conservation} requires at least one decay product to have opposite spin, precluding decay into the gapless branch alone. Decay becomes possible only at energies exceeding $\Delta_0$, where one decay product can transition into the gapped branch. Consequently, the gapless magnon branch in class-(i) magnets is universally stable below the energy gap $\Delta_0$ and protected by s-wave spin splitting.

\textbf{Class-(ii) magnets:} In antiferromagnets, which lack spin splitting, the magnon branches are degenerate: $\epsilon_+(\vec{k}) = \epsilon_-(\vec{k}) = \epsilon(\vec{k})$. Magnon stability at long wavelengths depends on the curvature of the dispersion relation. Consider the dispersion $\epsilon(\vec{k}) = v |\vec{k}| + \kappa |\vec{k}|^3$ where $v>0$ is the magnon velocity and $\kappa$ characterizes the leading nonlinearity. The sign of $\kappa$ determines the stability \cite{ZhitomirskyChernyshev2013}: For $\kappa>0$, the magnon dispersion has positive curvature, in which case Eq.~\eqref{eq:energy conservation} has solutions and magnons are unstable. If $\kappa<0$, Eq.~\eqref{eq:energy conservation} has no solutions and magnons are stable. Beyond the isotropic approximation, one has to account for lattice symmetries, that is, for $\kappa$ exhibiting directional dependence, resulting in magnon damping varying with orientation. Thus, class-(ii) magnets can support stable gapless magnons under specific conditions dictated by leading nonlinearities of the dispersion, but lack spin splitting to universally protect magnons from decay.

\begin{figure*}[t]
    \centering
    \includegraphics[width=\textwidth]{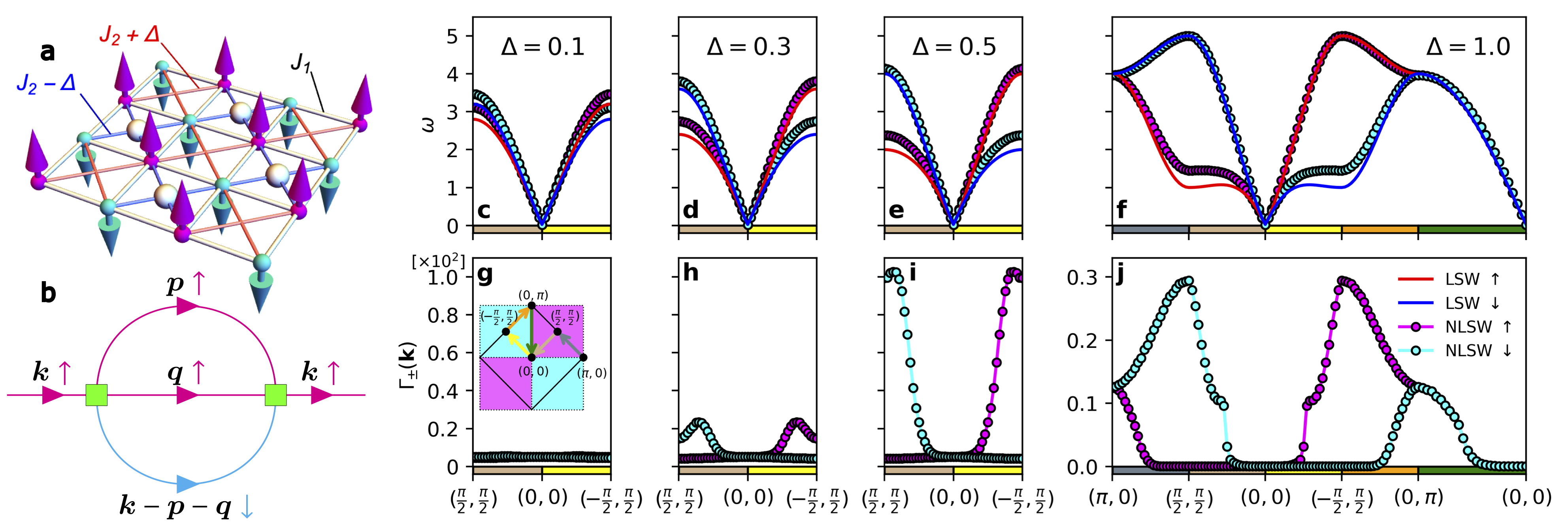}   
    \caption{\textbf{Nonlinear spin-wave analysis of a two-dimensional d-wave altermagnet.} (\textbf{a}) Schematic of the spin model for a two-dimensional d-wave altermagnet on a checkerboard lattice, described by the Hamiltonian in Eq.~\eqref{eq:model}. The system hosts a collinear $(\pi,\pi)$ N\'{e}el order with nearest-neighbor (white bonds) and next-nearest-neighbor (red and blue bonds) exchange interactions $J_1$ and $J_2\pm\Delta$, respectively. The checkerboard modulation---parametrized by $\Delta$---arises from nonmagnetic atoms (silver spheres) that alter the diagonal exchange, inducing a $d$-wave spin splitting of magnons. (\textbf{b}) Diagrammatic representation of the three-magnon scattering process, the leading mechanism for magnon decay. Due to momentum conservation (up to a reciprocal lattice vector), the sum of the momenta of the three intermediate states equals the initial momentum $\vec{k}$. Total spin conservation ensures that an initial single spin-$\uparrow$ magnon scatters into two spin-$\uparrow$ magnons and one spin-$\downarrow$ magnon, with a similar process for an initial spin-$\downarrow$ magnon. (\textbf{c-f}) Renormalized magnon spectrum (lines with dot markers) compared to the linear magnon spectrum (solid lines) for $J_1=1$, $J_2=-0.5$, $S=1/2$, and selected values of $\Delta$ as indicated. The altermagnetic splitting is most pronounced at $(\frac{\pi}{2}, \pm\frac{\pi}{2})$ and zero along the nodal planes, e.g., in the direction from $(0,0)$ to $(\pi,0)$. (\textbf{g-j}) Magnon damping $\Gamma_\pm({\vec{k}})$, highlighting the instability of the upper magnon branch. The inset in panel (\textbf{g}) shows the magnetic Brillouin zone with indicated high-symmetry paths. The color of these paths is used on the horizontal momentum axes in panels (\textbf{c-j}) to indicate direction. The cyan/magenta checkerboard pattern indicates the sign of the d-wave altermagnetic spin splitting.
    The legend in panel (\textbf{j}) distinguishes between nonlinear spin-wave (NLSW) and linear spin-wave (LSW) calculations.}
    \label{fig:model_and_spectra}
\end{figure*}

\textbf{Class-(iii) magnets:} 
We derive general insights into magnon decay in altermagnets by performing a low-energy expansion of the magnon dispersion in two dimensions. 
We consider an ideal long-wavelength altermagnetic magnon dispersion relation, in which crystallographic symmetries enter only via the altermagnetic splitting, given by
\begin{align}
    \epsilon_\pm(\vec{k})
    =
    v |\vec{k}| + \kappa |\vec{k}|^3 \pm \Delta  \sin\left( n \varphi_{\vec{k}} \right) |\vec{k}|^n,
    \label{eq:general-expansion}
\end{align}
where $v>0$ is the magnon velocity, $\kappa$ represents the leading (cubic) non-altermagnetic nonlinearity, $\tan \varphi_{\vec{k}} = k_y/k_x$, and $\Delta \ge 0$ parametrizes the altermagnetic splitting with d-wave ($n=2$), g-wave ($n=4$) or i-wave character ($n = 6$). The unconventional time-reversal symmetry breaking results in a gap $\epsilon_+(\vec{k}) - \epsilon_-(\vec{k}) = 2 \Delta  \sin\left( n \varphi_{\vec{k}} \right) |\vec{k}|^n$. Along a nodal plane (or line in 2D), e.g., $\vec{k} = (k_x,k_y) = (k,0)$, the splitting vanishes, suggesting that the class-(ii) curvature argument is relevant. Away from the nodal planes, the magnon bands split, creating novel decay channels, see \textbf{Fig.~\ref{fig:central-message}c}. 

We first discuss magnon stability along nodal planes. For $\kappa>0$, nodal-plane magnons decay as in class-(ii) magnets. For $\kappa<0$, magnons in class-(ii) magnets remain stable. However, consider the decay depicted in \textbf{Fig.~\ref{fig:decay3D}a}:
a nodal-plane magnon decays into a long-wavelength magnon and two nearby lower-branch magnons.
As shown in Supplementary Material I, this process obeys Eq.~\eqref{eq:energy conservation}, if $\Delta > \Delta^\star$, where
\begin{align}
    \Delta^\star = \sqrt{3} 2^{n-5/2} n^{-1} |\vec{k}|^{2-n} \sqrt{-\kappa} \sqrt{4v+3|\vec{k}|^2 \kappa} \quad \text{for} \quad \kappa < 0.
    \label{eq:deltastar}
\end{align}
As $|\vec{k}| \to 0$, $\Delta^\star$ diverges for $n=4$ (g-wave) and $n=6$ (i-wave), indicating that nodal-plane magnons are stable. For $n=2$ (d-wave), however, $\Delta^\star$ converges to
\begin{align}
    \Delta^\star_\text{d-wave} = \sqrt{-\frac{3}{2} v \kappa} \quad \text{for} \quad \kappa <0. \label{eq:criticalDeltadwave}
\end{align}
Therefore, d-wave magnets with $\kappa<0$ have unstable nodal-plane magnons if $\Delta > \Delta^\star_\text{d-wave}$. 

Next, we analyze the decay of the lower-branch magnon along a general direction, where it can decay into two lower-branch magnons and a spin-opposite magnon near the Goldstone mode (\textbf{Fig.~\ref{fig:decay3D}b}). For $\kappa < 0$, this decay process is kinematically forbidden. For $\kappa > 0$, it remains forbidden if
\begin{align}
    |\vec{k}|^{3-n} < \frac{2^{2-n}\left(2^n-2\right)}{3 \kappa}  \Delta \left|\sin( n \varphi_{\vec{k}}) \right| .
\end{align}
Thus, in d-wave altermagnets ($n=2$), lower-branch magnons have an ``island of stability'' at $|\vec{k}| < k^\star_\text{d-wave} \left|\sin\left( 2 \varphi_{\vec{k}} \right) \right|$, with
\begin{align}
    k^\star_\text{d-wave} = \frac{2 \Delta }{ 3\kappa }  \quad \text{for} \quad \kappa > 0. \label{eq:kstar}
\end{align}
No such solution exists for g-wave magnets ($n=4$) and i-wave magnets ($n=6$).

The upper magnon branch has two decay channels. In \textbf{Fig.~\ref{fig:decay3D}c}, it decays into a mode close to the origin and two modes to its left and right. This process is possible in d-wave magnets even for $\kappa<0$, as the altermagnetic splitting dominates over the cubic nonlinearity at long wavelengths. The process in \textbf{Fig.~\ref{fig:decay3D}d}, where it decays into a lower-branch magnon at approximately the same momentum and two upper-branch magnons around the origin, is always kinematically possible. Thus, the upper magnon branch in altermagnets is universally unstable.

\subsection*{\textbf{Nonlinear spin-wave theory of altermagnets}}

\begin{figure}[t]
    \centering
    \includegraphics[width=\columnwidth]{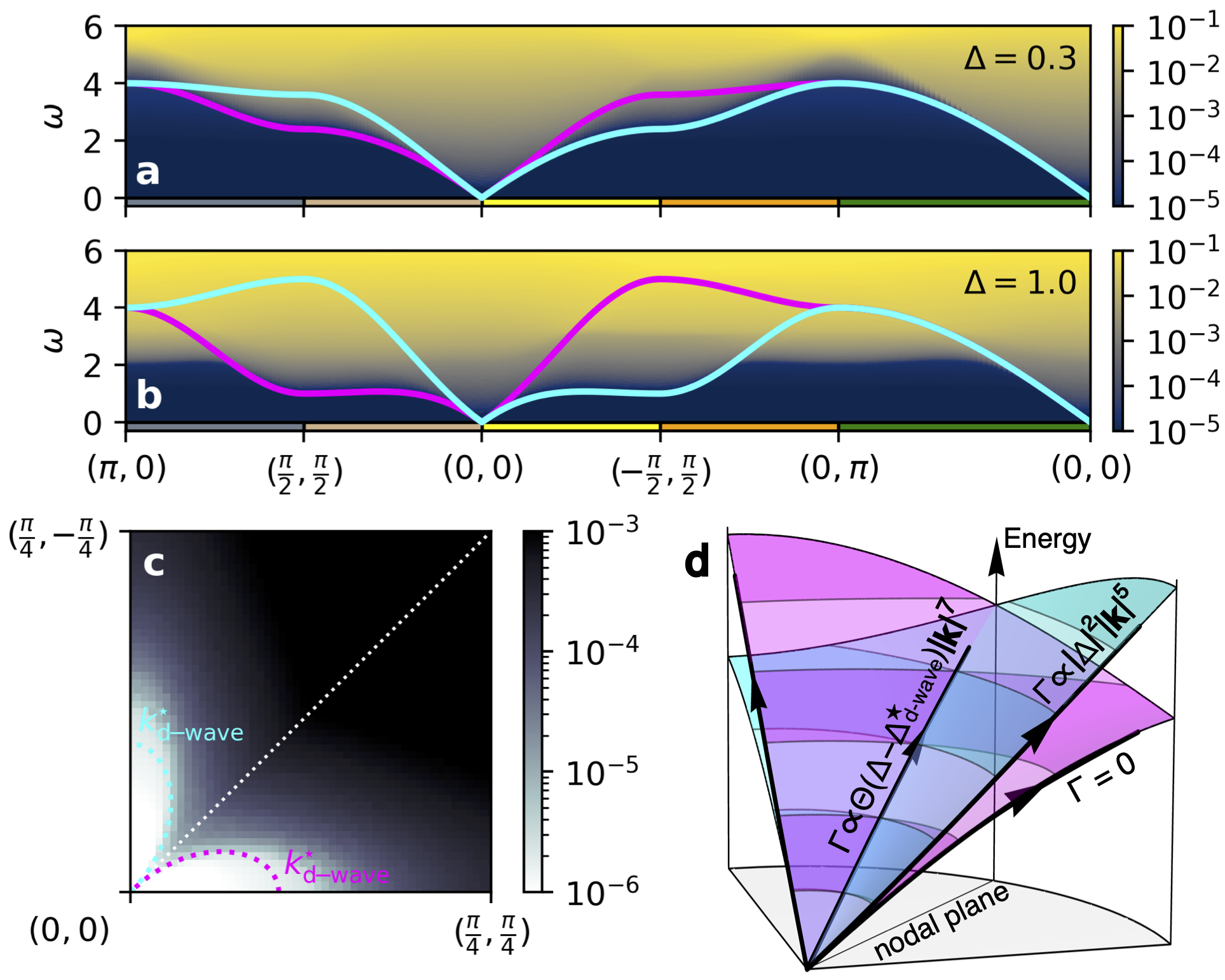}   
    \caption{
    \textbf{Magnon decay phase space analysis and long-wavelength scaling of magnon decay rate in d-wave altermagnets.} (\textbf{a}) Three-magnon density of states $\mathcal{D}_+(\omega,\vec{k})$, encoding the available decay phase space for the spin-$\uparrow$ magnon with energy $\epsilon_+(\vec{k})$ (magenta line) for moderate altermagnetic splitting ($\Delta=0.3$). 
    (\textbf{b}) Same as (\textbf{a}) but for $\Delta=1.0$, pushing the upper magnon branch deeper into the continuum. The colors on the horizontal momentum axes in (\textbf{a}) and (\textbf{b}) are identical to those in the Brillouin zone inset in \textbf{Fig.~\ref{fig:model_and_spectra}g}. Parameters used in (\textbf{a}) and (\textbf{b}) are $J_1=1$, $J_2=-0.5$, $S=1/2$ and $\Delta$ as indicated, realizing the case of a negative cubic nonlinearity: $\kappa <0$. (\textbf{c}) On-shell three-magnon density of states for the lower-energy magnon branch, $\mathcal{D}_\xi(\epsilon_\xi(\vec{k}),\vec{k})$, in vicinity of the Brillouin zone origin. $\xi$ is chosen so that the lower band is picked. The d-wave-shaped ``island of stability'' agrees with the indicated $k^\star_{\text{d-wave}}$ as obtained from the kinematic analysis in Eq.~\eqref{eq:kstar}. The nodal plane is indicated as a dotted diagonal line. Parameters are $J_1 = 1/10$, $J_2 = -1$, $S=1/2$, and $\Delta=0.5$, realizing the case of a positive cubic nonlinearity: $\kappa > 0$. 
    (\textbf{d}) Long-wavelength scaling of the magnon decay rate $\Gamma$ along selected directions. The upper magnon branch shows a characteristic $|\vec{k}|^5$ scaling away from the nodal planes that dominates potential subleading $|\vec{k}|^7$ contributions. Along the nodal planes, the $|\vec{k}|^7$ scaling is the leading contribution. The indicated scaling assumes two-dimensional d-wave altermagnets.
    }
    \label{fig:DOSs}
\end{figure}

To expand on the previous discussion of magnon decay, we consider a d-wave altermagnet on the checkerboard lattice (see \textbf{Fig.~\ref{fig:model_and_spectra}a}), performing a nonlinear spin-wave analysis. The spin Hamiltonian is
\begin{align}
    H 
    =  
    \sum_{\ev{\vec{r},\vec{r}'}}          
    J_1 \vec{S}_{\vec{r}}\cdot\vec{S}_{\vec{r}'}
    + 
    \sum_{\ev{\ev{\vec{r},\vec{r}'}}_{\pm}} 
    (J_2 \pm \Delta) 
    \vec{S}_\vec{r}\cdot\vec{S}_{\vec{r}'}.
    \label{eq:model}
\end{align}
Here, $J_1>0$ is the antiferromagnetic nearest-neighbor, and $J_2$ the second-nearest-neighbor exchange interaction. $\Delta>0$ represents the d-wave altermagnetic magnon band splitting.

Using a $1/S$ spin-wave expansion around the Néel-ordered ground state based on the Holstein-Primakoff transformation \cite{holsteinprimakoff1940}, we obtain the noninteracting single-magnon spectrum  $\epsilon_\pm(\vec{k})$
given in Appendix B. At low energies, the spectrum maps onto Eq.~\eqref{eq:general-expansion} with a rotated nodal plane and additional directional dependence of cubic nonlinearities. Up to order $1/S^2$, the renormalized spectrum is
$
    \tilde{\epsilon}_\pm(\vec{k}) = \epsilon_\pm(\vec{k}) + \delta \epsilon_\pm(\vec{k}) - \mathrm{i} \Gamma_\pm(\vec{k})
$,
where $\delta \epsilon_\pm(\vec{k})$ corrects the energy and 
\begin{align}
    \Gamma_\pm(\vec{k})
    =
    &\frac{\pi}{N_\textrm{muc}^2} \sum_{\vec{p}, \vec{q}} |W_\pm(\vec{k},\vec{p},\vec{q})|^2 \nonumber \\ 
    &\times \delta\left( \epsilon_\pm(\vec{k}) -  \epsilon_\pm(\vec{p}) - \epsilon_\pm(\vec{q}) - \epsilon_\mp(\vec{k}-\vec{p}- \vec{q}) \right),
    \label{eq:damping rate}
\end{align}
is the decay rate associated with the self-energy diagram in \textbf{Fig.~\ref{fig:model_and_spectra}b}. In Eq.~\eqref{eq:damping rate}, $N_\textrm{muc}$ is the number of magnetic unit cells and $W_\pm(\vec{k},\vec{p},\vec{q})$ is the interaction vertex (Supplementary Material II).  

The interacting magnon spectrum $\tilde{\epsilon}_\pm(\vec{k})$ is shown in \textbf{Figs.~\ref{fig:model_and_spectra}c-j} for $\kappa < 0$, exhibiting d-wave splitting that increases with $\Delta$. Many-body interactions stiffen both bands and cause damping of the upper magnon branch, see \textbf{Figs.~\ref{fig:model_and_spectra}g-j}. The damping grows with increasing $\Delta$, especially in the short-wavelength limit. When $\Delta > \Delta^\star_\text{d-wave}$ the damping is nonzero along the nodal planes (\textbf{Figs.~\ref{fig:model_and_spectra}h}), in agreement with kinematic predictions.

The damping rate $\Gamma_\pm(\vec{k})$ is related to the density of states $\mathcal{D}_\pm(\omega,\vec{k})$ of the three-magnon continuum with net spin $\pm 1$:
\begin{align}
    \mathcal{D}_\pm(\omega,\vec{k}) = \frac{1}{ N_\textrm{muc}^2} \sum_{\vec{p}, \vec{q} } \delta\left( \omega -  \epsilon_\pm(\vec{p}) - \epsilon_\pm(\vec{q}) - \epsilon_\mp(\vec{k}-\vec{p}- \vec{q}) \right).
    \label{eq:DOS}
\end{align}
When these continua overlap with the single-magnon branch of the same spin, that is, when $\mathcal{D}_\pm(\epsilon_\pm(\vec{k}),\vec{k}) \ne 0$, spontaneous decays are kinematically possible \cite{ZhitomirskyChernyshev2013}.
\textbf{Figure~\ref{fig:DOSs}a} shows $\mathcal{D}_+(\omega,\vec{k})$ together with $\epsilon_\pm(\vec{k})$ for $\kappa < 0$ and $\Delta = 0.3$. Clearly, the spin-$\uparrow$ magnon $\epsilon_+(\vec{k})$ is embedded within $\mathcal{D}_+(\omega,\vec{k})$ along $(0,0)$ to $(\frac{\pi}{2},-\frac{\pi}{2})$, allowing decays. In contrast, along $(0,0)$ to $(\frac{\pi}{2},\frac{\pi}{2})$ the continuum begins exactly at that branch's energy (due to the Goldstone mode), precluding decays. A similar discussion applies to the spin-$\downarrow$ magnon $\epsilon_-(\vec{k})$ and $\mathcal{D}_-(\omega,\vec{k})$ (not shown). Along the nodal-plane from $(0,0)$ to $(\pi,0)$, the continuum also begins exactly at the single-particle energies because $\Delta<\Delta^\star_\text{d-wave}$, ensuring stability.

For $\kappa < 0$ and $\Delta = 1.0$, the lower magnon branch has a local minimum at $(\pm\frac{\pi}{2},\frac{\pi}{2})$, causing a rapid increase in the three-particle density of states at approximately three times its energy, as shown in \textbf{Fig.~\ref{fig:DOSs}b}. This results in a sharp increase in $\Gamma(\vec{k})$ for the upper branch, in agreement with \textbf{Fig.~\ref{fig:model_and_spectra}j}. Since $\Delta > \Delta^\star_\text{d-wave}$, magnon energies are embedded within the continua along the nodal planes from $(0,0)$ to $(\pi,0)$, explaining the finite decay rate in \textbf{Fig.~\ref{fig:model_and_spectra}j}.

For $\kappa>0$, kinematics predicts an island of stability of size $k^\star_\text{d-wave}$ for the lower magnon branch around the origin. As shown in \textbf{Fig.~\ref{fig:DOSs}c}, the relevant continua exhibit a drop to zero with d-wave-shaped lobes. Near the origin, the density of states remains finite only along the nodal planes, where the altermagnetic splitting does not protect the lower branch from decay.

\subsection*{\textbf{Long-wavelength scaling of magnon decay rates in altermagnets}}

The presence of Goldstone modes raises the question of how $\Gamma_\pm(\vec{k})$ in Eq.~\eqref{eq:damping rate} scales with $|\vec{k}|$ as $|\vec{k}| \to 0$. This scaling depends on the decay phase space (denoted by $V(\vec{k})$), specifically the scaling of the three-magnon density of states $\mathcal{D}_\pm(\epsilon_\pm(\vec{k}),\vec{k})$, and that of $W_\pm(\vec{k},\vec{q},\vec{p})$.

At long wavelengths, the linear magnon spectrum $\epsilon(\vec{k}) = v |\vec{k}|$ is the reference point. A magnon at $\vec{k}$ is ``ready to decay'' into magnons at $\vec{p}$, $\vec{q}$, and $\vec{k}-\vec{p}-\vec{q}$, with any positive curvature activating decays in a threshold behavior, as in \textbf{Fig.~\ref{fig:decay3D}c}. 
Since the two momenta $\vec{p}=(p_x,p_y)$ and $\vec{q}=(q_x,q_y)$ have to be almost aligned with $\vec{k}$ to fulfill energy and momentum conservation, the four-dimensional decay phase space spanned by $\{p_x,p_y,q_x,q_y\}$ has a cigar shape.
In the two long directions, it is constrained by $|\vec{k}|$, and in the other two short directions by $|\vec{k}|^{3/2}$, yielding $V(\vec{k}) \propto |\vec{k}|^3$ for a positively curved magnon dispersion. In altermagnets with magnon dispersion $\epsilon_\pm(\vec{k}) = v |\vec{k}| \pm \Delta \sin(n\varphi_{\vec{k}}) |\vec{k}|^n $, this implies $V(\vec{k})$ of the upper branch is zero at $\Delta = 0$ and jumps to a finite value at $\Delta \ne 0$.

The threshold value of $\Delta$ becomes nonzero and $n$-dependent once additional nonlinearities are taken into account. 
When a negative cubic nonlinearity ($\kappa < 0$) is included as in Eq.~\eqref{eq:general-expansion}, $\Delta$ has to exceed $-C_n \kappa |\vec{k}|^{3-n}$ (with $C_n > 0$ being a constant) for the upper branch to decay at $|\vec{k}|$.
For d-wave magnets ($n=2$), $\Delta > -C_n \kappa |\vec{k}|$ for any $\kappa<0$ as long as $|\vec{k}|$ is small enough and their upper-branch magnons  have a decay phase space $V_{\text{d-wave}}(\vec{k}) \propto |\vec{k}|^3$.
In contrast, the threshold value for $\Delta$ diverges as $|\vec{k}| \to 0$ in g-wave and i-wave magnets, implying that their upper magnon branch does \textit{not} exhibit the $V(\vec{k}) \propto |\vec{k}|^3$ long-wavelength scaling for $\kappa < 0$. Instead, other higher-order-in-momentum mechanisms, e.g., the one in \textbf{Fig.~\ref{fig:decay3D}d}, have to be identified. 

With the relevant decay process for d-wave altermagnets identified, we plug the ``cigar-shaped'' scaling ($\propto |\vec{k}|$ for the two long and $\propto |\Delta|^{1/2} |\vec{k}|^{3/2}$ for the two short directions) for the respective components of $\vec{p}$ and $\vec{q}$ into $W_\pm(\vec{k},\vec{q},\vec{p})$ and take $|\vec{k}| \to 0$. We find $W_\pm(\vec{k},\vec{q},\vec{p}) \propto |\Delta| |\vec{k}| |\sin (2\varphi_\vec{k})|$, where we have restored the angular dependence. Therefore, as indicated in \textbf{Fig.~\ref{fig:DOSs}d}, the damping of the upper magnon branch is 
\begin{align}
    \Gamma_\text{upper}(\vec{k}) \propto |\Delta|^2 |\vec{k}|^5 |\sin (2\varphi_\vec{k})|^2 \quad  \text{as} \quad |\vec{k}| \to 0.
    \label{eq:ktothefive}
\end{align}

Magnons along the nodal planes are stable if $\Delta < \Delta^\star_\text{d-wave}$ but exhibit the decays shown in \textbf{Fig.~\ref{fig:decay3D}a} for $\Delta > \Delta^\star_\text{d-wave}$, with $\Delta^\star_\text{d-wave}$ given in Eq.~\eqref{eq:criticalDeltadwave}. In this case, the cigar-shaped decay space has a short-axes scaling of $|\vec{k}|^2$, such that the damping along the nodes is
\begin{align}
    \Gamma_\text{node}(\vec{k}) \propto \Theta \left( \Delta- \Delta^\star_\text{d-wave} \right) |\vec{k}|^7  \quad  \text{as} \quad |\vec{k}| \to 0,
\end{align}
where $\Theta$ indicates the Heaviside step function. Also, for $\kappa > 0$, an isotropic scaling $\Gamma(\vec{k}) \propto |\vec{k}|^7$ is expected \cite{Stephanovich2011}. 

Therefore, in d-wave altermagnets, the cubic nonlinearity is subleading at long-wavelengths and away from the nodal planes, giving rise to a unique scaling $\Gamma \propto |\vec{k}|^5$ in two dimensions. For further details, see Supplementary Material III and IV.

\subsection*{\textbf{Nonperturbative quantum simulations of altermagnets}}

\begin{figure*}
    \centering
    \includegraphics[width=0.9\textwidth]{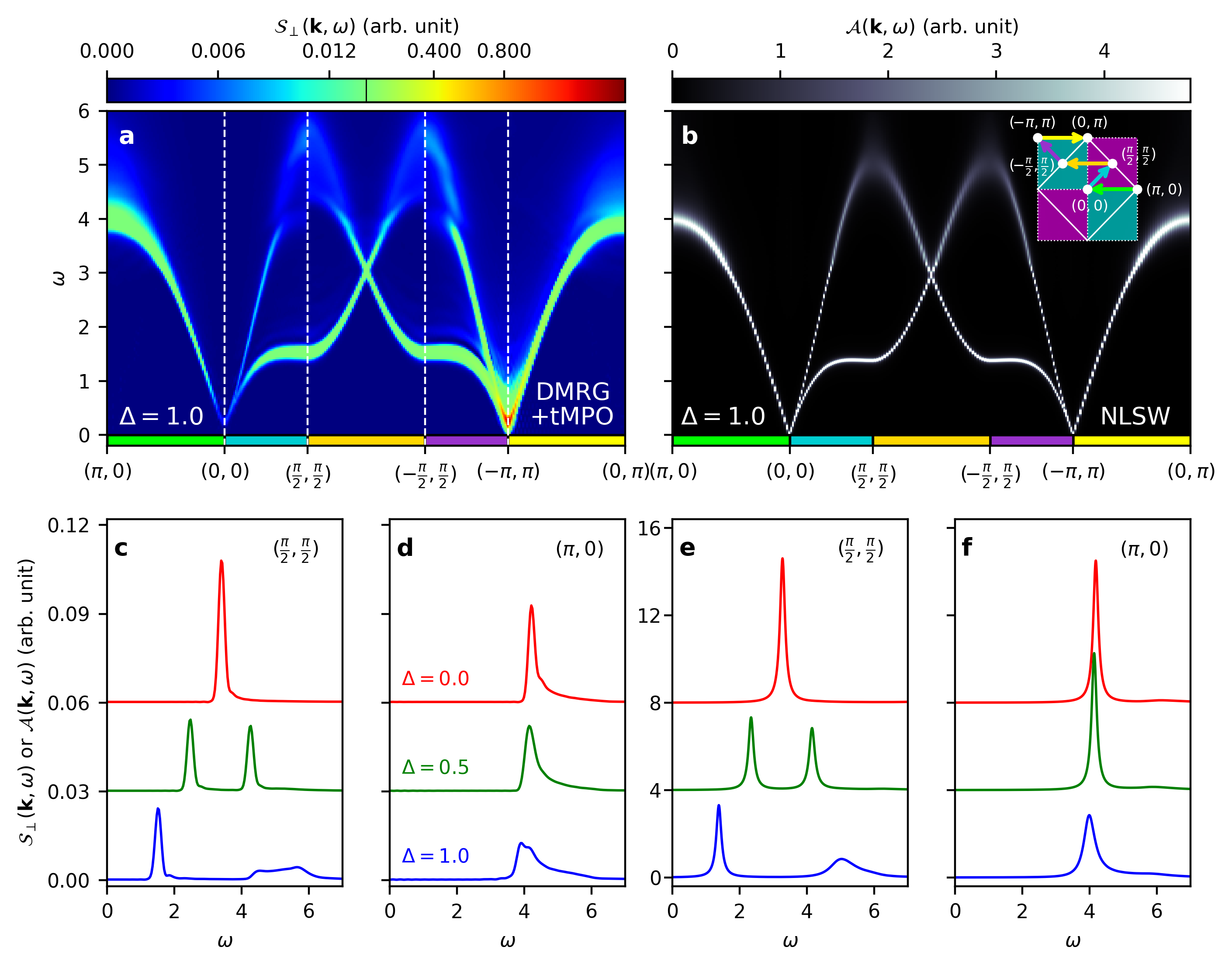}   
    \caption{\textbf{Comparison of nonperturbative and perturbative transverse quantum dynamics in two-dimensional d-wave altermagnets.} (\textbf{a}) The transverse dynamical structure factor $\mathcal{S}_\perp(\vec{k},\omega)$, as obtained from DMRG$+$tMPO, shows clear altermagnetic splitting along the path $(0,0)$ to $(\frac{\pi}{2},\frac{\pi}{2})$ in reciprocal space. Around this point, the spectral weight of the upper magnon branch gets almost completely wiped out due to hybridization with continua. A small XXZ anisotropy $\lambda=0.02$ was added for numerical stability. It causes a tiny gap of the Goldstone modes. Note the double linear color bar scale. Dashed white lines indicate points at which simulation results for different simulation geometries were patched together (see Appendix C). As the simulation on cylinders weakly breaks the exact four-fold rotational symmetry of the lattice, small discontinuities in the intensity can be identified. (\textbf{b}) The single-particle spectral function $\mathcal{A}(\vec{k},\omega)$ as obtained within perturbative nonlinear spin-wave (NLSW) theory of $1/S^2$ accuracy modulo additional off-shell corrections. The colors on the horizontal momentum axes in (\textbf{a}) and (\textbf{b}) match the colors of the high-symmetry paths indicated in the Brillouin zone in the inset of (\textbf{b}).
    (\textbf{c},\textbf{d}) Line cuts of the nonperturbative $\mathcal{S}_\perp(\vec{k},\omega)$ versus frequency at momenta $(\frac{\pi}{2},\frac{\pi}{2})$ and $(\pi,0)$, respectively.
    (\textbf{e},\textbf{f}) Line cuts of the perturbative $\mathcal{A}(\vec{k},\omega)$ versus frequency at momenta $(\frac{\pi}{2},\frac{\pi}{2})$ and $(\pi,0)$, respectively.
    Parameters read $J_1=1$, $J_2=-0.5$, $S=1/2$, and $\Delta$ as indicated.
    }
    \label{fig:dynstruc}
\end{figure*}

At low energies, where decay phase space and magnon-magnon interaction vertices are small, perturbation theory is reliable \cite{Harris1971, chernyshevzhitomirsky2009}. At higher energies, our calculations give $\Gamma / \epsilon \lesssim 0.1$ for $S=1/2$, indicating that magnons remain well-defined quasiparticles, validating the perturbative approach. However, small $S$, non-universal high-energy decays, and enhanced fluctuations in low-dimensional systems require nonperturbative methods for assessing magnon stability. We use density matrix renormalization group (DMRG) \cite{white_density_1992, schollwoeck2011} and matrix product operator time evolution (tMPO) \cite{zaletel_timeevolving_2015} to compute the dynamical spin structure factor $\mathcal{S}_{\mu\nu}(\vec{k}, \omega)$, as measured in inelastic neutron scattering (see Appendix C).

\textbf{Figure~\ref{fig:dynstruc}a} shows the transverse $\mathcal{S}_\perp(\vec{k},\omega) = \mathcal{S}_{xx}(\vec{k},\omega) + \mathcal{S}_{yy}(\vec{k},\omega)$ from nonperturbative simulations, revealing distinct magnon peaks. 
Vanishing spectral weight at $(0,0)$ is attributed to the disappearance of non-Bravais lattice form factors in the long-wavelength limit. While peak intensity at finite momenta is, in principle, finite due to nonzero form factors, it decreases significantly under the influence of magnon damping.
The lower mode remains sharp (within Gaussian filtering of width $\sigma_\omega=0.0896$), while the upper mode broadens along $(0,0) \to (\frac{\pi}{2},\frac{\pi}{2})$, aligning with perturbation theory. At $(\frac{\pi}{2},\frac{\pi}{2})$, its non-Lorentzian shape at large $\Delta$ signals strong nonperturbative effects.

We find good agreement between the nonperturbative spectrum in \textbf{Fig.~\ref{fig:dynstruc}a} and the $1/S^2$ off-shell single-particle spectral function $\mathcal{A}(\vec{k},\omega)$ in \textbf{Fig.~\ref{fig:dynstruc}b} (see Appendix A). Linecuts through $\mathcal{S}_\perp(\vec{k},\omega)$ and $\mathcal{A}(\vec{k},\omega)$ in \textbf{Fig.~\ref{fig:dynstruc}c,d} and \textbf{Fig.~\ref{fig:dynstruc}e,f} at  $(\frac{\pi}{2},\frac{\pi}{2})$ and $(\pi,0)$ show remarkable agreement. 
However, perturbation theory underestimates nonperturbative high-energy tails (compare \textbf{Fig.~\ref{fig:dynstruc}d} with \textbf{Fig.~\ref{fig:dynstruc}f}), possibly due to three-particle contributions, spinon continua \cite{DallaPiazza2014}, or Higgs resonances \cite{Powalski2015}. Capturing the full quantum dynamics of $S=1/2$ altermagnets may thus require incorporating amplitude oscillations of the order parameter, which lie beyond perturbation theory.

\section*{Discussion}

\subsection*{Experimental and theoretical implications}
The isotropic Heisenberg model describes quantum magnets when relativistic effects are negligible. In real materials, relativistic corrections such as magnetocrystalline anisotropy, $\lambda \sum_{\vec{r}, \vec{r}'} S_{\vec{r}}^z S_{\vec{r}'}^z $, influence magnon stability. For easy-axis anisotropy ($\lambda < 0$), relevant for altermagnetic candidates like MnF$_2$, CoF$_2$, and $\alpha$-Fe$_2$O$_3$ 
\cite{Smejkal2022altermagnets, SmejkalMagnons23, Hoyer2025}, both Goldstone modes acquire a gap scaling as $\sqrt{\lambda J}$. Beyond a critical anisotropy $\lambda^\star$, magnons become fully stable, a situation likely realized in MnF$_2$ \cite{Morano2024MnF2NoSplitting}. 
Experimental evidence for altermagnetic magnon spin splitting exists for MnTe \cite{Liu2024ChiralMagnonMnTe}, which shows a $2\,\text{meV}$ g-wave splitting over a $36\,\text{meV}$ bandwidth. However, its easy-plane anisotropy gaps one Goldstone mode by $4\,\text{meV}$ and violates spin conservation, requiring separate analysis. So do relativistic interactions such as Dzyaloshinskii-Moriya and pseudo-dipolar exchange (and field-induced canting) that enable order-$1/S$ two-magnon decay processes \cite{Mourigal2010, ZhitomirskyChernyshev2013, ChernyshevMaksimov2016, Winter2017, Hong2017,McClarty2018, McClarty2019, mook2021, Gohlke2023PRL}, which potentially overshadow intrinsic $1/S^2$ many-body effects characteristic of altermagnets.

Ideal platforms to study altermagnetic magnon decay are insulators to suppress electron-magnon damping \cite{SmejkalMagnons23, Costa2024AnisotropicDamping}, should minimize relativistic corrections, have a quasi-two-dimensional structure, and small spin quantum number $S$. Organic altermagnets such as $\kappa$-Cl are promising candidates, predicted to exhibit $4\,\text{meV}$ d-wave splitting over a $120\,\text{meV}$ bandwidth \cite{Naka2019}. We estimate $\Gamma \lesssim 1\,\mu$eV below the resolution of neutron spin-echo spectroscopy \cite{Keller2021}, which has previously measured magnon lifetimes \cite{Bayrakci2006, Nafradi2011, Chernyshev2012spinecho, Bayrakci2013}. However, $\kappa$-Cl-like systems with tenfold increased splitting show resolvable $\Gamma \approx 6\,\text{meV}$ (see Appendix D). With ongoing efforts in large-scale altermagnetic material discovery \cite{Guo2023LargeScaleAltermagnet, Bai2024AltermagnetReview, Sdequist2024TwoDimAlter}, suitable compounds are likely to emerge. Alternatively, altermagnetism could also be induced through strain \cite{Chakraborty2024strain}, twisting \cite{He2023twistingaltermagnet, Liu2024AltermagnetTwisted}, electric fields \cite{Mazin2023AMelectricfields}, or secondary order parameters \cite{Leeb2024AMorbitalOrdering}.


Altermagnets occupy a unique position in the broader context of spontaneous quasiparticle decay. Their \textit{continuous} SO$(2)$  symmetry makes spin-conserving three-particle decay the dominant instability, unlike other systems where two-particle decay prevails, such as phonons \cite{pitaevskii1959properties, Maradudin1962} or magnons in relativistic and noncollinear magnets \cite{ZhitomirskyChernyshev2013}. Note the difference to collinear magnets with a \textit{discrete} $C_2$ spin symmetry whose two-particle decay is also suppressed but their three-particle decays need not conserve spin \cite{Stephanovich2011, ZhitomirskyChernyshev2013, Khatua2023}. Notably, continuous spin symmetry does not always prevent two-particle decay---for instance, in quantum spin-gap magnets, two-triplon decay is allowed despite spin conservation \cite{Zhitomirski2006triplondecay}, requiring additional \textit{real}-space symmetries for suppression \cite{Gopalan1994, ZhitomirskyChernyshev2013}. This makes altermagnets a compelling platform to explore symmetry-constrained many-body dynamics.

In systems with multiple Goldstone modes, such as phonons  \cite{ZhitomirskyChernyshev2013}, decay from faster to slower modes usually depends on linear dispersion differences. In altermagnets, however, the mode splitting  is at least quadratic, fundamentally altering decay kinematics and giving rise to a unique ``island of stability'' in magnon lifetimes. These features underscore the exceptional many-body physics of altermagnets, marking a significant departure from conventional paradigms in quantum magnetism and quasiparticle decay.

\subsection*{Note added}
While preparing the manuscript we have become aware of independent complementary work on spontaneous magnon decay in d-wave altermagnets by Cichutek \textit{et al.} \cite{Cichutek25}, where an analytical long-wavelength analysis is presented for the checkerboard model.

\subsection*{Data availability}
All data and codes are available from the authors upon reasonable request.

\begin{acknowledgments}
    This work is funded by 
    the Deutsche Forschungsgemeinschaft (DFG, German Research Foundation) -- Project No.~504261060 (Emmy Noether Programme) and within the Transregional Collaborative Research Center TRR 173/3 - 268565370 ``Spin$+$X'' (Project A03), 
    by JSPS KAKENHI (Grants No.~22K14008, No.~24H02231, and No.~25H00611), 
    JST CREST (Grant No.~JPMJCR20T1), 
    and by Waseda University Grant for Special Research Project (Grant No.~2025C-133). 
    R.~E. was supported by a Grant-in-Aid for JSPS Fellows (Grant No.~23KJ2047) and the JSPS Overseas Challenge Program for Young Researchers FY2024.
    M.~G. acknowledges support by the Theory of Quantum Matter Unit of the Okinawa Institute of Science and Technology Graduate University (OIST). 
    The work of A.~L.~C. was supported by the U.S. Department of Energy, Office of Science, Basic Energy Sciences under Award No.~DE-SC0021221. 
    A part of numerical simulations was carried out at the Supercomputer Center, Institute of Solid State Physics, University of Tokyo. 
    M.~G. acknowledges support by the Scientific Computing section of the Research Support Division at OIST for providing the HPC resources.
\end{acknowledgments}


\section*{Appendix A: Spin-wave expansion for collinear isotropic magnets}
We assume that the classical ground state $|\text{CGS}\rangle$ of the nonrelativistic SO$(3)$-symmetric Heisenberg model in Eq.~\eqref{eq:HeisenbergModel} exhibits spontaneous symmetry breaking with collinear order, such that the magnet falls in one of the three classes mentioned in the main text. Without loss of generality, we fix the collinear order pointing along the $z$-direction, such that $|\text{CGS}\rangle$ becomes a product state of fully (positively or negatively) polarized eigenstates of the local spin-$z$ operator $S^z_{\vec{r}}$, i.e., the uniform state for ferromagnets and the Néel state for antiferromagnets and altermagnets. There is still a residual SO$(2)$ symmetry around the collinear axis, suggesting that the total spin-$z$ operator $S^z = \sum_{\vec{r}} S_{\vec{r}}^z$ commutes with $H$. However, $|\text{CGS}\rangle$ is only an eigenstate of $S^z$ but not of $H$. The situation is particularly pronounced for compensated magnets because their exact ground state is a fully isotropic spin singlet state \cite{anderson1997concepts}. However, as argued by Anderson \cite{anderson1997concepts}, the exact ground state is irrelevant because one can build a state with finite sublattice magnetizations as a superposition of states that are degenerate with the ground state in the thermodynamic limit. The point of a spin-wave expansion is then to approximate such a state perturbatively, starting from $|\text{CGS}\rangle$. Since $|\text{CGS}\rangle$ is an eigenstate of $S^z$, so are all excited states encountered in the spin-wave expansion, which are built upon $|\text{CGS}\rangle$ by acting on it with $S_{\vec{r}}^\pm = S_{\vec{r}}^x \pm \mathrm{i} S_{\vec{r}}^y$. As a particular result, magnons---obtained by a single action of $S_{\vec{r}}^\pm$---carry a definite spin quantum number of $\pm 1$ ($\hbar = 1$) relative to the ground state, and the interactions among them may not change the spin.

Gratifyingly, the above observation of a spin conservation does not have to be put into the spin-wave expansion by hand but rather emerges naturally when carrying out the expansion, as we will show in the following. We exclude the trivial case of ferromagnets. Being interested in the low-energy physics, we assume without loss of generality that there are only two sublattices, called A and B and carrying spins of length $S_\text{A}$ and $S_\text{B}$, such that there are no optical magnon branches. By means of the Holstein-Primakoff transformation \cite{holsteinprimakoff1940}, 
\begin{align}
    S^z_{\vec{r} \in \text{A}} = S_\text{A} - a^\dagger_{\vec{r}} a_{\vec{r}} 
    \quad &\text{and} \quad 
    S^z_{\vec{r} \in \text{B}} = - S_\text{B} + b^\dagger_{\vec{r}} b_{\vec{r}} 
    \\
    S^-_{\vec{r} \in \text{A}} = a^\dagger_{\vec{r}} \sqrt{2S_\text{A} - a^\dagger_{\vec{r}} a_{\vec{r}}} 
    \quad &\text{and} \quad 
    S^-_{\vec{r} \in \text{B}} =  \sqrt{2S_\text{B} - b^\dagger_{\vec{r}} b_{\vec{r}}} \, b_{\vec{r}}
    \label{eq:HPtrafo}
\end{align}
we map the spin operators onto sublattice bosons with annihilators $a_{\vec{r}}$ and $b_{\vec{r}}$ and creators $a^\dagger_{\vec{r}}$ and $b^\dagger_{\vec{r}}$. By expanding the square root in Eq.~\eqref{eq:HPtrafo}---assuming ground state expectation values of $a^\dagger_{\vec{r}} a_{\vec{r}}$ and $b^\dagger_{\vec{r}} b_{\vec{r}}$ small compared to $2S_\text{A/B}$---we obtain an infinite expansion $H = H_0 + H_2 + H_4 + H_6 + \ldots$, where the subscript denotes the power in bosonic operators. There are no terms with an odd power of bosons because of the residual SO$(2)$ symmetry around the collinear axis. (More generally, a less stringent requirement for the absence of odd terms in the spin-wave expansion of collinear magnets is a $C_2$ symmetry around the collinear axis \cite{Stephanovich2011, ZhitomirskyChernyshev2013, Khatua2023}. In this case, however, the even terms do not have to conserve spin.) $H_0$ is the classical ground state energy that is of no further interest. $H_2$ encodes the harmonic theory of noninteracting magnons, and $H_4$ and beyond captures magnon-magnon interactions. In momentum space, the harmonic theory is block diagonal,
\begin{align}
    H_2 = \frac{1}{2} \sum_{\vec{k} \in \text{BZ}} \left( a^\dagger_{\vec{k}}, b_{-\vec{k}}, b^\dagger_{\vec{k}}, a_{-\vec{k}} \right) \begin{pmatrix}
        h_+(\vec{k}) & 0 \\ 0 & h_-(\vec{k})
    \end{pmatrix}
    \begin{pmatrix}
        a_{\vec{k}} \\ b^\dagger_{-\vec{k}} \\ b_{\vec{k}} \\ a^\dagger_{-\vec{k}}
    \end{pmatrix}
    +
    \delta H_0,
\end{align}
where the two blocks correspond to opposite spin quantum numbers $\pm 1$. Here, $\delta H_0$ is a number providing a correction to the ground state energy, and the two-by-two Hamilton matrix blocks read
\begin{align}
    h_{\pm}(\vec{k}) = \begin{pmatrix}
        A_{\pm}(\vec{k})  & B_{\pm}(\vec{k})  \\
        B^\ast_{\pm}(\vec{k}) & C_{\pm}(\vec{k}) 
    \end{pmatrix},
\end{align}
where the entries $A_{\pm}(\vec{k})$, $B_{\pm}(\vec{k})$, and $C_{\pm}(\vec{k})$ depend on the specifics of the system.
After a Bogoliubov transformation, $H_2$ becomes diagonal,
\begin{align}
    H_2 = \delta H'_0 + \sum_{\vec{k} \in \text{BZ}} \left( \epsilon_+(\vec{k}) \alpha^\dagger_{\vec{k}} \alpha_{\vec{k}} + \epsilon_-(\vec{k}) \beta^\dagger_{\vec{k}} \beta_{\vec{k}} \right),
\end{align}
where $\epsilon_\pm(\vec{k})$ is the harmonic magnon energy of the spin-$\uparrow$ and spin-$\downarrow$ magnons, created by $\alpha^\dagger_{\vec{k}}$ and $\beta^\dagger_{\vec{k}}$, respectively. These bosons are the eigenmodes of the system when interactions are neglected. $\delta H'_0$ is the full correction to the classical ground state energy due to harmonic quantum fluctuations. 

The Bogoliubov vacuum $| \varnothing \rangle$, defined by $\alpha_{\vec{k}} | \varnothing \rangle = \beta_{\vec{k}} | \varnothing \rangle = 0$, is the leading-order quantum correction to the classical ground state $|\text{CGS}\rangle$. It can be considered a two-mode squeezed vacuum of those sublattice-boson Fock states that carry the same number of bosons on sublattices A and B \cite{Kamra2019AFMsqueeze}. $| \varnothing \rangle$ is an eigenstate of $S^z$ with the same eigenvalue as $|\text{CGS}\rangle$. Likewise, the single-magnon states 
$\alpha^\dagger_{\vec{k}} | \varnothing \rangle$ and 
$\beta^\dagger_{\vec{k}} | \varnothing \rangle$ are eigenstates of $S^z$ because they are two-mode squeezed states of those sublattice-boson Fock states that carry $n+1$ bosons on sublattice A and $n$ on sublattice B (or \textit{vice versa}) \cite{Kamra2019AFMsqueeze}. They carry spin relative to the vacuum identical to that of the sublattice-boson Fock states, that is $(n+1)-n = +1$ or $n-(n+1) = -1$.
The spin of magnons in collinear magnets is alternatively referred to as their handedness or chirality \cite{SmejkalMagnons23}, or polarization \cite{Nambu2020}. 

In terms of the bosons diagonalizing the harmonic theory, $\alpha$ and $\beta$, the leading magnon-magnon interactions are given by
\begin{align}
\begin{aligned}
    H_4 &= 
      W^{(1)} \alpha^\dagger \alpha^\dagger \alpha \alpha
    + W^{(2)} \beta^\dagger \beta^\dagger \beta \beta
    + W^{(3)} \alpha^\dagger \beta^\dagger \alpha \beta  \\
    &\quad
    + W^{(4)} \alpha^\dagger \alpha^\dagger \beta^\dagger \alpha
    + W^{(5)} \alpha^\dagger \beta^\dagger \beta^\dagger \beta
    + W^{(6)} \alpha^\dagger \alpha^\dagger \beta^\dagger \beta^\dagger \\
    &\quad
    + W^{(7)} \alpha^\dagger \alpha \alpha \beta
    + W^{(8)} \beta^\dagger \alpha \beta \beta
    + W^{(9)} \alpha \alpha \beta \beta,
    \label{eq:H4}
\end{aligned}
\end{align}
where we have suppressed momentum labels. The full expression in given in Supplementary Material II. Notably, none of the terms in Eq.~\eqref{eq:H4} change the total spin---the total spin of the destroyed bosons equals that of those created. While Ref.~\cite{GarciaGaitan2025} suggests the presence of terms like $\alpha^\dagger \alpha^\dagger \alpha^\dagger \beta$, such combinations do not actually appear. The same observation holds for $H_6$ and higher-order terms and is a consequence of spin conservation as discussed above.

To explore the effects of interactions in altermagnets, we set $S_\text{A} = S_\text{B} = S$ and follow the general strategy for many-body perturbation theory, e.g., see Refs.~\cite{Harris1971, Hamer1992ThirdOrderSpinWaveTheory, Syromyatnikov2010}. Formally, the small perturbative parameter of the spin-wave expansion is $1/S$, where $H_n$ is of order $S^{2-n/2}$. In the usual sense of many-body perturbation theory, we consider $H_2$ the unperturbed Hamiltonian piece and $V = V^{(1/S)} + V^{(1/S^2)}$ the perturbation. The order-$1/S$ perturbation $V^{(1/S)}$ contains $H_4$ and the order-$1/S^2$ perturbation $V^{(1/S^2)}$ contains $H_6$. For order-$1/S^2$ accuracy, we treat $V^{(1/S)}$ up to second-order perturbation theory and $V^{(1/S^2)}$ up to first-order perturbation theory.

The interaction-corrected magnon dispersion $\tilde{\epsilon}_\pm(\vec{k})$ is obtained from the poles of the retarded Green's function, $G_{\pm}(\vec{k},\omega) = [ \omega + \mathrm{i}0^+ - \epsilon_\pm(\vec{k}) - \Sigma^{(1/S)}_\pm(\vec{k},\omega) - \Sigma^{(1/S^2)}_\pm(\vec{k},\omega) ]^{-1}$, where the relevant zero-temperature single-magnon self-energies, $\Sigma^{(1/S)}_\pm(\vec{k},\omega)$ and $\Sigma^{(1/S^2)}_\pm(\vec{k},\omega)$, are given in Supplementary Material II. Within an on-shell calculation, the corrected spectrum is found to read
\begin{align}
    \tilde{\epsilon}_\pm(\vec{k}) = \epsilon_\pm(\vec{k}) + \delta \epsilon^{(1/S)}_\pm(\vec{k}) + \delta \epsilon^{(1/S^2)}_\pm(\vec{k}),
\end{align}
where $\delta \epsilon^{(1/S)}_\pm(\vec{k}) = \Sigma^{(1/S)}_\pm(\vec{k},\epsilon_\pm(\vec{k}))$ and $\delta \epsilon^{(1/S^2)}_\pm(\vec{k}) = \Sigma^{(1/S^2)}_\pm(\vec{k},\epsilon_\pm(\vec{k}))$.
At order $1/S$, there are only real corrections: $\delta \epsilon^{(1/S)}_\pm(\vec{k}) \in \mathbb{R}$. The leading contribution to the magnon lifetime $\tau_\pm(\vec{k}) = -1/\text{Im}[\delta \epsilon^{(1/S^2)}_\pm(\vec{k})]$ appears at order $1/S^2$ and is associated with the diagram in \textbf{Fig.~\ref{fig:model_and_spectra}b}. The corresponding  decay rate of magnons, $\Gamma_\pm(\vec{k}) = \tau^{-1}_\pm(\vec{k})$, is given in Eq.~\eqref{eq:damping rate}. It represents the half width at half maximum of the Lorentzian quasiparticle peak. We have explicitly verified that the real corrections to the spectrum do not gap out the Goldstone modes, that is, both $\delta \epsilon^{(1/S)}_\pm(\vec{k}) \to 0$ and $\mathrm{Re}\delta \epsilon^{(1/S^2)}_\pm(\vec{k}) \to 0$ as $k \to 0$.

For the off-shell calculation presented in \textbf{Fig.~\ref{fig:dynstruc}b}, we have plotted the single-particle spectral function $\mathcal{A}(\vec{k},\omega) = -\text{Im}[ G_+(\vec{k},\omega) + G_-(\vec{k},\omega)]/\pi$. The resulting spectrum is no longer consistent within $1/S$ and captures some nonperturbative effects, such as non-Lorenzian lineshapes. This inconcsistency in $1/S$ generates a problem for the Goldstone mode that gets pushed to negative energies. Numerically, we found it necessary to suppress one particular family of self-energy diagrams to avoid this problem; this neglection has almost no visible effects on the spectrum away from the Brillouin zone origin as the magnitude of these self-energies is tiny (see Supplementary Material II for further details).

\section*{Appendix B: Linear spin-waves in checkerboard d-wave altermagnet}
Applying spin-wave theory to the spin model in Eq.~\eqref{eq:model}, results in the following noninteracting magnon dispersion
\begin{align}
    \epsilon_\pm(\vec{k}) = S \left( \sqrt{A(\vec{k})^2 - B(\vec{k})^2} \pm \Delta(\vec{k}) \right),
\end{align}
where
\begin{align}
    A(\vec{k}) = 4J_1-4J_2+2J_2\left(\cos (k_x+k_y) + \cos (k_x-k_y) \right)
\end{align}
and 
\begin{align}
    B(\vec{k}) = 2J_1\left( \cos k_x + \cos k_y \right).
\end{align}
The d-wave splitting is given by $\epsilon_+(\vec{k}) - \epsilon_-(\vec{k}) = 2S\Delta(\vec{k})$, where
\begin{align}
    \Delta(\vec{k}) = - 2\Delta \left(\cos (k_x+k_y) - \cos (k_x-k_y) \right).   
\end{align}
In the long-wavelength limit, the dispersion is expanded as follows
\begin{align}
    \frac{\epsilon_\pm(\vec{k})}{S}
    \approx 
    v |\vec{k}| \pm 2\Delta \sin(2\varphi_{\vec{k}}) |\vec{k}|^2 + \left(\kappa + \kappa_{4\varphi} \cos(4\varphi_{\vec{k}})  \right) |\vec{k}|^3, 
\end{align}
where the nodal lines of the altermagnetic splitting are rotated by $\pi/4$ compared to Eq.~\eqref{eq:general-expansion}, and we have explicitly included in $\kappa_{4\varphi}$ the angular dependence of the cubic term. The expansion coefficients are related to $J_1$ and $J_2$ by the following relations:
\begin{align}
    v &= 2 \sqrt{2J_1(J_1-2J_2)}, \\
    \kappa &= \frac{-3 J_1^2 + 4 J_1 J_2 + 8 J_2^2}{8\sqrt{2J_1(J_1-2J_2)}}, \\
    \kappa_{4\varphi} &= \frac{J_1 (J_1 + 4 J_2) }{24 \sqrt{2J_1(J_1-2J_2)}}.
\end{align}
Throughout the text, we set $S=1/2$, assume units of meV, and explore two parameter sets: 

The first set of parameters is used in \textbf{Fig.~\ref{fig:model_and_spectra}}, \textbf{Fig.~\ref{fig:DOSs}a,b}, and \textbf{Fig.~\ref{fig:dynstruc}}: $J_1 = 1$ and $J_2 = -0.5$, such that $v = 4$, $\kappa = -3/16<0$, and $\kappa_{4\varphi} = -1/48$, realizing the negatively curved case with forbidden class-(ii) type decays. According to Eq.~\eqref{eq:criticalDeltadwave}, we find $\Delta^\star_\text{d-wave} = 3/(4\sqrt{2}) \approx 0.53$. 

The second set of parameters is used in \textbf{Fig.~\ref{fig:DOSs}c}: $J_1 = 1/10$ and $J_2 = -1$, such that $v = \sqrt{42}/5$, $\kappa = 757/(80 \sqrt{42})>0$, and $\kappa_{4\varphi} = -13/(80 \sqrt{42})$, realizing the positively curved case supporting class-(ii) type decays. According to Eq.~\eqref{eq:kstar}, we find $ k^\star_\text{d-wave} = 640/757 \times \sqrt{7/6} \approx 0.913 \Delta$.

For both parameter sets, $|\kappa_{4\varphi}| \ll |\kappa|$ and we expect that the results obtained from Eq.~\eqref{eq:general-expansion}, which neglected $\kappa_{4\varphi}$, provide a reasonable expectation.

\section*{Appendix C: Matrix product operator based time evolution}

\begin{figure}
    \centering
    \includegraphics[width=0.95\linewidth]{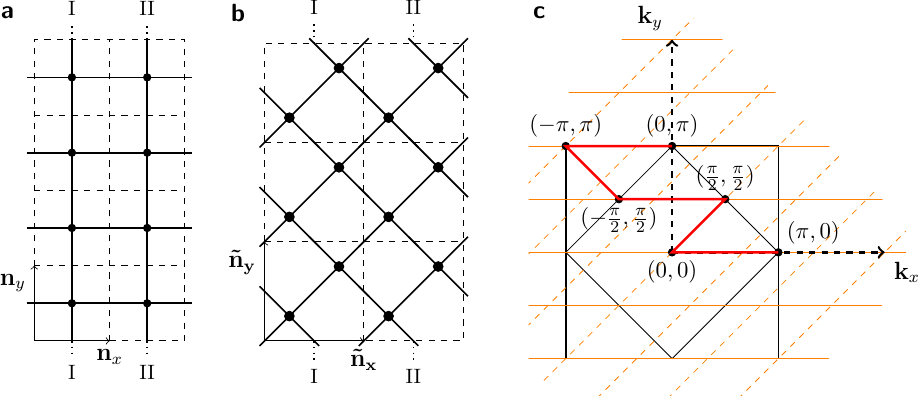}
    \caption{\textbf{Cylinder geometries and accessible momenta in reciprocal space in nonperturbative DMRG$+$tMPO simulations.}
        (\textbf{a}) Regular square lattice with a circumference of $L_y=4$ unit cells. 
        (\textbf{b}) $\pi/4$-rotated lattice with $L_y=3$.
        (\textbf{c}) Brillouin zone and accessible momenta cuts of regular (solid) and rotated lattice (dashed). The red line illustrates the Brillouin zone path used in \textbf{Fig.~\ref{fig:dynstruc}a}. The segment from $(-\frac{\pi}{2},\frac{\pi}{2})$ to $(-\pi,\pi)$ is taken from the path $(\frac{\pi}{2},\frac{\pi}{2})$ to $(\pi,\pi)$. Although these paths are not related by symmetry and have opposite altermagnetic splitting, the transverse dynamical structure factor $\mathcal S_\perp (\vec k,\omega)$ is identical.}
    \label{fig:cyl_geom_and_BZ}
\end{figure}

We perform nonperturbative quantum simulations to obtain dynamical correlations using infinite matrix-product states \cite{mcculloch_infinite_2008,phien_infinite_2012} and a time evolution based on matrix product operators \cite{zaletel_timeevolving_2015}. See also Ref.~\cite{gohlke_dynamics_JK_2017,Verresen2018AFM} for more details on the method.
MPS provide an efficient representation of quantum wave functions in one spatial dimension, but have been applied widely to two-dimensional systems by wrapping the lattice onto a cylinder and winding around the one-dimensional tensor train of the MPS and MPO. 
The finite circumference of the cylinder discretizes reciprocal momentum as $k_y = 2\pi/L_y$. 
On the other hand, reciprocal momenta $k_x$ parallel to the cylinder axis are very dense, resulting in lines of accessible momenta in reciprocal space.
We illustrate both lattice geometries, their periodic boundary around the cylinder, and the corresponding Brillouin zone in Fig.~\ref{fig:cyl_geom_and_BZ}.

First, we obtain the ground state in MPS form using infinite DMRG \cite{white_density_1992,mcculloch_infinite_2008,schollwoeck2011,phien_infinite_2012} and a bond dimension of $\chi=400$. 
Here, a small XXZ anisotropy $\lambda \sum_{\langle \vec{r},\vec{r}'\rangle } S^z_{\vec{r}} S^z_{\vec{r}'}$ of $\lambda=0.02$ improves accuracy and convergence by gapping out the Nambu-Goldstone mode, and ensuring a finite local moment of $\langle S^z_{\vec r_i}\rangle \approx \pm 0.28$.
After applying a local spin operator $S^{\gamma=\{x,y,z\}}_{\vec r_\text{center}}$ in the middle of the cylinder, we perform the iterative application of the time-evolution MPO \cite{zaletel_timeevolving_2015} with a time step of $dt = 0.125/7 \approx 0.01786$, while spin-spin correlations are measured each $\Delta t = 0.125$ up to a maximum time of $t_\text{max}=25$.
We allow $\chi$ to grow adaptively throughout the time-evolution with a cap set to $\chi_\text{max} = 600$ in order to limit required computational resources.

The objective is to compute the dynamical spin-structure factor $\mathcal S(\vec k,\omega)$ defined as,
\begin{equation}
    \mathcal S_{\gamma\gamma} (\bm{k},\omega) = N \int \mathrm{d}t ~ \mathrm{e}^{\mathrm{i} \omega t} \sum_{\vec r, \vec r'} \mathrm{e}^{\mathrm{i} \vec{k} \cdot (\vec{r} - \vec{r'})} ~ C^{\gamma\gamma}_{\vec r, \vec r'}(t) ~,
    \label{eq:tMPO_Skw}
\end{equation}
where $\gamma = \{x,y,z\}$ and $C^{\gamma\gamma}_{\vec r, \vec r'}(t) = \langle S^\gamma_{\vec r} (t) S^\gamma_{\vec r'} (0) \rangle$ are the dynamical spin-spin correlations.
The protocol is as follows:
(i) $C^{\gamma\gamma}_{\vec r, \vec r'}(t)$ is Fourier transformed in space providing $C^{\gamma\gamma}(\vec k,t)$,
(ii) $C^{\gamma\gamma}(\vec k,t)$ gets extended in time using dynamical mode decomposition (DMD) \cite{kaneko_dmd_2025,schmid_dmd_review_2022} and convoluted with a Gaussian of width $\sigma_t \approx 13.02$ to suppress ringing from the box function of the finite time window, 
and (iii) in doing a Fourier transform in time and coherent summations over the two sublattices of the magnetic unit cell, we obtain $\mathcal S_{\gamma\gamma}(\vec k,\omega)$ as in Eq.~\eqref{eq:tMPO_Skw}. 
The convolution with a Gaussian in (ii) results in a Gaussian broadening in $\mathcal S_{\gamma\gamma}(\bm k, \omega)$ of width $\sigma_\omega \approx 0.0768$. The final $\mathcal S_{\perp}(\bm k, \omega)= \mathcal S_{xx}(\bm k, \omega) + \mathcal S_{yy}(\bm k, \omega)$ is plotted in \textbf{Fig.~\ref{fig:dynstruc}a}.

Our simulations complement those of Ref.~\cite{GarciaGaitan2025}. Here, we remark on the difference to the approach taken in Ref.~\cite{GarciaGaitan2025} and its implications.
We compute the ground state on an infinite cylinder and perform the time-evolution on a finite cylinder segment embedded in an infinite cylinder. This is to be contrasted to the finite length, $L_x=80$, cylinder in Ref.~\cite{GarciaGaitan2025}. 
Therefore, we are less affected by finite-size effects. In turn, we add a finite XXZ anisotropy, $\lambda=0.02$, to control magnetization, convergence and computational resources, while retaining interpretability of the gap.
For comparison, Ref.~\cite{GarciaGaitan2025} has a spin gap of $\varepsilon \approx 0.5$, while at $\lambda = 0.02$ we obtain $\varepsilon \approx 0.23$.
Furthermore, the ground state of the isotropic, $\lambda=0$, Heisenberg model on a finite-width cylinder does not break $\mathrm{SO}(3)$ spontaneously and develops a paramagnetic, singlet-like ground state instead. As a result, $\mathcal{S}_{\mu\mu}(\vec{k},\omega)$ with $\mu = x,y,z$ is isotropic and picks up single-magnon and two-magnon contributions.
In our case, an XXZ anisotropy of $\lambda = 0.02$ ensures finite sublattice magnetization, which enables us to separate the transverse $\mathcal S_\perp (\vec k, \omega)$ from the longitudinal $\mathcal S_{zz} (\vec k, \omega)$, the latter of which we are presenting in Supplementary Material~V. However, as $\mathcal S_{zz} (\vec k, \omega)$ is proportional to the two-magnon sector rather than single particles and therefore does not directly reflect quasiparticle damping, we propose measurement of $\mathcal{S}_\perp(\vec{k},\omega)$ in experiments.

\section*{Appendix D: Magnon damping estimates for $\kappa$-Cl-like altermagnets}

\begin{figure}
    \centering
    \includegraphics[width=\columnwidth]{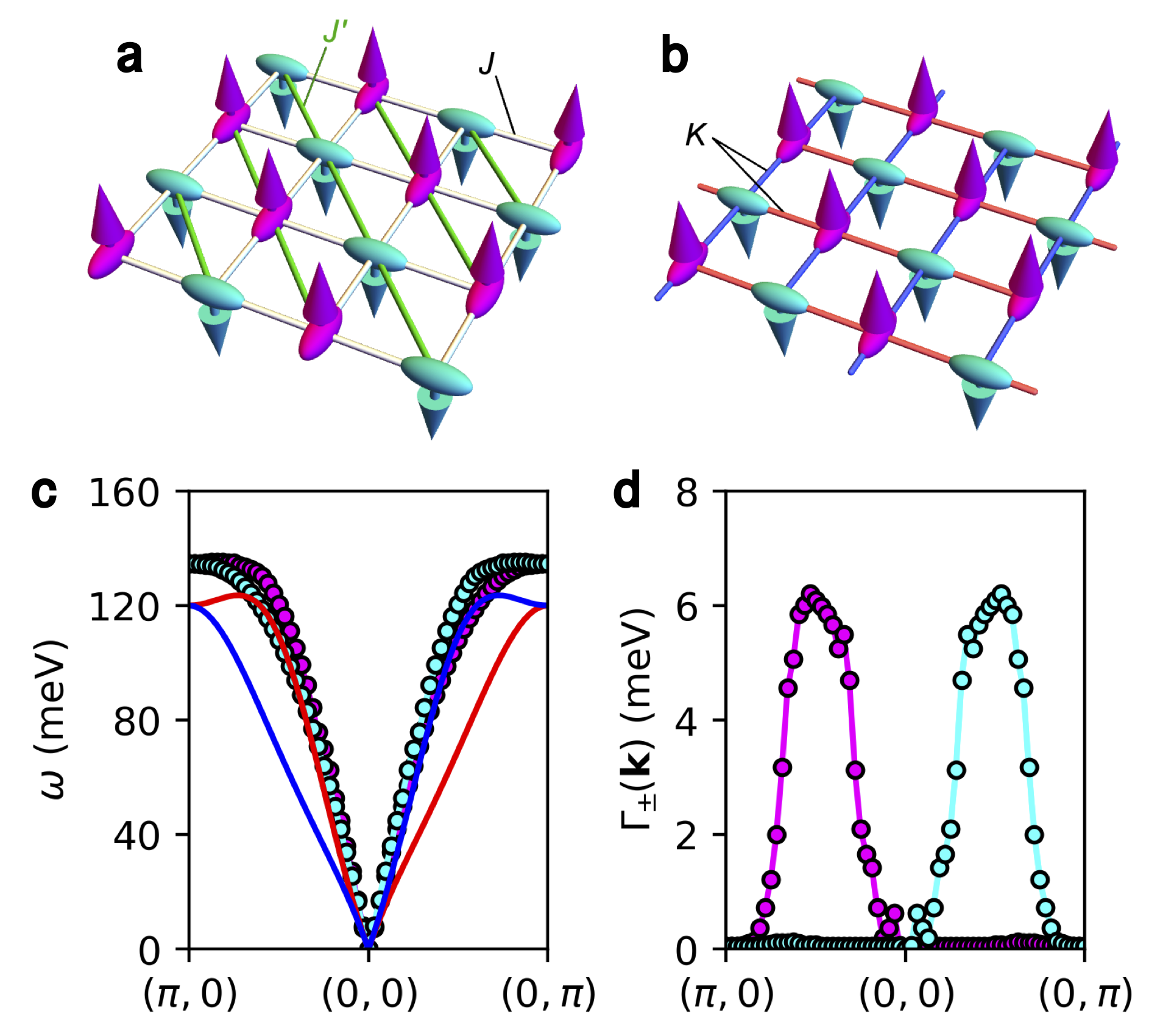}
    \caption{\textbf{Spontaneous magnon decay in $\kappa$-Cl-like organic altermagnets.} (\textbf{a,b}) Effective spin model of $\kappa$-Cl on the square lattice derived in Ref.~\cite{Naka2019}. Magenta and cyan arrows with ellipsoids respectively denote the collinear ground state $(\pi,\pi)$ N\'{e}el order. Directions of the ellipsoids correspond to the molecular orientations, which is the origin of altermagnetism in this material. White and green bonds in (\textbf{a}) indicate the nearest-neighbor and the next-nearest-neighbor exchange interactions $J$ and $J'$, respectively. In (\textbf{b}), due to the differing molecular orientations between the two sublattices, the third-nearest neighbor exchange interactions along the bonds parallel and perpendicular to the orientation direction are non-equivalent. Here, only the latter is denoted as $K$ and the former is neglected. (\textbf{c}) Renormalized magnon spectrum (lines with dot markers) compared to the linear magnon spectrum (solid lines) for $J=80\,$meV, $J'=20\,$meV, $S=1/2$ and $K = 20\,$meV. The value of $K$ sets the magnitude of the altermagnetic splitting and is approximately ten times larger than what is predicted for $\kappa$-Cl \cite{Naka2019}. (\textbf{d}) Magnon damping $\Gamma_\pm(\vec{k})$ along momentum directions with maximal band splitting.}
    \label{fig:kappa-Cl}
\end{figure}

$\kappa$-Cl is an effectively two-dimensional d-wave altermagnet, whose low-energy spin model was derived in Ref.~\cite{Naka2019}: 
\begin{align}
    H 
    =  
    \sum_{\ev{\vec{r},\vec{r}'}}          
    J \vec{S}_{\vec{r}}\cdot\vec{S}_{\vec{r}'}
    +
    \sum_{\ev{\ev{\vec{r},\vec{r}'}}}          
    J' \vec{S}_{\vec{r}}\cdot\vec{S}_{\vec{r}'}
    + 
    \sum_{\ev{\ev{\ev{\vec{r},\vec{r}'}}}} 
    K
    \vec{S}_\vec{r}\cdot\vec{S}_{\vec{r}'}.
    \label{eq:model_kappaCl}
\end{align}
Here, the local spin operators $\vec{S}_{\vec{r}}$ of length $S=1/2$ are associated with molecular dimers, which are indicated by ellipsoids in \textbf{Fig.~\ref{fig:kappa-Cl}a}. $J$ and $J'$ are the first and second-nearest-neighbor exchange interaction, respectively. $K$ is a third-nearest neighbor interaction along those third-neighbor bonds that align with the short axis of the dimers, see \textbf{Fig.~\ref{fig:kappa-Cl}b}. This third-nearest neighbor coupling causes the altermagnetic splitting in contrast to the model in Eq.~\eqref{eq:model}, where it derives from second-neighbor coupling. As a result, the magnon splitting is maximal at the midpoint of the line from $(0,0)$ to $(\pi,0)$ or to $(0,\pi)$, see \textbf{Fig.~\ref{fig:kappa-Cl}c}, but it is zero at the entire boundary of the Brillouin zone. By carrying out a nonlinear spin-wave analysis identical to the one outlined above, we find that for the parameters predicted in Ref.~\cite{Naka2019}, that is, $J=80\,$meV, $J'=20\,$meV, and $K = 2\,$meV, the maximal magnon damping is smaller than $1\,\mu$eV. However, if $K$ is increased by a factor of ten, we obtain a maximal damping of $\Gamma \sim 6\,$meV, as shown in \textbf{Fig.~\ref{fig:kappa-Cl}d}.

In our initial long-wavelength expansion of the magnon dispersion of altermagnets in  Eq.~\eqref{eq:general-expansion} we have neglected the crystallographic influence on the velocity $v$ and the cubic nonlinearity $\kappa$. Taking them into account for $\kappa\textrm{-Cl}$, we obtain
\begin{align}
    \epsilon^{\kappa\textrm{-Cl}}_\pm(\vec{k}) &= \nu\sqrt{1+(\nu_{2\varphi}|\nu_{2\varphi}|/\nu^2)\sin (2\varphi_\vec{k})} |\vec{k}| \nonumber \\
    &\quad + \frac{\kappa+\kappa_{2\varphi}\sin(2\varphi_\vec{k})+\kappa_{4\varphi}\cos(4\varphi_\vec{k})}{\sqrt{1+(\nu_{2\varphi}|\nu_{2\varphi}|/\nu^2)\sin (2\varphi_\vec{k})}} |\vec{k}|^3 \nonumber \\
    &\quad \pm \Delta \cos (2\varphi_\vec{k}) |\vec{k}|^2.
\end{align}
We find that $|\nu_{2\varphi}|\ll\nu$ and $|\kappa_{2\varphi}|$, $|\kappa_{4\varphi}|\ll|\kappa|$ are satisfied. The angular dependence of the velocity $\nu_{2\varphi}$ and the cubic nonlinearity $\kappa_{2\varphi}$ stems from the next-nearest-neighbor exchange interaction $J'$, which breaks the four-fold rotation symmetry of the square lattice. Since the crystal structure of $\kappa$-Cl belongs to the crystallographic Laue group $mmm$, which does not preserve four-fold rotational symmetry, both $\nu_{2\varphi}$ and $\kappa_{2\varphi}$ originate purely from lattice-geometric factors. We emphasize that a finite $\nu_{2\varphi}$ does not disrupt the scaling theory of $\Gamma$ in the long-wavelength limit.

We note that planar altermagnets belong to one of the following four crystallographic Laue groups: $mmm$, $4/m$, $4/mmm$, and $6/mmm$ \cite{Smejkal2022altermagnets}. Those belonging to $mmm$ are the only with nonzero angular corrections to the velocity, $\nu_{2\varphi} \neq 0$. In contrast, angular corrections to $\kappa$, namely $\kappa_{4\varphi}$ in d-wave and g-wave, and $\kappa_{6\varphi}$ in i-wave, are generically present. They have a qualitative effect on magnon damping if they modulate the sign of the cubic nonlinearity with direction.


\bibliography{bib}


\clearpage
\onecolumngrid
\appendix

\section{Kinematics of magnon decay in altermagnets}
\renewcommand{\theequation}{A.\arabic{equation}}

\subsection{Decay of nodal-plane magnons for $\kappa<0$}
\label{sec:decay-d-wave-nodal}
We recall from the main text that the low-energy expansion of altermagnets is given by
\begin{align}
    \epsilon_\pm(\bm{k})
    =
    v |\bm{k}| + \kappa |\bm{k}|^3 \pm \Delta  \sin\left( n \varphi_{\bm{k}} \right) |\bm{k}|^n,
    \label{eq:expansion}
\end{align}
where $v>0$ is the magnon velocity, $\kappa$ represents the leading (cubic) non-altermagnetic nonlinearity, $\tan \varphi_{\bm{k}} = k_y/k_x$, and $\Delta$ parametrizes the altermagnetic splitting with a d-wave ($n=2$), g-wave ($n=4$) or i-wave character ($n = 6$). Without loss of generality, we set $\Delta \ge 0$. Note that the expansion in Eq.~\eqref{eq:expansion} neglects the angular dependence of $v$ and $\kappa$ due to crystallographic symmetries.

We consider a nodal-plane magnon with momentum $\bm{k}=(k,0)$ decaying into three magnons with momenta $\bm{p}=(p_x,p_y)$, $\bm{q}=(q_x,q_y)$, and $\bm{k}-\bm{q}-\bm{p}=(0,0)$. One decay product is put in the Goldstone mode in order to make the energy of the decay products as small as possible. The other two decay products have net spin zero and belong to the lower magnon branch. We assume that their momenta have the same magnitude and are symmetrically tilted by angles $\pm \theta$ around the nodal plane ($x$-axis), such that
\begin{align}
    \bm{q} 
    = 
    \frac{k}{2 \cos \theta} 
    \begin{pmatrix}
        \cos \theta \\ \sin \theta
    \end{pmatrix}
    \quad
    \text{and}
    \quad
    \bm{p} 
    = 
    \frac{k}{2 \cos \theta} 
    \begin{pmatrix}
        \cos (-\theta) \\ \sin(-\theta)
    \end{pmatrix}.
    \label{eq:momentumansatz}
\end{align}
Plugging this ansatz for the momenta into the energy conservation condition,
\begin{align}
    \epsilon_+(\bm{k}) = \epsilon_-(\bm{q}) + \epsilon_+(\bm{p}) + \epsilon_+(\bm{k}-\bm{q}-\bm{p}),
\end{align}
yields an equation for $\theta$: 
\begin{align}
    0 = -k \left( v +\kappa k^2 \right) + \frac{k v}{\cos \theta} + \frac{k^3\kappa}{4 \cos^3 \theta} - 2 \Delta \left( \frac{k}{2\sin{\theta}} \right)^n \sin(n\theta).
\end{align}
A second-order Taylor expansion in $\theta$ results in the quadratic equation 
\begin{align}
    0 = -2^{4-n} k^n n \Delta \theta + 4 k v \theta^2 + 3 k^2 \kappa \left(\theta^2-2 \right),
\end{align}
with solutions
\begin{align}
    \theta = \frac{2^{3-n}k^n \Delta \pm \sqrt{4^{3-n}k^{2n} n^2 \Delta ^2 + 6 k^4 \kappa\left( 4v + 3 k^2 \kappa \right)}}{4 k v + 3 k^3 \kappa}.
    \label{eq:solution_theta}
\end{align}
These are real only if
\begin{align}
    4^{3-n}k^{2n} n^2 \Delta ^2 + 6 k^4 \kappa\left( 4v + 3 k^2 \kappa \right) \ge 0,
\end{align}
which provides the condition $\Delta \ge \Delta^\star$, where
\begin{align}
    \Delta^\star = \sqrt{3} \frac{2^{n-5/2}}{n k^{n-2}} \sqrt{-\kappa} \sqrt{4v+3k^2 \kappa},
\end{align}
as given in the main text. We emphasize that $\kappa < 0$.

\subsection{Decay of lower-branch magnons for $\kappa>0$}
We consider the decay of a lower-branch magnon with momentum $\bm{k}=k (\cos \rho, \sin \rho)$ along a general direction into three magnons with momenta $\bm{p}=(p_x,p_y)$, $\bm{q}=(q_x,q_y)$, and $\bm{k}-\bm{q}-\bm{p}=(0,0)$. Again, one decay product is put in the Goldstone mode in order to make the energy of the decay products as small as possible. It carries opposite spin. The other two decay products have the same spin as the decaying magnon. We make the ansatz that the momenta $\bm{q}$ and $\bm{p}$ have the same length and are symmetrically tilted away from the axis of the decaying magnon by an angle $\pm \theta$, that is,
\begin{align}
    \bm{q} 
    = 
    \frac{k}{2 \cos \theta} 
    \begin{pmatrix}
        \cos (\rho + \theta) \\ \sin (\rho + \theta)
    \end{pmatrix}
    \quad
    \text{and}
    \quad
    \bm{p} 
    = 
    \frac{k}{2 \cos \theta} 
    \begin{pmatrix}
        \cos (\rho - \theta) \\ \sin(\rho - \theta)
    \end{pmatrix}.
    \label{eq:momentumansatz2}
\end{align}
Whether $\epsilon_-(\bm{k})$ or $\epsilon_+(\bm{k})$ is the lower-magnon branch is determined by the angle $\rho$. We note that the ansatz in Eq.~\eqref{eq:momentumansatz2} neglects the absence of a reflection symmetry of the dispersion about a general direction. First, we plug it into energy conservation for a decaying spin-$\downarrow$ magnon,
\begin{align}
    \epsilon_-(\bm{k}) = \epsilon_-(\bm{q}) + \epsilon_-(\bm{p}) + \epsilon_+(\bm{k}-\bm{q}-\bm{p}),
\end{align}
and obtain
\begin{align}
    0 = -k \left( v +k^2 \kappa \right)
    + \frac{k v}{\cos \theta}
    + \frac{k^3 \kappa}{4 \cos^3 \theta}
    + \frac{\Delta}{2^n} \left( 2^n k^n - 2 \cos (n \rho) \left(  \frac{k}{\cos \theta}\right)^n \right) \sin( n \rho ).
\end{align}
A second-order Taylor expansion in $\theta$ results again in a quadratic equation,
\begin{align}
    0 = \frac{1}{2} kv\theta^2
    + \frac{3}{8} k^3 \left( \theta^2 - 2\right) \kappa
    + \frac{k^n}{2^n} \Delta \left( 2^n -2 (n-1) n \theta^2 \right) \sin( n \rho ).
\end{align}
Its solutions,
\begin{align}
    \theta = \pm \frac{
    \sqrt{6k^3 \kappa - 2^{3-n} (2^n -2) k^n \Delta \sin( n \rho )}
    }{
    \sqrt{ 4 kv + 3 k^3 \kappa - 2^{3-n} k^n n (n-1) \Delta \sin( n \rho ) }
    }
\end{align}
are real if
\begin{align}
    6k^3 \kappa - 2^{3-n} (2^n -2) k^n \Delta \sin( n \rho ) \ge 0.
\end{align}
The spin-$\downarrow$ magnon with energy $\epsilon_-(\bm{k})$ is the lower-energy branch for $\sin( n \rho )>0$. If the sine is negative, one has to look at the spin-$\uparrow$ magnon with energy $\epsilon_+(\bm{k})$ to derive a similar condition. Overall, the relevant condition for decays of the lower-energy magnon is
\begin{align}
    3k^3 \kappa - 2^{2-n} (2^n -2) k^n \Delta |\sin( n \rho )| \ge 0,
\end{align}
from which the expression in the main text is derived. We emphasize that $\kappa > 0$. 

\section{Linear and nonlinear spin-wave theory with Holstein-Primakoff transformation}
\renewcommand{\theequation}{B.\arabic{equation}}
Here, we present the a general introduction to linear and nonlinear spin-wave theory and explain how the main text results were obtained.

\subsection{Holstein-Primakoff transformation}
We expand the spin-$S$ operator ${\bm S}_{\bm r}$ at position ${\bm r}$ by using the Holstein-Primakoff transformation given by
\begin{equation}
    {\bm S}_{\bm r} \rightarrow \hat{\mathsf{S}}^1_{\bm r}{\bm e}^1_{\bm r} + \hat{\mathsf{S}}^2_{\bm r}{\bm e}^2_{\bm r} + \hat{\mathsf{S}}^0_{\bm r}{\bm e}^0_{\bm r},
    \label{eq:SpinOpExpantion}
\end{equation}
where
\begin{equation}
    \left\{
    \begin{aligned}
        \hat{\mathsf{S}}^+_{\bm r} &= \frac{\hat{\mathsf{S}}^1_{\bm r}+i\hat{\mathsf{S}}^2_{\bm r}}{\sqrt{2}} = \sqrt{S-\frac{\hat{a}^\dagger_{\bm r}\hat{a}_{\bm r}}{2}}\hat{a}_{\bm r} = \sqrt{S}\left( \hat{a}_{\bm r} - \frac{1}{4S}\hat{a}^\dagger_{\bm r}\hat{a}_{\bm r}\hat{a}_{\bm r} + \cdots \right), \\
        \hat{\mathsf{S}}^-_{\bm r} &= \frac{\hat{\mathsf{S}}^1_{\bm r}-i\hat{\mathsf{S}}^2_{\bm r}}{\sqrt{2}} = \hat{a}^\dagger_{\bm r}\sqrt{S-\frac{\hat{a}^\dagger_{\bm r}\hat{a}_{\bm r}}{2}} = \sqrt{S}\left( \hat{a}^\dagger_{\bm r} - \frac{1}{4S}\hat{a}^\dagger_{\bm r}\hat{a}^\dagger_{\bm r}\hat{a}_{\bm r} + \cdots \right), \\
        \hat{\mathsf{S}}^0_{\bm r} &= S-\hat{a}^\dagger_{\bm r}\hat{a}_{\bm r}.
        \label{eq:HPtransform}
    \end{aligned}
    \right.
\end{equation}
Here, $\hat{a}_{\bm r}$ $(\hat{a}^\dagger_{\bm r})$ denotes the magnon annihilation (creation) operator on the site ${\bm r}$. We have defined a pair of complex vectors on the site ${\bm r}$, namely ${\bm e}^\pm_{\bm r} = \left({\bm e}^1_{\bm r}\pm i{\bm e}^2_{\bm r}\right)/\sqrt{2}$, where ${\bm e}^1_{\bm r}$, ${\bm e}^2_{\bm r}$, and ${\bm e}^0_{\bm r}\parallel\ev{{\bm S}_{\bm r}}$ are real unit vectors and the relation ${\bm e}^1_{\bm r}\times{\bm e}^2_{\bm r} = {\bm e}^0_{\bm r}$ is satisfied. $\ev{{\bm S}_{\bm r}}$ denotes the expectation value of ${\bm S}_{\bm r}$ in the classical ground state. By expanding the square roots in Eq.~(\ref{eq:HPtransform}), one gets
\begin{equation}
\begin{aligned}
    {\bm S}_{\bm r} &\rightarrow \hat{\mathsf{S}}^1_{\bm r}{\bm e}^1_{\bm r} + \hat{\mathsf{S}}^2_{\bm r}{\bm e}^2_{\bm r} + \hat{\mathsf{S}}^0_{\bm r}{\bm e}^0_{\bm r} \\
    &= \frac{\hat{\mathsf{S}}^+_{\bm r}+\hat{\mathsf{S}}^-_{\bm r}}{\sqrt{2}}\frac{{\bm e}^+_{\bm r}+{\bm e}^-_{\bm r}}{\sqrt{2}} + \frac{\hat{\mathsf{S}}^+_{\bm r}-\hat{\mathsf{S}}^-_{\bm r}}{\sqrt{2}i}\frac{{\bm e}^+_{\bm r}-{\bm e}^-_{\bm r}}{\sqrt{2}i} + \hat{\mathsf{S}}^0_{\bm r}{\bm e}^0_{\bm r} \\
    &= \hat{\mathsf{S}}^+_{\bm r}{\bm e}^-_{\bm r} + \hat{\mathsf{S}}^-_{\bm r}{\bm e}^+_{\bm r} + \hat{\mathsf{S}}^0_{\bm r}{\bm e}^0_{\bm r} \\
    &= {\bm e}^-_{\bm r} \sqrt{S} \left( \underbrace{\hat{a}_{\bm r}}_{\rm I}           
     - \underbrace{\frac{1}{4S} \hat{a}^\dagger_{\bm r}\hat{a}_{\bm r}\hat{a}_{\bm r}}_{\rm II}
     - \frac{1}{32S^2} \hat{a}^\dagger_{\bm r}\hat{a}_{\bm r}\hat{a}_{\bm r}
     - \frac{1}{32S^2} \hat{a}^\dagger_{\bm r}\hat{a}^\dagger_{\bm r}\hat{a}_{\bm r}\hat{a}_{\bm r}\hat{a}_{\bm r}
     + \cdots \right) \\
    &\quad + {\bm e}^+_{\bm r} \sqrt{S} \left( \underbrace{\hat{a}^\dagger_{\bm r}}_{\rm III} 
     - \underbrace{\frac{1}{4S} \hat{a}^\dagger_{\bm r}\hat{a}^\dagger_{\bm r}\hat{a}_{\bm r}}_{\rm IV}
     - \frac{1}{32S^2} \hat{a}^\dagger_{\bm r}\hat{a}^\dagger_{\bm r}\hat{a}_{\bm r}
     - \frac{1}{32S^2} \hat{a}^\dagger_{\bm r}\hat{a}^\dagger_{\bm r}\hat{a}^\dagger_{\bm r}\hat{a}_{\bm r}\hat{a}_{\bm r}
     + \cdots \right) \\
    &\quad + {\bm e}^0_{\bm r} \left( S-\underbrace{\hat{a}^\dagger_{\bm r}\hat{a}_{\bm r}}_{\rm V} \right),
    \label{eq:S_HPtransform}
\end{aligned}
\end{equation}
where we have used roman numbers to indicate the most relevant leading and subleading terms of the expansion.

We apply Eq.~(\ref{eq:S_HPtransform}) to a general bilinear spin Hamiltonian and get
\begin{equation}
\begin{aligned}
    &\quad {\bm S}_{\bm r}^\top \mathbb{J}_{{\bm r},{\bm r}'} {\bm S}_{{\bm r}'} \\
    &= \mathcal{J}^{+-}_{{\bm r},{\bm r}'} \left( S\hat{a}^\dagger_{{\bm r}}\hat{a}_{{\bm r}'}
     - \frac{1}{4} \left(1+\frac{1}{8S}\right) \overbrace{\underbrace{\hat{a}^\dagger_{{\bm r}}\hat{a}^\dagger_{{\bm r}}\hat{a}_{{\bm r}}\hat{a}_{{\bm r}'}}_{{\rm IV}\times{\rm I}'}}^\mathit{1} 
     - \frac{1}{4} \left(1+\frac{1}{8S}\right) \overbrace{\underbrace{\hat{a}^\dagger_{{\bm r}}\hat{a}^\dagger_{{\bm r}'}\hat{a}_{{\bm r}'}\hat{a}_{{\bm r}'}}_{{\rm III}\times{\rm II}'}}^\mathit{2}
    \right. \\ &\quad\quad\quad\quad\quad\quad \left.
     - \frac{1}{32S} \hat{a}^\dagger_{{\bm r} }\hat{a}^\dagger_{{\bm r}'}\hat{a}^\dagger_{{\bm r}'}\hat{a}        _{{\bm r}'}\hat{a}        _{{\bm r}'}\hat{a}        _{{\bm r}'}
     - \frac{1}{32S} \hat{a}^\dagger_{{\bm r} }\hat{a}^\dagger_{{\bm r} }\hat{a}^\dagger_{{\bm r} }\hat{a}        _{{\bm r} }\hat{a}        _{{\bm r} }\hat{a}        _{{\bm r}'}
     + \frac{1}{16S} \hat{a}^\dagger_{{\bm r} }\hat{a}^\dagger_{{\bm r} }\hat{a}^\dagger_{{\bm r}'}\hat{a}        _{{\bm r} }\hat{a}        _{{\bm r}'}\hat{a}        _{{\bm r}'}
     + \cdots \right) \\
    &+ \mathcal{J}^{-+}_{{\bm r},{\bm r}'} \left( S\hat{a}_{{\bm r}}\hat{a}^\dagger_{{\bm r}'}
     - \frac{1}{4} \left(1+\frac{1}{8S}\right) \overbrace{\underbrace{\hat{a}^\dagger_{{\bm r}'}\hat{a}^\dagger_{{\bm r}'}\hat{a}_{{\bm r}}\hat{a}_{{\bm r}'}}_{{\rm IV }'\times{\rm I }}}^\mathit{4}
     - \frac{1}{4} \left(1+\frac{1}{8S}\right) \overbrace{\underbrace{\hat{a}^\dagger_{{\bm r} }\hat{a}^\dagger_{{\bm r}'}\hat{a}_{{\bm r}}\hat{a}_{{\bm r} }}_{{\rm III}'\times{\rm II}}}^\mathit{3} 
    \right. \\ &\quad\quad\quad\quad\quad\quad \left.
     - \frac{1}{32S} \hat{a}^\dagger_{{\bm r}'}\hat{a}^\dagger_{{\bm r}'}\hat{a}^\dagger_{{\bm r}'}\hat{a}        _{{\bm r} }\hat{a}        _{{\bm r}'}\hat{a}        _{{\bm r}'}
     - \frac{1}{32S} \hat{a}^\dagger_{{\bm r} }\hat{a}^\dagger_{{\bm r} }\hat{a}^\dagger_{{\bm r}'}\hat{a}        _{{\bm r} }\hat{a}        _{{\bm r} }\hat{a}        _{{\bm r} }
     + \frac{1}{16S} \hat{a}^\dagger_{{\bm r} }\hat{a}^\dagger_{{\bm r}'}\hat{a}^\dagger_{{\bm r}'}\hat{a}        _{{\bm r} }\hat{a}        _{{\bm r} }\hat{a}        _{{\bm r}'}
     + \cdots \right) \\
    &+\mathcal{J}^{00}_{{\bm r},{\bm r}'} \left( S^2 - S\hat{a}^\dagger_{{\bm r}}\hat{a}_{{\bm r}} - S\hat{a}^\dagger_{{\bm r}'}\hat{a}_{{\bm r}'}
     + \overbrace{\underbrace{\hat{a}^\dagger_{{\bm r}}\hat{a}^\dagger_{{\bm r}'}\hat{a}_{{\bm r}}\hat{a}_{{\bm r}'}}_{{\rm V}\times{\rm V}'}}^{\mathit{5}} \right) \\
    &+ \mathcal{J}^{++}_{{\bm r},{\bm r}'} \left( S\hat{a}^\dagger_{{\bm r}}\hat{a}^\dagger_{{\bm r}'}
     - \frac{1}{4} \left(1+\frac{1}{8S}\right) \overbrace{\underbrace{\hat{a}^\dagger_{{\bm r}}\hat{a}^\dagger_{{\bm r}}\hat{a}^\dagger_{{\bm r}'}\hat{a}_{{\bm r}}}_{{\rm III}'\times{\rm IV}}}^\mathit{6} 
     - \frac{1}{4} \left(1+\frac{1}{8S}\right) \overbrace{\underbrace{\hat{a}^\dagger_{{\bm r}}\hat{a}^\dagger_{{\bm r}'}\hat{a}^\dagger_{{\bm r}'}\hat{a}_{{\bm r}'}}_{{\rm III}\times{\rm IV}'}}^\mathit{7} 
    \right. \\ &\quad\quad\quad\quad\quad\quad \left.
     - \frac{1}{32S} \hat{a}^\dagger_{{\bm r} }\hat{a}^\dagger_{{\bm r}'}\hat{a}^\dagger_{{\bm r}'}\hat{a}^\dagger_{{\bm r}'}\hat{a}        _{{\bm r}'}\hat{a}        _{{\bm r}'}
     - \frac{1}{32S} \hat{a}^\dagger_{{\bm r} }\hat{a}^\dagger_{{\bm r} }\hat{a}^\dagger_{{\bm r} }\hat{a}^\dagger_{{\bm r}'}\hat{a}        _{{\bm r} }\hat{a}        _{{\bm r} }
     + \frac{1}{16S} \hat{a}^\dagger_{{\bm r} }\hat{a}^\dagger_{{\bm r} }\hat{a}^\dagger_{{\bm r}'}\hat{a}^\dagger_{{\bm r}'}\hat{a}        _{{\bm r} }\hat{a}        _{{\bm r}'}
     + \cdots \right) \\
    &+ \mathcal{J}^{--}_{{\bm r},{\bm r}'} \left( S\hat{a}_{{\bm r}}\hat{a}_{{\bm r}'}
     - \frac{1}{4} \left(1+\frac{1}{8S}\right) \overbrace{\underbrace{\hat{a}^\dagger_{{\bm r}}\hat{a}_{{\bm r}}\hat{a}_{{\bm r}}\hat{a}_{{\bm r}'}}_{{\rm II}\times{\rm I}'}}^\mathit{8} 
     - \frac{1}{4} \left(1+\frac{1}{8S}\right) \overbrace{\underbrace{\hat{a}^\dagger_{{\bm r}'}\hat{a}_{{\bm r}}\hat{a}_{{\bm r}'}\hat{a}_{{\bm r}'}}_{{\rm II}'\times{\rm I}}}^\mathit{9} 
     \right. \\ &\quad\quad\quad\quad\quad\quad \left.
     - \frac{1}{32S} \hat{a}^\dagger_{{\bm r}'}\hat{a}^\dagger_{{\bm r}'}\hat{a}        _{{\bm r} }\hat{a}        _{{\bm r}'}\hat{a}        _{{\bm r}'}\hat{a}        _{{\bm r}'}
     - \frac{1}{32S} \hat{a}^\dagger_{{\bm r} }\hat{a}^\dagger_{{\bm r} }\hat{a}        _{{\bm r} }\hat{a}        _{{\bm r} }\hat{a}        _{{\bm r} }\hat{a}        _{{\bm r}'}
     + \frac{1}{16S} \hat{a}^\dagger_{{\bm r} }\hat{a}^\dagger_{{\bm r}'}\hat{a}        _{{\bm r} }\hat{a}        _{{\bm r} }\hat{a}        _{{\bm r}'}\hat{a}        _{{\bm r}'}
     + \cdots \right) \\
    &+ \mathcal{J}^{+0}_{{\bm r},{\bm r}'} \left( \cdots \right)
    +  \mathcal{J}^{0+}_{{\bm r},{\bm r}'} \left( \cdots \right)
    +  \mathcal{J}^{-0}_{{\bm r},{\bm r}'} \left( \cdots \right)
    +  \mathcal{J}^{0-}_{{\bm r},{\bm r}'} \left( \cdots \right),
    \label{eq:HP_SJS}
\end{aligned}
\end{equation}
where $\mathcal{J}^{ab}_{{\bm r},{\bm r}'} \equiv \left({\bm e}^a_{\bm r}\right)^\top \mathbb{J}_{{\bm r},{\bm r}'} {\bm e}^b_{{\bm r}'} \ \ (a,b=+,-,0)$. Here, $\mathbb{J}_{{\bm r},{\bm r}'}$ is a $3\times 3$ matrix which determines the spin-spin interaction between two sites ${\bm r}$ and ${\bm r}'(\neq{\bm r})$. 
Roman numbers indicate the origin of the respective terms, that is, from which terms in the expansion in Eq.~(\ref{eq:S_HPtransform}) they are derived. For example, IV$\times$I' indicates that the IV contribution on site $\bm{r}$ got multiplied with the I contribution on site $\bm{r}'$. The nine different four-magnon terms are numbered by latin numbers for later reference.
We do not present the explicit form of the odd-magnon-number terms because $\mathcal{J}^{+0}_{{\bm r},{\bm r}'}=\mathcal{J}^{0+}_{{\bm r},{\bm r}'}=\mathcal{J}^{-0}_{{\bm r},{\bm r}'}=\mathcal{J}^{0-}_{{\bm r},{\bm r}'}=0$ in nonrelativistic collinear magnets. 

Next, we symmetrize the four-magnon term $\mathit{5}$ in Eq.~(\ref{eq:HP_SJS}) as follows:
\begin{equation}
    \overbrace{ \hat{a}^\dagger_{{\bm r}}\hat{a}^\dagger_{{\bm r}'}\hat{a}_{{\bm r}}\hat{a}_{{\bm r}'} }^{\mathit{5}}
    = \frac{1}{4} \overbrace{ \hat{a}^\dagger_{{\bm r}}\hat{a}^\dagger_{{\bm r}'}\hat{a}_{{\bm r}}\hat{a}_{{\bm r}'} }^{\mathit{5a}}
    + \frac{1}{4} \overbrace{ \hat{a}^\dagger_{{\bm r}}\hat{a}^\dagger_{{\bm r}'}\hat{a}_{{\bm r}'}\hat{a}_{{\bm r}} }^{\mathit{5b}}
    + \frac{1}{4} \overbrace{ \hat{a}^\dagger_{{\bm r}'}\hat{a}^\dagger_{{\bm r}}\hat{a}_{{\bm r}}\hat{a}_{{\bm r}'} }^{\mathit{5c}}
    + \frac{1}{4} \overbrace{ \hat{a}^\dagger_{{\bm r}'}\hat{a}^\dagger_{{\bm r}}\hat{a}_{{\bm r}'}\hat{a}_{{\bm r}} }^{\mathit{5d}}.
    \label{eq:HP4_SJS}
\end{equation}

We finally get the following series expansion of the general form of the bilinear spin Hamiltonian $\mathcal{H}=\sum_{{\bm r},{\bm r}'} {\bm S}_{\bm r}^\top \mathbb{J}_{{\bm r},{\bm r}'} {\bm S}_{{\bm r}'}$ $({\bm r}\neq{\bm r}')$ for nonrelativistic collinear magnets with respect to the inverse spin length $1/S$:
\begin{equation}
    \mathcal{H} = S^2\mathcal{H}^{(0)} + S\mathcal{H}^{(2)} + \mathcal{H}^{(4)} + S^{-1}\mathcal{H}^{(4\leftarrow6)} + S^{-1}\mathcal{H}^{(6)} + \cdots.
    \label{eq:MagnonHm1}
\end{equation}
Here $\mathcal{H}^{(n)}$ denotes the sum of $n$-magnon terms and $\mathcal{H}^{(0)} = \sum_{{\bm r},{\bm r}'} \mathcal{J}^{00}_{{\bm r},{\bm r}'}$ is the classical energy which is a constant. $\mathcal{H}^{(1)}$ vanishes because we consider (meta)stable classical order. 
For all nonrelativistic collinear magnets, odd-order terms $\mathcal{H}^{(2n+1)}$ $(n\in\mathbb{N}, \ n\geq1)$ vanishes because $\mathcal{J}^{+0}_{{\bm r},{\bm r}'}=\mathcal{J}^{0+}_{{\bm r},{\bm r}'}=\mathcal{J}^{-0}_{{\bm r},{\bm r}'}=\mathcal{J}^{0-}_{{\bm r},{\bm r}'}=0$ is satisfied. $\mathcal{H}^{(4\leftarrow6)}$ stems from the normal-ordering process of the series-expanded square root in Eq.~(\ref{eq:HPtransform}).
\subsection{Definition of the Fourier transform}
We use the definition of the Fourier transform between the real-space magnetic unit cell coordinate ${\bm R}$ and the crystal momentum ${\bm k}$ given by
\begin{equation}
    \hat{a}_{{\bm k},\alpha} = \frac{1}{\sqrt{N_{\rm muc}}} \sum_{\bm R} \hat{a}_{{\bm R}+{\bm r}_\alpha} e^{i{\bm k}\cdot\left({\bm R}+{\bm r}_\alpha\right)}, \ \hat{a}^\dagger_{{\bm k},\alpha} = \frac{1}{\sqrt{N_{\rm muc}}} \sum_{\bm R} \hat{a}^\dagger_{{\bm R}+{\bm r}_\alpha} e^{-i{\bm k}\cdot\left({\bm R}+{\bm r}_\alpha\right)}.
    \label{eq:Fourier}
\end{equation}
$N_\mathrm{muc}$ is the number of magnetic unit cells, i.e., the number of sampled ${\bm k}$-points in a magnetic Brillouin zone, and ${\bm r}_\alpha$ denotes the {\it internal} coordinate of the real-space position of the $\alpha$-th sublattice in a magnetic Brillouin zone. \par
\subsection{Linear spin-wave theory}
As the first step, we consider the bilinear Hamiltonian piece $\mathcal{H}^{(2)}$, for which the matrix representation is given by $\mathcal{H}^{(2)} = \sum_{\bm k} \hat{\phi}^\dagger_{\bm k} H^{(2)}({\bm k}) \hat{\phi}_{\bm k}$ with
\begin{equation}
\begin{aligned}
    H^{(2)}({\bm k}) = \left( 
    \begin{array}{cc}
        D({\bm k}) & O({\bm k}) \\
        O^*(-{\bm k}) & ^\top D(-{\bm k})
    \end{array}
    \right)
\end{aligned}
\end{equation}
where $D({\bm k})$ and $O({\bm k})$ are $N_\mathrm{sub}\times N_\mathrm{sub}$-dimensional matrices whose elements are given by
\begin{equation}
\begin{aligned}
    \left[ D({\bm k}) \right]_{\alpha\beta} &= \tilde{\mathcal{J}}^{+-}_{{\bm k},\alpha\beta} - \delta_{\alpha\beta} \sum_{\gamma=1}^{N_\mathrm{sub}} \tilde{\mathcal{J}}^{00}_{{\bm 0},\alpha\gamma}, \\
    \left[ O({\bm k}) \right]_{\alpha\beta} &= \tilde{\mathcal{J}}^{++}_{{\bm k},\alpha\beta}.
\end{aligned}
\end{equation}
Note that $N_\textrm{sub}$ is the number of sublattices in a magnetic unit cell. $\hat{\phi}_{\bm k}$ denotes the Nambu space basis whose explicit expression is given by
\begin{equation}
    \hat{\phi}_{\bm k} = ^\top \left( \hat{a}_{{\bm k},1},\hat{a}_{{\bm k},2},\cdots,\hat{a}_{{\bm k},N_\mathrm{sub}},\hat{a}^\dagger_{{\bm k},1},\hat{a}^\dagger_{-{\bm k},2},\cdots,\hat{a}^\dagger_{-{\bm k},N_\mathrm{sub}} \right).
\end{equation}
We diagonalize $\hat{\phi}^\dagger_{\bm k}H^{(2)}({\bm k})\hat{\phi}_{\bm k}$ by using a paraunitary matrix $T_{\bm k}$. The diagonalization is explicitly given by
\begin{equation}
    \hat{\phi}^\dagger_{\bm k} \left(T^\dagger_{\bm k}\right)^{-1} T^\dagger_{\bm k} H^{(2)}({\bm k}) T_{\bm k} T_{\bm k}^{-1} \hat{\phi}_{\bm k} = \hat{\varphi}^\dagger_{\bm k} E({\bm k}) \hat{\varphi}_{\bm k},
\end{equation}
where
\begin{gather}
    E({\bm k}) = \mathrm{diag}\left( \varepsilon_1({\bm k}),\varepsilon_2({\bm k}),\cdots,\varepsilon_{N_\textrm{sub}}({\bm k}),\varepsilon_1(-{\bm k}),\varepsilon_2(-{\bm k}),\cdots,\varepsilon_{N_\mathrm{sub}}(-{\bm k}) \right), \\
    \hat{\varphi}_{\bm k} = ^\top\left( \hat{b}_{{\bm k},1},\hat{b}_{{\bm k},2},\cdots,\hat{b}_{{\bm k},N_\mathrm{sub}},\hat{b}^\dagger_{-{\bm k},1},\hat{b}^\dagger_{-{\bm k},2},\cdots,\hat{b}^\dagger_{-{\bm k},N_\mathrm{sub}} \right), \\
    T^\dagger_{\bm k} H^{(2)}({\bm k}) T_{\bm k} = E({\bm k}), \\
    \hat{\phi}_{\bm k} = T_{\bm k}\hat{\varphi}_{\bm k}.
\end{gather}
Here $\varepsilon_\nu({{\bm k}})$, $\hat{b}_{{\bm k},\nu}$, and $\hat{b}^\dagger_{{\bm k},\nu}$ denote energy, annihilation operator, and creation operator of a magnon with band index $\nu$ and crystal momentum ${\bm k}$, respectively. The paraunitary matrix $T_{\bm k}$ can be represented by using two $N_\mathrm{sub}\times N_\mathrm{sub}$-dimensional matrices $U_{\bm k}$ and $V_{\bm k}$ as follows:
\begin{equation}
    T_{\bm k} = \left( \begin{array}{cc}
        U_{{\bm k}} & V^*_{-{\bm k}} \\
        V_{{\bm k}} & U^*_{-{\bm k}}
    \end{array} \right).
\end{equation}
By using the matrix elements of $U_{\bm k}$ and $V_{\bm k}$, namely $U_{{\bm k},\alpha\nu}=\left[ U_{\bm k} \right]_{\alpha\nu}$ and $V_{{\bm k},\alpha\nu}=\left[ V_{\bm k} \right]_{\alpha\nu}$, the diagonalization, i.e., Bogoliubov transformation, is given by
\begin{equation}
            \hat{a}_{ {\bm k},\alpha} = \sum_{\nu=1}^{N_\mathrm{sub}} \left[   U_{ {\bm k},\alpha\nu}        \hat{b}_{ {\bm k},\nu} + V^*_{-{\bm k},\alpha\nu}\hat{b}^\dagger_{-{\bm k},\nu} \right], 
            \qquad
    \hat{a}^\dagger_{-{\bm k},\alpha} = \sum_{\nu=1}^{N_\mathrm{sub}} \left[ U^*_{-{\bm k},\alpha\nu}\hat{b}^\dagger_{-{\bm k},\nu} +   V_{ {\bm k},\alpha\nu}        \hat{b}_{ {\bm k},\nu} \right].
\end{equation}
\subsection{Many-body perturbation theory}
We start with the magnon Hamiltonian $\mathcal{H}-S^2\mathcal{H}^{(0)}$ where $\mathcal{H}$ is given by Eq.~(\ref{eq:MagnonHm1}), and then split it into the nonperturbative part $S\mathcal{H}_0$ and the perturbative part $S\mathcal{V}$, respectively given by
\begin{align}
    \mathcal{H}_0 &= \mathcal{H}^{(2)}, \\
    \mathcal{V}   &= S^{-1}\mathcal{H}^{(4)} + S^{-2}\mathcal{H}^{(4\leftarrow6)} + S^{-2}\mathcal{H}^{(6)} + \cdots.
\end{align}
In the framework of many-body perturbation theory, the interacting (Matsubara) one-magnon Green's function is given by
\begin{equation}
    \mathcal{G}_{{\bm k},mn}(\tau) = \sum_{j=0}^\infty \mathcal{G}^{(j)}_{{\bm k},mn}(\tau) = -\sum_{j=0}^\infty \left( -\frac{1}{\hbar} \right)^j \frac{1}{j!} \int_0^{\beta\hbar}d\tau_1 \cdots \int_0^{\beta\hbar}d\tau_j \ev{\mathcal{T}_\tau \left[ \mathcal{V}(\tau_1)\cdots\mathcal{V}(\tau_j) \hat{b}_{{\bm k}m}(\tau)\hat{b}^\dagger_{{\bm k}n}(0) \right]}_{(0)}^{\rm connected}.
\end{equation}
The zeroth-order term of this sum is the noninteracting Green's function $\mathcal{G}^{(0)}_{{\bm k},mn}(\tau)=-\ev{\mathcal{T}_\tau\hat{b}_{{\bm k}m}(\tau)\hat{b}^\dagger_{{\bm k}n}(0)}_{(0)}$. $\beta=(k_{\rm B}T)^{-1}$ is inverse temperature ($k_{\rm B}$ being Boltzmann constant). $\mathcal{T}_\tau$ is the time-ordering operator of the imaginary times $\tau_j$. The ensemble average $\ev{\cdot}_{(0)}^{\rm connected}$ is taken with respect to the nonperturbative term $\mathcal{H}_0$ over connected diagrams. \par
\begin{figure}[htpb]
    \includegraphics[width=17.0cm]{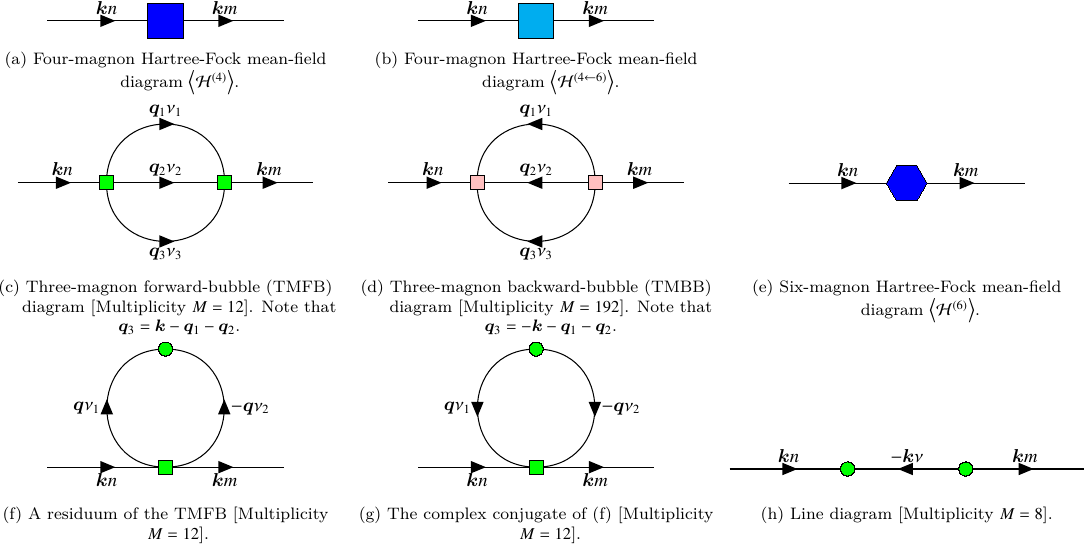}
    \caption{Feynman diagrams corresponding to corrections up to the $1/S^2$ order (The diagram (a) corresponds to the $1/S$ order correction and the diagrams (b)-(h) correspond to the $1/S^2$ order corrections).}
    \label{fig:diagrams-NLSW}
\end{figure}

To calculate the higher-order ensemble averages $\ev{\mathcal{T}_\tau \left[ \mathcal{V}(\tau_1)\cdots\mathcal{V}(\tau_j) \hat{b}_{{\bm k}m}(\tau)\hat{b}^\dagger_{{\bm k}n}(0) \right]}_{(0)}^{\rm connected}$ with $j\geq2$, it is convenient to represent $\mathcal{V}$ in terms of the diagonalized operators $\hat{b}_{{\bm k},\nu}$ and $\hat{b}^\dagger_{{\bm k},\nu}$. In Sec.~\ref{sec:fourmagnon}, we present $\mathcal{H}^{(4)}$ in the language of normal-ordered $\left(\hat{b}_{{\bm k},\nu},\hat{b}^\dagger_{{\bm k},\nu}\right)$. It is the only term relevant to the higher-order ensemble averages up to order $1/S^2$. By applying the Wick's theorem, we get eight diagrams presented in Fig.~\ref{fig:diagrams-NLSW} which have all contributions up to the $1/S^2$ order.
The order $1/S$ and $1/S^2$ self-energies are respectively given by
\begin{align}
    \Sigma^{(1/S)}_{{\bm k},mn} &= \left[ T^\dagger_{\bm k} \mathcal{H}^{(4)}_{\rm HF}({\bm k}) T_{\bm k} \right]_{mn}, \\
    \Sigma^{(1/S^2)}_{{\bm k},mn}(i\omega_s) &= \frac{1}{S}\left[ T^\dagger_{\bm k} \left( \mathcal{H}^{(4\leftarrow 6)}_{\rm HF}({\bm k}) + \mathcal{H}^{(6)}_{\rm HF}({\bm k}) \right) T_{\bm k} \right]_{mn} + \Sigma_{{\bm k},mn}^{\rm TMFB}(i\omega_s) + \Sigma_{{\bm k},mn}^{\rm TMBB}(i\omega_s) + \Sigma_{{\bm k},mn}^{{\rm RTMFB}} + \Sigma_{{\bm k},mn}^{{\rm RTMFB}^*} + \Sigma_{{\bm k},mn}^{{\rm Line}}(i\omega_s).
\end{align}
We show the explicit form of the Hartree-Fock mean field $\mathcal{H}^{(4)}_{\rm HF}({\bm k})$ and short notes for $\mathcal{H}^{(4\leftarrow 6)}_{\rm HF}({\bm k})$ and $\mathcal{H}^{(6)}_{\rm HF}({\bm k})$ in Sec.~\ref{sec:HartreeFockMF}.
The explicit expressions of the self-energies associated with the diagrams presented in Figs.~\ref{fig:diagrams-NLSW}(c), \ref{fig:diagrams-NLSW}(d), and \ref{fig:diagrams-NLSW}(f)-(h) are respectively given by (at zero temperature)
\begin{align}
    \Sigma_{{\bm k},mn}^{\rm TMFB}(i\omega_s) &=  \frac{6}{\hbar^2}  \sum_{{\bm q}_1,{\bm q}_2} \sum_{\nu_1,\nu_2,\nu_3} \frac{ \tilde{\mathcal{C}}_{{\bm q}_1,{\bm q}_2,{\bm k}-{\bm q}_1-{\bm q}_2\leftarrow{\bm k}}^{\nu_1,\nu_2,\nu_3\leftarrow n} \left( \tilde{\mathcal{C}}_{{\bm q}_1,{\bm q}_2,{\bm k}-{\bm q}_1-{\bm q}_2\leftarrow{\bm k}}^{\nu_1,\nu_2,\nu_3\leftarrow m} \right)^*}{i\omega_s-\varepsilon_{\nu_1}({\bm q}_1)-\varepsilon_{\nu_2}({\bm q}_2)-\varepsilon_{\nu_3}({\bm k}-{\bm q}_1-{\bm q}_2)}, \\
    \Sigma_{{\bm k},mn}^{\rm TMBB}(i\omega_s) &= -\frac{96}{\hbar^2} \sum_{{\bm q}_1,{\bm q}_2} \sum_{\nu_1,\nu_2,\nu_3} \frac{ \left( \tilde{\mathcal{A}}_{{\bm q}_1,{\bm q}_2,-{\bm k}-{\bm q}_1-{\bm q}_2,{\bm k}}^{\nu_1,\nu_2,\nu_3,n} \right)^* \tilde{\mathcal{A}}_{{\bm q}_1,{\bm q}_2,-{\bm k}-{\bm q}_1-{\bm q}_2,{\bm k}}^{\nu_1,\nu_2,\nu_3,m} }{i\omega_s+\varepsilon_{\nu_1}({\bm q}_1)+\varepsilon_{\nu_2}({\bm q}_2)+\varepsilon_{\nu_3}({\bm k}-{\bm q}_1-{\bm q}_2)}, \\
    \Sigma_{{\bm k},mn}^{{\rm RTMFB}  } &= -\frac{6}{\hbar^2} \sum_{\bm q} \sum_{\nu_1,\nu_2} \frac{ \tilde{\mathcal{C}}_{{\bm k},{\bm q},-{\bm q}\leftarrow{\bm k}}^{m,\nu_1,\nu_2\leftarrow n} \left(\tilde{\mathcal{R}}_{{\bm q},-{\bm q}}^{\nu_1,\nu_2}\right)^* }{\varepsilon_{\nu_1}({\bm q})+\varepsilon_{\nu_2}(-{\bm q})}, \\
    \Sigma_{{\bm k},mn}^{{\rm RTMFB}^*} &= -\frac{6}{\hbar^2} \sum_{\bm q} \sum_{\nu_1,\nu_2} \frac{ \left(\tilde{\mathcal{C}}_{{\bm k},{\bm q},-{\bm q}\leftarrow{\bm k}}^{n,\nu_1,\nu_2\leftarrow m}\right)^* \tilde{\mathcal{R}}_{{\bm q},-{\bm q}}^{\nu_1,\nu_2} }{\varepsilon_{\nu_1}({\bm q})+\varepsilon_{\nu_2}(-{\bm q})}, \\
    \Sigma_{{\bm k},mn}^{{\rm Line}}(i\omega_s) &= -\frac{4}{\hbar^2} \sum_\nu \frac{ \left( \tilde{\mathcal{R}}_{{\bm k},-{\bm k}}^{n,\nu} \right)^* \tilde{\mathcal{R}}_{{\bm k},-{\bm k}}^{m,\nu} }{i\omega_s+\varepsilon_\nu(-{\bm k})}.
\end{align}
Most individual self-energies in gapless systems with type-I (linearly dispersive) Goldstone modes (GMs) at the $\Gamma$ point exhibit unphysical divergences due to the divergences of the Bogoliubov rotation matrix elements $\left\{U_{{\bm k}=\Gamma,\alpha\nu=\text{GM}}\right\}$ and $\left\{V_{{\bm k}=\Gamma,\alpha\nu=\text{GM}}\right\}$. However, as long as all self-energies contributing to the order of $(1/S)^i$ (where $i$ is an arbitrary positive integer) are consistently treated using on-shell approximations, these unphysical divergences fully cancel out. Explicitly, we note that the self-energy diagrams shown in Figs.~\ref{fig:diagrams-NLSW}(c)-(f) individually exhibit unphysical divergences. However, when taking the sum of self-energies in Figs.~\ref{fig:diagrams-NLSW}(c), (d), and (e), the divergencies cancel out. Similarly, there is no divergence in the sum of self-energies in Figs.~\ref{fig:diagrams-NLSW}(f), (g), and (h).
\par Off-shell treatments of the frequency-dependent self-energies can be problematic because they are not consistent with the strict $1/S^2$ order. This inconsistency can bring about uncompensated divergencies in the vicinity of the Goldstone mode. We actually observe spurious poles induced by the line diagram self-energy given by Fig.~\ref{fig:diagrams-NLSW}(h), and thus, we neglect the set of diagrams given by Figs.~\ref{fig:diagrams-NLSW}(f),~\ref{fig:diagrams-NLSW}(g), and~\ref{fig:diagrams-NLSW}(h) in the off-shell self-energy calculations. This treatment is justified because spectral corrections stemming from these diagrams are tiny ($\sim 0.1\%$) compared to those from the diagrams shown in Figs.~\ref{fig:diagrams-NLSW}(a)-(e).
\subsection{Four-magnon vertices}
\label{sec:fourmagnon}
We present the four-magnon Hamiltonian $\mathcal{H}^{(4)}$ in the language of the Bogoliubov-transformed magnon operators $\hat{b}^\dagger_{{\bm q}\nu}$ and $\hat{b}_{{\bm q}\nu}$ as follows:
\begin{equation}
    \mathcal{H}^{(4)} = \tilde{\mathcal{H}}^{(22)} + \tilde{\mathcal{H}}^{(31)} + \tilde{\mathcal{H}}^{(13)} + \tilde{\mathcal{H}}^{(31{\rm R})} + \tilde{\mathcal{H}}^{(13{\rm R})} + \tilde{\mathcal{H}}^{(40)} + \tilde{\mathcal{H}}^{(04)},
\end{equation}
where $\tilde{\mathcal{H}}^{(ab)}$ consists of $a$ creation operators and $b$ annihilation operators. $\tilde{\mathcal{H}}^{(31{\rm R})}$ and $\tilde{\mathcal{H}}^{(13{\rm R})}$ are the residual terms of $\tilde{\mathcal{H}}^{(31)}$ and $\tilde{\mathcal{H}}^{(13)}$ after taking the normal ordering.
\subsubsection{Three-out-one-in term $\tilde{\mathcal{H}}^{(31)}$}
\begin{figure}[htpb]
    \includegraphics[width=4.0cm]{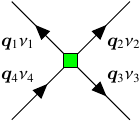}
    \caption{Symmetrized 3-1 vertex $\tilde{\mathcal{C}}_{{\bm q}_1,{\bm q}_2,{\bm q}_3\leftarrow{\bm q}_4}^{\nu_1,\nu_2,\nu_3\leftarrow\nu_4}$.}
\end{figure}
The explicit form of $\tilde{\mathcal{H}}^{(31)}$ is given by
\begin{equation}
    \tilde{\mathcal{H}}^{(31)} = \sum_{{\bm q}_1,{\bm q}_2,{\bm q}_3,{\bm q}_4} \sum_{\nu_1,\nu_2,\nu_3,\nu_4} 
    \tilde{\mathcal{C}}_{{\bm q}_1,{\bm q}_2,{\bm q}_3\leftarrow{\bm q}_4}^{\nu_1,\nu_2,\nu_3\leftarrow\nu_4} \
    \hat{b}^\dagger_{{\bm q}_1,\nu_1} \hat{b}^\dagger_{{\bm q}_2,\nu_2} \hat{b}^\dagger_{{\bm q}_3,\nu_3} \hat{b}_{{\bm q}_4,\nu_4},
\end{equation}
where $\tilde{\mathcal{C}}_{{\bm q}_1,{\bm q}_2,{\bm q}_3\leftarrow{\bm q}_4}^{\nu_1,\nu_2,\nu_3\leftarrow\nu_4}$ is the symmetrized three-out-one-in vertex given by, 
\begin{equation}
    \tilde{\mathcal{C}}_{{\bm q}_1,{\bm q}_2,{\bm q}_3\leftarrow{\bm q}_4}^{\nu_1,\nu_2,\nu_3\leftarrow\nu_4} = \frac{1}{2} \frac{1}{3!} \left(
      \mathcal{C}_{{\bm q}_1,{\bm q}_2,{\bm q}_3\leftarrow{\bm q}_4}^{\nu_1,\nu_2,\nu_3\leftarrow\nu_4}
    + \mathcal{C}_{{\bm q}_1,{\bm q}_3,{\bm q}_2\leftarrow{\bm q}_4}^{\nu_1,\nu_3,\nu_2\leftarrow\nu_4}
    + \mathcal{C}_{{\bm q}_2,{\bm q}_3,{\bm q}_1\leftarrow{\bm q}_4}^{\nu_2,\nu_3,\nu_1\leftarrow\nu_4}
    + \mathcal{C}_{{\bm q}_2,{\bm q}_1,{\bm q}_3\leftarrow{\bm q}_4}^{\nu_2,\nu_1,\nu_3\leftarrow\nu_4}
    + \mathcal{C}_{{\bm q}_3,{\bm q}_1,{\bm q}_2\leftarrow{\bm q}_4}^{\nu_3,\nu_1,\nu_2\leftarrow\nu_4}
    + \mathcal{C}_{{\bm q}_3,{\bm q}_2,{\bm q}_1\leftarrow{\bm q}_4}^{\nu_3,\nu_2,\nu_1\leftarrow\nu_4} \right).
\end{equation}
The factor $1/2$ is introduced to cancel double counting of each spin-spin interaction, and the factor $1/3!$ comes from the number of symmetrized components.
Note that $\tilde{\mathcal{H}}^{(13)}$ is given by the hermitian conjugate of $\tilde{\mathcal{H}}^{(31)}$. To get the explicit expression of the unsymmetrized vertex $\mathcal{C}_{{\bm q}_1,{\bm q}_2,{\bm q}_3\leftarrow{\bm q}_4}^{\nu_1,\nu_2,\nu_3\leftarrow\nu_4}$, we need to pick up the normal-ordered three-out-one-in part of the four magnon terms $\mathit{1},\mathit{2},\cdots,\mathit{9}$ in Eqs.~(\ref{eq:HP_SJS}) and (\ref{eq:HP4_SJS}). We list all contributions in the following:
\begin{equation}
\begin{aligned}
    \quad \mathit{1.} &\quad \sum_{{\bm r},{\bm r}'} \mathcal{J}^{+-}_{{\bm r},{\bm r}'} \hat{a}^\dagger_{{\bm r}}\hat{a}^\dagger_{{\bm r}}\hat{a}_{{\bm r}}\hat{a}_{{\bm r}'} \\
    &= \frac{1}{N_{\rm muc}} \sum_{\nu_1\nu_2\nu_3\nu_4} \sum_{\alpha\beta\gamma\delta} \sum_{{\bm q}_1,{\bm q}_2,{\bm q}_3,{\bm q}_4} \delta_{\alpha\beta}\delta_{\alpha\gamma} \tilde{\mathcal{J}}^{+-}_{{\bm q}_4,\gamma\delta}  \delta\left( {\bm q}_1+{\bm q}_2-{\bm q}_3-{\bm q}_4 \right) \\
    &\quad\quad \times 
    \left( U^*_{{\bm q}_1,\alpha\nu_1} \hat{b}^\dagger_{{\bm q}_1,\nu_1} +   V_{-{\bm q}_1,\alpha\nu_1}         \hat{b}_{-{\bm q}_1,\nu_1} \right)
    \left( U^*_{{\bm q}_2,\beta \nu_2} \hat{b}^\dagger_{{\bm q}_2,\nu_2} +   V_{-{\bm q}_2,\beta \nu_2}         \hat{b}_{-{\bm q}_2,\nu_2} \right) \\
    &\quad\quad \times
    \left(   U_{{\bm q}_3,\gamma\nu_3}         \hat{b}_{{\bm q}_3,\nu_3} + V^*_{-{\bm q}_3,\gamma\nu_3} \hat{b}^\dagger_{-{\bm q}_3,\nu_3} \right)
    \left(   U_{{\bm q}_4,\delta\nu_4}         \hat{b}_{{\bm q}_4,\nu_4} + V^*_{-{\bm q}_4,\delta\nu_4} \hat{b}^\dagger_{-{\bm q}_4,\nu_4} \right) \\
    &\rightarrow \frac{1}{N_{\rm muc}} \sum_{{\bm q}_1,{\bm q}_2,{\bm q}_3,{\bm q}_4} \sum_{\nu_1\nu_2\nu_3\nu_4} \sum_{\alpha\beta\gamma\delta} \delta_{\alpha\beta}\delta_{\alpha\gamma} \delta\left( {\bm q}_1+{\bm q}_2+{\bm q}_3-{\bm q}_4 \right) \\
    &\quad \times \left[ 
        \tilde{\mathcal{J}}^{+-}_{ {\bm q}_4,\gamma\delta} U^*_{{\bm q}_1,\alpha\nu_1}U^*_{{\bm q}_2,\beta \nu_2}V^*_{{\bm q}_3,\gamma\nu_3}U_{{\bm q}_4,\delta\nu_4}
      + \tilde{\mathcal{J}}^{+-}_{-{\bm q}_3,\gamma\delta} V_{{\bm q}_4,\alpha\nu_4}U^*_{{\bm q}_1,\beta \nu_1}V^*_{{\bm q}_2,\gamma\nu_2}V^*_{{\bm q}_3,\delta\nu_3}
    \right. \\ &\quad\quad \left.
      + \tilde{\mathcal{J}}^{+-}_{-{\bm q}_1,\gamma\delta} U^*_{{\bm q}_2,\alpha\nu_2}U^*_{{\bm q}_3,\beta \nu_3}U_{{\bm q}_4,\gamma\nu_4}V^*_{{\bm q}_1,\delta\nu_1}
      + \tilde{\mathcal{J}}^{+-}_{-{\bm q}_2,\gamma\delta} U^*_{{\bm q}_3,\alpha\nu_3}V_{{\bm q}_4,\beta \nu_4}V^*_{{\bm q}_1,\gamma\nu_1}V^*_{{\bm q}_2,\delta\nu_2}
    \right] 
    \hat{b}^\dagger_{{\bm q}_1,\nu_1}\hat{b}^\dagger_{{\bm q}_2,\nu_2}\hat{b}^\dagger_{{\bm q}_3,\nu_3}\hat{b}_{{\bm q}_4,\nu_4},
\end{aligned}
\end{equation}
\begin{equation}
\begin{aligned}
    \mathit{2}. &\quad \sum_{{\bm r},{\bm r}'}\mathcal{J}^{+-}_{{\bm r},{\bm r}'} \hat{a}^\dagger_{{\bm r}}\hat{a}^\dagger_{{\bm r}'}\hat{a}_{{\bm r}'}\hat{a}_{{\bm r}'} \\
    &\rightarrow \frac{1}{N_{\rm muc}} \sum_{{\bm q}_1,{\bm q}_2,{\bm q}_3,{\bm q}_4} \sum_{\nu_1\nu_2\nu_3\nu_4} \sum_{\alpha\beta\gamma\delta} 
    \delta_{\beta\delta}\delta_{\gamma\delta} 
    \delta\left( {\bm q}_1+{\bm q}_2+{\bm q}_3-{\bm q}_4 \right) \\
    &\quad\quad \times
    \left[
          \tilde{\mathcal{J}}^{+-}_{ {\bm q}_1,\alpha\beta } U^*_{{\bm q}_1,\alpha\nu_1}U^*_{{\bm q}_2,\beta \nu_2}V^*_{{\bm q}_3,\gamma\nu_3}U_{{\bm q}_4,\delta\nu_4}
        + \tilde{\mathcal{J}}^{+-}_{-{\bm q}_4,\alpha\beta } V_{{\bm q}_4,\alpha\nu_4}U^*_{{\bm q}_1,\beta \nu_1}V^*_{{\bm q}_2,\gamma\nu_2}V^*_{{\bm q}_3,\delta\nu_3}
    \right. \\ &\quad\quad\quad \left.
        + \tilde{\mathcal{J}}^{+-}_{ {\bm q}_2,\alpha\beta } U^*_{{\bm q}_2,\alpha\nu_2}U^*_{{\bm q}_3,\beta \nu_3}U_{{\bm q}_4,\gamma\nu_4}V^*_{{\bm q}_1,\delta\nu_1}
        + \tilde{\mathcal{J}}^{+-}_{ {\bm q}_3,\alpha\beta } U^*_{{\bm q}_3,\alpha\nu_3}V_{{\bm q}_4,\beta \nu_4}V^*_{{\bm q}_1,\gamma\nu_1}V^*_{{\bm q}_2,\delta\nu_2}
    \right] 
    \hat{b}^\dagger_{{\bm q}_1,\nu_1} \hat{b}^\dagger_{{\bm q}_2,\nu_2} \hat{b}^\dagger_{{\bm q}_3,\nu_3} \hat{b}_{{\bm q}_4,\nu_4},
\end{aligned}
\end{equation}
\begin{equation}
\begin{aligned}
    \mathit{3}. &\quad \sum_{{\bm r},{\bm r}'}\mathcal{J}^{-+}_{{\bm r},{\bm r}'} \hat{a}^\dagger_{{\bm r}}\hat{a}^\dagger_{{\bm r}'}\hat{a}_{{\bm r}}\hat{a}_{{\bm r}} \\
    &\rightarrow \frac{1}{N_{\rm muc}} \sum_{{\bm q}_1,{\bm q}_2,{\bm q}_3,{\bm q}_4} \sum_{\nu_1\nu_2\nu_3\nu_4} \sum_{\alpha\beta\gamma\delta} 
    \delta_{\alpha\gamma}\delta_{\alpha\delta} 
    \delta\left( {\bm q}_1+{\bm q}_2+{\bm q}_3-{\bm q}_4 \right) \\
    &\quad\quad \times 
    \left[
          \tilde{\mathcal{J}}^{-+}_{-{\bm q}_2,\alpha\beta } U^*_{{\bm q}_1,\alpha\nu_1}U^*_{{\bm q}_2,\beta \nu_2}V^*_{{\bm q}_3,\gamma\nu_3}U_{{\bm q}_4,\delta\nu_4}
        + \tilde{\mathcal{J}}^{-+}_{-{\bm q}_1,\alpha\beta } V_{{\bm q}_4,\alpha\nu_4}U^*_{{\bm q}_1,\beta \nu_1}V^*_{{\bm q}_2,\gamma\nu_2}V^*_{{\bm q}_3,\delta\nu_3}
    \right. \\ &\quad\quad\quad \left.
        + \tilde{\mathcal{J}}^{-+}_{-{\bm q}_3,\alpha\beta } U^*_{{\bm q}_2,\alpha\nu_2}U^*_{{\bm q}_3,\beta \nu_3}U_{{\bm q}_4,\gamma\nu_4}V^*_{{\bm q}_1,\delta\nu_1}
        + \tilde{\mathcal{J}}^{-+}_{ {\bm q}_4,\alpha\beta } U^*_{{\bm q}_3,\alpha\nu_3}V_{{\bm q}_4,\beta \nu_4}V^*_{{\bm q}_1,\gamma\nu_1}V^*_{{\bm q}_2,\delta\nu_2}
    \right] 
    \hat{b}^\dagger_{{\bm q}_1,\nu_1} \hat{b}^\dagger_{{\bm q}_2,\nu_2} \hat{b}^\dagger_{{\bm q}_3,\nu_3} \hat{b}_{{\bm q}_4,\nu_4},
\end{aligned}
\end{equation}
\begin{equation}
\begin{aligned}
    \mathit{4}. &\quad \sum_{{\bm r},{\bm r}'}\mathcal{J}^{-+}_{{\bm r},{\bm r}'} \hat{a}^\dagger_{{\bm r}'}\hat{a}^\dagger_{{\bm r}'}\hat{a}_{{\bm r}}\hat{a}_{{\bm r}'} \\
    &\rightarrow \frac{1}{N_{\rm muc}} \sum_{{\bm q}_1,{\bm q}_2,{\bm q}_3,{\bm q}_4} \sum_{\nu_1\nu_2\nu_3\nu_4} \sum_{\alpha\beta\gamma\delta} 
    \delta_{\alpha\beta}\delta_{\alpha\delta} 
    \delta\left( {\bm q}_1+{\bm q}_2+{\bm q}_3-{\bm q}_4 \right) \\
    &\quad\quad \times 
    \left[
          \tilde{\mathcal{J}}^{-+}_{ {\bm q}_3,\gamma\delta} U^*_{{\bm q}_1,\alpha\nu_1}U^*_{{\bm q}_2,\beta \nu_2}V^*_{{\bm q}_3,\gamma\nu_3}U_{{\bm q}_4,\delta\nu_4}
        + \tilde{\mathcal{J}}^{-+}_{ {\bm q}_2,\gamma\delta} V_{{\bm q}_4,\alpha\nu_4}U^*_{{\bm q}_1,\beta \nu_1}V^*_{{\bm q}_2,\gamma\nu_2}V^*_{{\bm q}_3,\delta\nu_3}
    \right. \\ &\quad\quad\quad \left.
        + \tilde{\mathcal{J}}^{-+}_{-{\bm q}_4,\gamma\delta} U^*_{{\bm q}_2,\alpha\nu_2}U^*_{{\bm q}_3,\beta \nu_3}U_{{\bm q}_4,\gamma\nu_4}V^*_{{\bm q}_1,\delta\nu_1}
        + \tilde{\mathcal{J}}^{-+}_{ {\bm q}_1,\gamma\delta} U^*_{{\bm q}_3,\alpha\nu_3}V_{{\bm q}_4,\beta \nu_4}V^*_{{\bm q}_1,\gamma\nu_1}V^*_{{\bm q}_2,\delta\nu_2}
    \right] 
    \hat{b}^\dagger_{{\bm q}_1,\nu_1} \hat{b}^\dagger_{{\bm q}_2,\nu_2} \hat{b}^\dagger_{{\bm q}_3,\nu_3} \hat{b}_{{\bm q}_4,\nu_4},
\end{aligned}
\end{equation}
\begin{equation}
\begin{aligned}
    \mathit{5a}. &\quad \sum_{{\bm r},{\bm r}'}\mathcal{J}^{00}_{{\bm r},{\bm r}'} \hat{a}^\dagger_{{\bm r}}\hat{a}^\dagger_{{\bm r}'}\hat{a}_{{\bm r}}\hat{a}_{{\bm r}'} \\
    &\rightarrow \frac{1}{N_{\rm muc}} \sum_{{\bm q}_1,{\bm q}_2,{\bm q}_3,{\bm q}_4} \sum_{\nu_1\nu_2\nu_3\nu_4} \sum_{\alpha\beta\gamma\delta} 
    \delta_{\alpha\gamma}\delta_{\beta\delta} 
    \delta\left( {\bm q}_1+{\bm q}_2+{\bm q}_3-{\bm q}_4 \right) \\
    &\quad\quad \times 
    \left[
          \tilde{\mathcal{J}}^{00}_{ {\bm q}_1+{\bm q}_3,\alpha\beta } U^*_{{\bm q}_1,\alpha\nu_1}U^*_{{\bm q}_2,\beta \nu_2}V^*_{{\bm q}_3,\gamma\nu_3}U_{{\bm q}_4,\delta\nu_4}
        + \tilde{\mathcal{J}}^{00}_{-{\bm q}_4+{\bm q}_2,\alpha\beta } V_{{\bm q}_4,\alpha\nu_4}U^*_{{\bm q}_1,\beta \nu_1}V^*_{{\bm q}_2,\gamma\nu_2}V^*_{{\bm q}_3,\delta\nu_3}
    \right. \\ &\quad\quad\quad \left.
        + \tilde{\mathcal{J}}^{00}_{ {\bm q}_2-{\bm q}_4,\alpha\beta } U^*_{{\bm q}_2,\alpha\nu_2}U^*_{{\bm q}_3,\beta \nu_3}U_{{\bm q}_4,\gamma\nu_4}V^*_{{\bm q}_1,\delta\nu_1}
        + \tilde{\mathcal{J}}^{00}_{ {\bm q}_3+{\bm q}_1,\alpha\beta } U^*_{{\bm q}_3,\alpha\nu_3}V_{{\bm q}_4,\beta \nu_4}V^*_{{\bm q}_1,\gamma\nu_1}V^*_{{\bm q}_2,\delta\nu_2}
    \right] 
    \hat{b}^\dagger_{{\bm q}_1,\nu_1} \hat{b}^\dagger_{{\bm q}_2,\nu_2} \hat{b}^\dagger_{{\bm q}_3,\nu_3} \hat{b}_{{\bm q}_4,\nu_4},
\end{aligned}
\end{equation}
\begin{equation}
\begin{aligned}
    \mathit{5b}. &\quad \sum_{{\bm r},{\bm r}'}\mathcal{J}^{00}_{{\bm r},{\bm r}'} \hat{a}^\dagger_{{\bm r}}\hat{a}^\dagger_{{\bm r}'}\hat{a}_{{\bm r}'}\hat{a}_{{\bm r}} \\
    &\rightarrow \frac{1}{N_{\rm muc}} \sum_{{\bm q}_1,{\bm q}_2,{\bm q}_3,{\bm q}_4} \sum_{\nu_1\nu_2\nu_3\nu_4} \sum_{\alpha\beta\gamma\delta} 
    \delta_{\alpha\delta}\delta_{\beta\gamma} 
    \delta\left( {\bm q}_1+{\bm q}_2+{\bm q}_3-{\bm q}_4 \right) \\
    &\quad\quad \times 
    \left[
          \tilde{\mathcal{J}}^{00}_{ {\bm q}_1-{\bm q}_4,\alpha\beta } U^*_{{\bm q}_1,\alpha\nu_1}U^*_{{\bm q}_2,\beta \nu_2}V^*_{{\bm q}_3,\gamma\nu_3}U_{{\bm q}_4,\delta\nu_4}
        + \tilde{\mathcal{J}}^{00}_{-{\bm q}_4+{\bm q}_3,\alpha\beta } V_{{\bm q}_4,\alpha\nu_4}U^*_{{\bm q}_1,\beta \nu_1}V^*_{{\bm q}_2,\gamma\nu_2}V^*_{{\bm q}_3,\delta\nu_3}
    \right. \\ &\quad\quad\quad \left.
        + \tilde{\mathcal{J}}^{00}_{ {\bm q}_2+{\bm q}_1,\alpha\beta } U^*_{{\bm q}_2,\alpha\nu_2}U^*_{{\bm q}_3,\beta \nu_3}U_{{\bm q}_4,\gamma\nu_4}V^*_{{\bm q}_1,\delta\nu_1}
        + \tilde{\mathcal{J}}^{00}_{ {\bm q}_3+{\bm q}_2,\alpha\beta } U^*_{{\bm q}_3,\alpha\nu_3}V_{{\bm q}_4,\beta \nu_4}V^*_{{\bm q}_1,\gamma\nu_1}V^*_{{\bm q}_2,\delta\nu_2}
    \right] 
    \hat{b}^\dagger_{{\bm q}_1,\nu_1} \hat{b}^\dagger_{{\bm q}_2,\nu_2} \hat{b}^\dagger_{{\bm q}_3,\nu_3} \hat{b}_{{\bm q}_4,\nu_4},
\end{aligned}
\end{equation}
\begin{equation}
\begin{aligned}
    \mathit{5c}. &\quad \sum_{{\bm r},{\bm r}'}\mathcal{J}^{00}_{{\bm r},{\bm r}'} \hat{a}^\dagger_{{\bm r}'}\hat{a}^\dagger_{{\bm r}}\hat{a}_{{\bm r}}\hat{a}_{{\bm r}'} \\
    &\rightarrow \frac{1}{N_{\rm muc}} \sum_{{\bm q}_1,{\bm q}_2,{\bm q}_3,{\bm q}_4} \sum_{\nu_1\nu_2\nu_3\nu_4} \sum_{\alpha\beta\gamma\delta} 
    \delta_{\alpha\delta}\delta_{\beta\gamma} 
    \delta\left( {\bm q}_1+{\bm q}_2+{\bm q}_3-{\bm q}_4 \right) \\
    &\quad\quad \times 
    \left[
          \tilde{\mathcal{J}}^{00}_{ {\bm q}_2+{\bm q}_3,\beta\alpha } U^*_{{\bm q}_1,\alpha\nu_1}U^*_{{\bm q}_2,\beta \nu_2}V^*_{{\bm q}_3,\gamma\nu_3}U_{{\bm q}_4,\delta\nu_4}
        + \tilde{\mathcal{J}}^{00}_{ {\bm q}_1+{\bm q}_2,\beta\alpha } V_{{\bm q}_4,\alpha\nu_4}U^*_{{\bm q}_1,\beta \nu_1}V^*_{{\bm q}_2,\gamma\nu_2}V^*_{{\bm q}_3,\delta\nu_3}
    \right. \\ &\quad\quad\quad \left.
        + \tilde{\mathcal{J}}^{00}_{ {\bm q}_3-{\bm q}_4,\beta\alpha } U^*_{{\bm q}_2,\alpha\nu_2}U^*_{{\bm q}_3,\beta \nu_3}U_{{\bm q}_4,\gamma\nu_4}V^*_{{\bm q}_1,\delta\nu_1}
        + \tilde{\mathcal{J}}^{00}_{-{\bm q}_4+{\bm q}_1,\beta\alpha } U^*_{{\bm q}_3,\alpha\nu_3}V_{{\bm q}_4,\beta \nu_4}V^*_{{\bm q}_1,\gamma\nu_1}V^*_{{\bm q}_2,\delta\nu_2}
    \right] 
    \hat{b}^\dagger_{{\bm q}_1,\nu_1} \hat{b}^\dagger_{{\bm q}_2,\nu_2} \hat{b}^\dagger_{{\bm q}_3,\nu_3} \hat{b}_{{\bm q}_4,\nu_4},
\end{aligned}
\end{equation}
\begin{equation}
\begin{aligned}
    \mathit{5d}. &\quad \sum_{{\bm r},{\bm r}'}\mathcal{J}^{00}_{{\bm r},{\bm r}'} \hat{a}^\dagger_{{\bm r}'}\hat{a}^\dagger_{{\bm r}}\hat{a}_{{\bm r}'}\hat{a}_{{\bm r}} \\
    &\rightarrow \frac{1}{N_{\rm muc}} \sum_{{\bm q}_1,{\bm q}_2,{\bm q}_3,{\bm q}_4} \sum_{\nu_1\nu_2\nu_3\nu_4} \sum_{\alpha\beta\gamma\delta} 
    \delta_{\alpha\gamma}\delta_{\beta\delta} 
    \delta\left( {\bm q}_1+{\bm q}_2+{\bm q}_3-{\bm q}_4 \right) \\
    &\quad\quad \times 
    \left[
          \tilde{\mathcal{J}}^{00}_{ {\bm q}_2-{\bm q}_4,\beta\alpha } U^*_{{\bm q}_1,\alpha\nu_1}U^*_{{\bm q}_2,\beta \nu_2}V^*_{{\bm q}_3,\gamma\nu_3}U_{{\bm q}_4,\delta\nu_4}
        + \tilde{\mathcal{J}}^{00}_{ {\bm q}_1+{\bm q}_3,\beta\alpha } V_{{\bm q}_4,\alpha\nu_4}U^*_{{\bm q}_1,\beta \nu_1}V^*_{{\bm q}_2,\gamma\nu_2}V^*_{{\bm q}_3,\delta\nu_3}
    \right. \\ &\quad\quad\quad \left.
        + \tilde{\mathcal{J}}^{00}_{ {\bm q}_3+{\bm q}_1,\beta\alpha } U^*_{{\bm q}_2,\alpha\nu_2}U^*_{{\bm q}_3,\beta \nu_3}U_{{\bm q}_4,\gamma\nu_4}V^*_{{\bm q}_1,\delta\nu_1}
        + \tilde{\mathcal{J}}^{00}_{-{\bm q}_4+{\bm q}_2,\beta\alpha } U^*_{{\bm q}_3,\alpha\nu_3}V_{{\bm q}_4,\beta \nu_4}V^*_{{\bm q}_1,\gamma\nu_1}V^*_{{\bm q}_2,\delta\nu_2}
    \right] 
    \hat{b}^\dagger_{{\bm q}_1,\nu_1} \hat{b}^\dagger_{{\bm q}_2,\nu_2} \hat{b}^\dagger_{{\bm q}_3,\nu_3} \hat{b}_{{\bm q}_4,\nu_4},
\end{aligned}
\end{equation}
\begin{equation}
\begin{aligned}
    \mathit{6}. &\quad \sum_{{\bm r},{\bm r}'} \mathcal{J}^{++}_{{\bm r},{\bm r}'} \hat{a}^\dagger_{{\bm r}}\hat{a}^\dagger_{{\bm r}}\hat{a}^\dagger_{{\bm r}'}\hat{a}_{{\bm r}} \\
    &= \frac{1}{N_{\rm muc}} \sum_{\nu_1\nu_2\nu_3\nu_4} \sum_{\alpha\beta\gamma\delta} \sum_{{\bm q}_1,{\bm q}_2,{\bm q}_3,{\bm q}_4} \delta_{\alpha\beta}\delta_{\alpha\delta} \tilde{\mathcal{J}}^{++}_{-{\bm q}_3,\alpha\gamma} \delta\left( {\bm q}_1+{\bm q}_2+{\bm q}_3-{\bm q}_4 \right) \\
    &\quad\quad \times
    \left( U^*_{{\bm q}_1,\alpha\nu_1} \hat{b}^\dagger_{{\bm q}_1,\nu_1} + V_{-{\bm q}_1,\alpha\nu_1}   \hat{b}_{-{\bm q}_1,\nu_1}         \right)
    \left( U^*_{{\bm q}_2,\beta \nu_2} \hat{b}^\dagger_{{\bm q}_2,\nu_2} + V_{-{\bm q}_2,\beta \nu_2}   \hat{b}_{-{\bm q}_2,\nu_2}         \right) \\
    &\quad\quad \times
    \left( U^*_{{\bm q}_3,\gamma\nu_3} \hat{b}^\dagger_{{\bm q}_3,\nu_3} + V_{-{\bm q}_3,\gamma\nu_3}   \hat{b}_{-{\bm q}_3,\nu_3}         \right)
    \left( U_{{\bm q}_4,\delta\nu_4}   \hat{b}_{{\bm q}_4,\nu_4}         + V^*_{-{\bm q}_4,\delta\nu_4} \hat{b}^\dagger_{-{\bm q}_4,\nu_4} \right) \\
    &\rightarrow \frac{1}{N_{\rm muc}} \sum_{{\bm q}_1,{\bm q}_2,{\bm q}_3,{\bm q}_4} \sum_{\nu_1\nu_2\nu_3\nu_4} \sum_{\alpha\beta\gamma\delta} \delta_{\alpha\beta}\delta_{\alpha\delta} \delta\left( {\bm q}_1+{\bm q}_2+{\bm q}_3-{\bm q}_4 \right) \\
    &\quad\quad \times
    \left[
          \tilde{\mathcal{J}}^{++}_{-{\bm q}_3,\alpha\gamma} U^*_{{\bm q}_1,\alpha\nu_1}U^*_{{\bm q}_2,\beta \nu_2}U^*_{{\bm q}_3,\gamma\nu_3}U_{{\bm q}_4,\delta\nu_4}
        + \tilde{\mathcal{J}}^{++}_{-{\bm q}_2,\alpha\gamma} V_{{\bm q}_4,\alpha\nu_4}U^*_{{\bm q}_1,\beta \nu_1}U^*_{{\bm q}_2,\gamma\nu_2}V^*_{{\bm q}_3,\delta\nu_3}
    \right. \\ &\quad\quad\quad \left.
        + \tilde{\mathcal{J}}^{++}_{ {\bm q}_4,\alpha\gamma} U^*_{{\bm q}_2,\alpha\nu_2}U^*_{{\bm q}_3,\beta \nu_3}V_{{\bm q}_4,\gamma\nu_4}V^*_{{\bm q}_1,\delta\nu_1}
        + \tilde{\mathcal{J}}^{++}_{-{\bm q}_1,\alpha\gamma} U^*_{{\bm q}_3,\alpha\nu_3}V_{{\bm q}_4,\beta \nu_4}U^*_{{\bm q}_1,\gamma\nu_1}V^*_{{\bm q}_2,\delta\nu_2}
    \right]
    \hat{b}^\dagger_{{\bm q}_1,\nu_1} \hat{b}^\dagger_{{\bm q}_2,\nu_2} \hat{b}^\dagger_{{\bm q}_3,\nu_3} \hat{b}_{{\bm q}_4,\nu_4},
\end{aligned}
\end{equation}
\begin{equation}
\begin{aligned}
    \mathit{7}. &\quad \sum_{{\bm r},{\bm r}'} \mathcal{J}^{++}_{{\bm r},{\bm r}'} \hat{a}^\dagger_{{\bm r}}\hat{a}^\dagger_{{\bm r}'}\hat{a}^\dagger_{{\bm r}'}\hat{a}_{{\bm r}'} \\
    &\rightarrow \frac{1}{N_{\rm muc}} \sum_{{\bm q}_1,{\bm q}_2,{\bm q}_3,{\bm q}_4} \sum_{\nu_1\nu_2\nu_3\nu_4} \sum_{\alpha\beta\gamma\delta} \delta_{\beta\gamma}\delta_{\beta\delta} \delta\left( {\bm q}_1+{\bm q}_2+{\bm q}_3-{\bm q}_4 \right) \\
    &\quad\quad \times
    \left[
          \tilde{\mathcal{J}}^{++}_{ {\bm q}_1,\alpha\delta} U^*_{{\bm q}_1,\alpha\nu_1}U^*_{{\bm q}_2,\beta \nu_2}U^*_{{\bm q}_3,\gamma\nu_3}U_{{\bm q}_4,\delta\nu_4}
        + \tilde{\mathcal{J}}^{++}_{-{\bm q}_4,\alpha\delta} V_{{\bm q}_4,\alpha\nu_4}U^*_{{\bm q}_1,\beta \nu_1}U^*_{{\bm q}_2,\gamma\nu_2}V^*_{{\bm q}_3,\delta\nu_3}
    \right. \\ &\quad\quad\quad \left.
        + \tilde{\mathcal{J}}^{++}_{ {\bm q}_2,\alpha\delta} U^*_{{\bm q}_2,\alpha\nu_2}U^*_{{\bm q}_3,\beta \nu_3}V_{{\bm q}_4,\gamma\nu_4}V^*_{{\bm q}_1,\delta\nu_1}
        + \tilde{\mathcal{J}}^{++}_{ {\bm q}_3,\alpha\delta} U^*_{{\bm q}_3,\alpha\nu_3}V_{{\bm q}_4,\beta \nu_4}U^*_{{\bm q}_1,\gamma\nu_1}V^*_{{\bm q}_2,\delta\nu_2}
    \right]
    \hat{b}^\dagger_{{\bm q}_1,\nu_1} \hat{b}^\dagger_{{\bm q}_2,\nu_2} \hat{b}^\dagger_{{\bm q}_3,\nu_3} \hat{b}_{{\bm q}_4,\nu_4},
\end{aligned}
\end{equation}
\begin{equation}
\begin{aligned}
    \mathit{8}. &\quad \sum_{{\bm r},{\bm r}'} \mathcal{J}^{--}_{{\bm r},{\bm r}'} \hat{a}^\dagger_{{\bm r}}\hat{a}_{{\bm r}}\hat{a}_{{\bm r}}\hat{a}_{{\bm r}'} \\
    &= \frac{1}{N_{\rm muc}} \sum_{\nu_1\nu_2\nu_3\nu_4} \sum_{\alpha\beta\gamma\delta} \sum_{{\bm q_1},{\bm q}_2,{\bm q}_3,{\bm q}_4} \delta_{\alpha\beta}\delta_{\alpha\gamma} \tilde{\mathcal{J}}^{--}_{{\bm q}_4,\alpha\delta} \delta({\bm q}_1-{\bm q}_2-{\bm q}_3-{\bm q}_4) \\
    &\quad\quad \times 
    \left( U^*_{{\bm q}_1,\alpha\nu_1} \hat{b}^\dagger_{{\bm q}_1,\nu_1} + V_{-{\bm q}_1,\alpha\nu_1}   \hat{b}_{-{\bm q}_1,\nu_1}         \right)
    \left( U_{{\bm q}_2,\beta \nu_2}   \hat{b}_{{\bm q}_2,\nu_2}         + V^*_{-{\bm q}_2,\beta \nu_2} \hat{b}^\dagger_{-{\bm q}_2,\nu_2} \right) \\
    &\quad\quad \times
    \left( U_{{\bm q}_3,\gamma\nu_3}   \hat{b}_{{\bm q}_3,\nu_3}         + V^*_{-{\bm q}_3,\gamma\nu_3} \hat{b}^\dagger_{-{\bm q}_3,\nu_3} \right)
    \left( U_{{\bm q}_4,\delta\nu_4}   \hat{b}_{{\bm q}_4,\nu_4}         + V^*_{-{\bm q}_4,\delta\nu_4} \hat{b}^\dagger_{-{\bm q}_4,\nu_4} \right) \\
    &\rightarrow \frac{1}{N_{\rm muc}} \sum_{{\bm q_1},{\bm q}_2,{\bm q}_3,{\bm q}_4} \sum_{\nu_1\nu_2\nu_3\nu_4} \sum_{\alpha\beta\gamma\delta} \delta_{\alpha\beta}\delta_{\alpha\gamma} \delta({\bm q}_1+{\bm q}_2+{\bm q}_3-{\bm q}_4) \\
    &\quad\quad \times
    \left[
          \tilde{\mathcal{J}}^{--}_{ {\bm q}_4,\alpha\delta} U^*_{{\bm q}_1,\alpha\nu_1}V^*_{{\bm q}_2,\beta \nu_2}V^*_{{\bm q}_3,\gamma\nu_3}U_{{\bm q}_4,\delta\nu_4}
        + \tilde{\mathcal{J}}^{--}_{-{\bm q}_3,\alpha\delta} V_{{\bm q}_4,\alpha\nu_4}V^*_{{\bm q}_1,\beta \nu_1}V^*_{{\bm q}_2,\gamma\nu_2}V^*_{{\bm q}_3,\delta\nu_3}
    \right. \\ &\quad\quad\quad \left.
        + \tilde{\mathcal{J}}^{--}_{-{\bm q}_1,\alpha\delta} U^*_{{\bm q}_2,\alpha\nu_2}V^*_{{\bm q}_3,\beta \nu_3}U_{{\bm q}_4,\gamma\nu_4}V^*_{{\bm q}_1,\delta\nu_1}
        + \tilde{\mathcal{J}}^{--}_{-{\bm q}_2,\alpha\delta} U^*_{{\bm q}_3,\alpha\nu_3}U_{{\bm q}_4,\beta \nu_4}V^*_{{\bm q}_1,\gamma\nu_1}V^*_{{\bm q}_2,\delta\nu_2}
    \right]
    \hat{b}^\dagger_{{\bm q}_1,\nu_1} \hat{b}^\dagger_{{\bm q}_2,\nu_2} \hat{b}^\dagger_{{\bm q}_3,\nu_3} \hat{b}_{{\bm q}_4,\nu_4},
\end{aligned}
\end{equation}
\begin{equation}
\begin{aligned}
    \mathit{9}. &\quad \sum_{{\bm r},{\bm r}'} \mathcal{J}^{--}_{{\bm r},{\bm r}'} \hat{a}^\dagger_{{\bm r}'}\hat{a}_{{\bm r}}\hat{a}_{{\bm r}'}\hat{a}_{{\bm r}'} \\
    &\rightarrow \frac{1}{N_{\rm muc}} \sum_{{\bm q_1},{\bm q}_2,{\bm q}_3,{\bm q}_4} \sum_{\nu_1\nu_2\nu_3\nu_4} \sum_{\alpha\beta\gamma\delta} \delta_{\alpha\gamma}\delta_{\alpha\delta} \delta({\bm q}_1+{\bm q}_2+{\bm q}_3-{\bm q}_4) \\
    &\quad\quad \times
    \left[
          \tilde{\mathcal{J}}^{--}_{ {\bm q}_2,\beta\alpha } U^*_{{\bm q}_1,\alpha\nu_1}V^*_{{\bm q}_2,\beta \nu_2}V^*_{{\bm q}_3,\gamma\nu_3}U_{{\bm q}_4,\delta\nu_4}
        + \tilde{\mathcal{J}}^{--}_{ {\bm q}_1,\beta\alpha } V_{{\bm q}_4,\alpha\nu_4}V^*_{{\bm q}_1,\beta \nu_1}V^*_{{\bm q}_2,\gamma\nu_2}V^*_{{\bm q}_3,\delta\nu_3}
    \right. \\ &\quad\quad\quad \left.
        + \tilde{\mathcal{J}}^{--}_{ {\bm q}_3,\beta\alpha } U^*_{{\bm q}_2,\alpha\nu_2}V^*_{{\bm q}_3,\beta \nu_3}U_{{\bm q}_4,\gamma\nu_4}V^*_{{\bm q}_1,\delta\nu_1}
        + \tilde{\mathcal{J}}^{--}_{-{\bm q}_4,\beta\alpha } U^*_{{\bm q}_3,\alpha\nu_3}U_{{\bm q}_4,\beta \nu_4}V^*_{{\bm q}_1,\gamma\nu_1}V^*_{{\bm q}_2,\delta\nu_2}
    \right]
    \hat{b}^\dagger_{{\bm q}_1,\nu_1} \hat{b}^\dagger_{{\bm q}_2,\nu_2} \hat{b}^\dagger_{{\bm q}_3,\nu_3} \hat{b}_{{\bm q}_4,\nu_4}.
\end{aligned}
\end{equation}
Taken together, the explicit expression of the unsymmetrized 3-1 vertex is given by
\begin{equation}
\begin{aligned}
    &\quad \ \mathcal{C}_{{\bm q}_1,{\bm q}_2,{\bm q}_3\leftarrow{\bm q}_4}^{\nu_1,\nu_2,\nu_3\leftarrow\nu_4} \\
    &= \frac{1}{4N_{\rm muc}} \sum_{\alpha,\beta,\gamma,\delta=1}^{N_{\rm sub}} \delta({\bm q}_1+{\bm q}_2+{\bm q}_3-{\bm q}_4) \\
    &\quad\quad \times \left[ 
        - \delta_{\alpha\beta }\delta_{\alpha\gamma} \tilde{\mathcal{J}}^{+-}_{ {\bm q}_4,\gamma\delta} U^*_{{\bm q}_1,\alpha\nu_1}U^*_{{\bm q}_2,\beta \nu_2}V^*_{{\bm q}_3,\gamma\nu_3}U_{{\bm q}_4,\delta\nu_4}
        - \delta_{\alpha\beta }\delta_{\alpha\gamma} \tilde{\mathcal{J}}^{+-}_{-{\bm q}_3,\gamma\delta} V_{{\bm q}_4,\alpha\nu_4}U^*_{{\bm q}_1,\beta \nu_1}V^*_{{\bm q}_2,\gamma\nu_2}V^*_{{\bm q}_3,\delta\nu_3}
    \right. \\ &\quad\quad\quad
        - \delta_{\alpha\beta }\delta_{\alpha\gamma} \tilde{\mathcal{J}}^{+-}_{-{\bm q}_1,\gamma\delta} U^*_{{\bm q}_2,\alpha\nu_2}U^*_{{\bm q}_3,\beta \nu_3}U_{{\bm q}_4,\gamma\nu_4}V^*_{{\bm q}_1,\delta\nu_1}
        - \delta_{\alpha\beta }\delta_{\alpha\gamma} \tilde{\mathcal{J}}^{+-}_{-{\bm q}_2,\gamma\delta} U^*_{{\bm q}_3,\alpha\nu_3}V_{{\bm q}_4,\beta \nu_4}V^*_{{\bm q}_1,\gamma\nu_1}V^*_{{\bm q}_2,\delta\nu_2}
    \\ &\quad\quad\quad
        - \delta_{\beta\delta }\delta_{\gamma\delta} \tilde{\mathcal{J}}^{+-}_{ {\bm q}_1,\alpha\beta } U^*_{{\bm q}_1,\alpha\nu_1}U^*_{{\bm q}_2,\beta \nu_2}V^*_{{\bm q}_3,\gamma\nu_3}U_{{\bm q}_4,\delta\nu_4}
        - \delta_{\beta\delta }\delta_{\gamma\delta} \tilde{\mathcal{J}}^{+-}_{-{\bm q}_4,\alpha\beta } V_{{\bm q}_4,\alpha\nu_4}U^*_{{\bm q}_1,\beta \nu_1}V^*_{{\bm q}_2,\gamma\nu_2}V^*_{{\bm q}_3,\delta\nu_3}
    \\ &\quad\quad\quad
        - \delta_{\beta\delta }\delta_{\gamma\delta} \tilde{\mathcal{J}}^{+-}_{ {\bm q}_2,\alpha\beta } U^*_{{\bm q}_2,\alpha\nu_2}U^*_{{\bm q}_3,\beta \nu_3}U_{{\bm q}_4,\gamma\nu_4}V^*_{{\bm q}_1,\delta\nu_1}
        - \delta_{\beta\delta }\delta_{\gamma\delta} \tilde{\mathcal{J}}^{+-}_{ {\bm q}_3,\alpha\beta } U^*_{{\bm q}_3,\alpha\nu_3}V_{{\bm q}_4,\beta \nu_4}V^*_{{\bm q}_1,\gamma\nu_1}V^*_{{\bm q}_2,\delta\nu_2}
    \\ &\quad\quad\quad
        - \delta_{\alpha\gamma}\delta_{\alpha\delta} \tilde{\mathcal{J}}^{-+}_{-{\bm q}_2,\alpha\beta } U^*_{{\bm q}_1,\alpha\nu_1}U^*_{{\bm q}_2,\beta \nu_2}V^*_{{\bm q}_3,\gamma\nu_3}U_{{\bm q}_4,\delta\nu_4}
        - \delta_{\alpha\gamma}\delta_{\alpha\delta} \tilde{\mathcal{J}}^{-+}_{-{\bm q}_1,\alpha\beta } V_{{\bm q}_4,\alpha\nu_4}U^*_{{\bm q}_1,\beta \nu_1}V^*_{{\bm q}_2,\gamma\nu_2}V^*_{{\bm q}_3,\delta\nu_3}
    \\ &\quad\quad\quad
        - \delta_{\alpha\gamma}\delta_{\alpha\delta} \tilde{\mathcal{J}}^{-+}_{-{\bm q}_3,\alpha\beta } U^*_{{\bm q}_2,\alpha\nu_2}U^*_{{\bm q}_3,\beta \nu_3}U_{{\bm q}_4,\gamma\nu_4}V^*_{{\bm q}_1,\delta\nu_1}
        - \delta_{\alpha\gamma}\delta_{\alpha\delta} \tilde{\mathcal{J}}^{-+}_{ {\bm q}_4,\alpha\beta } U^*_{{\bm q}_3,\alpha\nu_3}V_{{\bm q}_4,\beta \nu_4}V^*_{{\bm q}_1,\gamma\nu_1}V^*_{{\bm q}_2,\delta\nu_2}
    \\ &\quad\quad\quad
        - \delta_{\alpha\beta }\delta_{\alpha\delta} \tilde{\mathcal{J}}^{-+}_{ {\bm q}_3,\gamma\delta} U^*_{{\bm q}_1,\alpha\nu_1}U^*_{{\bm q}_2,\beta \nu_2}V^*_{{\bm q}_3,\gamma\nu_3}U_{{\bm q}_4,\delta\nu_4}
        - \delta_{\alpha\beta }\delta_{\alpha\delta} \tilde{\mathcal{J}}^{-+}_{ {\bm q}_2,\gamma\delta} V_{{\bm q}_4,\alpha\nu_4}U^*_{{\bm q}_1,\beta \nu_1}V^*_{{\bm q}_2,\gamma\nu_2}V^*_{{\bm q}_3,\delta\nu_3}
    \\ &\quad\quad\quad
        - \delta_{\alpha\beta }\delta_{\alpha\delta} \tilde{\mathcal{J}}^{-+}_{-{\bm q}_4,\gamma\delta} U^*_{{\bm q}_2,\alpha\nu_2}U^*_{{\bm q}_3,\beta \nu_3}U_{{\bm q}_4,\gamma\nu_4}V^*_{{\bm q}_1,\delta\nu_1}
        - \delta_{\alpha\beta }\delta_{\alpha\delta} \tilde{\mathcal{J}}^{-+}_{ {\bm q}_1,\gamma\delta} U^*_{{\bm q}_3,\alpha\nu_3}V_{{\bm q}_4,\beta \nu_4}V^*_{{\bm q}_1,\gamma\nu_1}V^*_{{\bm q}_2,\delta\nu_2}
    \\ &\quad\quad\quad
        + \delta_{\alpha\gamma}\delta_{\beta\delta } \tilde{\mathcal{J}}^{00}_{ {\bm q}_1+{\bm q}_3,\alpha\beta } U^*_{{\bm q}_1,\alpha\nu_1}U^*_{{\bm q}_2,\beta \nu_2}V^*_{{\bm q}_3,\gamma\nu_3}U_{{\bm q}_4,\delta\nu_4}
        + \delta_{\alpha\gamma}\delta_{\beta\delta } \tilde{\mathcal{J}}^{00}_{-{\bm q}_4+{\bm q}_2,\alpha\beta } V_{{\bm q}_4,\alpha\nu_4}U^*_{{\bm q}_1,\beta \nu_1}V^*_{{\bm q}_2,\gamma\nu_2}V^*_{{\bm q}_3,\delta\nu_3}
    \\ &\quad\quad\quad
        + \delta_{\alpha\gamma}\delta_{\beta\delta } \tilde{\mathcal{J}}^{00}_{ {\bm q}_2-{\bm q}_4,\alpha\beta } U^*_{{\bm q}_2,\alpha\nu_2}U^*_{{\bm q}_3,\beta \nu_3}U_{{\bm q}_4,\gamma\nu_4}V^*_{{\bm q}_1,\delta\nu_1}
        + \delta_{\alpha\gamma}\delta_{\beta\delta } \tilde{\mathcal{J}}^{00}_{ {\bm q}_3+{\bm q}_1,\alpha\beta } U^*_{{\bm q}_3,\alpha\nu_3}V_{{\bm q}_4,\beta \nu_4}V^*_{{\bm q}_1,\gamma\nu_1}V^*_{{\bm q}_2,\delta\nu_2}
    \\ &\quad\quad\quad
        + \delta_{\alpha\delta}\delta_{\beta\gamma } \tilde{\mathcal{J}}^{00}_{ {\bm q}_1-{\bm q}_4,\alpha\beta } U^*_{{\bm q}_1,\alpha\nu_1}U^*_{{\bm q}_2,\beta \nu_2}V^*_{{\bm q}_3,\gamma\nu_3}U_{{\bm q}_4,\delta\nu_4}
        + \delta_{\alpha\delta}\delta_{\beta\gamma } \tilde{\mathcal{J}}^{00}_{-{\bm q}_4+{\bm q}_3,\alpha\beta } V_{{\bm q}_4,\alpha\nu_4}U^*_{{\bm q}_1,\beta \nu_1}V^*_{{\bm q}_2,\gamma\nu_2}V^*_{{\bm q}_3,\delta\nu_3}
    \\ &\quad\quad\quad
        + \delta_{\alpha\delta}\delta_{\beta\gamma } \tilde{\mathcal{J}}^{00}_{ {\bm q}_2+{\bm q}_1,\alpha\beta } U^*_{{\bm q}_2,\alpha\nu_2}U^*_{{\bm q}_3,\beta \nu_3}U_{{\bm q}_4,\gamma\nu_4}V^*_{{\bm q}_1,\delta\nu_1}
        + \delta_{\alpha\delta}\delta_{\beta\gamma } \tilde{\mathcal{J}}^{00}_{ {\bm q}_3+{\bm q}_2,\alpha\beta } U^*_{{\bm q}_3,\alpha\nu_3}V_{{\bm q}_4,\beta \nu_4}V^*_{{\bm q}_1,\gamma\nu_1}V^*_{{\bm q}_2,\delta\nu_2}
    \\ &\quad\quad\quad
        + \delta_{\alpha\delta}\delta_{\beta\gamma } \tilde{\mathcal{J}}^{00}_{ {\bm q}_2+{\bm q}_3,\beta\alpha } U^*_{{\bm q}_1,\alpha\nu_1}U^*_{{\bm q}_2,\beta \nu_2}V^*_{{\bm q}_3,\gamma\nu_3}U_{{\bm q}_4,\delta\nu_4}
        + \delta_{\alpha\delta}\delta_{\beta\gamma } \tilde{\mathcal{J}}^{00}_{ {\bm q}_1+{\bm q}_2,\beta\alpha } V_{{\bm q}_4,\alpha\nu_4}U^*_{{\bm q}_1,\beta \nu_1}V^*_{{\bm q}_2,\gamma\nu_2}V^*_{{\bm q}_3,\delta\nu_3}
    \\ &\quad\quad\quad
        + \delta_{\alpha\delta}\delta_{\beta\gamma } \tilde{\mathcal{J}}^{00}_{ {\bm q}_3-{\bm q}_4,\beta\alpha } U^*_{{\bm q}_2,\alpha\nu_2}U^*_{{\bm q}_3,\beta \nu_3}U_{{\bm q}_4,\gamma\nu_4}V^*_{{\bm q}_1,\delta\nu_1}
        + \delta_{\alpha\delta}\delta_{\beta\gamma } \tilde{\mathcal{J}}^{00}_{-{\bm q}_4+{\bm q}_1,\beta\alpha } U^*_{{\bm q}_3,\alpha\nu_3}V_{{\bm q}_4,\beta \nu_4}V^*_{{\bm q}_1,\gamma\nu_1}V^*_{{\bm q}_2,\delta\nu_2}
    \\ &\quad\quad\quad
        + \delta_{\alpha\gamma}\delta_{\beta\delta } \tilde{\mathcal{J}}^{00}_{ {\bm q}_2-{\bm q}_4,\beta\alpha } U^*_{{\bm q}_1,\alpha\nu_1}U^*_{{\bm q}_2,\beta \nu_2}V^*_{{\bm q}_3,\gamma\nu_3}U_{{\bm q}_4,\delta\nu_4}
        + \delta_{\alpha\gamma}\delta_{\beta\delta } \tilde{\mathcal{J}}^{00}_{ {\bm q}_1+{\bm q}_3,\beta\alpha } V_{{\bm q}_4,\alpha\nu_4}U^*_{{\bm q}_1,\beta \nu_1}V^*_{{\bm q}_2,\gamma\nu_2}V^*_{{\bm q}_3,\delta\nu_3}
    \\ &\quad\quad\quad
        + \delta_{\alpha\gamma}\delta_{\beta\delta } \tilde{\mathcal{J}}^{00}_{ {\bm q}_3+{\bm q}_1,\beta\alpha } U^*_{{\bm q}_2,\alpha\nu_2}U^*_{{\bm q}_3,\beta \nu_3}U_{{\bm q}_4,\gamma\nu_4}V^*_{{\bm q}_1,\delta\nu_1}
        + \delta_{\alpha\gamma}\delta_{\beta\delta } \tilde{\mathcal{J}}^{00}_{-{\bm q}_4+{\bm q}_2,\beta\alpha } U^*_{{\bm q}_3,\alpha\nu_3}V_{{\bm q}_4,\beta \nu_4}V^*_{{\bm q}_1,\gamma\nu_1}V^*_{{\bm q}_2,\delta\nu_2}
    \\ &\quad\quad\quad
        - \delta_{\alpha\beta }\delta_{\alpha\delta} \tilde{\mathcal{J}}^{++}_{-{\bm q}_3,\alpha\gamma} U^*_{{\bm q}_1,\alpha\nu_1}U^*_{{\bm q}_2,\beta \nu_2}U^*_{{\bm q}_3,\gamma\nu_3}U_{{\bm q}_4,\delta\nu_4}
        - \delta_{\alpha\beta }\delta_{\alpha\delta} \tilde{\mathcal{J}}^{++}_{-{\bm q}_2,\alpha\gamma} V_{{\bm q}_4,\alpha\nu_4}U^*_{{\bm q}_1,\beta \nu_1}U^*_{{\bm q}_2,\gamma\nu_2}V^*_{{\bm q}_3,\delta\nu_3}
    \\ &\quad\quad\quad
        - \delta_{\alpha\beta }\delta_{\alpha\delta} \tilde{\mathcal{J}}^{++}_{ {\bm q}_4,\alpha\gamma} U^*_{{\bm q}_2,\alpha\nu_2}U^*_{{\bm q}_3,\beta \nu_3}V_{{\bm q}_4,\gamma\nu_4}V^*_{{\bm q}_1,\delta\nu_1}
        - \delta_{\alpha\beta }\delta_{\alpha\delta} \tilde{\mathcal{J}}^{++}_{-{\bm q}_1,\alpha\gamma} U^*_{{\bm q}_3,\alpha\nu_3}V_{{\bm q}_4,\beta \nu_4}U^*_{{\bm q}_1,\gamma\nu_1}V^*_{{\bm q}_2,\delta\nu_2}
    \\ &\quad\quad\quad
        - \delta_{\beta\gamma }\delta_{\beta\delta } \tilde{\mathcal{J}}^{++}_{ {\bm q}_1,\alpha\delta} U^*_{{\bm q}_1,\alpha\nu_1}U^*_{{\bm q}_2,\beta \nu_2}U^*_{{\bm q}_3,\gamma\nu_3}U_{{\bm q}_4,\delta\nu_4}
        - \delta_{\beta\gamma }\delta_{\beta\delta } \tilde{\mathcal{J}}^{++}_{-{\bm q}_4,\alpha\delta} V_{{\bm q}_4,\alpha\nu_4}U^*_{{\bm q}_1,\beta \nu_1}U^*_{{\bm q}_2,\gamma\nu_2}V^*_{{\bm q}_3,\delta\nu_3}
    \\ &\quad\quad\quad
        - \delta_{\beta\gamma }\delta_{\beta\delta } \tilde{\mathcal{J}}^{++}_{ {\bm q}_2,\alpha\delta} U^*_{{\bm q}_2,\alpha\nu_2}U^*_{{\bm q}_3,\beta \nu_3}V_{{\bm q}_4,\gamma\nu_4}V^*_{{\bm q}_1,\delta\nu_1}
        - \delta_{\beta\gamma }\delta_{\beta\delta } \tilde{\mathcal{J}}^{++}_{ {\bm q}_3,\alpha\delta} U^*_{{\bm q}_3,\alpha\nu_3}V_{{\bm q}_4,\beta \nu_4}U^*_{{\bm q}_1,\gamma\nu_1}V^*_{{\bm q}_2,\delta\nu_2}
    \\ &\quad\quad\quad
        - \delta_{\alpha\beta }\delta_{\alpha\gamma} \tilde{\mathcal{J}}^{--}_{ {\bm q}_4,\alpha\delta} U^*_{{\bm q}_1,\alpha\nu_1}V^*_{{\bm q}_2,\beta \nu_2}V^*_{{\bm q}_3,\gamma\nu_3}U_{{\bm q}_4,\delta\nu_4}
        - \delta_{\alpha\beta }\delta_{\alpha\gamma} \tilde{\mathcal{J}}^{--}_{-{\bm q}_3,\alpha\delta} V_{{\bm q}_4,\alpha\nu_4}V^*_{{\bm q}_1,\beta \nu_1}V^*_{{\bm q}_2,\gamma\nu_2}V^*_{{\bm q}_3,\delta\nu_3}
    \\ &\quad\quad\quad
        - \delta_{\alpha\beta }\delta_{\alpha\gamma} \tilde{\mathcal{J}}^{--}_{-{\bm q}_1,\alpha\delta} U^*_{{\bm q}_2,\alpha\nu_2}V^*_{{\bm q}_3,\beta \nu_3}U_{{\bm q}_4,\gamma\nu_4}V^*_{{\bm q}_1,\delta\nu_1}
        - \delta_{\alpha\beta }\delta_{\alpha\gamma} \tilde{\mathcal{J}}^{--}_{-{\bm q}_2,\alpha\delta} U^*_{{\bm q}_3,\alpha\nu_3}U_{{\bm q}_4,\beta \nu_4}V^*_{{\bm q}_1,\gamma\nu_1}V^*_{{\bm q}_2,\delta\nu_2}
    \\ &\quad\quad\quad
        - \delta_{\alpha\gamma}\delta_{\alpha\delta} \tilde{\mathcal{J}}^{--}_{ {\bm q}_2,\beta\alpha } U^*_{{\bm q}_1,\alpha\nu_1}V^*_{{\bm q}_2,\beta \nu_2}V^*_{{\bm q}_3,\gamma\nu_3}U_{{\bm q}_4,\delta\nu_4}
        - \delta_{\alpha\gamma}\delta_{\alpha\delta} \tilde{\mathcal{J}}^{--}_{ {\bm q}_1,\beta\alpha } V_{{\bm q}_4,\alpha\nu_4}V^*_{{\bm q}_1,\beta \nu_1}V^*_{{\bm q}_2,\gamma\nu_2}V^*_{{\bm q}_3,\delta\nu_3}
    \\ &\quad\quad\quad \left.
        - \delta_{\alpha\gamma}\delta_{\alpha\delta} \tilde{\mathcal{J}}^{--}_{ {\bm q}_3,\beta\alpha } U^*_{{\bm q}_2,\alpha\nu_2}V^*_{{\bm q}_3,\beta \nu_3}U_{{\bm q}_4,\gamma\nu_4}V^*_{{\bm q}_1,\delta\nu_1}
        - \delta_{\alpha\gamma}\delta_{\alpha\delta} \tilde{\mathcal{J}}^{--}_{-{\bm q}_4,\beta\alpha } U^*_{{\bm q}_3,\alpha\nu_3}U_{{\bm q}_4,\beta \nu_4}V^*_{{\bm q}_1,\gamma\nu_1}V^*_{{\bm q}_2,\delta\nu_2}
    \right].
\end{aligned}
\end{equation}
\subsubsection{Normal-ordered residua $\tilde{\mathcal{H}}^{(31{\rm R})}$}
\begin{figure}[htpb]
    \includegraphics[width=3.0cm]{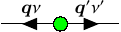}
    \caption{Symmetrized residuum of the 3-1 vertex $\tilde{\mathcal{R}}_{{\bm q},{\bm q}'}^{\nu,\nu'}$.}
\end{figure}
The explicit form of $\tilde{\mathcal{H}}^{(31{\rm R})}$ is given by
\begin{equation}
    \mathcal{H}^{(31{\rm R})} = \sum_{{\bm k},{\bm k}'} \sum_{\nu,\nu'} \tilde{\mathcal{R}}_{{\bm k},{\bm k}'}^{\nu,\nu'} \hat{b}^\dagger_{{\bm k},\nu}\hat{b}^\dagger_{{\bm k}',\nu'},
\end{equation}
where $ \tilde{\mathcal{R}}_{{\bm k},{\bm k}'}^{\nu,\nu'}$ is the symmetrized residuum of the three-out-one-in vertex given by,
\begin{equation}
    \tilde{\mathcal{R}}_{{\bm k},{\bm k}'}^{\nu,\nu'} = \frac{1}{2} \frac{1}{3!} \sum_{\bm q}\sum_\mu\left(
          \mathcal{R}_{{\bm q} ,{\bm k} ,{\bm k}'\leftarrow{\bm q}}^{\mu ,\nu ,\nu'\leftarrow\mu}
        + \mathcal{R}_{{\bm q} ,{\bm k}',{\bm k} \leftarrow{\bm q}}^{\mu ,\nu',\nu \leftarrow\mu}
        + \mathcal{R}_{{\bm k}',{\bm q} ,{\bm k} \leftarrow{\bm q}}^{\nu',\mu ,\nu \leftarrow\mu}
        + \mathcal{R}_{{\bm k}',{\bm k} ,{\bm q} \leftarrow{\bm q}}^{\nu',\nu ,\mu \leftarrow\mu}
        + \mathcal{R}_{{\bm k} ,{\bm k}',{\bm q} \leftarrow{\bm q}}^{\nu ,\nu',\mu \leftarrow\mu}
        + \mathcal{R}_{{\bm k} ,{\bm q} ,{\bm k}'\leftarrow{\bm q}}^{\nu ,\mu ,\nu'\leftarrow\mu}
    \right).
\end{equation}
Note that $\tilde{\mathcal{H}}^{(13{\rm R})}$ is given by the hermitian conjugate of $\tilde{\mathcal{H}}^{(31{\rm R})}$. To get the explicit expression of the unsymmetrized vertex $\mathcal{R}_{{\bm q}_1 ,{\bm q}_2 ,{\bm q}_3\leftarrow{\bm q}_4}^{\nu_1 ,\nu_2 ,\nu_3\leftarrow\nu_4}$, we need to pick up the normal-ordered residua of the three-out-one-in part of the four magnon terms $\mathit{1},\mathit{2},\cdots,\mathit{9}$ in Eqs.~(\ref{eq:HP_SJS}) and (\ref{eq:HP4_SJS}). We list all contributions in the following:
\begin{equation}
\begin{aligned}
    \quad \mathit{1.} &\quad \sum_{{\bm r},{\bm r}'} \mathcal{J}^{+-}_{{\bm r},{\bm r}'} \hat{a}^\dagger_{{\bm r}}\hat{a}^\dagger_{{\bm r}}\hat{a}_{{\bm r}}\hat{a}_{{\bm r}'} \\
    &= \frac{1}{N_{\rm muc}} \sum_{\nu_1\nu_2\nu_3\nu_4} \sum_{\alpha\beta\gamma\delta} \sum_{{\bm q}_1,{\bm q}_2,{\bm q}_3,{\bm q}_4} \delta_{\alpha\beta}\delta_{\alpha\gamma} \tilde{\mathcal{J}}^{+-}_{{\bm q}_4,\gamma\delta}  \delta\left( {\bm q}_1+{\bm q}_2-{\bm q}_3-{\bm q}_4 \right) \\
    &\quad\quad \times 
    \left( U^*_{{\bm q}_1,\alpha\nu_1} \hat{b}^\dagger_{{\bm q}_1,\nu_1} +   V_{-{\bm q}_1,\alpha\nu_1}         \hat{b}_{-{\bm q}_1,\nu_1} \right)
    \left( U^*_{{\bm q}_2,\beta \nu_2} \hat{b}^\dagger_{{\bm q}_2,\nu_2} +   V_{-{\bm q}_2,\beta \nu_2}         \hat{b}_{-{\bm q}_2,\nu_2} \right) \\
    &\quad\quad \times
    \left(   U_{{\bm q}_3,\gamma\nu_3}         \hat{b}_{{\bm q}_3,\nu_3} + V^*_{-{\bm q}_3,\gamma\nu_3} \hat{b}^\dagger_{-{\bm q}_3,\nu_3} \right)
    \left(   U_{{\bm q}_4,\delta\nu_4}         \hat{b}_{{\bm q}_4,\nu_4} + V^*_{-{\bm q}_4,\delta\nu_4} \hat{b}^\dagger_{-{\bm q}_4,\nu_4} \right) \\
    &\rightarrow \frac{1}{N_{\rm muc}} \sum_{{\bm q}_1,{\bm q}_2,{\bm q}_3,{\bm q}_4} \sum_{\nu_1\nu_2\nu_3\nu_4} \sum_{\alpha\beta\gamma\delta} \delta_{\alpha\beta}\delta_{\alpha\gamma} \delta\left( {\bm q}_1+{\bm q}_2+{\bm q}_3-{\bm q}_4 \right)
    \times \sum_{{\bm k}',{\bm q}_a,{\bm q}_b} \sum_{\nu'\nu_a\nu_b} \delta_{{\bm q}_1,{\bm k}'}\delta_{\nu_1,\nu'} \delta_{{\bm q}_2,{\bm q}_a}\delta_{\nu_2,\nu_a} \delta_{{\bm q}_3,{\bm q}_b}\delta_{\nu_3,\nu_b} \delta_{{\bm q}_4,{\bm k}'}\delta_{\nu_4,\nu'} \\
    &\quad \times \left[ 
          \tilde{\mathcal{J}}^{+-}_{-{\bm q}_3,\gamma\delta} V_{{\bm q}_4,\alpha\nu_4}U^*_{{\bm q}_1,\beta \nu_1}V^*_{{\bm q}_2,\gamma\nu_2}V^*_{{\bm q}_3,\delta\nu_3}
        + \tilde{\mathcal{J}}^{+-}_{-{\bm q}_3,\gamma\delta} V_{{\bm q}_4,\alpha\nu_4}U^*_{{\bm q}_2,\beta \nu_2}V^*_{{\bm q}_1,\gamma\nu_1}V^*_{{\bm q}_3,\delta\nu_3}
    \right. \\ &\quad\quad
        + \tilde{\mathcal{J}}^{+-}_{-{\bm q}_1,\gamma\delta} V_{{\bm q}_4,\alpha\nu_4}U^*_{{\bm q}_2,\beta \nu_2}V^*_{{\bm q}_3,\gamma\nu_3}V^*_{{\bm q}_1,\delta\nu_1}
        + \tilde{\mathcal{J}}^{+-}_{-{\bm q}_1,\gamma\delta} U^*_{{\bm q}_2,\alpha\nu_2}U^*_{{\bm q}_3,\beta \nu_3}U_{{\bm q}_4,\gamma\nu_4}V^*_{{\bm q}_1,\delta\nu_1}
    \\ &\quad\quad \left.
        + \tilde{\mathcal{J}}^{+-}_{-{\bm q}_3,\gamma\delta} U^*_{{\bm q}_2,\alpha\nu_2}V_{{\bm q}_4,\beta \nu_4}V^*_{{\bm q}_1,\gamma\nu_1}V^*_{{\bm q}_3,\delta\nu_3}
        + \tilde{\mathcal{J}}^{+-}_{-{\bm q}_1,\gamma\delta} U^*_{{\bm q}_3,\alpha\nu_3}V_{{\bm q}_4,\beta \nu_4}V^*_{{\bm q}_2,\gamma\nu_2}V^*_{{\bm q}_1,\delta\nu_1}
    \right] \hat{b}^\dagger_{{\bm q}_a,\nu_a}\hat{b}^\dagger_{{\bm q}_b,\nu_b},
\end{aligned}
\end{equation}
\begin{equation}
\begin{aligned}
    \mathit{2}. &\quad \sum_{{\bm r},{\bm r}'}\mathcal{J}^{+-}_{{\bm r},{\bm r}'} \hat{a}^\dagger_{{\bm r}}\hat{a}^\dagger_{{\bm r}'}\hat{a}_{{\bm r}'}\hat{a}_{{\bm r}'} \\
    &\rightarrow \frac{1}{N_{\rm muc}} \sum_{{\bm q}_1,{\bm q}_2,{\bm q}_3,{\bm q}_4} \sum_{\nu_1\nu_2\nu_3\nu_4} \sum_{\alpha\beta\gamma\delta} 
    \delta_{\beta\delta}\delta_{\gamma\delta} 
    \delta\left( {\bm q}_1+{\bm q}_2+{\bm q}_3-{\bm q}_4 \right)
    \times \sum_{{\bm k}',{\bm q}_a,{\bm q}_b} \sum_{\nu'\nu_a\nu_b} \delta_{{\bm q}_1,{\bm k}'}\delta_{\nu_1,\nu'} \delta_{{\bm q}_2,{\bm q}_a}\delta_{\nu_2,\nu_a} \delta_{{\bm q}_3,{\bm q}_b}\delta_{\nu_3,\nu_b} \delta_{{\bm q}_4,{\bm k}'}\delta_{\nu_4,\nu'} \\
    &\quad \times \left[ 
          \tilde{\mathcal{J}}^{+-}_{-{\bm q}_4,\alpha\beta } V_{{\bm q}_4,\alpha\nu_4}U^*_{{\bm q}_1,\beta \nu_1}V^*_{{\bm q}_2,\gamma\nu_2}V^*_{{\bm q}_3,\delta\nu_3}
        + \tilde{\mathcal{J}}^{+-}_{-{\bm q}_4,\alpha\beta } V_{{\bm q}_4,\alpha\nu_4}U^*_{{\bm q}_2,\beta \nu_2}V^*_{{\bm q}_1,\gamma\nu_1}V^*_{{\bm q}_3,\delta\nu_3}
    \right. \\ &\quad\quad
        + \tilde{\mathcal{J}}^{+-}_{-{\bm q}_4,\alpha\beta } V_{{\bm q}_4,\alpha\nu_4}U^*_{{\bm q}_2,\beta \nu_2}V^*_{{\bm q}_3,\gamma\nu_3}V^*_{{\bm q}_1,\delta\nu_1}
        + \tilde{\mathcal{J}}^{+-}_{ {\bm q}_2,\alpha\beta } U^*_{{\bm q}_2,\alpha\nu_2}U^*_{{\bm q}_3,\beta \nu_3}U_{{\bm q}_4,\gamma\nu_4}V^*_{{\bm q}_1,\delta\nu_1}
    \\ &\quad\quad \left.
        + \tilde{\mathcal{J}}^{+-}_{ {\bm q}_2,\alpha\beta } U^*_{{\bm q}_2,\alpha\nu_2}V_{{\bm q}_4,\beta \nu_4}V^*_{{\bm q}_1,\gamma\nu_1}V^*_{{\bm q}_3,\delta\nu_3}
        + \tilde{\mathcal{J}}^{+-}_{ {\bm q}_3,\alpha\beta } U^*_{{\bm q}_3,\alpha\nu_3}V_{{\bm q}_4,\beta \nu_4}V^*_{{\bm q}_2,\gamma\nu_2}V^*_{{\bm q}_1,\delta\nu_1}
    \right] \hat{b}^\dagger_{{\bm q}_a,\nu_a}\hat{b}^\dagger_{{\bm q}_b,\nu_b},
\end{aligned}
\end{equation}
\begin{equation}
\begin{aligned}
    \mathit{3}. &\quad \sum_{{\bm r},{\bm r}'}\mathcal{J}^{-+}_{{\bm r},{\bm r}'} \hat{a}^\dagger_{{\bm r}}\hat{a}^\dagger_{{\bm r}'}\hat{a}_{{\bm r}}\hat{a}_{{\bm r}} \\
    &\rightarrow \frac{1}{N_{\rm muc}} \sum_{{\bm q}_1,{\bm q}_2,{\bm q}_3,{\bm q}_4} \sum_{\nu_1\nu_2\nu_3\nu_4} \sum_{\alpha\beta\gamma\delta} 
    \delta_{\alpha\gamma}\delta_{\alpha\delta} 
    \delta\left( {\bm q}_1+{\bm q}_2+{\bm q}_3-{\bm q}_4 \right)
    \times \sum_{{\bm k}',{\bm q}_a,{\bm q}_b} \sum_{\nu'\nu_a\nu_b} \delta_{{\bm q}_1,{\bm k}'}\delta_{\nu_1,\nu'} \delta_{{\bm q}_2,{\bm q}_a}\delta_{\nu_2,\nu_a} \delta_{{\bm q}_3,{\bm q}_b}\delta_{\nu_3,\nu_b} \delta_{{\bm q}_4,{\bm k}'}\delta_{\nu_4,\nu'} \\
    &\quad \times \left[ 
          \tilde{\mathcal{J}}^{-+}_{-{\bm q}_1,\alpha\beta } V_{{\bm q}_4,\alpha\nu_4}U^*_{{\bm q}_1,\beta \nu_1}V^*_{{\bm q}_2,\gamma\nu_2}V^*_{{\bm q}_3,\delta\nu_3}
        + \tilde{\mathcal{J}}^{-+}_{-{\bm q}_2,\alpha\beta } V_{{\bm q}_4,\alpha\nu_4}U^*_{{\bm q}_2,\beta \nu_2}V^*_{{\bm q}_1,\gamma\nu_1}V^*_{{\bm q}_3,\delta\nu_3}
    \right. \\ &\quad\quad
        + \tilde{\mathcal{J}}^{-+}_{-{\bm q}_2,\alpha\beta } V_{{\bm q}_4,\alpha\nu_4}U^*_{{\bm q}_2,\beta \nu_2}V^*_{{\bm q}_3,\gamma\nu_3}V^*_{{\bm q}_1,\delta\nu_1}
        + \tilde{\mathcal{J}}^{-+}_{-{\bm q}_3,\alpha\beta } U^*_{{\bm q}_2,\alpha\nu_2}U^*_{{\bm q}_3,\beta \nu_3}U_{{\bm q}_4,\gamma\nu_4}V^*_{{\bm q}_1,\delta\nu_1}
    \\ &\quad\quad \left.
        + \tilde{\mathcal{J}}^{-+}_{ {\bm q}_4,\alpha\beta } U^*_{{\bm q}_2,\alpha\nu_2}V_{{\bm q}_4,\beta \nu_4}V^*_{{\bm q}_1,\gamma\nu_1}V^*_{{\bm q}_3,\delta\nu_3}
        + \tilde{\mathcal{J}}^{-+}_{ {\bm q}_4,\alpha\beta } U^*_{{\bm q}_3,\alpha\nu_3}V_{{\bm q}_4,\beta \nu_4}V^*_{{\bm q}_2,\gamma\nu_2}V^*_{{\bm q}_1,\delta\nu_1}
    \right] \hat{b}^\dagger_{{\bm q}_a,\nu_a}\hat{b}^\dagger_{{\bm q}_b,\nu_b},
\end{aligned}
\end{equation}
\begin{equation}
\begin{aligned}
    \mathit{4}. &\quad \sum_{{\bm r},{\bm r}'}\mathcal{J}^{-+}_{{\bm r},{\bm r}'} \hat{a}^\dagger_{{\bm r}'}\hat{a}^\dagger_{{\bm r}'}\hat{a}_{{\bm r}}\hat{a}_{{\bm r}'} \\
    &\rightarrow \frac{1}{N_{\rm muc}} \sum_{{\bm q}_1,{\bm q}_2,{\bm q}_3,{\bm q}_4} \sum_{\nu_1\nu_2\nu_3\nu_4} \sum_{\alpha\beta\gamma\delta} 
    \delta_{\alpha\beta}\delta_{\alpha\delta} 
    \delta\left( {\bm q}_1+{\bm q}_2+{\bm q}_3-{\bm q}_4 \right) 
    \times \sum_{{\bm k}',{\bm q}_a,{\bm q}_b} \sum_{\nu'\nu_a\nu_b} \delta_{{\bm q}_1,{\bm k}'}\delta_{\nu_1,\nu'} \delta_{{\bm q}_2,{\bm q}_a}\delta_{\nu_2,\nu_a} \delta_{{\bm q}_3,{\bm q}_b}\delta_{\nu_3,\nu_b} \delta_{{\bm q}_4,{\bm k}'}\delta_{\nu_4,\nu'} \\
    &\quad \times \left[ 
          \tilde{\mathcal{J}}^{-+}_{ {\bm q}_2,\gamma\delta} V_{{\bm q}_4,\alpha\nu_4}U^*_{{\bm q}_1,\beta \nu_1}V^*_{{\bm q}_2,\gamma\nu_2}V^*_{{\bm q}_3,\delta\nu_3}
        + \tilde{\mathcal{J}}^{-+}_{ {\bm q}_1,\gamma\delta} V_{{\bm q}_4,\alpha\nu_4}U^*_{{\bm q}_2,\beta \nu_2}V^*_{{\bm q}_1,\gamma\nu_1}V^*_{{\bm q}_3,\delta\nu_3}
    \right. \\ &\quad\quad
        + \tilde{\mathcal{J}}^{-+}_{ {\bm q}_3,\gamma\delta} V_{{\bm q}_4,\alpha\nu_4}U^*_{{\bm q}_2,\beta \nu_2}V^*_{{\bm q}_3,\gamma\nu_3}V^*_{{\bm q}_1,\delta\nu_1}
        + \tilde{\mathcal{J}}^{-+}_{-{\bm q}_4,\gamma\delta} U^*_{{\bm q}_2,\alpha\nu_2}U^*_{{\bm q}_3,\beta \nu_3}U_{{\bm q}_4,\gamma\nu_4}V^*_{{\bm q}_1,\delta\nu_1}
    \\ &\quad\quad \left.
        + \tilde{\mathcal{J}}^{-+}_{ {\bm q}_1,\gamma\delta} U^*_{{\bm q}_2,\alpha\nu_2}V_{{\bm q}_4,\beta \nu_4}V^*_{{\bm q}_1,\gamma\nu_1}V^*_{{\bm q}_3,\delta\nu_3}
        + \tilde{\mathcal{J}}^{-+}_{ {\bm q}_2,\gamma\delta} U^*_{{\bm q}_3,\alpha\nu_3}V_{{\bm q}_4,\beta \nu_4}V^*_{{\bm q}_2,\gamma\nu_2}V^*_{{\bm q}_1,\delta\nu_1}
    \right] \hat{b}^\dagger_{{\bm q}_a,\nu_a}\hat{b}^\dagger_{{\bm q}_b,\nu_b},
\end{aligned}
\end{equation}
\begin{equation}
\begin{aligned}
    \mathit{5a}. &\quad \sum_{{\bm r},{\bm r}'}\mathcal{J}^{00}_{{\bm r},{\bm r}'} \hat{a}^\dagger_{{\bm r}}\hat{a}^\dagger_{{\bm r}'}\hat{a}_{{\bm r}}\hat{a}_{{\bm r}'} \\
    &\rightarrow \frac{1}{N_{\rm muc}} \sum_{{\bm q}_1,{\bm q}_2,{\bm q}_3,{\bm q}_4} \sum_{\nu_1\nu_2\nu_3\nu_4} \sum_{\alpha\beta\gamma\delta} 
    \delta_{\alpha\gamma}\delta_{\beta\delta} 
    \delta\left( {\bm q}_1+{\bm q}_2+{\bm q}_3-{\bm q}_4 \right)
    \times \sum_{{\bm k}',{\bm q}_a,{\bm q}_b} \sum_{\nu'\nu_a\nu_b} \delta_{{\bm q}_1,{\bm k}'}\delta_{\nu_1,\nu'} \delta_{{\bm q}_2,{\bm q}_a}\delta_{\nu_2,\nu_a} \delta_{{\bm q}_3,{\bm q}_b}\delta_{\nu_3,\nu_b} \delta_{{\bm q}_4,{\bm k}'}\delta_{\nu_4,\nu'} \\
    &\quad \times \left[ 
          \tilde{\mathcal{J}}^{00}_{-{\bm q}_4+{\bm q}_2,\alpha\beta } V_{{\bm q}_4,\alpha\nu_4}U^*_{{\bm q}_1,\beta \nu_1}V^*_{{\bm q}_2,\gamma\nu_2}V^*_{{\bm q}_3,\delta\nu_3}
        + \tilde{\mathcal{J}}^{00}_{-{\bm q}_4+{\bm q}_1,\alpha\beta } V_{{\bm q}_4,\alpha\nu_4}U^*_{{\bm q}_2,\beta \nu_2}V^*_{{\bm q}_1,\gamma\nu_1}V^*_{{\bm q}_3,\delta\nu_3}
    \right. \\ &\quad\quad
        + \tilde{\mathcal{J}}^{00}_{-{\bm q}_4+{\bm q}_3,\alpha\beta } V_{{\bm q}_4,\alpha\nu_4}U^*_{{\bm q}_2,\beta \nu_2}V^*_{{\bm q}_3,\gamma\nu_3}V^*_{{\bm q}_1,\delta\nu_1}
        + \tilde{\mathcal{J}}^{00}_{ {\bm q}_2-{\bm q}_4,\alpha\beta } U^*_{{\bm q}_2,\alpha\nu_2}U^*_{{\bm q}_3,\beta \nu_3}U_{{\bm q}_4,\gamma\nu_4}V^*_{{\bm q}_1,\delta\nu_1}
    \\ &\quad\quad \left.
        + \tilde{\mathcal{J}}^{00}_{ {\bm q}_2+{\bm q}_3,\alpha\beta } U^*_{{\bm q}_2,\alpha\nu_2}V_{{\bm q}_4,\beta \nu_4}V^*_{{\bm q}_1,\gamma\nu_1}V^*_{{\bm q}_3,\delta\nu_3}
        + \tilde{\mathcal{J}}^{00}_{ {\bm q}_3+{\bm q}_2,\alpha\beta } U^*_{{\bm q}_3,\alpha\nu_3}V_{{\bm q}_4,\beta \nu_4}V^*_{{\bm q}_2,\gamma\nu_2}V^*_{{\bm q}_1,\delta\nu_1}
    \right] \hat{b}^\dagger_{{\bm q}_a,\nu_a}\hat{b}^\dagger_{{\bm q}_b,\nu_b},
\end{aligned}
\end{equation}
\begin{equation}
\begin{aligned}
    \mathit{5b}. &\quad \sum_{{\bm r},{\bm r}'}\mathcal{J}^{00}_{{\bm r},{\bm r}'} \hat{a}^\dagger_{{\bm r}}\hat{a}^\dagger_{{\bm r}'}\hat{a}_{{\bm r}'}\hat{a}_{{\bm r}} \\
    &\rightarrow \frac{1}{N_{\rm muc}} \sum_{{\bm q}_1,{\bm q}_2,{\bm q}_3,{\bm q}_4} \sum_{\nu_1\nu_2\nu_3\nu_4} \sum_{\alpha\beta\gamma\delta} 
    \delta_{\alpha\delta}\delta_{\beta\gamma} 
    \delta\left( {\bm q}_1+{\bm q}_2+{\bm q}_3-{\bm q}_4 \right)
    \times \sum_{{\bm k}',{\bm q}_a,{\bm q}_b} \sum_{\nu'\nu_a\nu_b} \delta_{{\bm q}_1,{\bm k}'}\delta_{\nu_1,\nu'} \delta_{{\bm q}_2,{\bm q}_a}\delta_{\nu_2,\nu_a} \delta_{{\bm q}_3,{\bm q}_b}\delta_{\nu_3,\nu_b} \delta_{{\bm q}_4,{\bm k}'}\delta_{\nu_4,\nu'} \\
    &\quad \times \left[ 
          \tilde{\mathcal{J}}^{00}_{-{\bm q}_4+{\bm q}_3,\alpha\beta } V_{{\bm q}_4,\alpha\nu_4}U^*_{{\bm q}_1,\beta \nu_1}V^*_{{\bm q}_2,\gamma\nu_2}V^*_{{\bm q}_3,\delta\nu_3}
        + \tilde{\mathcal{J}}^{00}_{-{\bm q}_4+{\bm q}_3,\alpha\beta } V_{{\bm q}_4,\alpha\nu_4}U^*_{{\bm q}_2,\beta \nu_2}V^*_{{\bm q}_1,\gamma\nu_1}V^*_{{\bm q}_3,\delta\nu_3}
    \right. \\ &\quad\quad
        + \tilde{\mathcal{J}}^{00}_{-{\bm q}_4+{\bm q}_1,\alpha\beta } V_{{\bm q}_4,\alpha\nu_4}U^*_{{\bm q}_2,\beta \nu_2}V^*_{{\bm q}_3,\gamma\nu_3}V^*_{{\bm q}_1,\delta\nu_1}
        + \tilde{\mathcal{J}}^{00}_{ {\bm q}_2+{\bm q}_1,\alpha\beta } U^*_{{\bm q}_2,\alpha\nu_2}U^*_{{\bm q}_3,\beta \nu_3}U_{{\bm q}_4,\gamma\nu_4}V^*_{{\bm q}_1,\delta\nu_1}
    \\ &\quad\quad \left.
        + \tilde{\mathcal{J}}^{00}_{ {\bm q}_2+{\bm q}_3,\alpha\beta } U^*_{{\bm q}_2,\alpha\nu_2}V_{{\bm q}_4,\beta \nu_4}V^*_{{\bm q}_1,\gamma\nu_1}V^*_{{\bm q}_3,\delta\nu_3}
        + \tilde{\mathcal{J}}^{00}_{ {\bm q}_3+{\bm q}_1,\alpha\beta } U^*_{{\bm q}_3,\alpha\nu_3}V_{{\bm q}_4,\beta \nu_4}V^*_{{\bm q}_2,\gamma\nu_2}V^*_{{\bm q}_1,\delta\nu_1}
    \right] \hat{b}^\dagger_{{\bm q}_a,\nu_a}\hat{b}^\dagger_{{\bm q}_b,\nu_b},
\end{aligned}
\end{equation}
\begin{equation}
\begin{aligned}
    \mathit{5c}. &\quad \sum_{{\bm r},{\bm r}'}\mathcal{J}^{00}_{{\bm r},{\bm r}'} \hat{a}^\dagger_{{\bm r}'}\hat{a}^\dagger_{{\bm r}}\hat{a}_{{\bm r}}\hat{a}_{{\bm r}'} \\
    &\rightarrow \frac{1}{N_{\rm muc}} \sum_{{\bm q}_1,{\bm q}_2,{\bm q}_3,{\bm q}_4} \sum_{\nu_1\nu_2\nu_3\nu_4} \sum_{\alpha\beta\gamma\delta} 
    \delta_{\alpha\delta}\delta_{\beta\gamma} 
    \delta\left( {\bm q}_1+{\bm q}_2+{\bm q}_3-{\bm q}_4 \right)
    \times \sum_{{\bm k}',{\bm q}_a,{\bm q}_b} \sum_{\nu'\nu_a\nu_b} \delta_{{\bm q}_1,{\bm k}'}\delta_{\nu_1,\nu'} \delta_{{\bm q}_2,{\bm q}_a}\delta_{\nu_2,\nu_a} \delta_{{\bm q}_3,{\bm q}_b}\delta_{\nu_3,\nu_b} \delta_{{\bm q}_4,{\bm k}'}\delta_{\nu_4,\nu'} \\
    &\quad \times \left[ 
          \tilde{\mathcal{J}}^{00}_{ {\bm q}_1+{\bm q}_2,\beta\alpha } V_{{\bm q}_4,\alpha\nu_4}U^*_{{\bm q}_1,\beta \nu_1}V^*_{{\bm q}_2,\gamma\nu_2}V^*_{{\bm q}_3,\delta\nu_3}
        + \tilde{\mathcal{J}}^{00}_{ {\bm q}_2+{\bm q}_1,\beta\alpha } V_{{\bm q}_4,\alpha\nu_4}U^*_{{\bm q}_2,\beta \nu_2}V^*_{{\bm q}_1,\gamma\nu_1}V^*_{{\bm q}_3,\delta\nu_3}
    \right. \\ &\quad\quad
        + \tilde{\mathcal{J}}^{00}_{ {\bm q}_2+{\bm q}_3,\beta\alpha } V_{{\bm q}_4,\alpha\nu_4}U^*_{{\bm q}_2,\beta \nu_2}V^*_{{\bm q}_3,\gamma\nu_3}V^*_{{\bm q}_1,\delta\nu_1}
        + \tilde{\mathcal{J}}^{00}_{ {\bm q}_3-{\bm q}_4,\beta\alpha } U^*_{{\bm q}_2,\alpha\nu_2}U^*_{{\bm q}_3,\beta \nu_3}U_{{\bm q}_4,\gamma\nu_4}V^*_{{\bm q}_1,\delta\nu_1}
    \\ &\quad\quad \left.
        + \tilde{\mathcal{J}}^{00}_{-{\bm q}_4+{\bm q}_1,\beta\alpha } U^*_{{\bm q}_2,\alpha\nu_2}V_{{\bm q}_4,\beta \nu_4}V^*_{{\bm q}_1,\gamma\nu_1}V^*_{{\bm q}_3,\delta\nu_3}
        + \tilde{\mathcal{J}}^{00}_{-{\bm q}_4+{\bm q}_2,\beta\alpha } U^*_{{\bm q}_3,\alpha\nu_3}V_{{\bm q}_4,\beta \nu_4}V^*_{{\bm q}_2,\gamma\nu_2}V^*_{{\bm q}_1,\delta\nu_1}
    \right] \hat{b}^\dagger_{{\bm q}_a,\nu_a}\hat{b}^\dagger_{{\bm q}_b,\nu_b},
\end{aligned}
\end{equation}
\begin{equation}
\begin{aligned}
    \mathit{5d}. &\quad \sum_{{\bm r},{\bm r}'}\mathcal{J}^{00}_{{\bm r},{\bm r}'} \hat{a}^\dagger_{{\bm r}'}\hat{a}^\dagger_{{\bm r}}\hat{a}_{{\bm r}'}\hat{a}_{{\bm r}} \\
    &\rightarrow \frac{1}{N_{\rm muc}} \sum_{{\bm q}_1,{\bm q}_2,{\bm q}_3,{\bm q}_4} \sum_{\nu_1\nu_2\nu_3\nu_4} \sum_{\alpha\beta\gamma\delta} 
    \delta_{\alpha\gamma}\delta_{\beta\delta} 
    \delta\left( {\bm q}_1+{\bm q}_2+{\bm q}_3-{\bm q}_4 \right)
    \times \sum_{{\bm k}',{\bm q}_a,{\bm q}_b} \sum_{\nu'\nu_a\nu_b} \delta_{{\bm q}_1,{\bm k}'}\delta_{\nu_1,\nu'} \delta_{{\bm q}_2,{\bm q}_a}\delta_{\nu_2,\nu_a} \delta_{{\bm q}_3,{\bm q}_b}\delta_{\nu_3,\nu_b} \delta_{{\bm q}_4,{\bm k}'}\delta_{\nu_4,\nu'} \\
    &\quad \times \left[ 
          \tilde{\mathcal{J}}^{00}_{ {\bm q}_1+{\bm q}_3,\beta\alpha } V_{{\bm q}_4,\alpha\nu_4}U^*_{{\bm q}_1,\beta \nu_1}V^*_{{\bm q}_2,\gamma\nu_2}V^*_{{\bm q}_3,\delta\nu_3}
        + \tilde{\mathcal{J}}^{00}_{ {\bm q}_2+{\bm q}_3,\beta\alpha } V_{{\bm q}_4,\alpha\nu_4}U^*_{{\bm q}_2,\beta \nu_2}V^*_{{\bm q}_1,\gamma\nu_1}V^*_{{\bm q}_3,\delta\nu_3}
    \right. \\ &\quad\quad
        + \tilde{\mathcal{J}}^{00}_{ {\bm q}_2+{\bm q}_1,\beta\alpha } V_{{\bm q}_4,\alpha\nu_4}U^*_{{\bm q}_2,\beta \nu_2}V^*_{{\bm q}_3,\gamma\nu_3}V^*_{{\bm q}_1,\delta\nu_1}
        + \tilde{\mathcal{J}}^{00}_{ {\bm q}_3+{\bm q}_1,\beta\alpha } U^*_{{\bm q}_2,\alpha\nu_2}U^*_{{\bm q}_3,\beta \nu_3}U_{{\bm q}_4,\gamma\nu_4}V^*_{{\bm q}_1,\delta\nu_1}
    \\ &\quad\quad \left.
        + \tilde{\mathcal{J}}^{00}_{-{\bm q}_4+{\bm q}_3,\beta\alpha } U^*_{{\bm q}_2,\alpha\nu_2}V_{{\bm q}_4,\beta \nu_4}V^*_{{\bm q}_1,\gamma\nu_1}V^*_{{\bm q}_3,\delta\nu_3}
        + \tilde{\mathcal{J}}^{00}_{-{\bm q}_4+{\bm q}_1,\beta\alpha } U^*_{{\bm q}_3,\alpha\nu_3}V_{{\bm q}_4,\beta \nu_4}V^*_{{\bm q}_2,\gamma\nu_2}V^*_{{\bm q}_1,\delta\nu_1}
    \right] \hat{b}^\dagger_{{\bm q}_a,\nu_a}\hat{b}^\dagger_{{\bm q}_b,\nu_b},
\end{aligned}
\end{equation}
\begin{equation}
\begin{aligned}
    \mathit{6}. &\quad \sum_{{\bm r},{\bm r}'} \mathcal{J}^{++}_{{\bm r},{\bm r}'} \hat{a}^\dagger_{{\bm r}}\hat{a}^\dagger_{{\bm r}}\hat{a}^\dagger_{{\bm r}'}\hat{a}_{{\bm r}} \\
    &= \frac{1}{N_{\rm muc}} \sum_{\nu_1\nu_2\nu_3\nu_4} \sum_{\alpha\beta\gamma\delta} \sum_{{\bm q}_1,{\bm q}_2,{\bm q}_3,{\bm q}_4} \delta_{\alpha\beta}\delta_{\alpha\delta} \tilde{\mathcal{J}}^{++}_{-{\bm q}_3,\alpha\gamma} \delta\left( {\bm q}_1+{\bm q}_2+{\bm q}_3-{\bm q}_4 \right) \\
    &\quad\quad \times
    \left( U^*_{{\bm q}_1,\alpha\nu_1} \hat{b}^\dagger_{{\bm q}_1,\nu_1} + V_{-{\bm q}_1,\alpha\nu_1}   \hat{b}_{-{\bm q}_1,\nu_1}         \right)
    \left( U^*_{{\bm q}_2,\beta \nu_2} \hat{b}^\dagger_{{\bm q}_2,\nu_2} + V_{-{\bm q}_2,\beta \nu_2}   \hat{b}_{-{\bm q}_2,\nu_2}         \right) \\
    &\quad\quad \times
    \left( U^*_{{\bm q}_3,\gamma\nu_3} \hat{b}^\dagger_{{\bm q}_3,\nu_3} + V_{-{\bm q}_3,\gamma\nu_3}   \hat{b}_{-{\bm q}_3,\nu_3}         \right)
    \left( U_{{\bm q}_4,\delta\nu_4}   \hat{b}_{{\bm q}_4,\nu_4}         + V^*_{-{\bm q}_4,\delta\nu_4} \hat{b}^\dagger_{-{\bm q}_4,\nu_4} \right) \\
    &\rightarrow \frac{1}{N_{\rm muc}} \sum_{{\bm q}_1,{\bm q}_2,{\bm q}_3,{\bm q}_4} \sum_{\nu_1\nu_2\nu_3\nu_4} \sum_{\alpha\beta\gamma\delta} \delta_{\alpha\beta}\delta_{\alpha\delta} \delta\left( {\bm q}_1+{\bm q}_2+{\bm q}_3-{\bm q}_4 \right)
    \times \sum_{{\bm k}',{\bm q}_a,{\bm q}_b} \sum_{\nu'\nu_a\nu_b} \delta_{{\bm q}_1,{\bm k}'}\delta_{\nu_1,\nu'} \delta_{{\bm q}_2,{\bm q}_a}\delta_{\nu_2,\nu_a} \delta_{{\bm q}_3,{\bm q}_b}\delta_{\nu_3,\nu_b} \delta_{{\bm q}_4,{\bm k}'}\delta_{\nu_4,\nu'} \\
    &\quad\quad \times
    \left[
          \tilde{\mathcal{J}}^{++}_{-{\bm q}_2,\alpha\gamma} V_{{\bm q}_4,\alpha\nu_4}U^*_{{\bm q}_1,\beta \nu_1}U^*_{{\bm q}_2,\gamma\nu_2}V^*_{{\bm q}_3,\delta\nu_3}
        + \tilde{\mathcal{J}}^{++}_{-{\bm q}_1,\alpha\gamma} V_{{\bm q}_4,\alpha\nu_4}U^*_{{\bm q}_2,\beta \nu_2}U^*_{{\bm q}_1,\gamma\nu_1}V^*_{{\bm q}_3,\delta\nu_3} 
    \right. \\ &\quad\quad\quad
        + \tilde{\mathcal{J}}^{++}_{-{\bm q}_3,\alpha\gamma} V_{{\bm q}_4,\alpha\nu_4}U^*_{{\bm q}_2,\beta \nu_2}U^*_{{\bm q}_3,\gamma\nu_3}V^*_{{\bm q}_1,\delta\nu_1}
        + \tilde{\mathcal{J}}^{++}_{ {\bm q}_4,\alpha\gamma} U^*_{{\bm q}_2,\alpha\nu_2}U^*_{{\bm q}_3,\beta \nu_3}V_{{\bm q}_4,\gamma\nu_4}V^*_{{\bm q}_1,\delta\nu_1}
    \\ &\quad\quad\quad \left.
        + \tilde{\mathcal{J}}^{++}_{-{\bm q}_1,\alpha\gamma} U^*_{{\bm q}_2,\alpha\nu_2}V_{{\bm q}_4,\beta \nu_4}U^*_{{\bm q}_1,\gamma\nu_1}V^*_{{\bm q}_3,\delta\nu_3}
        + \tilde{\mathcal{J}}^{++}_{-{\bm q}_2,\alpha\gamma} U^*_{{\bm q}_3,\alpha\nu_3}V_{{\bm q}_4,\beta \nu_4}U^*_{{\bm q}_2,\gamma\nu_2}V^*_{{\bm q}_1,\delta\nu_1}
    \right] \hat{b}^\dagger_{{\bm q}_a,\nu_a}\hat{b}^\dagger_{{\bm q}_b,\nu_b},
\end{aligned}
\end{equation}
\begin{equation}
\begin{aligned}
    \mathit{7}. &\quad \sum_{{\bm r},{\bm r}'} \mathcal{J}^{++}_{{\bm r},{\bm r}'} \hat{a}^\dagger_{{\bm r}}\hat{a}^\dagger_{{\bm r}'}\hat{a}^\dagger_{{\bm r}'}\hat{a}_{{\bm r}'} \\
    &\rightarrow \frac{1}{N_{\rm muc}} \sum_{{\bm q}_1,{\bm q}_2,{\bm q}_3,{\bm q}_4} \sum_{\nu_1\nu_2\nu_3\nu_4} \sum_{\alpha\beta\gamma\delta} \delta_{\beta\gamma}\delta_{\beta\delta} \delta\left( {\bm q}_1+{\bm q}_2+{\bm q}_3-{\bm q}_4 \right)
    \times \sum_{{\bm k}',{\bm q}_a,{\bm q}_b} \sum_{\nu'\nu_a\nu_b} \delta_{{\bm q}_1,{\bm k}'}\delta_{\nu_1,\nu'} \delta_{{\bm q}_2,{\bm q}_a}\delta_{\nu_2,\nu_a} \delta_{{\bm q}_3,{\bm q}_b}\delta_{\nu_3,\nu_b} \delta_{{\bm q}_4,{\bm k}'}\delta_{\nu_4,\nu'} \\
    &\quad\quad \times
    \left[
          \tilde{\mathcal{J}}^{++}_{-{\bm q}_4,\alpha\delta} V_{{\bm q}_4,\alpha\nu_4}U^*_{{\bm q}_1,\beta \nu_1}U^*_{{\bm q}_2,\gamma\nu_2}V^*_{{\bm q}_3,\delta\nu_3}
        + \tilde{\mathcal{J}}^{++}_{-{\bm q}_4,\alpha\delta} V_{{\bm q}_4,\alpha\nu_4}U^*_{{\bm q}_2,\beta \nu_2}U^*_{{\bm q}_1,\gamma\nu_1}V^*_{{\bm q}_3,\delta\nu_3} 
    \right. \\ &\quad\quad\quad
        + \tilde{\mathcal{J}}^{++}_{-{\bm q}_4,\alpha\delta} V_{{\bm q}_4,\alpha\nu_4}U^*_{{\bm q}_2,\beta \nu_2}U^*_{{\bm q}_3,\gamma\nu_3}V^*_{{\bm q}_1,\delta\nu_1}
        + \tilde{\mathcal{J}}^{++}_{ {\bm q}_2,\alpha\delta} U^*_{{\bm q}_2,\alpha\nu_2}U^*_{{\bm q}_3,\beta \nu_3}V_{{\bm q}_4,\gamma\nu_4}V^*_{{\bm q}_1,\delta\nu_1}
    \\ &\quad\quad\quad \left.
        + \tilde{\mathcal{J}}^{++}_{ {\bm q}_2,\alpha\delta} U^*_{{\bm q}_2,\alpha\nu_2}V_{{\bm q}_4,\beta \nu_4}U^*_{{\bm q}_1,\gamma\nu_1}V^*_{{\bm q}_3,\delta\nu_3}
        + \tilde{\mathcal{J}}^{++}_{ {\bm q}_3,\alpha\delta} U^*_{{\bm q}_3,\alpha\nu_3}V_{{\bm q}_4,\beta \nu_4}U^*_{{\bm q}_2,\gamma\nu_2}V^*_{{\bm q}_1,\delta\nu_1}
    \right] \hat{b}^\dagger_{{\bm q}_a,\nu_a}\hat{b}^\dagger_{{\bm q}_b,\nu_b},
\end{aligned}
\end{equation}
\begin{equation}
\begin{aligned}
    \mathit{8}. &\quad \sum_{{\bm r},{\bm r}'} \mathcal{J}^{--}_{{\bm r},{\bm r}'} \hat{a}^\dagger_{{\bm r}}\hat{a}_{{\bm r}}\hat{a}_{{\bm r}}\hat{a}_{{\bm r}'} \\
    &= \frac{1}{N_{\rm muc}} \sum_{\nu_1\nu_2\nu_3\nu_4} \sum_{\alpha\beta\gamma\delta} \sum_{{\bm q_1},{\bm q}_2,{\bm q}_3,{\bm q}_4} \delta_{\alpha\beta}\delta_{\alpha\gamma} \tilde{\mathcal{J}}^{--}_{{\bm q}_4,\alpha\delta} \delta({\bm q}_1-{\bm q}_2-{\bm q}_3-{\bm q}_4) \\
    &\quad\quad \times 
    \left( U^*_{{\bm q}_1,\alpha\nu_1} \hat{b}^\dagger_{{\bm q}_1,\nu_1} + V_{-{\bm q}_1,\alpha\nu_1}   \hat{b}_{-{\bm q}_1,\nu_1}         \right)
    \left( U_{{\bm q}_2,\beta \nu_2}   \hat{b}_{{\bm q}_2,\nu_2}         + V^*_{-{\bm q}_2,\beta \nu_2} \hat{b}^\dagger_{-{\bm q}_2,\nu_2} \right) \\
    &\quad\quad \times
    \left( U_{{\bm q}_3,\gamma\nu_3}   \hat{b}_{{\bm q}_3,\nu_3}         + V^*_{-{\bm q}_3,\gamma\nu_3} \hat{b}^\dagger_{-{\bm q}_3,\nu_3} \right)
    \left( U_{{\bm q}_4,\delta\nu_4}   \hat{b}_{{\bm q}_4,\nu_4}         + V^*_{-{\bm q}_4,\delta\nu_4} \hat{b}^\dagger_{-{\bm q}_4,\nu_4} \right) \\
    &\rightarrow \frac{1}{N_{\rm muc}} \sum_{{\bm q_1},{\bm q}_2,{\bm q}_3,{\bm q}_4} \sum_{\nu_1\nu_2\nu_3\nu_4} \sum_{\alpha\beta\gamma\delta} \delta_{\alpha\beta}\delta_{\alpha\gamma} \delta({\bm q}_1+{\bm q}_2+{\bm q}_3-{\bm q}_4)
    \times \sum_{{\bm k}',{\bm q}_a,{\bm q}_b} \sum_{\nu'\nu_a\nu_b} \delta_{{\bm q}_1,{\bm k}'}\delta_{\nu_1,\nu'} \delta_{{\bm q}_2,{\bm q}_a}\delta_{\nu_2,\nu_a} \delta_{{\bm q}_3,{\bm q}_b}\delta_{\nu_3,\nu_b} \delta_{{\bm q}_4,{\bm k}'}\delta_{\nu_4,\nu'} \\
    &\quad\quad \times
    \left[
          \tilde{\mathcal{J}}^{--}_{-{\bm q}_3,\alpha\delta} V_{{\bm q}_4,\alpha\nu_4}V^*_{{\bm q}_1,\beta \nu_1}V^*_{{\bm q}_2,\gamma\nu_2}V^*_{{\bm q}_3,\delta\nu_3}
        + \tilde{\mathcal{J}}^{--}_{-{\bm q}_3,\alpha\delta} V_{{\bm q}_4,\alpha\nu_4}V^*_{{\bm q}_2,\beta \nu_2}V^*_{{\bm q}_1,\gamma\nu_1}V^*_{{\bm q}_3,\delta\nu_3}
    \right. \\ &\quad\quad\quad
        + \tilde{\mathcal{J}}^{--}_{-{\bm q}_1,\alpha\delta} V_{{\bm q}_4,\alpha\nu_4}V^*_{{\bm q}_2,\beta \nu_2}V^*_{{\bm q}_3,\gamma\nu_3}V^*_{{\bm q}_1,\delta\nu_1}
        + \tilde{\mathcal{J}}^{--}_{-{\bm q}_1,\alpha\delta} U^*_{{\bm q}_2,\alpha\nu_2}V^*_{{\bm q}_3,\beta \nu_3}U_{{\bm q}_4,\gamma\nu_4}V^*_{{\bm q}_1,\delta\nu_1}
    \\ &\quad\quad\quad \left.
        + \tilde{\mathcal{J}}^{--}_{-{\bm q}_3,\alpha\delta} U^*_{{\bm q}_2,\alpha\nu_2}U_{{\bm q}_4,\beta \nu_4}V^*_{{\bm q}_1,\gamma\nu_1}V^*_{{\bm q}_3,\delta\nu_3}
        + \tilde{\mathcal{J}}^{--}_{-{\bm q}_1,\alpha\delta} U^*_{{\bm q}_3,\alpha\nu_3}U_{{\bm q}_4,\beta \nu_4}V^*_{{\bm q}_2,\gamma\nu_2}V^*_{{\bm q}_1,\delta\nu_1}
    \right] \hat{b}^\dagger_{{\bm q}_a,\nu_a}\hat{b}^\dagger_{{\bm q}_b,\nu_b},
\end{aligned}
\end{equation}
\begin{equation}
\begin{aligned}
    \mathit{9}. &\quad \sum_{{\bm r},{\bm r}'} \mathcal{J}^{--}_{{\bm r},{\bm r}'} \hat{a}^\dagger_{{\bm r}'}\hat{a}_{{\bm r}}\hat{a}_{{\bm r}'}\hat{a}_{{\bm r}'} \\
    &\rightarrow \frac{1}{N_{\rm muc}} \sum_{{\bm q_1},{\bm q}_2,{\bm q}_3,{\bm q}_4} \sum_{\nu_1\nu_2\nu_3\nu_4} \sum_{\alpha\beta\gamma\delta} \delta_{\alpha\gamma}\delta_{\alpha\delta} \delta({\bm q}_1+{\bm q}_2+{\bm q}_3-{\bm q}_4)\times \sum_{{\bm k}',{\bm q}_a,{\bm q}_b} \sum_{\nu'\nu_a\nu_b} \delta_{{\bm q}_1,{\bm k}'}\delta_{\nu_1,\nu'} \delta_{{\bm q}_2,{\bm q}_a}\delta_{\nu_2,\nu_a} \delta_{{\bm q}_3,{\bm q}_b}\delta_{\nu_3,\nu_b} \delta_{{\bm q}_4,{\bm k}'}\delta_{\nu_4,\nu'} \\
    &\quad\quad \times
    \left[
          \tilde{\mathcal{J}}^{--}_{ {\bm q}_1,\beta\alpha } V_{{\bm q}_4,\alpha\nu_4}V^*_{{\bm q}_1,\beta \nu_1}V^*_{{\bm q}_2,\gamma\nu_2}V^*_{{\bm q}_3,\delta\nu_3}
        + \tilde{\mathcal{J}}^{--}_{ {\bm q}_2,\beta\alpha } V_{{\bm q}_4,\alpha\nu_4}V^*_{{\bm q}_2,\beta \nu_2}V^*_{{\bm q}_1,\gamma\nu_1}V^*_{{\bm q}_3,\delta\nu_3}
    \right. \\ &\quad\quad\quad
        + \tilde{\mathcal{J}}^{--}_{ {\bm q}_2,\beta\alpha } V_{{\bm q}_4,\alpha\nu_4}V^*_{{\bm q}_2,\beta \nu_2}V^*_{{\bm q}_3,\gamma\nu_3}V^*_{{\bm q}_1,\delta\nu_1}
        + \tilde{\mathcal{J}}^{--}_{ {\bm q}_3,\beta\alpha } U^*_{{\bm q}_2,\alpha\nu_2}V^*_{{\bm q}_3,\beta \nu_3}U_{{\bm q}_4,\gamma\nu_4}V^*_{{\bm q}_1,\delta\nu_1}
    \\ &\quad\quad\quad \left.
        + \tilde{\mathcal{J}}^{--}_{-{\bm q}_4,\beta\alpha } U^*_{{\bm q}_2,\alpha\nu_2}U_{{\bm q}_4,\beta \nu_4}V^*_{{\bm q}_1,\gamma\nu_1}V^*_{{\bm q}_3,\delta\nu_3}
        + \tilde{\mathcal{J}}^{--}_{-{\bm q}_4,\beta\alpha } U^*_{{\bm q}_3,\alpha\nu_3}U_{{\bm q}_4,\beta \nu_4}V^*_{{\bm q}_2,\gamma\nu_2}V^*_{{\bm q}_1,\delta\nu_1}.
    \right] \hat{b}^\dagger_{{\bm q}_a,\nu_a}\hat{b}^\dagger_{{\bm q}_b,\nu_b}.
\end{aligned}
\end{equation}
Taken together, the explicit expression of the unsymmetrized residual 3-1 vertex is given by
\begin{equation}
\begin{aligned}
    &\quad \ \mathcal{R}_{{\bm q}_1,{\bm q}_2,{\bm q}_3\leftarrow{\bm q}_4}^{\nu_1,\nu_2,\nu_3\leftarrow\nu_4} \\
    &= \frac{1}{4N_{\rm muc}} \sum_{\alpha,\beta,\gamma,\delta=1}^{N_{\rm sub}} \delta({\bm q}_1+{\bm q}_2+{\bm q}_3-{\bm q}_4) \\
    &\quad\quad \times \left[ 
        - \delta_{\alpha\beta }\delta_{\alpha\gamma} \tilde{\mathcal{J}}^{+-}_{-{\bm q}_3,\gamma\delta} V_{{\bm q}_4,\alpha\nu_4}U^*_{{\bm q}_1,\beta \nu_1}V^*_{{\bm q}_2,\gamma\nu_2}V^*_{{\bm q}_3,\delta\nu_3}
        - \delta_{\alpha\beta }\delta_{\alpha\gamma} \tilde{\mathcal{J}}^{+-}_{-{\bm q}_3,\gamma\delta} V_{{\bm q}_4,\alpha\nu_4}U^*_{{\bm q}_2,\beta \nu_2}V^*_{{\bm q}_1,\gamma\nu_1}V^*_{{\bm q}_3,\delta\nu_3}
    \right. \\ &\quad\quad\quad
        - \delta_{\alpha\beta }\delta_{\alpha\gamma} \tilde{\mathcal{J}}^{+-}_{-{\bm q}_1,\gamma\delta} V_{{\bm q}_4,\alpha\nu_4}U^*_{{\bm q}_2,\beta \nu_2}V^*_{{\bm q}_3,\gamma\nu_3}V^*_{{\bm q}_1,\delta\nu_1}
        - \delta_{\alpha\beta }\delta_{\alpha\gamma} \tilde{\mathcal{J}}^{+-}_{-{\bm q}_1,\gamma\delta} U^*_{{\bm q}_2,\alpha\nu_2}U^*_{{\bm q}_3,\beta \nu_3}U_{{\bm q}_4,\gamma\nu_4}V^*_{{\bm q}_1,\delta\nu_1}
    \\ &\quad\quad\quad
        - \delta_{\alpha\beta }\delta_{\alpha\gamma} \tilde{\mathcal{J}}^{+-}_{-{\bm q}_3,\gamma\delta} U^*_{{\bm q}_2,\alpha\nu_2}V_{{\bm q}_4,\beta \nu_4}V^*_{{\bm q}_1,\gamma\nu_1}V^*_{{\bm q}_3,\delta\nu_3}
        - \delta_{\alpha\beta }\delta_{\alpha\gamma} \tilde{\mathcal{J}}^{+-}_{-{\bm q}_1,\gamma\delta} U^*_{{\bm q}_3,\alpha\nu_3}V_{{\bm q}_4,\beta \nu_4}V^*_{{\bm q}_2,\gamma\nu_2}V^*_{{\bm q}_1,\delta\nu_1}
    \\ &\quad\quad\quad
        - \delta_{\beta\delta }\delta_{\gamma\delta} \tilde{\mathcal{J}}^{+-}_{-{\bm q}_4,\alpha\beta } V_{{\bm q}_4,\alpha\nu_4}U^*_{{\bm q}_1,\beta \nu_1}V^*_{{\bm q}_2,\gamma\nu_2}V^*_{{\bm q}_3,\delta\nu_3}
        - \delta_{\beta\delta }\delta_{\gamma\delta} \tilde{\mathcal{J}}^{+-}_{-{\bm q}_4,\alpha\beta } V_{{\bm q}_4,\alpha\nu_4}U^*_{{\bm q}_2,\beta \nu_2}V^*_{{\bm q}_1,\gamma\nu_1}V^*_{{\bm q}_3,\delta\nu_3}
    \\ &\quad\quad\quad
        - \delta_{\beta\delta }\delta_{\gamma\delta} \tilde{\mathcal{J}}^{+-}_{-{\bm q}_4,\alpha\beta } V_{{\bm q}_4,\alpha\nu_4}U^*_{{\bm q}_2,\beta \nu_2}V^*_{{\bm q}_3,\gamma\nu_3}V^*_{{\bm q}_1,\delta\nu_1}
        - \delta_{\beta\delta }\delta_{\gamma\delta} \tilde{\mathcal{J}}^{+-}_{ {\bm q}_2,\alpha\beta } U^*_{{\bm q}_2,\alpha\nu_2}U^*_{{\bm q}_3,\beta \nu_3}U_{{\bm q}_4,\gamma\nu_4}V^*_{{\bm q}_1,\delta\nu_1}
    \\ &\quad\quad\quad
        - \delta_{\beta\delta }\delta_{\gamma\delta} \tilde{\mathcal{J}}^{+-}_{ {\bm q}_2,\alpha\beta } U^*_{{\bm q}_2,\alpha\nu_2}V_{{\bm q}_4,\beta \nu_4}V^*_{{\bm q}_1,\gamma\nu_1}V^*_{{\bm q}_3,\delta\nu_3}
        - \delta_{\beta\delta }\delta_{\gamma\delta} \tilde{\mathcal{J}}^{+-}_{ {\bm q}_3,\alpha\beta } U^*_{{\bm q}_3,\alpha\nu_3}V_{{\bm q}_4,\beta \nu_4}V^*_{{\bm q}_2,\gamma\nu_2}V^*_{{\bm q}_1,\delta\nu_1}
    \\ &\quad\quad\quad
        - \delta_{\alpha\gamma}\delta_{\alpha\delta} \tilde{\mathcal{J}}^{-+}_{-{\bm q}_1,\alpha\beta } V_{{\bm q}_4,\alpha\nu_4}U^*_{{\bm q}_1,\beta \nu_1}V^*_{{\bm q}_2,\gamma\nu_2}V^*_{{\bm q}_3,\delta\nu_3}
        - \delta_{\alpha\gamma}\delta_{\alpha\delta} \tilde{\mathcal{J}}^{-+}_{-{\bm q}_2,\alpha\beta } V_{{\bm q}_4,\alpha\nu_4}U^*_{{\bm q}_2,\beta \nu_2}V^*_{{\bm q}_1,\gamma\nu_1}V^*_{{\bm q}_3,\delta\nu_3}
    \\ &\quad\quad\quad
        - \delta_{\alpha\gamma}\delta_{\alpha\delta} \tilde{\mathcal{J}}^{-+}_{-{\bm q}_2,\alpha\beta } V_{{\bm q}_4,\alpha\nu_4}U^*_{{\bm q}_2,\beta \nu_2}V^*_{{\bm q}_3,\gamma\nu_3}V^*_{{\bm q}_1,\delta\nu_1}
        - \delta_{\alpha\gamma}\delta_{\alpha\delta} \tilde{\mathcal{J}}^{-+}_{-{\bm q}_3,\alpha\beta } U^*_{{\bm q}_2,\alpha\nu_2}U^*_{{\bm q}_3,\beta \nu_3}U_{{\bm q}_4,\gamma\nu_4}V^*_{{\bm q}_1,\delta\nu_1}
    \\ &\quad\quad\quad
        - \delta_{\alpha\gamma}\delta_{\alpha\delta} \tilde{\mathcal{J}}^{-+}_{ {\bm q}_4,\alpha\beta } U^*_{{\bm q}_2,\alpha\nu_2}V_{{\bm q}_4,\beta \nu_4}V^*_{{\bm q}_1,\gamma\nu_1}V^*_{{\bm q}_3,\delta\nu_3}
        - \delta_{\alpha\gamma}\delta_{\alpha\delta} \tilde{\mathcal{J}}^{-+}_{ {\bm q}_4,\alpha\beta } U^*_{{\bm q}_3,\alpha\nu_3}V_{{\bm q}_4,\beta \nu_4}V^*_{{\bm q}_2,\gamma\nu_2}V^*_{{\bm q}_1,\delta\nu_1}
    \\ &\quad\quad\quad
        - \delta_{\alpha\beta }\delta_{\alpha\delta} \tilde{\mathcal{J}}^{-+}_{ {\bm q}_2,\gamma\delta} V_{{\bm q}_4,\alpha\nu_4}U^*_{{\bm q}_1,\beta \nu_1}V^*_{{\bm q}_2,\gamma\nu_2}V^*_{{\bm q}_3,\delta\nu_3}
        - \delta_{\alpha\beta }\delta_{\alpha\delta} \tilde{\mathcal{J}}^{-+}_{ {\bm q}_1,\gamma\delta} V_{{\bm q}_4,\alpha\nu_4}U^*_{{\bm q}_2,\beta \nu_2}V^*_{{\bm q}_1,\gamma\nu_1}V^*_{{\bm q}_3,\delta\nu_3}
    \\ &\quad\quad\quad
        - \delta_{\alpha\beta }\delta_{\alpha\delta} \tilde{\mathcal{J}}^{-+}_{ {\bm q}_3,\gamma\delta} V_{{\bm q}_4,\alpha\nu_4}U^*_{{\bm q}_2,\beta \nu_2}V^*_{{\bm q}_3,\gamma\nu_3}V^*_{{\bm q}_1,\delta\nu_1}
        - \delta_{\alpha\beta }\delta_{\alpha\delta} \tilde{\mathcal{J}}^{-+}_{-{\bm q}_4,\gamma\delta} U^*_{{\bm q}_2,\alpha\nu_2}U^*_{{\bm q}_3,\beta \nu_3}U_{{\bm q}_4,\gamma\nu_4}V^*_{{\bm q}_1,\delta\nu_1}
    \\ &\quad\quad\quad
        - \delta_{\alpha\beta }\delta_{\alpha\delta} \tilde{\mathcal{J}}^{-+}_{ {\bm q}_1,\gamma\delta} U^*_{{\bm q}_2,\alpha\nu_2}V_{{\bm q}_4,\beta \nu_4}V^*_{{\bm q}_1,\gamma\nu_1}V^*_{{\bm q}_3,\delta\nu_3}
        - \delta_{\alpha\beta }\delta_{\alpha\delta} \tilde{\mathcal{J}}^{-+}_{ {\bm q}_2,\gamma\delta} U^*_{{\bm q}_3,\alpha\nu_3}V_{{\bm q}_4,\beta \nu_4}V^*_{{\bm q}_2,\gamma\nu_2}V^*_{{\bm q}_1,\delta\nu_1}
    \\ &\quad\quad\quad
        + \delta_{\alpha\gamma}\delta_{\beta\delta } \tilde{\mathcal{J}}^{00}_{-{\bm q}_4+{\bm q}_2,\alpha\beta } V_{{\bm q}_4,\alpha\nu_4}U^*_{{\bm q}_1,\beta \nu_1}V^*_{{\bm q}_2,\gamma\nu_2}V^*_{{\bm q}_3,\delta\nu_3}
        + \delta_{\alpha\gamma}\delta_{\beta\delta } \tilde{\mathcal{J}}^{00}_{-{\bm q}_4+{\bm q}_1,\alpha\beta } V_{{\bm q}_4,\alpha\nu_4}U^*_{{\bm q}_2,\beta \nu_2}V^*_{{\bm q}_1,\gamma\nu_1}V^*_{{\bm q}_3,\delta\nu_3}
    \\ &\quad\quad\quad
        + \delta_{\alpha\gamma}\delta_{\beta\delta } \tilde{\mathcal{J}}^{00}_{-{\bm q}_4+{\bm q}_3,\alpha\beta } V_{{\bm q}_4,\alpha\nu_4}U^*_{{\bm q}_2,\beta \nu_2}V^*_{{\bm q}_3,\gamma\nu_3}V^*_{{\bm q}_1,\delta\nu_1}
        + \delta_{\alpha\gamma}\delta_{\beta\delta } \tilde{\mathcal{J}}^{00}_{ {\bm q}_2-{\bm q}_4,\alpha\beta } U^*_{{\bm q}_2,\alpha\nu_2}U^*_{{\bm q}_3,\beta \nu_3}U_{{\bm q}_4,\gamma\nu_4}V^*_{{\bm q}_1,\delta\nu_1}
    \\ &\quad\quad\quad
        + \delta_{\alpha\gamma}\delta_{\beta\delta } \tilde{\mathcal{J}}^{00}_{ {\bm q}_2+{\bm q}_3,\alpha\beta } U^*_{{\bm q}_2,\alpha\nu_2}V_{{\bm q}_4,\beta \nu_4}V^*_{{\bm q}_1,\gamma\nu_1}V^*_{{\bm q}_3,\delta\nu_3}
        + \delta_{\alpha\gamma}\delta_{\beta\delta } \tilde{\mathcal{J}}^{00}_{ {\bm q}_3+{\bm q}_2,\alpha\beta } U^*_{{\bm q}_3,\alpha\nu_3}V_{{\bm q}_4,\beta \nu_4}V^*_{{\bm q}_2,\gamma\nu_2}V^*_{{\bm q}_1,\delta\nu_1}
    \\ &\quad\quad\quad
        + \delta_{\alpha\delta}\delta_{\beta\gamma } \tilde{\mathcal{J}}^{00}_{-{\bm q}_4+{\bm q}_3,\alpha\beta } V_{{\bm q}_4,\alpha\nu_4}U^*_{{\bm q}_1,\beta \nu_1}V^*_{{\bm q}_2,\gamma\nu_2}V^*_{{\bm q}_3,\delta\nu_3}
        + \delta_{\alpha\delta}\delta_{\beta\gamma } \tilde{\mathcal{J}}^{00}_{-{\bm q}_4+{\bm q}_3,\alpha\beta } V_{{\bm q}_4,\alpha\nu_4}U^*_{{\bm q}_2,\beta \nu_2}V^*_{{\bm q}_1,\gamma\nu_1}V^*_{{\bm q}_3,\delta\nu_3}
    \\ &\quad\quad\quad
        + \delta_{\alpha\delta}\delta_{\beta\gamma } \tilde{\mathcal{J}}^{00}_{-{\bm q}_4+{\bm q}_1,\alpha\beta } V_{{\bm q}_4,\alpha\nu_4}U^*_{{\bm q}_2,\beta \nu_2}V^*_{{\bm q}_3,\gamma\nu_3}V^*_{{\bm q}_1,\delta\nu_1}
        + \delta_{\alpha\delta}\delta_{\beta\gamma } \tilde{\mathcal{J}}^{00}_{ {\bm q}_2+{\bm q}_1,\alpha\beta } U^*_{{\bm q}_2,\alpha\nu_2}U^*_{{\bm q}_3,\beta \nu_3}U_{{\bm q}_4,\gamma\nu_4}V^*_{{\bm q}_1,\delta\nu_1}
    \\ &\quad\quad\quad
        + \delta_{\alpha\delta}\delta_{\beta\gamma } \tilde{\mathcal{J}}^{00}_{ {\bm q}_2+{\bm q}_3,\alpha\beta } U^*_{{\bm q}_2,\alpha\nu_2}V_{{\bm q}_4,\beta \nu_4}V^*_{{\bm q}_1,\gamma\nu_1}V^*_{{\bm q}_3,\delta\nu_3}
        + \delta_{\alpha\delta}\delta_{\beta\gamma } \tilde{\mathcal{J}}^{00}_{ {\bm q}_3+{\bm q}_1,\alpha\beta } U^*_{{\bm q}_3,\alpha\nu_3}V_{{\bm q}_4,\beta \nu_4}V^*_{{\bm q}_2,\gamma\nu_2}V^*_{{\bm q}_1,\delta\nu_1}
    \\ &\quad\quad\quad
        + \delta_{\alpha\delta}\delta_{\beta\gamma } \tilde{\mathcal{J}}^{00}_{ {\bm q}_1+{\bm q}_2,\beta\alpha } V_{{\bm q}_4,\alpha\nu_4}U^*_{{\bm q}_1,\beta \nu_1}V^*_{{\bm q}_2,\gamma\nu_2}V^*_{{\bm q}_3,\delta\nu_3}
        + \delta_{\alpha\delta}\delta_{\beta\gamma } \tilde{\mathcal{J}}^{00}_{ {\bm q}_2+{\bm q}_1,\beta\alpha } V_{{\bm q}_4,\alpha\nu_4}U^*_{{\bm q}_2,\beta \nu_2}V^*_{{\bm q}_1,\gamma\nu_1}V^*_{{\bm q}_3,\delta\nu_3}
    \\ &\quad\quad\quad
        + \delta_{\alpha\delta}\delta_{\beta\gamma } \tilde{\mathcal{J}}^{00}_{ {\bm q}_2+{\bm q}_3,\beta\alpha } V_{{\bm q}_4,\alpha\nu_4}U^*_{{\bm q}_2,\beta \nu_2}V^*_{{\bm q}_3,\gamma\nu_3}V^*_{{\bm q}_1,\delta\nu_1}
        + \delta_{\alpha\delta}\delta_{\beta\gamma } \tilde{\mathcal{J}}^{00}_{ {\bm q}_3-{\bm q}_4,\beta\alpha } U^*_{{\bm q}_2,\alpha\nu_2}U^*_{{\bm q}_3,\beta \nu_3}U_{{\bm q}_4,\gamma\nu_4}V^*_{{\bm q}_1,\delta\nu_1}
    \\ &\quad\quad\quad
        + \delta_{\alpha\delta}\delta_{\beta\gamma } \tilde{\mathcal{J}}^{00}_{-{\bm q}_4+{\bm q}_1,\beta\alpha } U^*_{{\bm q}_2,\alpha\nu_2}V_{{\bm q}_4,\beta \nu_4}V^*_{{\bm q}_1,\gamma\nu_1}V^*_{{\bm q}_3,\delta\nu_3}
        + \delta_{\alpha\delta}\delta_{\beta\gamma } \tilde{\mathcal{J}}^{00}_{-{\bm q}_4+{\bm q}_2,\beta\alpha } U^*_{{\bm q}_3,\alpha\nu_3}V_{{\bm q}_4,\beta \nu_4}V^*_{{\bm q}_2,\gamma\nu_2}V^*_{{\bm q}_1,\delta\nu_1}
    \\ &\quad\quad\quad
        + \delta_{\alpha\gamma}\delta_{\beta\delta } \tilde{\mathcal{J}}^{00}_{ {\bm q}_1+{\bm q}_3,\beta\alpha } V_{{\bm q}_4,\alpha\nu_4}U^*_{{\bm q}_1,\beta \nu_1}V^*_{{\bm q}_2,\gamma\nu_2}V^*_{{\bm q}_3,\delta\nu_3}
        + \delta_{\alpha\gamma}\delta_{\beta\delta } \tilde{\mathcal{J}}^{00}_{ {\bm q}_2+{\bm q}_3,\beta\alpha } V_{{\bm q}_4,\alpha\nu_4}U^*_{{\bm q}_2,\beta \nu_2}V^*_{{\bm q}_1,\gamma\nu_1}V^*_{{\bm q}_3,\delta\nu_3}
    \\ &\quad\quad\quad
        + \delta_{\alpha\gamma}\delta_{\beta\delta } \tilde{\mathcal{J}}^{00}_{ {\bm q}_2+{\bm q}_1,\beta\alpha } V_{{\bm q}_4,\alpha\nu_4}U^*_{{\bm q}_2,\beta \nu_2}V^*_{{\bm q}_3,\gamma\nu_3}V^*_{{\bm q}_1,\delta\nu_1}
        + \delta_{\alpha\gamma}\delta_{\beta\delta } \tilde{\mathcal{J}}^{00}_{ {\bm q}_3+{\bm q}_1,\beta\alpha } U^*_{{\bm q}_2,\alpha\nu_2}U^*_{{\bm q}_3,\beta \nu_3}U_{{\bm q}_4,\gamma\nu_4}V^*_{{\bm q}_1,\delta\nu_1}
    \\ &\quad\quad\quad
        + \delta_{\alpha\gamma}\delta_{\beta\delta } \tilde{\mathcal{J}}^{00}_{-{\bm q}_4+{\bm q}_3,\beta\alpha } U^*_{{\bm q}_2,\alpha\nu_2}V_{{\bm q}_4,\beta \nu_4}V^*_{{\bm q}_1,\gamma\nu_1}V^*_{{\bm q}_3,\delta\nu_3}
        + \delta_{\alpha\gamma}\delta_{\beta\delta } \tilde{\mathcal{J}}^{00}_{-{\bm q}_4+{\bm q}_1,\beta\alpha } U^*_{{\bm q}_3,\alpha\nu_3}V_{{\bm q}_4,\beta \nu_4}V^*_{{\bm q}_2,\gamma\nu_2}V^*_{{\bm q}_1,\delta\nu_1}
    \\ &\quad\quad\quad
        - \delta_{\alpha\beta }\delta_{\alpha\delta} \tilde{\mathcal{J}}^{++}_{-{\bm q}_2,\alpha\gamma} V_{{\bm q}_4,\alpha\nu_4}U^*_{{\bm q}_1,\beta \nu_1}U^*_{{\bm q}_2,\gamma\nu_2}V^*_{{\bm q}_3,\delta\nu_3}
        - \delta_{\alpha\beta }\delta_{\alpha\delta} \tilde{\mathcal{J}}^{++}_{-{\bm q}_1,\alpha\gamma} V_{{\bm q}_4,\alpha\nu_4}U^*_{{\bm q}_2,\beta \nu_2}U^*_{{\bm q}_1,\gamma\nu_1}V^*_{{\bm q}_3,\delta\nu_3} 
    \\ &\quad\quad\quad
        - \delta_{\alpha\beta }\delta_{\alpha\delta} \tilde{\mathcal{J}}^{++}_{-{\bm q}_3,\alpha\gamma} V_{{\bm q}_4,\alpha\nu_4}U^*_{{\bm q}_2,\beta \nu_2}U^*_{{\bm q}_3,\gamma\nu_3}V^*_{{\bm q}_1,\delta\nu_1}
        - \delta_{\alpha\beta }\delta_{\alpha\delta} \tilde{\mathcal{J}}^{++}_{ {\bm q}_4,\alpha\gamma} U^*_{{\bm q}_2,\alpha\nu_2}U^*_{{\bm q}_3,\beta \nu_3}V_{{\bm q}_4,\gamma\nu_4}V^*_{{\bm q}_1,\delta\nu_1}
    \\ &\quad\quad\quad
        - \delta_{\alpha\beta }\delta_{\alpha\delta} \tilde{\mathcal{J}}^{++}_{-{\bm q}_1,\alpha\gamma} U^*_{{\bm q}_2,\alpha\nu_2}V_{{\bm q}_4,\beta \nu_4}U^*_{{\bm q}_1,\gamma\nu_1}V^*_{{\bm q}_3,\delta\nu_3}
        - \delta_{\alpha\beta }\delta_{\alpha\delta} \tilde{\mathcal{J}}^{++}_{-{\bm q}_2,\alpha\gamma} U^*_{{\bm q}_3,\alpha\nu_3}V_{{\bm q}_4,\beta \nu_4}U^*_{{\bm q}_2,\gamma\nu_2}V^*_{{\bm q}_1,\delta\nu_1}
    \\ &\quad\quad\quad
        - \delta_{\beta\gamma }\delta_{\beta\delta } \tilde{\mathcal{J}}^{++}_{-{\bm q}_4,\alpha\delta} V_{{\bm q}_4,\alpha\nu_4}U^*_{{\bm q}_1,\beta \nu_1}U^*_{{\bm q}_2,\gamma\nu_2}V^*_{{\bm q}_3,\delta\nu_3}
        - \delta_{\beta\gamma }\delta_{\beta\delta } \tilde{\mathcal{J}}^{++}_{-{\bm q}_4,\alpha\delta} V_{{\bm q}_4,\alpha\nu_4}U^*_{{\bm q}_2,\beta \nu_2}U^*_{{\bm q}_1,\gamma\nu_1}V^*_{{\bm q}_3,\delta\nu_3} 
    \\ &\quad\quad\quad
        - \delta_{\beta\gamma }\delta_{\beta\delta } \tilde{\mathcal{J}}^{++}_{-{\bm q}_4,\alpha\delta} V_{{\bm q}_4,\alpha\nu_4}U^*_{{\bm q}_2,\beta \nu_2}U^*_{{\bm q}_3,\gamma\nu_3}V^*_{{\bm q}_1,\delta\nu_1}
        - \delta_{\beta\gamma }\delta_{\beta\delta } \tilde{\mathcal{J}}^{++}_{ {\bm q}_2,\alpha\delta} U^*_{{\bm q}_2,\alpha\nu_2}U^*_{{\bm q}_3,\beta \nu_3}V_{{\bm q}_4,\gamma\nu_4}V^*_{{\bm q}_1,\delta\nu_1}
    \\ &\quad\quad\quad
        - \delta_{\beta\gamma }\delta_{\beta\delta } \tilde{\mathcal{J}}^{++}_{ {\bm q}_2,\alpha\delta} U^*_{{\bm q}_2,\alpha\nu_2}V_{{\bm q}_4,\beta \nu_4}U^*_{{\bm q}_1,\gamma\nu_1}V^*_{{\bm q}_3,\delta\nu_3}
        - \delta_{\beta\gamma }\delta_{\beta\delta } \tilde{\mathcal{J}}^{++}_{ {\bm q}_3,\alpha\delta} U^*_{{\bm q}_3,\alpha\nu_3}V_{{\bm q}_4,\beta \nu_4}U^*_{{\bm q}_2,\gamma\nu_2}V^*_{{\bm q}_1,\delta\nu_1}
    \\ &\quad\quad\quad
        - \delta_{\alpha\beta }\delta_{\alpha\gamma} \tilde{\mathcal{J}}^{--}_{-{\bm q}_3,\alpha\delta} V_{{\bm q}_4,\alpha\nu_4}V^*_{{\bm q}_1,\beta \nu_1}V^*_{{\bm q}_2,\gamma\nu_2}V^*_{{\bm q}_3,\delta\nu_3}
        - \delta_{\alpha\beta }\delta_{\alpha\gamma} \tilde{\mathcal{J}}^{--}_{-{\bm q}_3,\alpha\delta} V_{{\bm q}_4,\alpha\nu_4}V^*_{{\bm q}_2,\beta \nu_2}V^*_{{\bm q}_1,\gamma\nu_1}V^*_{{\bm q}_3,\delta\nu_3}
    \\ &\quad\quad\quad
        - \delta_{\alpha\beta }\delta_{\alpha\gamma} \tilde{\mathcal{J}}^{--}_{-{\bm q}_1,\alpha\delta} V_{{\bm q}_4,\alpha\nu_4}V^*_{{\bm q}_2,\beta \nu_2}V^*_{{\bm q}_3,\gamma\nu_3}V^*_{{\bm q}_1,\delta\nu_1}
        - \delta_{\alpha\beta }\delta_{\alpha\gamma} \tilde{\mathcal{J}}^{--}_{-{\bm q}_1,\alpha\delta} U^*_{{\bm q}_2,\alpha\nu_2}V^*_{{\bm q}_3,\beta \nu_3}U_{{\bm q}_4,\gamma\nu_4}V^*_{{\bm q}_1,\delta\nu_1}
    \\ &\quad\quad\quad
        - \delta_{\alpha\beta }\delta_{\alpha\gamma} \tilde{\mathcal{J}}^{--}_{-{\bm q}_3,\alpha\delta} U^*_{{\bm q}_2,\alpha\nu_2}U_{{\bm q}_4,\beta \nu_4}V^*_{{\bm q}_1,\gamma\nu_1}V^*_{{\bm q}_3,\delta\nu_3}
        - \delta_{\alpha\beta }\delta_{\alpha\gamma} \tilde{\mathcal{J}}^{--}_{-{\bm q}_1,\alpha\delta} U^*_{{\bm q}_3,\alpha\nu_3}U_{{\bm q}_4,\beta \nu_4}V^*_{{\bm q}_2,\gamma\nu_2}V^*_{{\bm q}_1,\delta\nu_1}
    \\ &\quad\quad\quad
        - \delta_{\alpha\gamma}\delta_{\alpha\delta} \tilde{\mathcal{J}}^{--}_{ {\bm q}_1,\beta\alpha } V_{{\bm q}_4,\alpha\nu_4}V^*_{{\bm q}_1,\beta \nu_1}V^*_{{\bm q}_2,\gamma\nu_2}V^*_{{\bm q}_3,\delta\nu_3}
        - \delta_{\alpha\gamma}\delta_{\alpha\delta} \tilde{\mathcal{J}}^{--}_{ {\bm q}_2,\beta\alpha } V_{{\bm q}_4,\alpha\nu_4}V^*_{{\bm q}_2,\beta \nu_2}V^*_{{\bm q}_1,\gamma\nu_1}V^*_{{\bm q}_3,\delta\nu_3}
    \\ &\quad\quad\quad
        - \delta_{\alpha\gamma}\delta_{\alpha\delta} \tilde{\mathcal{J}}^{--}_{ {\bm q}_2,\beta\alpha } V_{{\bm q}_4,\alpha\nu_4}V^*_{{\bm q}_2,\beta \nu_2}V^*_{{\bm q}_3,\gamma\nu_3}V^*_{{\bm q}_1,\delta\nu_1}
        - \delta_{\alpha\gamma}\delta_{\alpha\delta} \tilde{\mathcal{J}}^{--}_{ {\bm q}_3,\beta\alpha } U^*_{{\bm q}_2,\alpha\nu_2}V^*_{{\bm q}_3,\beta \nu_3}U_{{\bm q}_4,\gamma\nu_4}V^*_{{\bm q}_1,\delta\nu_1}
    \\ &\quad\quad\quad \left.
        - \delta_{\alpha\gamma}\delta_{\alpha\delta} \tilde{\mathcal{J}}^{--}_{-{\bm q}_4,\beta\alpha } U^*_{{\bm q}_2,\alpha\nu_2}U_{{\bm q}_4,\beta \nu_4}V^*_{{\bm q}_1,\gamma\nu_1}V^*_{{\bm q}_3,\delta\nu_3}
        - \delta_{\alpha\gamma}\delta_{\alpha\delta} \tilde{\mathcal{J}}^{--}_{-{\bm q}_4,\beta\alpha } U^*_{{\bm q}_3,\alpha\nu_3}U_{{\bm q}_4,\beta \nu_4}V^*_{{\bm q}_2,\gamma\nu_2}V^*_{{\bm q}_1,\delta\nu_1}
    \right].
\end{aligned}
\end{equation}
\subsubsection{Four-out-zero-in term $\tilde{\mathcal{H}}^{(40)}$}
\begin{figure}[htpb]
    \includegraphics[width=4.0cm]{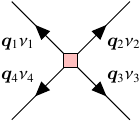}
    \caption{Symmetrized 4-0 vertex $\tilde{\mathcal{A}}_{{\bm q}_1,{\bm q}_2,{\bm q}_3,{\bm q}_4}^{\nu_1,\nu_2,\nu_3,\nu_4}$.}
\end{figure}
The explicit form of $\tilde{\mathcal{H}}^{(40)}$ is given by
\begin{equation}
    \mathcal{H}^{(40)} = \sum_{{\bm q}_1,{\bm q}_2,{\bm q}_3,{\bm q}_4} \sum_{\nu_1,\nu_2,\nu_3,\nu_4} 
    \tilde{\mathcal{A}}_{{\bm q}_1,{\bm q}_2,{\bm q}_3,{\bm q}_4}^{\nu_1,\nu_2,\nu_3,\nu_4} \
    \hat{b}^\dagger_{{\bm q}_1,\nu_1} \hat{b}^\dagger_{{\bm q}_2,\nu_2} \hat{b}^\dagger_{{\bm q}_3,\nu_3}\hat{b}^\dagger_{{\bm q}_4,\nu_4}
\end{equation}
where $\tilde{\mathcal{A}}_{{\bm q}_1,{\bm q}_2,{\bm q}_3,{\bm q}_4}^{\nu_1,\nu_2,\nu_3,\nu_4}$ is the symmetrized four-out-zero-in vertex given by,
\begin{equation}
\begin{aligned}
    \tilde{\mathcal{A}}_{{\bm q}_1,{\bm q}_2,{\bm q}_3,{\bm q}_4}^{\nu_1,\nu_2,\nu_3,\nu_4} &= \frac{1}{2} \frac{1}{4!} \left(
      \mathcal{A}_{{\bm q}_1,{\bm q}_2,{\bm q}_3,{\bm q}_4}^{\nu_1,\nu_2,\nu_3,\nu_4}
    + \mathcal{A}_{{\bm q}_1,{\bm q}_2,{\bm q}_4,{\bm q}_3}^{\nu_1,\nu_2,\nu_4,\nu_3}
    + \mathcal{A}_{{\bm q}_1,{\bm q}_3,{\bm q}_2,{\bm q}_4}^{\nu_1,\nu_3,\nu_2,\nu_4}
    + \mathcal{A}_{{\bm q}_1,{\bm q}_3,{\bm q}_4,{\bm q}_2}^{\nu_1,\nu_3,\nu_4,\nu_2}
    + \mathcal{A}_{{\bm q}_1,{\bm q}_4,{\bm q}_2,{\bm q}_3}^{\nu_1,\nu_4,\nu_2,\nu_3}
    + \mathcal{A}_{{\bm q}_1,{\bm q}_4,{\bm q}_3,{\bm q}_2}^{\nu_1,\nu_4,\nu_3,\nu_2} \right. \\
    &\quad\quad
    + \mathcal{A}_{{\bm q}_2,{\bm q}_1,{\bm q}_3,{\bm q}_4}^{\nu_2,\nu_1,\nu_3,\nu_4}
    + \mathcal{A}_{{\bm q}_2,{\bm q}_1,{\bm q}_4,{\bm q}_3}^{\nu_2,\nu_1,\nu_4,\nu_3}
    + \mathcal{A}_{{\bm q}_2,{\bm q}_3,{\bm q}_1,{\bm q}_4}^{\nu_2,\nu_3,\nu_1,\nu_4}
    + \mathcal{A}_{{\bm q}_2,{\bm q}_3,{\bm q}_4,{\bm q}_1}^{\nu_2,\nu_3,\nu_4,\nu_1}
    + \mathcal{A}_{{\bm q}_2,{\bm q}_4,{\bm q}_1,{\bm q}_3}^{\nu_2,\nu_4,\nu_1,\nu_3}
    + \mathcal{A}_{{\bm q}_2,{\bm q}_4,{\bm q}_3,{\bm q}_1}^{\nu_2,\nu_4,\nu_3,\nu_1} \\
    &\quad\quad
    + \mathcal{A}_{{\bm q}_3,{\bm q}_1,{\bm q}_2,{\bm q}_4}^{\nu_3,\nu_1,\nu_2,\nu_4}
    + \mathcal{A}_{{\bm q}_3,{\bm q}_1,{\bm q}_4,{\bm q}_2}^{\nu_3,\nu_1,\nu_4,\nu_2}
    + \mathcal{A}_{{\bm q}_3,{\bm q}_2,{\bm q}_1,{\bm q}_4}^{\nu_3,\nu_2,\nu_1,\nu_4}
    + \mathcal{A}_{{\bm q}_3,{\bm q}_2,{\bm q}_4,{\bm q}_1}^{\nu_3,\nu_2,\nu_4,\nu_1}
    + \mathcal{A}_{{\bm q}_3,{\bm q}_4,{\bm q}_1,{\bm q}_2}^{\nu_3,\nu_4,\nu_1,\nu_2}
    + \mathcal{A}_{{\bm q}_3,{\bm q}_4,{\bm q}_2,{\bm q}_1}^{\nu_3,\nu_4,\nu_2,\nu_1} \\
    &\quad\quad \left.
    + \mathcal{A}_{{\bm q}_4,{\bm q}_1,{\bm q}_2,{\bm q}_3}^{\nu_4,\nu_1,\nu_2,\nu_3}
    + \mathcal{A}_{{\bm q}_4,{\bm q}_1,{\bm q}_3,{\bm q}_2}^{\nu_4,\nu_1,\nu_3,\nu_2}
    + \mathcal{A}_{{\bm q}_4,{\bm q}_2,{\bm q}_1,{\bm q}_3}^{\nu_4,\nu_2,\nu_1,\nu_3}
    + \mathcal{A}_{{\bm q}_4,{\bm q}_2,{\bm q}_3,{\bm q}_1}^{\nu_4,\nu_2,\nu_3,\nu_1}
    + \mathcal{A}_{{\bm q}_4,{\bm q}_3,{\bm q}_1,{\bm q}_2}^{\nu_4,\nu_3,\nu_1,\nu_2}
    + \mathcal{A}_{{\bm q}_4,{\bm q}_3,{\bm q}_2,{\bm q}_1}^{\nu_4,\nu_3,\nu_2,\nu_1} \right).
\end{aligned}
\end{equation}
Note that $\tilde{\mathcal{H}}^{(04)}$ is given by the hermitian conjugate of $\tilde{\mathcal{H}}^{(40)}$. Once again, to get the explicit form of the unsymmetrized vertex $\mathcal{A}_{{\bm q}_1,{\bm q}_2,{\bm q}_3,{\bm q}_4}^{\nu_1,\nu_2,\nu_3,\nu_4}$, we need to pick up the four-out-zero-in part of the four magnon terms $\mathit{1},\mathit{2},\cdots,\mathit{9}$ in Eqs.~(\ref{eq:HP_SJS}) and (\ref{eq:HP4_SJS}). We list all contributions in the following:
\begin{equation}
\begin{aligned}
    \mathit{1.} &\quad \sum_{{\bm r},{\bm r}'} \mathcal{J}^{+-}_{{\bm r},{\bm r}'} \hat{a}^\dagger_{{\bm r}}\hat{a}^\dagger_{{\bm r}}\hat{a}_{{\bm r}}\hat{a}_{{\bm r}'} \\
    &= \frac{1}{N_{\rm muc}} \sum_{\nu_1\nu_2\nu_3\nu_4} \sum_{\alpha\beta\gamma\delta} \sum_{{\bm q}_1,{\bm q}_2,{\bm q}_3,{\bm q}_4} \delta_{\alpha\beta}\delta_{\alpha\gamma} \tilde{\mathcal{J}}^{+-}_{{\bm q}_4,\gamma\delta}  \delta\left( {\bm q}_1+{\bm q}_2-{\bm q}_3-{\bm q}_4 \right) \\
    &\quad\quad \times 
    \left( U^*_{{\bm q}_1,\alpha\nu_1} \hat{b}^\dagger_{{\bm q}_1,\nu_1} +   V_{-{\bm q}_1,\alpha\nu_1}         \hat{b}_{-{\bm q}_1,\nu_1} \right)
    \left( U^*_{{\bm q}_2,\beta \nu_2} \hat{b}^\dagger_{{\bm q}_2,\nu_2} +   V_{-{\bm q}_2,\beta \nu_2}         \hat{b}_{-{\bm q}_2,\nu_2} \right) \\
    &\quad\quad \times
    \left(   U_{{\bm q}_3,\gamma\nu_3}         \hat{b}_{{\bm q}_3,\nu_3} + V^*_{-{\bm q}_3,\gamma\nu_3} \hat{b}^\dagger_{-{\bm q}_3,\nu_3} \right)
    \left(   U_{{\bm q}_4,\delta\nu_4}         \hat{b}_{{\bm q}_4,\nu_4} + V^*_{-{\bm q}_4,\delta\nu_4} \hat{b}^\dagger_{-{\bm q}_4,\nu_4} \right) \\
    &\rightarrow \frac{1}{N} \sum_{{\bm q}_1,{\bm q}_2,{\bm q}_3,{\bm q}_4} \sum_{\nu_1\nu_2\nu_3\nu_4} \sum_{\alpha\beta\gamma\delta} \delta_{\alpha\beta}\delta_{\alpha\gamma} \tilde{\mathcal{J}}^{+-}_{-{\bm q}_4,\gamma\delta}  \delta\left( {\bm q}_1+{\bm q}_2+{\bm q}_3+{\bm q}_4 \right) U^*_{{\bm q}_1,\alpha\nu_1}U^*_{{\bm q}_2,\beta \nu_2}V^*_{{\bm q}_3,\gamma\nu_3}V^*_{{\bm q}_4,\delta\nu_4} \hat{b}^\dagger_{{\bm q}_1,\nu_1}\hat{b}^\dagger_{{\bm q}_2,\nu_2}\hat{b}^\dagger_{{\bm q}_3,\nu_3}\hat{b}^\dagger_{{\bm q}_4,\nu_4},
\end{aligned}
\end{equation}
\begin{equation}
\begin{aligned}
    \mathit{2.} &\quad \sum_{{\bm r},{\bm r}'}\mathcal{J}^{+-}_{{\bm r},{\bm r}'} \hat{a}^\dagger_{{\bm r}}\hat{a}^\dagger_{{\bm r}'}\hat{a}_{{\bm r}'}\hat{a}_{{\bm r}'} \\
    &\rightarrow \frac{1}{N_{\rm muc}} \sum_{{\bm q}_1,{\bm q}_2,{\bm q}_3,{\bm q}_4} \sum_{\nu_1\nu_2\nu_3\nu_4} \sum_{\alpha\beta\gamma\delta} \delta_{\beta\delta}\delta_{\gamma\delta} \tilde{\mathcal{J}}^{+-}_{{\bm q}_1,\alpha\beta} \delta\left( {\bm q}_1+{\bm q}_2+{\bm q}_3+{\bm q}_4 \right) U^*_{{\bm q}_1,\alpha\nu_1}U^*_{{\bm q}_2,\beta \nu_2}V^*_{{\bm q}_3,\gamma\nu_3}V^*_{{\bm q}_4,\delta\nu_4} \hat{b}^\dagger_{{\bm q}_1,\nu_1}\hat{b}^\dagger_{{\bm q}_2,\nu_2}\hat{b}^\dagger_{{\bm q}_3,\nu_3}\hat{b}^\dagger_{{\bm q}_4,\nu_4},
\end{aligned}
\end{equation}
\begin{equation}
\begin{aligned}
    \mathit{3.} &\quad \sum_{{\bm r},{\bm r}'}\mathcal{J}^{-+}_{{\bm r},{\bm r}'} \hat{a}^\dagger_{{\bm r}}\hat{a}^\dagger_{{\bm r}'}\hat{a}_{{\bm r}}\hat{a}_{{\bm r}} \\
    &\rightarrow \frac{1}{N_{\rm muc}} \sum_{{\bm q}_1,{\bm q}_2,{\bm q}_3,{\bm q}_4} \sum_{\nu_1\nu_2\nu_3\nu_4} \sum_{\alpha\beta\gamma\delta} \delta_{\alpha\gamma}\delta_{\alpha\delta} \tilde{\mathcal{J}}^{-+}_{-{\bm q}_2,\alpha\beta} \delta\left( {\bm q}_1+{\bm q}_2+{\bm q}_3+{\bm q}_4 \right) U^*_{{\bm q}_1,\alpha\nu_1}U^*_{{\bm q}_2,\beta \nu_2}V^*_{{\bm q}_3,\gamma\nu_3}V^*_{{\bm q}_4,\delta\nu_4} \hat{b}^\dagger_{{\bm q}_1,\nu_1}\hat{b}^\dagger_{{\bm q}_2,\nu_2}\hat{b}^\dagger_{{\bm q}_3,\nu_3}\hat{b}^\dagger_{{\bm q}_4,\nu_4},
\end{aligned}
\end{equation}
\begin{equation}
\begin{aligned}
    \mathit{4.} &\quad \sum_{{\bm r},{\bm r}'}\mathcal{J}^{-+}_{{\bm r},{\bm r}'} \hat{a}^\dagger_{{\bm r}'}\hat{a}^\dagger_{{\bm r}'}\hat{a}_{{\bm r}}\hat{a}_{{\bm r}'} \\
    &\rightarrow \frac{1}{N_{\rm muc}} \sum_{{\bm q}_1,{\bm q}_2,{\bm q}_3,{\bm q}_4} \sum_{\nu_1\nu_2\nu_3\nu_4} \sum_{\alpha\beta\gamma\delta} \delta_{\alpha\beta}\delta_{\alpha\delta} \tilde{\mathcal{J}}^{-+}_{{\bm q}_3,\gamma\delta} \delta\left( {\bm q}_1+{\bm q}_2+{\bm q}_3+{\bm q}_4 \right) U^*_{{\bm q}_1,\alpha\nu_1}U^*_{{\bm q}_2,\beta \nu_2}V^*_{{\bm q}_3,\gamma\nu_3}V^*_{{\bm q}_4,\delta\nu_4} \hat{b}^\dagger_{{\bm q}_1,\nu_1}\hat{b}^\dagger_{{\bm q}_2,\nu_2}\hat{b}^\dagger_{{\bm q}_3,\nu_3}\hat{b}^\dagger_{{\bm q}_4,\nu_4},
\end{aligned}
\end{equation}
\begin{equation}
\begin{aligned}
    \mathit{5a.} &\quad \sum_{{\bm r},{\bm r}'}\mathcal{J}^{00}_{{\bm r},{\bm r}'} \hat{a}^\dagger_{{\bm r}}\hat{a}^\dagger_{{\bm r}'}\hat{a}_{{\bm r}}\hat{a}_{{\bm r}'} \\
    &\rightarrow \frac{1}{N_{\rm muc}} \sum_{{\bm q}_1,{\bm q}_2,{\bm q}_3,{\bm q}_4} \sum_{\nu_1\nu_2\nu_3\nu_4} \sum_{\alpha\beta\gamma\delta} \delta_{\alpha\gamma}\delta_{\beta\delta} \tilde{\mathcal{J}}^{00}_{{\bm q}_1+{\bm q}_3,\alpha\beta} \delta\left( {\bm q}_1+{\bm q}_2+{\bm q}_3+{\bm q}_4 \right) U^*_{{\bm q}_1,\alpha\nu_1}U^*_{{\bm q}_2,\beta \nu_2}V^*_{{\bm q}_3,\gamma\nu_3}V^*_{{\bm q}_4,\delta\nu_4} \hat{b}^\dagger_{{\bm q}_1,\nu_1}\hat{b}^\dagger_{{\bm q}_2,\nu_2}\hat{b}^\dagger_{{\bm q}_3,\nu_3}\hat{b}^\dagger_{{\bm q}_4,\nu_4},
\end{aligned}
\end{equation}
\begin{equation}
\begin{aligned}
    \mathit{5b.} &\quad \sum_{{\bm r},{\bm r}'}\mathcal{J}^{00}_{{\bm r},{\bm r}'} \hat{a}^\dagger_{{\bm r}}\hat{a}^\dagger_{{\bm r}'}\hat{a}_{{\bm r}'}\hat{a}_{{\bm r}} \\
    &\rightarrow \frac{1}{N_{\rm muc}} \sum_{{\bm q}_1,{\bm q}_2,{\bm q}_3,{\bm q}_4} \sum_{\nu_1\nu_2\nu_3\nu_4} \sum_{\alpha\beta\gamma\delta} \delta_{\alpha\delta}\delta_{\beta\gamma} \tilde{\mathcal{J}}^{00}_{{\bm q}_1+{\bm q}_4,\alpha\beta} \delta\left( {\bm q}_1+{\bm q}_2+{\bm q}_3+{\bm q}_4 \right) U^*_{{\bm q}_1,\alpha\nu_1}U^*_{{\bm q}_2,\beta \nu_2}V^*_{{\bm q}_3,\gamma\nu_3}V^*_{{\bm q}_4,\delta\nu_4} \hat{b}^\dagger_{{\bm q}_1,\nu_1}\hat{b}^\dagger_{{\bm q}_2,\nu_2}\hat{b}^\dagger_{{\bm q}_3,\nu_3}\hat{b}^\dagger_{{\bm q}_4,\nu_4},
\end{aligned}
\end{equation}
\begin{equation}
\begin{aligned}
    \mathit{5c.} &\quad \sum_{{\bm r},{\bm r}'}\mathcal{J}^{00}_{{\bm r},{\bm r}'} \hat{a}^\dagger_{{\bm r}'}\hat{a}^\dagger_{{\bm r}}\hat{a}_{{\bm r}}\hat{a}_{{\bm r}'} \\
    &\rightarrow \frac{1}{N_{\rm muc}} \sum_{{\bm q}_1,{\bm q}_2,{\bm q}_3,{\bm q}_4} \sum_{\nu_1\nu_2\nu_3\nu_4} \sum_{\alpha\beta\gamma\delta} \delta_{\alpha\delta}\delta_{\beta\gamma} \tilde{\mathcal{J}}^{00}_{{\bm q}_2+{\bm q}_3,\beta\alpha} \delta\left( {\bm q}_1+{\bm q}_2+{\bm q}_3+{\bm q}_4 \right) U^*_{{\bm q}_1,\alpha\nu_1}U^*_{{\bm q}_2,\beta \nu_2}V^*_{{\bm q}_3,\gamma\nu_3}V^*_{{\bm q}_4,\delta\nu_4} \hat{b}^\dagger_{{\bm q}_1,\nu_1}\hat{b}^\dagger_{{\bm q}_2,\nu_2}\hat{b}^\dagger_{{\bm q}_3,\nu_3}\hat{b}^\dagger_{{\bm q}_4,\nu_4},
\end{aligned}
\end{equation}
\begin{equation}
\begin{aligned}
    \mathit{5d.} &\quad \sum_{{\bm r},{\bm r}'}\mathcal{J}^{00}_{{\bm r},{\bm r}'} \hat{a}^\dagger_{{\bm r}'}\hat{a}^\dagger_{{\bm r}}\hat{a}_{{\bm r}'}\hat{a}_{{\bm r}} \\
    &\rightarrow \frac{1}{N_{\rm muc}} \sum_{{\bm q}_1,{\bm q}_2,{\bm q}_3,{\bm q}_4} \sum_{\nu_1\nu_2\nu_3\nu_4} \sum_{\alpha\beta\gamma\delta} \delta_{\alpha\gamma}\delta_{\beta\delta} \tilde{\mathcal{J}}^{00}_{{\bm q}_2+{\bm q}_4,\beta\alpha} \delta\left( {\bm q}_1+{\bm q}_2+{\bm q}_3+{\bm q}_4 \right) U^*_{{\bm q}_1,\alpha\nu_1}U^*_{{\bm q}_2,\beta \nu_2}V^*_{{\bm q}_3,\gamma\nu_3}V^*_{{\bm q}_4,\delta\nu_4} \hat{b}^\dagger_{{\bm q}_1,\nu_1}\hat{b}^\dagger_{{\bm q}_2,\nu_2}\hat{b}^\dagger_{{\bm q}_3,\nu_3}\hat{b}^\dagger_{{\bm q}_4,\nu_4},
\end{aligned}
\end{equation}
\begin{equation}
\begin{aligned}
    \mathit{6}. &\quad \sum_{{\bm r},{\bm r}'} \mathcal{J}^{++}_{{\bm r},{\bm r}'} \hat{a}^\dagger_{{\bm r}}\hat{a}^\dagger_{{\bm r}}\hat{a}^\dagger_{{\bm r}'}\hat{a}_{{\bm r}} \\
    &= \frac{1}{N_{\rm muc}} \sum_{\nu_1\nu_2\nu_3\nu_4} \sum_{\alpha\beta\gamma\delta} \sum_{{\bm q}_1,{\bm q}_2,{\bm q}_3,{\bm q}_4} \delta_{\alpha\beta}\delta_{\alpha\delta} \tilde{\mathcal{J}}^{++}_{-{\bm q}_3,\alpha\gamma} \delta\left( {\bm q}_1+{\bm q}_2+{\bm q}_3-{\bm q}_4 \right) \\
    &\quad\quad \times
    \left( U^*_{{\bm q}_1,\alpha\nu_1} \hat{b}^\dagger_{{\bm q}_1,\nu_1} + V_{-{\bm q}_1,\alpha\nu_1}   \hat{b}_{-{\bm q}_1,\nu_1}         \right)
    \left( U^*_{{\bm q}_2,\beta \nu_2} \hat{b}^\dagger_{{\bm q}_2,\nu_2} + V_{-{\bm q}_2,\beta \nu_2}   \hat{b}_{-{\bm q}_2,\nu_2}         \right) \\
    &\quad\quad \times
    \left( U^*_{{\bm q}_3,\gamma\nu_3} \hat{b}^\dagger_{{\bm q}_3,\nu_3} + V_{-{\bm q}_3,\gamma\nu_3}   \hat{b}_{-{\bm q}_3,\nu_3}         \right)
    \left( U_{{\bm q}_4,\delta\nu_4}   \hat{b}_{{\bm q}_4,\nu_4}         + V^*_{-{\bm q}_4,\delta\nu_4} \hat{b}^\dagger_{-{\bm q}_4,\nu_4} \right) \\
    &\rightarrow \frac{1}{N_{\rm muc}} \sum_{{\bm q}_1,{\bm q}_2,{\bm q}_3,{\bm q}_4} \sum_{\nu_1\nu_2\nu_3\nu_4} \sum_{\alpha\beta\gamma\delta} \delta_{\alpha\beta}\delta_{\alpha\delta} \tilde{\mathcal{J}}^{++}_{-{\bm q}_3,\alpha\gamma} \delta\left( {\bm q}_1+{\bm q}_2+{\bm q}_3+{\bm q}_4 \right) U^*_{{\bm q}_1,\alpha\nu_1}U^*_{{\bm q}_2,\beta \nu_2}U^*_{{\bm q}_3,\gamma\nu_3}V^*_{{\bm q}_4,\delta\nu_4} \hat{b}^\dagger_{{\bm q}_1,\nu_1}\hat{b}^\dagger_{{\bm q}_2,\nu_2}\hat{b}^\dagger_{{\bm q}_3,\nu_3}\hat{b}^\dagger_{{\bm q}_4,\nu_4},
\end{aligned}
\end{equation}
\begin{equation}
\begin{aligned}
    \mathit{7}. &\quad \sum_{{\bm r},{\bm r}'} \mathcal{J}^{++}_{{\bm r},{\bm r}'} \hat{a}^\dagger_{{\bm r}}\hat{a}^\dagger_{{\bm r}'}\hat{a}^\dagger_{{\bm r}'}\hat{a}_{{\bm r}'} \\
    &\rightarrow \frac{1}{N_{\rm muc}} \sum_{{\bm q}_1,{\bm q}_2,{\bm q}_3,{\bm q}_4} \sum_{\nu_1\nu_2\nu_3\nu_4} \sum_{\alpha\beta\gamma\delta} \delta_{\beta\gamma}\delta_{\beta\delta} \tilde{\mathcal{J}}^{++}_{{\bm q}_1,\alpha\delta} \delta\left( {\bm q}_1+{\bm q}_2+{\bm q}_3+{\bm q}_4 \right) U^*_{{\bm q}_1,\alpha\nu_1}U^*_{{\bm q}_2,\beta \nu_2}U^*_{{\bm q}_3,\gamma\nu_3}V^*_{{\bm q}_4,\delta\nu_4} \hat{b}^\dagger_{{\bm q}_1,\nu_1}\hat{b}^\dagger_{{\bm q}_2,\nu_2}\hat{b}^\dagger_{{\bm q}_3,\nu_3}\hat{b}^\dagger_{{\bm q}_4,\nu_4},
\end{aligned}
\end{equation}
\begin{equation}
\begin{aligned}
    \mathit{8}. &\quad \sum_{{\bm r},{\bm r}'} \mathcal{J}^{--}_{{\bm r},{\bm r}'} \hat{a}^\dagger_{{\bm r}}\hat{a}_{{\bm r}}\hat{a}_{{\bm r}}\hat{a}_{{\bm r}'} \\
    &= \frac{1}{N_{\rm muc}} \sum_{\nu_1\nu_2\nu_3\nu_4} \sum_{\alpha\beta\gamma\delta} \sum_{{\bm q_1},{\bm q}_2,{\bm q}_3,{\bm q}_4} \delta_{\alpha\beta}\delta_{\alpha\gamma} \tilde{\mathcal{J}}^{--}_{{\bm q}_4,\alpha\delta} \delta({\bm q}_1-{\bm q}_2-{\bm q}_3-{\bm q}_4) \\
    &\quad\quad \times 
    \left( U^*_{{\bm q}_1,\alpha\nu_1} \hat{b}^\dagger_{{\bm q}_1,\nu_1} + V_{-{\bm q}_1,\alpha\nu_1}   \hat{b}_{-{\bm q}_1,\nu_1}         \right)
    \left( U_{{\bm q}_2,\beta \nu_2}   \hat{b}_{{\bm q}_2,\nu_2}         + V^*_{-{\bm q}_2,\beta \nu_2} \hat{b}^\dagger_{-{\bm q}_2,\nu_2} \right) \\
    &\quad\quad \times
    \left( U_{{\bm q}_3,\gamma\nu_3}   \hat{b}_{{\bm q}_3,\nu_3}         + V^*_{-{\bm q}_3,\gamma\nu_3} \hat{b}^\dagger_{-{\bm q}_3,\nu_3} \right)
    \left( U_{{\bm q}_4,\delta\nu_4}   \hat{b}_{{\bm q}_4,\nu_4}         + V^*_{-{\bm q}_4,\delta\nu_4} \hat{b}^\dagger_{-{\bm q}_4,\nu_4} \right) \\
    &\rightarrow \frac{1}{N_{\rm muc}} \sum_{{\bm q_1},{\bm q}_2,{\bm q}_3,{\bm q}_4} \sum_{\nu_1\nu_2\nu_3\nu_4} \sum_{\alpha\beta\gamma\delta} \delta_{\alpha\beta}\delta_{\alpha\gamma} \tilde{\mathcal{J}}^{--}_{-{\bm q}_4,\alpha\delta} \delta({\bm q}_1+{\bm q}_2+{\bm q}_3+{\bm q}_4)  U^*_{{\bm q}_1,\alpha\nu_1}V^*_{{\bm q}_2,\beta \nu_2}V^*_{{\bm q}_3,\gamma\nu_3}V^*_{{\bm q}_4,\delta\nu_4} \hat{b}^\dagger_{{\bm q}_1,\nu_1}\hat{b}^\dagger_{{\bm q}_2,\nu_2}\hat{b}^\dagger_{{\bm q}_3,\nu_3}\hat{b}^\dagger_{{\bm q}_4,\nu_4},
\end{aligned}
\end{equation}
\begin{equation}
\begin{aligned}
    \mathit{9}. &\quad \sum_{{\bm r},{\bm r}'} \mathcal{J}^{--}_{{\bm r},{\bm r}'} \hat{a}^\dagger_{{\bm r}'}\hat{a}_{{\bm r}}\hat{a}_{{\bm r}'}\hat{a}_{{\bm r}'} \\
    &\rightarrow \frac{1}{N_{\rm muc}} \sum_{{\bm q}_1,{\bm q}_2,{\bm q}_3,{\bm q}_4} \sum_{\nu_1\nu_2\nu_3\nu_4} \sum_{\alpha\beta\gamma\delta} \delta_{\alpha\gamma}\delta_{\alpha\delta} \tilde{\mathcal{J}}^{--}_{{\bm q}_2,\beta\alpha} \delta\left( {\bm q}_1+{\bm q}_2+{\bm q}_3+{\bm q}_4 \right) U^*_{{\bm q}_1,\alpha\nu_1}V^*_{{\bm q}_2,\beta \nu_2}V^*_{{\bm q}_3,\gamma\nu_3}V^*_{{\bm q}_4,\delta\nu_4} \hat{b}^\dagger_{{\bm q}_1,\nu_1}\hat{b}^\dagger_{{\bm q}_2,\nu_2}\hat{b}^\dagger_{{\bm q}_3,\nu_3}\hat{b}^\dagger_{{\bm q}_4,\nu_4}.
\end{aligned}
\end{equation}
Taken together, the explicit expression of the unsymmetrized 4-0 vertex is given by
\begin{equation}
\begin{aligned}
    \mathcal{A}_{{\bm q}_1,{\bm q}_2,{\bm q}_3,{\bm q}_4}^{\nu_1,\nu_2,\nu_3,\nu_4}
    &=  \frac{1}{4N_{\rm muc}} \sum_{\alpha,\beta,\gamma,\delta=1}^{N_{\rm sub}} \delta({\bm q}_1+{\bm q}_2+{\bm q}_3+{\bm q}_4) \\
    &\quad\quad \left[ \left(
          \delta_{\alpha\gamma}\delta_{\beta\delta } \tilde{\mathcal{J}}^{00}_{ {\bm q}_1+{\bm q}_3,\alpha\beta }
        + \delta_{\alpha\delta}\delta_{\beta\gamma } \tilde{\mathcal{J}}^{00}_{ {\bm q}_1+{\bm q}_4,\alpha\beta }
        + \delta_{\alpha\delta}\delta_{\beta\gamma } \tilde{\mathcal{J}}^{00}_{ {\bm q}_2+{\bm q}_3,\beta\alpha }
        + \delta_{\alpha\gamma}\delta_{\beta\delta } \tilde{\mathcal{J}}^{00}_{ {\bm q}_2+{\bm q}_4,\beta\alpha } \right.
    \right. \\ &\quad\quad\quad \left. \left.
        - \delta_{\alpha\beta }\delta_{\alpha\gamma} \tilde{\mathcal{J}}^{+-}_{-{\bm q}_4,\gamma\delta} 
        - \delta_{\beta\delta }\delta_{\gamma\delta} \tilde{\mathcal{J}}^{+-}_{ {\bm q}_1,\alpha\beta }
        - \delta_{\alpha\gamma}\delta_{\alpha\delta} \tilde{\mathcal{J}}^{-+}_{-{\bm q}_2,\alpha\beta }
        - \delta_{\alpha\beta }\delta_{\alpha\delta} \tilde{\mathcal{J}}^{-+}_{ {\bm q}_3,\gamma\delta} \right)
    U^*_{{\bm q}_1,\alpha\nu_1}U^*_{{\bm q}_2,\beta \nu_2}V^*_{{\bm q}_3,\gamma\nu_3}V^*_{{\bm q}_4,\delta\nu_4}
    \right.
    \\ &\quad\quad\quad - \left(
          \delta_{\alpha\beta }\delta_{\alpha\delta} \tilde{\mathcal{J}}^{++}_{-{\bm q}_3,\alpha\gamma}
        + \delta_{\beta\gamma }\delta_{\beta\delta } \tilde{\mathcal{J}}^{++}_{ {\bm q}_1,\alpha\delta}
        \right) U^*_{{\bm q}_1,\alpha\nu_1}U^*_{{\bm q}_2,\beta \nu_2}U^*_{{\bm q}_3,\gamma\nu_3}V^*_{{\bm q}_4,\delta\nu_4} 
    \\ &\quad\quad\quad \left. - \left(
          \delta_{\alpha\beta }\delta_{\alpha\gamma} \tilde{\mathcal{J}}^{--}_{-{\bm q}_4,\alpha\delta}
        + \delta_{\alpha\gamma}\delta_{\alpha\delta} \tilde{\mathcal{J}}^{--}_{ {\bm q}_2,\beta\alpha } \right) U^*_{{\bm q}_1,\alpha\nu_1}V^*_{{\bm q}_2,\beta \nu_2}V^*_{{\bm q}_3,\gamma\nu_3}V^*_{{\bm q}_4,\delta\nu_4}
    \right].
\end{aligned}
\end{equation}
\subsection{Hartree-Fock mean-fields}
\label{sec:HartreeFockMF}
\subsubsection{Four-magnon mean-fields}
We divide the four-magnon Hamiltonian $\mathcal{H}^{(4)}$ in terms of the numbers of created and annihilated magnons in the language of $\hat{a}$ and $\hat{a}^\dagger$ operators, given by
\begin{equation}
    \mathcal{H}^{(4)} = \mathcal{H}^{(22)} + \mathcal{H}^{(31)} + \mathcal{H}^{(13)}.
    \label{eq:H4_decomposed_aa}
\end{equation}
Note that $\mathcal{H}^{(13)} = \left[ \mathcal{H}^{(31)} \right]^\dagger$. According to Eqs.~(\ref{eq:HP_SJS}) and (\ref{eq:HP4_SJS}), each term in Eq.~(\ref{eq:H4_decomposed_aa}) is explicitly given by
\begin{align}
\begin{aligned}
    \mathcal{H}^{(22)} &= \sum_{\alpha\beta\gamma\delta} \sum_{{\bm q}_1,{\bm q}_2,{\bm q}_3,{\bm q}_4} \mathfrak{Q}_{{\bm q_1},{\bm q}_2\leftrightarrow{\bm q}_3,{\bm q}_4}^{\alpha,\beta\leftrightarrow\gamma,\delta} 
    \hat{a}^\dagger_{{\bm q}_1,\alpha}\hat{a}^\dagger_{{\bm q}_2,\beta}\hat{a}_{{\bm q}_3,\gamma}\hat{a}_{{\bm q}_4,\delta}, \\
    \mathfrak{Q}_{{\bm q_1},{\bm q}_2\leftrightarrow{\bm q}_3,{\bm q}_4}^{\alpha,\beta,\leftrightarrow\gamma,\delta} 
    &= \frac{1}{2} \frac{1}{4N_{\rm muc}} \delta({\bm q_1}+{\bm q}_2-{\bm q}_3-{\bm q}_4) \\
    &\quad\quad \times \left[ \left\{ 
        \delta_{\alpha\gamma}\delta_{\beta\delta}\tilde{\mathcal{J}}^{00}_{{\bm q}_1-{\bm q}_3,\alpha\beta} + 
        \delta_{\alpha\delta}\delta_{\beta\gamma}\tilde{\mathcal{J}}^{00}_{{\bm q}_1-{\bm q}_4,\alpha\beta} + 
        \delta_{\alpha\delta}\delta_{\beta\gamma}\tilde{\mathcal{J}}^{00}_{{\bm q}_2-{\bm q}_3,\beta\alpha} + 
        \delta_{\alpha\gamma}\delta_{\beta\delta}\tilde{\mathcal{J}}^{00}_{{\bm q}_2-{\bm q}_4,\beta\alpha} 
    \right\} \right. \\ 
    &\quad\quad\quad\quad 
    \left. - \left\{ 
        \delta_{\alpha\beta} \delta_{\alpha\gamma}\tilde{\mathcal{J}}^{+-}_{ {\bm q}_4,\gamma\delta} + 
        \delta_{\beta\delta} \delta_{\gamma\delta}\tilde{\mathcal{J}}^{+-}_{ {\bm q}_1,\alpha\beta}  + 
        \delta_{\alpha\gamma}\delta_{\alpha\delta}\tilde{\mathcal{J}}^{-+}_{-{\bm q}_2,\alpha\beta}  + 
        \delta_{\alpha\beta} \delta_{\alpha\delta}\tilde{\mathcal{J}}^{-+}_{-{\bm q}_3,\gamma\delta} 
    \right\} \right],
    \label{eq:H22_explicit}
\end{aligned}
\end{align}
and
\begin{align}
\begin{aligned}
    \mathcal{H}^{(31)} &=  \sum_{\alpha\beta\gamma\delta} \sum_{{\bm q}_1,{\bm q}_2,{\bm q}_3,{\bm q}_4} \mathfrak{C}_{{\bm q_1},{\bm q}_2,{\bm q}_3\leftarrow{\bm q}_4}^{\alpha,\beta,\gamma\leftarrow\delta} 
    \hat{a}^\dagger_{{\bm q}_1,\alpha}\hat{a}^\dagger_{{\bm q}_2,\beta}\hat{a}^\dagger_{{\bm q}_3,\gamma}\hat{a}_{{\bm q}_4,\delta}, \\
    \mathfrak{C}_{{\bm q_1},{\bm q}_2,{\bm q}_3\leftarrow{\bm q}_4}^{\alpha,\beta,\gamma\leftarrow\delta} 
    &= - \frac{1}{2} \frac{1}{4N_{\rm muc}} \delta({\bm q_1}+{\bm q}_2+{\bm q}_3-{\bm q}_4) \left[ 
          \delta_{\alpha\beta}\delta_{\alpha\delta}\tilde{\mathcal{J}}^{++}_{-{\bm q}_3,\alpha\gamma} 
        + \delta_{\beta\gamma}\delta_{\beta\delta }\tilde{\mathcal{J}}^{++}_{{\bm q}_1,\alpha\delta} \right].
    \label{eq:H31_explicit}
\end{aligned}
\end{align}
In the following, we consider the Hartree-Fock decomposition of $\mathcal{H}^{(4)}$. First, we present the bilinear expectation values of $\hat{b}$ and $\hat{b}^\dagger$ operators, given by
\begin{equation}
    \left\{
    \begin{array}{c}
        \ev{\hat{b}^\dagger_{{\bm k}\nu}\hat{b}_{{\bm k}'\nu'}} = \delta_{{\bm k},{\bm k}'}\delta_{\nu\nu'} n^{(0)}_{\rm B}(\varepsilon_\nu({\bm k})), \\
        \ev{\hat{b}_{{\bm k}\nu}\hat{b}^\dagger_{{\bm k}'\nu'}} = \delta_{{\bm k},{\bm k}'}\delta_{\nu\nu'} \left(1+n^{(0)}_{\rm B}(\varepsilon_\nu({\bm k}))\right), \\
        \ev{\hat{b}^\dagger_{{\bm k}\nu}\hat{b}^\dagger_{{\bm k}'\nu'}} = \ev{\hat{b}_{{\bm k}\nu}\hat{b}_{{\bm k}'\nu'}} = 0,
    \end{array}
    \right.
\end{equation}
where $n_\textrm{B}^{(0)}(x)=1/(e^{x/k_\textrm{B}T}-1)$ is the Bose distribution function. Using them, the bilinear expectation values of $\hat{a}$ and $\hat{a}^\dagger$ operators can be written as
\begin{align}
\begin{aligned}
    \ev{\hat{a}^\dagger_{{\bm k}\alpha}\hat{a}_{{\bm k}'\beta}} &= 
    \delta_{{\bm k},{\bm k}'} \sum_\nu 
    \left[ U^*_{{\bm k},\alpha\nu}U_{{\bm k},\beta\nu} \ev{\hat{b}^\dagger_{{\bm k}\nu}\hat{b}_{{\bm k}\nu}} + V_{-{\bm k},\alpha\nu}V^*_{-{\bm k},\beta\nu} \ev{\hat{b}_{-{\bm k}\nu}\hat{b}^\dagger_{-{\bm k}\nu}} \right] \\
    &= \delta_{{\bm k},{\bm k}'} \sum_\nu 
    \left[ U^*_{{\bm k},\alpha\nu}U_{{\bm k},\beta\nu} n^{(0)}_{\rm B}(\varepsilon_\nu({\bm k})) + V_{-{\bm k},\alpha\nu}V^*_{-{\bm k},\beta\nu} \left( 1+n^{(0)}_{\rm B}(\varepsilon_\nu(-{\bm k})) \right) \right],
\end{aligned}
\end{align}
and
\begin{align}
\begin{aligned}
    \ev{\hat{a}^\dagger_{{\bm k}\alpha}\hat{a}^\dagger_{{\bm k}'\beta}} &= 
    \delta_{{\bm k},-{\bm k}'} \sum_\nu 
    \left[ U^*_{{\bm k},\alpha\nu}V_{{\bm k},\beta\nu} \ev{\hat{b}^\dagger_{{\bm k}\nu}\hat{b}_{{\bm k}\nu}} + V_{-{\bm k},\alpha\nu}U^*_{-{\bm k},\beta\nu} \ev{\hat{b}_{-{\bm k}\nu}\hat{b}^\dagger_{-{\bm k}\nu}} \right] \\
    &= \delta_{{\bm k},-{\bm k}'} \sum_\nu 
    \left[ U^*_{{\bm k},\alpha\nu}V_{{\bm k},\beta\nu} n^{(0)}_{\rm B}(\varepsilon_\nu({\bm k})) + V_{-{\bm k},\alpha\nu}U^*_{-{\bm k},\beta\nu} \left( 1+n^{(0)}_{\rm B}(\varepsilon_\nu(-{\bm k})) \right) \right],
\end{aligned}
\end{align}
\begin{align}
\begin{aligned}
    \ev{\hat{a}_{{\bm k}\alpha}\hat{a}_{{\bm k}'\beta}} &= 
    \delta_{{\bm k},-{\bm k}'} \sum_\nu 
    \left[ U_{{\bm k},\alpha\nu}V^*_{{\bm k},\beta\nu} \ev{\hat{b}_{{\bm k}\nu}\hat{b}^\dagger_{{\bm k}\nu}} + V^*_{-{\bm k},\alpha\nu}U_{-{\bm k},\beta\nu} \ev{\hat{b}^\dagger_{-{\bm k}\nu}\hat{b}_{-{\bm k}\nu}} \right] \\
    &= \delta_{{\bm k},-{\bm k}'} \sum_\nu 
    \left[ U_{{\bm k},\alpha\nu}V^*_{{\bm k},\beta\nu} \left( 1+n^{(0)}_{\rm B}(\varepsilon_\nu({\bm k})) \right) + V^*_{-{\bm k},\alpha\nu}U_{-{\bm k},\beta\nu} n^{(0)}_{\rm B}(\varepsilon_\nu(-{\bm k})) \right].
\end{aligned}
\end{align}
A mean-field decoupling of the quartic term with two creation and two annihilation operators is given by
\begin{align}
\begin{aligned}
    \hat{a}^\dagger_{{\bm q}_1,\alpha}\hat{a}^\dagger_{{\bm q}_2,\beta}\hat{a}_{{\bm q}_3,\gamma}\hat{a}_{{\bm q}_4,\delta} &\approx 
    \ev{ \hat{a}^\dagger_{{\bm q}_1,\alpha} \hat{a}_{{\bm q}_3,\gamma} } \hat{a}^\dagger_{{\bm q}_2,\beta } \hat{a}_{{\bm q}_4,\delta} +
    \ev{ \hat{a}^\dagger_{{\bm q}_2,\beta } \hat{a}_{{\bm q}_4,\delta} } \hat{a}^\dagger_{{\bm q}_1,\alpha} \hat{a}_{{\bm q}_3,\gamma} \\
    &+
    \ev{ \hat{a}^\dagger_{{\bm q}_1,\alpha} \hat{a}_{{\bm q}_4,\delta} } \hat{a}^\dagger_{{\bm q}_2,\beta } \hat{a}_{{\bm q}_3,\gamma} +
    \ev{ \hat{a}^\dagger_{{\bm q}_2,\beta } \hat{a}_{{\bm q}_3,\gamma} } \hat{a}^\dagger_{{\bm q}_1,\alpha} \hat{a}_{{\bm q}_4,\delta} \\
    &+
    \ev{ \hat{a}^\dagger_{{\bm q}_1,\alpha}\hat{a}^\dagger_{{\bm q}_2,\beta} } \hat{a}_{{\bm q}_3,\gamma}\hat{a}_{{\bm q}_4,\delta} +
    \ev{ \hat{a}_{{\bm q}_3,\gamma}\hat{a}_{{\bm q}_4,\delta} } \hat{a}^\dagger_{{\bm q}_1,\alpha}\hat{a}^\dagger_{{\bm q}_2,\beta}.
    \label{eq:MFdecoupling_22}
\end{aligned}
\end{align}
Substituting Eq.~(\ref{eq:MFdecoupling_22}) into Eq.~(\ref{eq:H22_explicit}), we get
\begin{align}
\begin{aligned}
    \mathcal{H}^{(22)}
    &= \sum_{{\bm q}_1,{\bm q}_2,{\bm q}_3,{\bm q}_4} \sum_{\alpha\beta\gamma\delta} \delta\left({\bm q}_1+{\bm q}_2-{\bm q}_3-{\bm q}_4\right) \mathfrak{Q}_{{\bm q}_1,{\bm q}_2\leftrightarrow{\bm q}_3,{\bm q}_4}^{\alpha,\beta\leftrightarrow\gamma,\delta} \hat{a}^\dagger_{{\bm q}_1,\alpha}\hat{a}^\dagger_{{\bm q}_2,\beta}\hat{a}_{{\bm q}_3,\gamma}\hat{a}_{{\bm q}_4,\delta} \\
    &\approx \sum_{{\bm k}}\sum_{{\bm p}} \sum_{\alpha\beta\gamma\delta} \left[
        \underbrace{
        \mathfrak{Q}_{{\bm p},{\bm k}\leftrightarrow{\bm p},{\bm k}}^{\alpha,\beta\leftrightarrow\gamma,\delta} 
        \ev{\hat{a}^\dagger_{{\bm p},\alpha}\hat{a}_{{\bm p},\gamma}} \hat{a}^\dagger_{{\bm k},\beta }\hat{a}_{{\bm k},\delta} +
        \mathfrak{Q}_{{\bm k},{\bm p}\leftrightarrow{\bm k},{\bm p}}^{\alpha,\beta\leftrightarrow\gamma,\delta}
        \ev{\hat{a}^\dagger_{{\bm p},\beta }\hat{a}_{{\bm p},\delta}} \hat{a}^\dagger_{{\bm k},\alpha}\hat{a}_{{\bm k},\gamma}}_{\hat{H}^{(22)}_{\rm N}} \right. \\
    &\quad\quad\quad\quad\quad\quad +
        \underbrace{
        \mathfrak{Q}_{{\bm p},{\bm k}\leftrightarrow{\bm k},{\bm p}}^{\alpha,\beta\leftrightarrow\gamma,\delta} 
        \ev{\hat{a}^\dagger_{{\bm p},\alpha}\hat{a}_{{\bm p},\delta}} \hat{a}^\dagger_{{\bm k},\beta }\hat{a}_{{\bm k},\gamma} + 
        \mathfrak{Q}_{{\bm k},{\bm p}\leftrightarrow{\bm p},{\bm k}}^{\alpha,\beta\leftrightarrow\gamma,\delta} 
        \ev{\hat{a}^\dagger_{{\bm p},\beta }\hat{a}_{{\bm p},\gamma}} \hat{a}^\dagger_{{\bm k},\alpha}\hat{a}_{{\bm k},\delta}}_{\hat{H}^{(22)}_{\rm N}}  \\
    &\quad\quad\quad\quad\quad\quad \left. +
        \underbrace{
        \mathfrak{Q}_{{\bm p},-{\bm p}\leftrightarrow-{\bm k},{\bm k}}^{\alpha,\beta\leftrightarrow\gamma,\delta} 
        \ev{\hat{a}^\dagger_{{\bm p},\alpha}\hat{a}^\dagger_{-{\bm p},\beta }} \hat{a}_{-{\bm k},\gamma}\hat{a}_{{\bm k},\delta}}_{\hat{H}^{(22)}_{\rm AA}} + 
        \underbrace{
        \mathfrak{Q}_{{\bm k},-{\bm k}\leftrightarrow{\bm p},-{\bm p}}^{\alpha,\beta\leftrightarrow\gamma,\delta} 
        \ev{\hat{a}_{{\bm p},\gamma}\hat{a}_{-{\bm p},\delta}} \hat{a}^\dagger_{{\bm k},\alpha}\hat{a}^\dagger_{-{\bm k},\beta }}_{\hat{H}^{(22)}_{\rm AC}}
    \right].
\end{aligned}
\end{align}
A mean-field decoupling of the quartic term with three creation and one annihilation operator is given by
\begin{align}
\begin{aligned}
    \hat{a}^\dagger_{{\bm q}_1,\alpha}\hat{a}^\dagger_{{\bm q}_2,\beta}\hat{a}^\dagger_{{\bm q}_3,\gamma}\hat{a}_{{\bm q}_4,\delta} &\approx 
    \ev{ \hat{a}^\dagger_{{\bm q}_1,\alpha} \hat{a}_{{\bm q}_4,\delta} } \hat{a}^\dagger_{{\bm q}_2,\beta } \hat{a}^\dagger_{{\bm q}_3,\gamma} +
    \ev{ \hat{a}^\dagger_{{\bm q}_2,\beta } \hat{a}^\dagger_{{\bm q}_3,\gamma} } \hat{a}^\dagger_{{\bm q}_1,\alpha} \hat{a}_{{\bm q}_4,\delta} \\
    &+
    \ev{ \hat{a}^\dagger_{{\bm q}_2,\beta } \hat{a}_{{\bm q}_4,\delta} } \hat{a}^\dagger_{{\bm q}_1,\alpha} \hat{a}^\dagger_{{\bm q}_3,\gamma} +
    \ev{ \hat{a}^\dagger_{{\bm q}_1,\alpha} \hat{a}^\dagger_{{\bm q}_3,\gamma} } \hat{a}^\dagger_{{\bm q}_2,\beta } \hat{a}_{{\bm q}_4,\delta} \\
    &+
    \ev{ \hat{a}^\dagger_{{\bm q}_3,\gamma} \hat{a}_{{\bm q}_4,\delta} } \hat{a}^\dagger_{{\bm q}_1,\alpha} \hat{a}^\dagger_{{\bm q}_2,\beta } +
    \ev{ \hat{a}^\dagger_{{\bm q}_1,\alpha} \hat{a}^\dagger_{{\bm q}_2,\beta } } \hat{a}^\dagger_{{\bm q}_3,\gamma} \hat{a}_{{\bm q}_4,\delta}.
    \label{eq:MFdecoupling_31}
\end{aligned}
\end{align}
Substituting Eq.~(\ref{eq:MFdecoupling_31}) into Eq.~(\ref{eq:H31_explicit}), we get
\begin{align}
\begin{aligned}
    \mathcal{H}^{(31)}
    &= \sum_{{\bm q}_1,{\bm q}_2,{\bm q}_3,{\bm q}_4} \sum_{\alpha\beta\gamma\delta} \delta\left({\bm q}_1+{\bm q}_2+{\bm q}_3-{\bm q}_4\right)  \mathfrak{C}_{{\bm q}_1,{\bm q}_2,{\bm q}_3\leftarrow{\bm q}_4}^{\alpha,\beta,\gamma\leftarrow\delta} \hat{a}^\dagger_{{\bm q}_1,\alpha}\hat{a}^\dagger_{{\bm q}_2,\beta}\hat{a}^\dagger_{{\bm q}_3,\gamma}\hat{a}_{{\bm q}_4,\delta} \\
    &\approx \sum_{{\bm k}}\sum_{{\bm p}} \sum_{\alpha\beta\gamma\delta} \left[
        \underbrace{
        \mathfrak{C}_{{\bm p},{\bm k},-{\bm k}\leftarrow{\bm p}}^{\alpha,\beta,\gamma\leftarrow\delta} \ev{ \hat{a}^\dagger_{{\bm p},\alpha} \hat{a}_{{\bm p},\delta} } \hat{a}^\dagger_{{\bm k},\beta } \hat{a}^\dagger_{-{\bm k},\gamma}}_{\hat{H}^{(31)}_{\rm AC}} +
        \underbrace{
        \mathfrak{C}_{{\bm k},{\bm p},-{\bm p}\leftarrow{\bm k}}^{\alpha,\beta,\gamma\leftarrow\delta} \ev{ \hat{a}^\dagger_{{\bm p},\beta } \hat{a}^\dagger_{-{\bm p},\gamma} } \hat{a}^\dagger_{{\bm k},\alpha} \hat{a}_{{\bm k},\delta}}_{\hat{H}^{(31)}_{\rm N}} \right. \\
    &\quad\quad\quad\quad\quad\quad + 
        \underbrace{
        \mathfrak{C}_{{\bm k},{\bm p},-{\bm k}\leftarrow{\bm p}}^{\alpha,\beta,\gamma\leftarrow\delta} \ev{ \hat{a}^\dagger_{{\bm p},\beta } \hat{a}_{{\bm p},\delta} } \hat{a}^\dagger_{{\bm k},\alpha} \hat{a}^\dagger_{-{\bm k},\gamma}}_{\hat{H}^{(31)}_{\rm AC}} + 
        \underbrace{
        \mathfrak{C}_{{\bm p},{\bm k},-{\bm p}\leftarrow{\bm k}}^{\alpha,\beta,\gamma\leftarrow\delta} \ev{ \hat{a}^\dagger_{{\bm p},\alpha} \hat{a}^\dagger_{-{\bm p},\gamma} } \hat{a}^\dagger_{{\bm k},\beta } \hat{a}_{{\bm k},\delta}}_{\hat{H}^{(31)}_{\rm N}} \\
    &\quad\quad\quad\quad\quad\quad \left. +
        \underbrace{
        \mathfrak{C}_{{\bm k},-{\bm k},{\bm p}\leftarrow{\bm p}}^{\alpha,\beta,\gamma\leftarrow\delta} \ev{ \hat{a}^\dagger_{{\bm p},\gamma} \hat{a}_{{\bm p},\delta} } \hat{a}^\dagger_{{\bm k},\alpha} \hat{a}^\dagger_{-{\bm k},\beta }}_{\hat{H}^{(31)}_{\rm AC}} +
        \underbrace{
        \mathfrak{C}_{{\bm p},-{\bm p},{\bm k}\leftarrow{\bm k}}^{\alpha,\beta,\gamma\leftarrow\delta} \ev{ \hat{a}^\dagger_{{\bm p},\alpha} \hat{a}^\dagger_{-{\bm p},\beta } } \hat{a}^\dagger_{{\bm k},\gamma} \hat{a}_{{\bm k},\delta}}_{\hat{H}^{(31)}_{\rm N}}
    \right].
\end{aligned}
\end{align}
As a result, the matrix representation of the four-magnon Hartree-Fock mean-fields in the Nambu space is given by $\ev{\mathcal{H}^{(4)}} = \sum_{\bm k} \hat{\phi}^\dagger_{\bm k} \mathcal{H}^{(4)}_{\rm HF}({\bm k}) \hat{\phi}_{\bm k}$ with
\begin{equation}
    \mathcal{H}^{(4)}_{\rm HF}({\bm k}) = \left(
        \begin{array}{cc}
            \hat{H}^{(22)}_{\rm N }({\bm k}) + \hat{H}^{(31)}_{\rm N }({\bm k}) + \hat{H}^{(13)}_{\rm N }({\bm k}) &
            \hat{H}^{(22)}_{\rm AC}({\bm k}) + \left[ \hat{H}^{(22)}_{\rm AA}({\bm k}) \right]^\dagger + \hat{H}^{(31)}_{\rm AC}({\bm k}) + \left[ \hat{H}^{(13)}_{\rm AA}({\bm k}) \right]^\dagger \\
            \left[ \hat{H}^{(22)}_{\rm AC}({\bm k}) \right]^\dagger + \hat{H}^{(22)}_{\rm AA}({\bm k}) + \left[ \hat{H}^{(31)}_{\rm AC}({\bm k}) \right]^\dagger + \hat{H}^{(13)}_{\rm AA}({\bm k}) &
            \left[ \hat{H}^{(22)}_{\rm N }({-\bm k}) \right]^\top + \left[ \hat{H}^{(31)}_{\rm N }(-{\bm k}) \right]^\top + \left[ \hat{H}^{(13)}_{\rm N }(-{\bm k}) \right]^\top
        \end{array}
    \right).
\end{equation}
Note that $\hat{H}^{(31)}_{\rm N}({\bm k})=\left[ \hat{H}^{(13)}_{\rm N}({\bm k}) \right]^\dagger$ and $\hat{H}^{(31)}_{\rm AC}({\bm k})=\left[ \hat{H}^{(13)}_{\rm AA}({\bm k}) \right]^\dagger$ are $N_\textrm{sub}\times N_\textrm{sub}$ matrices.

\par Another four magnon term stemming from normal-ordered square root $\mathcal{H}^{(4\leftarrow 6)}$ is given by
\begin{equation}
    \mathcal{H}^{(4\leftarrow 6)} = 
    \mathcal{H}^{(4\leftarrow 6,22)} + 
    \mathcal{H}^{(4\leftarrow 6,31)} +
    \mathcal{H}^{(4\leftarrow 6,13)}.
    \label{eq:H4n6_decomposed_aa}
\end{equation}
Note that $\mathcal{H}^{(4\leftarrow 6,13)}=\left[ \mathcal{H}^{(4\leftarrow 6,31)} \right]^\dagger$. According to Eq.~(\ref{eq:HP_SJS}), each term in Eq.~(\ref{eq:H4n6_decomposed_aa}) is given by
\begin{align}
\begin{aligned}
    \mathcal{H}^{(4\leftarrow6,22)} &= \sum_{\alpha\beta\gamma\delta} \sum_{{\bm q}_1,{\bm q}_2,{\bm q}_3,{\bm q}_4} \mathtt{Q}_{{\bm q_1},{\bm q}_2\leftrightarrow{\bm q}_3,{\bm q}_4}^{\alpha,\beta\leftrightarrow\gamma,\delta} 
    \hat{a}^\dagger_{{\bm q}_1,\alpha}\hat{a}^\dagger_{{\bm q}_2,\beta}\hat{a}_{{\bm q}_3,\gamma}\hat{a}_{{\bm q}_4,\delta}, \\
    \mathtt{Q}_{{\bm q_1},{\bm q}_2\leftrightarrow{\bm q}_3,{\bm q}_4}^{\alpha,\beta,\leftrightarrow\gamma,\delta} 
    &= \frac{1}{2} \frac{1}{32N_{\rm muc}} \delta({\bm q_1}+{\bm q}_2-{\bm q}_3-{\bm q}_4) \left(
        \delta_{\alpha\beta} \delta_{\alpha\gamma}\tilde{\mathcal{J}}^{+-}_{ {\bm q}_4,\gamma\delta} + 
        \delta_{\beta\delta} \delta_{\gamma\delta}\tilde{\mathcal{J}}^{+-}_{ {\bm q}_1,\alpha\beta}  + 
        \delta_{\alpha\gamma}\delta_{\alpha\delta}\tilde{\mathcal{J}}^{-+}_{-{\bm q}_2,\alpha\beta}  + 
        \delta_{\alpha\beta} \delta_{\alpha\delta}\tilde{\mathcal{J}}^{-+}_{-{\bm q}_3,\gamma\delta} 
    \right),
\end{aligned}
\end{align}
and
\begin{align}
\begin{aligned}
    \mathcal{H}^{(4\leftarrow 6,31)} &=  -\frac{1}{8} \sum_{\alpha\beta\gamma\delta} \sum_{{\bm q}_1,{\bm q}_2,{\bm q}_3,{\bm q}_4} \mathfrak{C}_{{\bm q_1},{\bm q}_2,{\bm q}_3\leftarrow{\bm q}_4}^{\alpha,\beta,\gamma\leftarrow\delta} 
    \hat{a}^\dagger_{{\bm q}_1,\alpha}\hat{a}^\dagger_{{\bm q}_2,\beta}\hat{a}^\dagger_{{\bm q}_3,\gamma}\hat{a}_{{\bm q}_4,\delta}.
\end{aligned}
\end{align}
Note that the explicit form of $\mathfrak{C}_{{\bm q_1},{\bm q}_2,{\bm q}_3\leftarrow{\bm q}_4}^{\alpha,\beta,\gamma\leftarrow\delta}$ is given in Eq.~(\ref{eq:H31_explicit}). Mean-field decoupling of $\mathcal{H}^{(4\leftarrow 6)}$ is obtained with almost the same procedures as those for $\mathcal{H}^{(4)}$ discussed earlier in this subsection.
\subsubsection{Six-magnon mean-fields}
We divide the six-magnon Hamiltonian $\mathcal{H}^{(6)}$ in terms of the numbers of created and annihilated magnons in the language of $\hat{a}$ and $\hat{a}^\dagger$ operators, given by
\begin{equation}
    \mathcal{H}^{(6)} = \mathcal{H}^{(33)} + \mathcal{H}^{(42)} + \mathcal{H}^{(24)}.
    \label{eq:H6_decomposed_aa}
\end{equation}
Note that $\mathcal{H}^{(24)} = \left[ \mathcal{H}^{(42)} \right]^\dagger$. According to Eqs.~(\ref{eq:HP_SJS}) and (\ref{eq:HP4_SJS}), each term in Eq.~(\ref{eq:H6_decomposed_aa}) is explicitly given by
\begin{align}
\begin{aligned}
    \mathcal{H}^{(33)} &= \sum_{\chi_1\chi_2\chi_3\chi_4\chi_5\chi_6} \sum_{{\bm q}_1,{\bm q}_2,{\bm q}_3,{\bm q}_4,{\bm q}_5,{\bm q}_6} 
    \mathfrak{K}_{{\bm q_1},{\bm q}_2,{\bm q}_3\leftrightarrow{\bm q}_4,{\bm q}_5,{\bm q}_6}^{\chi_1,\chi_2,\chi_3\leftrightarrow\chi_4,\chi_5,\chi_6} 
    \hat{a}^\dagger_{{\bm q}_1,\chi_1}\hat{a}^\dagger_{{\bm q}_2,\chi_2}\hat{a}^\dagger_{{\bm q}_3,\chi_3}\hat{a}_{{\bm q}_4,\chi_4}\hat{a}_{{\bm q}_5,\chi_5}\hat{a}_{{\bm q}_6,\chi_6}, \\
    \mathfrak{K}_{{\bm q_1},{\bm q}_2,{\bm q}_3\leftrightarrow{\bm q}_4,{\bm q}_5,{\bm q}_6}^{\chi_1,\chi_2,\chi_3\leftrightarrow\chi_4,\chi_5,\chi_6}
    &= \frac{1}{2} \frac{1}{32N_{\rm muc}^2} \delta({\bm q_1}+{\bm q}_2+{\bm q}_3-{\bm q}_4-{\bm q}_5-{\bm q}_6) \\
    &\quad\quad \times \left[ \left\{ 
        -  \delta_{\chi_2\chi_3\chi_4\chi_5\chi_6}                \tilde{\mathcal{J}}^{+-}_{ {\bm q}_1                    ,\chi_1\chi_2}
        -  \delta_{\chi_1\chi_2\chi_3\chi_4\chi_5}                \tilde{\mathcal{J}}^{+-}_{ {\bm q}_6                    ,\chi_1\chi_6}
        + 2\delta_{\chi_1\chi_2\chi_4}\delta_{\chi_3\chi_5\chi_6} \tilde{\mathcal{J}}^{+-}_{ {\bm q}_1+{\bm q}_2-{\bm q}_4,\chi_1\chi_3}
    \right\} \right. \\ 
    &\quad\quad\quad\quad 
    \left. + \left\{ 
        -  \delta_{\chi_1\chi_2\chi_3\chi_5\chi_6}                \tilde{\mathcal{J}}^{-+}_{-{\bm q}_4                    ,\chi_4\chi_1}
        -  \delta_{\chi_1\chi_2\chi_4\chi_5\chi_6}                \tilde{\mathcal{J}}^{-+}_{-{\bm q}_3                    ,\chi_1\chi_3}
        + 2\delta_{\chi_1\chi_4\chi_5}\delta_{\chi_2\chi_3\chi_6} \tilde{\mathcal{J}}^{-+}_{ {\bm q}_1-{\bm q}_4-{\bm q}_5,\chi_1\chi_2}
    \right\} \right],
\end{aligned}
\end{align}
\begin{align}
\begin{aligned}
    \mathcal{H}^{(42)} &= \sum_{\chi_1\chi_2\chi_3\chi_4\chi_5\chi_6} \sum_{{\bm q}_1,{\bm q}_2,{\bm q}_3,{\bm q}_4,{\bm q}_5,{\bm q}_6} 
    \mathfrak{S}_{{\bm q_1},{\bm q}_2,{\bm q}_3,{\bm q}_4\leftarrow{\bm q}_5,{\bm q}_6}^{\chi_1,\chi_2,\chi_3,\chi_4\leftarrow\chi_5,\chi_6} 
    \hat{a}^\dagger_{{\bm q}_1,\chi_1}\hat{a}^\dagger_{{\bm q}_2,\chi_2}\hat{a}^\dagger_{{\bm q}_3,\chi_3}\hat{a}^\dagger_{{\bm q}_4,\chi_4}\hat{a}_{{\bm q}_5,\chi_5}\hat{a}_{{\bm q}_6,\chi_6}, \\
    \mathfrak{S}_{{\bm q_1},{\bm q}_2,{\bm q}_3,{\bm q}_4\leftarrow{\bm q}_5,{\bm q}_6}^{\chi_1,\chi_2,\chi_3,\chi_4\leftarrow\chi_5,\chi_6}
    &= \frac{1}{2} \frac{1}{32N_{\rm muc}^2} \delta({\bm q_1}+{\bm q}_2+{\bm q}_3+{\bm q}_4-{\bm q}_5-{\bm q}_6) \\
    &\quad\quad \times \left[
        -  \delta_{\chi_2\chi_3\chi_4\chi_5\chi_6}                \tilde{\mathcal{J}}^{++}_{ {\bm q}_1                    ,\chi_1\chi_2}
        -  \delta_{\chi_1\chi_2\chi_3\chi_5\chi_6}                \tilde{\mathcal{J}}^{++}_{-{\bm q}_4                    ,\chi_1\chi_4}
        + 2\delta_{\chi_1\chi_2\chi_5}\delta_{\chi_3\chi_4\chi_6} \tilde{\mathcal{J}}^{++}_{ {\bm q}_1+{\bm q}_2-{\bm q}_5,\chi_1\chi_3}
    \right], \\ 
\end{aligned}
\end{align}
where $\delta_{\chi_1\chi_2\cdots\chi_n}=\delta_{\chi_1\chi_2}\delta_{\chi_2\chi_3}\cdots\delta_{\chi_{n-1}\chi_n}$. Regarding the mean-field decoupling of these six-magnon terms, we have to consider 90($={}_6C_2 \cdot {}_4C_2$) different ways of decoupling for each of $\hat{a}^\dagger\hat{a}^\dagger\hat{a}^\dagger\hat{a}\hat{a}\hat{a}$ and $\hat{a}^\dagger\hat{a}^\dagger\hat{a}^\dagger\hat{a}^\dagger\hat{a}\hat{a}$ type terms. To conduct this bothersome decoupling efficiently, we developed a numerical algorithm to take these 90 different mean-filed pairings automatically. 
\section{Nonlinear spin-wave theory with Dyson-Maleev transformation}
\renewcommand{\theequation}{C.\arabic{equation}}
\subsection{Dyson-Maleev transformation}
In this section, instead of the Holstein-Primakoff transformation, let us consider the Dyson-Maleev transformation. We will use the Dyson-Maleev transformation to derive the long-wavelength scaling of the magnon damping 
in Sec.~\ref{sec:scaling}.
\par We consider the collinear two-sublattice N\'{e}el order [also called $(\pi,\pi)$-order], which is the classical ground state of the $J_1-J_2-\Delta$ Heisenberg model on the checkerboard lattice given in Eq.~(8) in the main text. The explicit expression of the spin Hamiltonian is given by
\begin{align}
\begin{aligned}
    H
    &=        \frac{J_1}{2} \sum_{{\bm r}\in{\rm A}} {\bm S}_{\bm r} \cdot{\bm S}_{{\bm r}+{\bm e}_x} 
            + \frac{J_1}{2} \sum_{{\bm r}\in{\rm A}} {\bm S}_{\bm r} \cdot{\bm S}_{{\bm r}-{\bm e}_x}
            + \frac{J_1}{2} \sum_{{\bm r}\in{\rm B}} {\bm S}_{\bm r} \cdot{\bm S}_{{\bm r}+{\bm e}_x} 
            + \frac{J_1}{2} \sum_{{\bm r}\in{\rm B}} {\bm S}_{\bm r} \cdot{\bm S}_{{\bm r}-{\bm e}_x} \\
    &\quad  + \frac{J_1}{2} \sum_{{\bm r}\in{\rm A}} {\bm S}_{\bm r} \cdot{\bm S}_{{\bm r}+{\bm e}_y}
            + \frac{J_1}{2} \sum_{{\bm r}\in{\rm A}} {\bm S}_{\bm r} \cdot{\bm S}_{{\bm r}-{\bm e}_y}
            + \frac{J_1}{2} \sum_{{\bm r}\in{\rm B}} {\bm S}_{\bm r} \cdot{\bm S}_{{\bm r}+{\bm e}_y}
            + \frac{J_1}{2} \sum_{{\bm r}\in{\rm B}} {\bm S}_{\bm r} \cdot{\bm S}_{{\bm r}-{\bm e}_y} \\
    &\quad  + \frac{J_2+\Delta}{2} \sum_{{\bm r}\in{\rm A}} {\bm S}_{\bm r} \cdot{\bm S}_{{\bm r}+{\bm e}_x+{\bm e}_y}
            + \frac{J_2+\Delta}{2} \sum_{{\bm r}\in{\rm A}} {\bm S}_{\bm r} \cdot{\bm S}_{{\bm r}-{\bm e}_x-{\bm e}_y}
            + \frac{J_2-\Delta}{2} \sum_{{\bm r}\in{\rm B}} {\bm S}_{\bm r} \cdot{\bm S}_{{\bm r}+{\bm e}_x+{\bm e}_y}
            + \frac{J_2-\Delta}{2} \sum_{{\bm r}\in{\rm B}} {\bm S}_{\bm r} \cdot{\bm S}_{{\bm r}-{\bm e}_x-{\bm e}_y} \\
    &\quad  + \frac{J_2-\Delta}{2} \sum_{{\bm r}\in{\rm A}} {\bm S}_{\bm r} \cdot{\bm S}_{{\bm r}+{\bm e}_x-{\bm e}_y}
            + \frac{J_2-\Delta}{2} \sum_{{\bm r}\in{\rm A}} {\bm S}_{\bm r} \cdot{\bm S}_{{\bm r}-{\bm e}_x+{\bm e}_y}
            + \frac{J_2+\Delta}{2} \sum_{{\bm r}\in{\rm B}} {\bm S}_{\bm r} \cdot{\bm S}_{{\bm r}+{\bm e}_x-{\bm e}_y} 
            + \frac{J_2+\Delta}{2} \sum_{{\bm r}\in{\rm B}} {\bm S}_{\bm r} \cdot{\bm S}_{{\bm r}-{\bm e}_x+{\bm e}_y}.
\end{aligned}
\end{align}
We refer to the up-spin (down-spin) sites as sublattice A(B). Similar to the Holstein-Primakoff transformation, we start from Eq.~(\ref{eq:SpinOpExpantion}). However, we now assume explicitly that there are only two sublattices, for which the Dyson-Maleev transformation is given by 
\begin{equation}
    \left\{ \begin{aligned}
        \hat{\mathsf{S}}^+_{{\bm r},{\rm A}} &\rightarrow \sqrt{S} \left( a_{\bm r} - \frac{1}{2S}a^\dagger_{\bm r}a_{\bm r}a_{\bm r} \right) \\
        \hat{\mathsf{S}}^-_{{\bm r},{\rm A}} &\rightarrow \sqrt{S} a^\dagger_{\bm r} \\
        \hat{\mathsf{S}}^0_{{\bm r},{\rm A}} &\rightarrow S - a^\dagger_{\bm r}a_{\bm r}
    \end{aligned} \right.
\end{equation}
for the sublattice A, and
\begin{equation}
    \left\{ \begin{aligned}
        \hat{\mathsf{S}}^+_{{\bm r},{\rm B}} &\rightarrow \sqrt{S} b_{\bm r} \\
        \hat{\mathsf{S}}^-_{{\bm r},{\rm B}} &\rightarrow \sqrt{S} \left( b^\dagger_{\bm r} - \frac{1}{2S}b^\dagger_{\bm r}b^\dagger_{\bm r}b_{\bm r} \right) \\
        \hat{\mathsf{S}}^0_{{\bm r},{\rm B}} &\rightarrow S - b^\dagger_{\bm r}b_{\bm r}
    \end{aligned} \right.
\end{equation} 
for the sublattice B. Thus, the spin-spin interactions in the altermagnetic Heisenberg model can be written as
\begin{align}
\begin{aligned}
    &\quad\quad {\bm S}_{{\bm r},{\rm A}} \cdot {\bm S}_{{\bm r}',{\rm B}} \\
    &= \left( \hat{\mathsf{S}}^+_{{\bm r} ,{\rm A}} {\bm e}^-_{\rm A} + \hat{\mathsf{S}}^-_{{\bm r} ,{\rm A}} {\bm e}^+_{\rm A} + \hat{\mathsf{S}}^0_{{\bm r} ,{\rm A}} {\bm e}^0_{\rm A} \right)
        \left( \hat{\mathsf{S}}^+_{{\bm r}',{\rm B}} {\bm e}^-_{\rm B} + \hat{\mathsf{S}}^-_{{\bm r}',{\rm B}} {\bm e}^+_{\rm B} + \hat{\mathsf{S}}^0_{{\bm r}',{\rm B}} {\bm e}^0_{\rm B} \right) \\
    &= Sa_{\bm r}b_{{\bm r}'} - \frac{1}{2} a^\dagger_{\bm r}a_{\bm r}a_{\bm r}b_{{\bm r}'} + Sa^\dagger_{\bm r}b^\dagger_{{\bm r}'} - \frac{1}{2}a^\dagger_{\bm r}b^\dagger_{{\bm r}'}b^\dagger_{{\bm r}'}b_{{\bm r}'}
    - S^2 + Sa^\dagger_{\bm r}a_{\bm r} + Sb^\dagger_{{\bm r}'}b_{{\bm r}'} - a^\dagger_{\bm r}a_{\bm r}b^\dagger_{{\bm r}'}b_{{\bm r}'},
    \label{eq:DM_SS1}
\end{aligned}
\end{align}
\begin{align}
\begin{aligned}
    &\quad\quad {\bm S}_{{\bm r},{\rm B}} \cdot {\bm S}_{{\bm r}',{\rm A}} \\
    &= \left( \hat{\mathsf{S}}^+_{{\bm r} ,{\rm B}} {\bm e}^-_{\rm B} + \hat{\mathsf{S}}^-_{{\bm r} ,{\rm B}} {\bm e}^+_{\rm B} + \hat{\mathsf{S}}^0_{{\bm r} ,{\rm B}} {\bm e}^0_{\rm B} \right)
        \left( \hat{\mathsf{S}}^+_{{\bm r}',{\rm A}} {\bm e}^-_{\rm A} + \hat{\mathsf{S}}^-_{{\bm r}',{\rm A}} {\bm e}^+_{\rm A} + \hat{\mathsf{S}}^0_{{\bm r}',{\rm A}} {\bm e}^0_{\rm A} \right) \\
    &= Sb_{\bm r}a_{{\bm r}'} - \frac{1}{2} b_{\bm r}a^\dagger_{{\bm r}'}a_{{\bm r}'}a_{{\bm r}'} + Sb^\dagger_{\bm r}a^\dagger_{{\bm r}'} - \frac{1}{2}b^\dagger_{\bm r}b^\dagger_{\bm r}b_{\bm r}a^\dagger_{{\bm r}'}
    - S^2 + Sb^\dagger_{\bm r}b_{\bm r} + Sa^\dagger_{{\bm r}'}a_{{\bm r}'} - b^\dagger_{\bm r}b_{\bm r}a^\dagger_{{\bm r}'}a_{{\bm r}'},
    \label{eq:DM_SS2}
\end{aligned}
\end{align}
\begin{align}
\begin{aligned}
    &\quad\quad {\bm S}_{{\bm r},{\rm A}} \cdot {\bm S}_{{\bm r}',{\rm A}} \\
    &= \left( \hat{\mathsf{S}}^+_{{\bm r} ,{\rm A}} {\bm e}^-_{\rm A} + \hat{\mathsf{S}}^-_{{\bm r} ,{\rm A}} {\bm e}^+_{\rm A} + \hat{\mathsf{S}}^0_{{\bm r} ,{\rm A}} {\bm e}^0_{\rm A} \right)
        \left( \hat{\mathsf{S}}^+_{{\bm r}',{\rm A}} {\bm e}^-_{\rm A} + \hat{\mathsf{S}}^-_{{\bm r}',{\rm A}} {\bm e}^+_{\rm A} + \hat{\mathsf{S}}^0_{{\bm r}',{\rm A}} {\bm e}^0_{\rm A} \right) \\
    &= Sa_{\bm r}a^\dagger_{{\bm r}'} - \frac{1}{2}a^\dagger_{\bm r}a_{\bm r}a_{\bm r}a^\dagger_{{\bm r}'} + Sa^\dagger_{\bm r}a_{{\bm r}'} - \frac{1}{2}a^\dagger_{\bm r}a^\dagger_{{\bm r}'}a_{{\bm r}'}a_{{\bm r}'}
    + S^2 - Sa^\dagger_{\bm r}a_{\bm r} - Sa^\dagger_{{\bm r}'}a_{{\bm r}'} + a^\dagger_{\bm r}a_{\bm r}a^\dagger_{{\bm r}'}a_{{\bm r}'},
    \label{eq:DM_SS3}
\end{aligned}
\end{align}
\begin{align}
\begin{aligned}
    &\quad\quad {\bm S}_{{\bm r},{\rm B}} \cdot {\bm S}_{{\bm r}',{\rm B}} \\
    &= \left( \hat{\mathsf{S}}^+_{{\bm r} ,{\rm B}} {\bm e}^-_{\rm B} + \hat{\mathsf{S}}^-_{{\bm r} ,{\rm B}} {\bm e}^+_{\rm B} + \hat{\mathsf{S}}^0_{{\bm r} ,{\rm B}} {\bm e}^0_{\rm B} \right)
        \left( \hat{\mathsf{S}}^+_{{\bm r}',{\rm B}} {\bm e}^-_{\rm B} + \hat{\mathsf{S}}^-_{{\bm r}',{\rm B}} {\bm e}^+_{\rm B} + \hat{\mathsf{S}}^0_{{\bm r}',{\rm B}} {\bm e}^0_{\rm B} \right) \\
    &= Sb_{\bm r}b^\dagger_{{\bm r}'} - \frac{1}{2}b_{\bm r}b^\dagger_{{\bm r}'}b^\dagger_{{\bm r}'}b_{{\bm r}'} + Sb^\dagger_{\bm r}b_{{\bm r}'} - \frac{1}{2}b^\dagger_{\bm r}b^\dagger_{\bm r}b_{\bm r}b_{{\bm r}'}
    - Sb^\dagger_{\bm r}b_{\bm r} - Sb^\dagger_{{\bm r}'}b_{{\bm r}'} + b^\dagger_{\bm r}b_{\bm r}b^\dagger_{{\bm r}'}b_{{\bm r}'},
    \label{eq:DM_SS4}
\end{aligned}
\end{align}
where ${\bm r}\neq{\bm r}'$. It is worth noting that Eqs.~(\ref{eq:DM_SS1}),~(\ref{eq:DM_SS2}),~(\ref{eq:DM_SS3}), and~(\ref{eq:DM_SS4}) consist only of two- and four-magnon terms, and there are no six-magnon or higher-order terms. Regarding the $\mathcal{O}(1/S^2)$ corrections, recall that in the Holstein-Primakoff transformation for gapless systems with type-I Goldstone modes, the unphysical divergencies in the self-energies illustrated in Figs.~\ref{fig:diagrams-NLSW}(c)-(e) are completely compensated among themselves. The same holds true for the Dyson-Maleev transformation. Since there are no six-magnon terms, there are no corrections from six-magnon Hartree-Fock mean fields shown in Fig.~\ref{fig:diagrams-NLSW}(e). Therefore, the individual divergences of the two bubble diagrams---three-magnon forward and backward bubbles---always cancel each other out. 

\par Within the linear spin-wave approximation, the bilinear eigenvalue problem obtained by the Dyson-Maleev transformation is identical to that obtained by the Holstein-Primakoff transformation. As discussed in the Method section in the main text, the diagonalized bilinear Hamiltonian is given by
\begin{equation}
    H_2 = \delta H'_0 + \sum_{{\bm k}\in\text{BZ}} \left( \epsilon_+({\bm k})\alpha^\dagger_{\bm k}\alpha_{\bm k} + \epsilon_-({\bm k})\beta^\dagger_{\bm k}\beta_{\bm k} \right),
\end{equation}
where the linear spin-wave dispersion relation $\epsilon_\pm({\bm k})$ is given by
\begin{equation}
    \epsilon_{\pm}({\bm k})/S = \sqrt{A({\bm k})^2-B({\bm k})^2} \pm \Delta({\bm k}).
\end{equation}
Note that $A({\bm k})=4J_1-4J_2+2J_2\left( \cos (k_x+k_y)+\cos (k_x-k_y) \right)$, $B({\bm k})=2J_1\left( \cos k_x + \cos k_y \right)$, $\Delta({\bm k})=2\Delta\left( \cos (k_x+k_y) - \cos (k_x-k_y) \right)$, and the Bogoliubov rotation is given by
\begin{align}
    \left\{
    \begin{aligned}
        a_{\bm k} &= \alpha_{ {\bm k}} \cosh\frac{X_{\bm k}}{2} +  \beta^\dagger_{-{\bm k}}\sinh\frac{X_{\bm k}}{2}, \\
        b_{\bm k} &=  \beta_{ {\bm k}} \cosh\frac{X_{\bm k}}{2} + \alpha^\dagger_{-{\bm k}}\sinh\frac{X_{\bm k}}{2}, \\
        a^\dagger_{\bm k} &= \alpha^\dagger_{ {\bm k}} \cosh\frac{X_{\bm k}}{2} +  \beta_{-{\bm k}}\sinh\frac{X_{\bm k}}{2} ,\\
        b^\dagger_{\bm k} &=  \beta^\dagger_{ {\bm k}} \cosh\frac{X_{\bm k}}{2} + \alpha_{-{\bm k}}\sinh\frac{X_{\bm k}}{2},
    \end{aligned} \right.
    \label{eq:BogoliubovRotation_DM}
\end{align}
where
\begin{align}
    \cos \frac{X_{\bm q}}{2} &=  {\frac{1}{\sqrt{2}}\left( 1+\frac{A({\bm q})}{\sqrt{A({\bm q})^2-B({\bm q})^2}} \right)^{1/2}}, \\
    \sin \frac{X_{\bm q}}{2} &= -\frac{B({\bm q})}{\sqrt{2}\sqrt{A({\bm q})^2-B({\bm q})^2} \left( \sqrt{A({\bm q})^2-B({\bm q})^2} + A({\bm q}) \right)}.
\end{align}
\subsection{Nonlinear spin-wave theory}
The four-magnon term $H_4$ is explicitly given by
\begin{align}
\begin{aligned}
    H_4
    &= \frac{1}{N_\text{muc}} \sum_{{\bm q}_1,{\bm q}_2,{\bm q}_3,{\bm q}_4} \left[ \right. \\
    &\quad -                            J_1\cos\left\{{\bm q}_4\cdot {\bm e}_x \right\} a^\dagger_{{\bm q}_1}a        _{{\bm q}_2}a        _{{\bm q}_3}b        _{{\bm q}_4} \delta\left({\bm q}_1-{\bm q_2}-{\bm q}_3-{\bm q}_4\right) \\
    &\quad -                            J_1\cos\left\{{\bm q}_1\cdot {\bm e}_x \right\} a^\dagger_{{\bm q}_1}b^\dagger_{{\bm q}_2}b^\dagger_{{\bm q}_3}b        _{{\bm q}_4} \delta\left({\bm q}_1+{\bm q_2}+{\bm q}_3-{\bm q}_4\right) \\
    &\quad - 2J_{1}\cos\left\{\left({\bm q}_4-{\bm q}_2\right)\cdot {\bm e}_x \right\} a^\dagger_{{\bm q}_1}b^\dagger_{{\bm q}_2}a        _{{\bm q}_3}b        _{{\bm q}_4} \delta\left({\bm q}_1+{\bm q_2}-{\bm q}_3-{\bm q}_4\right) \\
    &\quad -                            J_1\cos\left\{{\bm q}_4\cdot {\bm e}_y \right\} a^\dagger_{{\bm q}_1}a        _{{\bm q}_2}a        _{{\bm q}_3}b        _{{\bm q}_4} \delta\left({\bm q}_1-{\bm q_2}-{\bm q}_3-{\bm q}_4\right) \\
    &\quad -                            J_1\cos\left\{{\bm q}_1\cdot {\bm e}_y \right\} a^\dagger_{{\bm q}_1}b^\dagger_{{\bm q}_2}b^\dagger_{{\bm q}_3}b        _{{\bm q}_4} \delta\left({\bm q}_1+{\bm q_2}+{\bm q}_3-{\bm q}_4\right) \\
    &\quad - 2J_{1}\cos\left\{\left({\bm q}_4-{\bm q}_2\right)\cdot {\bm e}_y \right\} a^\dagger_{{\bm q}_1}b^\dagger_{{\bm q}_2}a        _{{\bm q}_3}b        _{{\bm q}_4} \delta\left({\bm q}_1+{\bm q_2}-{\bm q}_3-{\bm q}_4\right) \\
    &\quad -                        (J_2+\Delta)\cos\left\{{\bm q}_1\cdot({\bm e}_x+{\bm e}_y)\right\} a^\dagger_{{\bm q}_1}a^\dagger_{{\bm q}_2}a        _{{\bm q}_3}a        _{{\bm q}_4} \delta\left({\bm q}_1+{\bm q_2}-{\bm q}_3-{\bm q}_4\right) \\
    &\quad + (J_2+\Delta)\cos\left\{\left({\bm q}_4-{\bm q}_2\right)\cdot({\bm e}_x+{\bm e}_y)\right\} a^\dagger_{{\bm q}_1}a^\dagger_{{\bm q}_2}a        _{{\bm q}_3}a        _{{\bm q}_4} \delta\left({\bm q}_1+{\bm q_2}-{\bm q}_3-{\bm q}_4\right) \\
    &\quad -                        (J_2-\Delta)\cos\left\{{\bm q}_1\cdot({\bm e}_x-{\bm e}_y)\right\} a^\dagger_{{\bm q}_1}a^\dagger_{{\bm q}_2}a        _{{\bm q}_3}a        _{{\bm q}_4} \delta\left({\bm q}_1+{\bm q_2}-{\bm q}_3-{\bm q}_4\right) \\
    &\quad + (J_2-\Delta)\cos\left\{\left({\bm q}_4-{\bm q}_2\right)\cdot({\bm e}_x-{\bm e}_y)\right\} a^\dagger_{{\bm q}_1}a^\dagger_{{\bm q}_2}a        _{{\bm q}_3}a        _{{\bm q}_4} \delta\left({\bm q}_1+{\bm q_2}-{\bm q}_3-{\bm q}_4\right) \\
    &\quad -                        (J_2-\Delta)\cos\left\{{\bm q}_4\cdot({\bm e}_x+{\bm e}_y)\right\} b^\dagger_{{\bm q}_1}b^\dagger_{{\bm q}_2}b        _{{\bm q}_3}b        _{{\bm q}_4} \delta\left({\bm q}_1+{\bm q_2}-{\bm q}_3-{\bm q}_4\right) \\
    &\quad + (J_2-\Delta)\cos\left\{\left({\bm q}_4-{\bm q}_2\right)\cdot({\bm e}_x+{\bm e}_y)\right\} b^\dagger_{{\bm q}_1}b^\dagger_{{\bm q}_2}b        _{{\bm q}_3}b        _{{\bm q}_4} \delta\left({\bm q}_1+{\bm q_2}-{\bm q}_3-{\bm q}_4\right) \\
    &\quad -                        (J_2+\Delta)\cos\left\{{\bm q}_4\cdot({\bm e}_x-{\bm e}_y)\right\} b^\dagger_{{\bm q}_1}b^\dagger_{{\bm q}_2}b        _{{\bm q}_3}b        _{{\bm q}_4} \delta\left({\bm q}_1+{\bm q_2}-{\bm q}_3-{\bm q}_4\right) \\
    &\quad + (J_2+\Delta)\cos\left\{\left({\bm q}_4-{\bm q}_2\right)\cdot({\bm e}_x-{\bm e}_y)\right\} b^\dagger_{{\bm q}_1}b^\dagger_{{\bm q}_2}b        _{{\bm q}_3}b        _{{\bm q}_4} \delta\left({\bm q}_1+{\bm q_2}-{\bm q}_3-{\bm q}_4\right) \left. \right].
    \label{eq:H4DM}
\end{aligned}
\end{align}
We apply the Bogoliubov rotation given by Eq.~(\ref{eq:BogoliubovRotation_DM}) to $H_4$ given by Eq.~(\ref{eq:H4DM}) and then pick up all normal-ordered four-magnon terms. The result is summarized as follows:
\begin{equation}
\begin{aligned}
    \tilde{H}_4 = \frac{1}{N_\text{muc}} \sum_{{\bm q}_1,{\bm q}_2,{\bm q}_3,{\bm q}_4} & \left[ \left\{
      W^{(1)}_{{\bm q}_1,{\bm q}_2\leftarrow{\bm q}_3,{\bm q}_4} \alpha_{{\bm q}_1}^\dagger \alpha_{{\bm q}_2}^\dagger \alpha_{{\bm q}_3} \alpha_{{\bm q}_4}
    + W^{(2)}_{{\bm q}_1,{\bm q}_2\leftarrow{\bm q}_3,{\bm q}_4} \beta_{{\bm q}_1}^\dagger \beta_{{\bm q}_2}^\dagger \beta_{{\bm q}_3} \beta_{{\bm q}_4}
    + W^{(3)}_{{\bm q}_1,{\bm q}_2\leftarrow{\bm q}_3,{\bm q}_4} \alpha_{{\bm q}_1}^\dagger \beta_{{\bm q}_2}^\dagger \alpha_{{\bm q}_3} \beta_{{\bm q}_4} \right\} \delta({\bm q}_1+{\bm q}_2-{\bm q}_3-{\bm q}_4) \right. \\
    &\quad + \left\{
      W^{(4)}_{{\bm q}_1,{\bm q}_2,{\bm q}_3\leftarrow{\bm q}_4} \alpha_{{\bm q}_1}^\dagger \alpha_{{\bm q}_2}^\dagger \beta_{{\bm q}_3}^\dagger \alpha_{{\bm q}_4}
    + W^{(5)}_{{\bm q}_1,{\bm q}_2,{\bm q}_3\leftarrow{\bm q}_4} \alpha_{{\bm q}_1}^\dagger \beta_{{\bm q}_2}^\dagger \beta_{{\bm q}_3}^\dagger \beta_{{\bm q}_4} \right\} \delta({\bm q}_1+{\bm q}_2+{\bm q}_3-{\bm q}_4) \\
    &\quad
    + W^{(6)}_{{\bm q}_1,{\bm q}_2,{\bm q}_3,{\bm q}_4} \alpha_{{\bm q}_1}^\dagger \alpha_{{\bm q}_2}^\dagger \beta_{{\bm q}_3}^\dagger \beta_{{\bm q}_4}^\dagger \delta({\bm q}_1+{\bm q}_2+{\bm q}_3+{\bm q}_4) \\
    &\quad + \left\{
      W^{(7)}_{{\bm q}_1\leftarrow{\bm q}_2,{\bm q}_3,{\bm q}_4} \alpha_{{\bm q}_1}^\dagger \alpha_{{\bm q}_2} \alpha_{{\bm q}_3} \beta_{{\bm q}_4}
    + W^{(8)}_{{\bm q}_1\leftarrow{\bm q}_2,{\bm q}_3,{\bm q}_4} \beta_{{\bm q}_1}^\dagger \alpha_{{\bm q}_2} \beta_{{\bm q}_3} \beta_{{\bm q}_4} \right\} \delta({\bm q}_1-{\bm q}_2-{\bm q}_3-{\bm q}_4) \\
    &\quad \left.
    + W^{(9)}_{{\bm q}_1,{\bm q}_2,{\bm q}_3,{\bm q}_4} \alpha_{{\bm q}_1} \alpha_{{\bm q}_2} \beta_{{\bm q}_3} \beta_{{\bm q}_4} \delta({\bm q}_1+{\bm q}_2+{\bm q}_3+{\bm q}_4) \right].
\end{aligned}
\end{equation}
The expressions of the vertices, for example, $W^{(4)}_{{\bm q}_1,{\bm q}_2,{\bm q}_3\leftarrow{\bm q}_4}$ and $W^{(7)}_{{\bm q}_1\leftarrow{\bm q}_2,{\bm q}_3,{\bm q}_4}$, are given by
\begin{align}
\begin{aligned}
    W^{(4)}_{{\bm q}_1,{\bm q}_2,{\bm q}_3\leftarrow{\bm q}_4} 
    &= \frac{1}{2} \left[
    -\frac{J_1}{2}    \mathsf{F}^{\rm NN}_{-{\bm q}_2} \mathsf{c}_{{\bm q}_1}\mathsf{s}_{{\bm q}_2}\mathsf{s}_{{\bm q}_3}\mathsf{c}_{{\bm q}_4}
    -\frac{J_1}{2}    \mathsf{F}^{\rm NN}_{ {\bm q}_1} \mathsf{c}_{{\bm q}_1}\mathsf{s}_{{\bm q}_2}\mathsf{c}_{{\bm q}_3}\mathsf{s}_{{\bm q}_4}
    -\frac{J_{1}}{2} \mathsf{F}^{\rm NN}_{-{\bm q}_2-{\bm q}_3} \mathsf{c}_{{\bm q}_1}\mathsf{s}_{{\bm q}_2}\mathsf{c}_{{\bm q}_3}\mathsf{c}_{{\bm q}_4}
    -\frac{J_{1}}{2} \mathsf{F}^{\rm NN}_{-{\bm q}_2+{\bm q}_4} \mathsf{c}_{{\bm q}_1}\mathsf{s}_{{\bm q}_2}\mathsf{s}_{{\bm q}_3}\mathsf{s}_{{\bm q}_4} \right. \\
    &\quad\quad
    -\left\{\frac{J_2+\Delta}{2}\left(\mathsf{F}^{{\rm NNN}x}_{ {\bm q}_1}-\mathsf{F}^{{\rm NNN}x}_{-{\bm q}_3-{\bm q}_2}\right) + \frac{J_2-\Delta}{2}\left(\mathsf{F}^{{\rm NNN}y}_{ {\bm q}_1}-\mathsf{F}^{{\rm NNN}y}_{-{\bm q}_3-{\bm q}_2}\right)\right\}
    \mathsf{c}_{{\bm q}_1}\mathsf{c}_{{\bm q}_2}\mathsf{s}_{{\bm q}_3}\mathsf{c}_{{\bm q}_4} \\
    &\quad\quad
    -\left\{\frac{J_2+\Delta}{2}\left(\mathsf{F}^{{\rm NNN}x}_{ {\bm q}_1}-\mathsf{F}^{{\rm NNN}x}_{ {\bm q}_4-{\bm q}_2}\right) + \frac{J_2-\Delta}{2}\left(\mathsf{F}^{{\rm NNN}y}_{ {\bm q}_1}-\mathsf{F}^{{\rm NNN}y}_{ {\bm q}_4-{\bm q}_2}\right)\right\}
    \mathsf{c}_{{\bm q}_1}\mathsf{c}_{{\bm q}_2}\mathsf{s}_{{\bm q}_3}\mathsf{c}_{{\bm q}_4} \\
    &\quad\quad
    -\left\{\frac{J_2-\Delta}{2}\left(\mathsf{F}^{{\rm NNN}x}_{-{\bm q}_2}-\mathsf{F}^{{\rm NNN}x}_{-{\bm q}_2+{\bm q}_4}\right) + \frac{J_2+\Delta}{2}\left(\mathsf{F}^{{\rm NNN}y}_{-{\bm q}_2}-\mathsf{F}^{{\rm NNN}y}_{-{\bm q}_2+{\bm q}_4}\right)\right\}
    \mathsf{s}_{{\bm q}_1}\mathsf{s}_{{\bm q}_2}\mathsf{c}_{{\bm q}_3}\mathsf{s}_{{\bm q}_4} \\
    &\quad\quad \left.
    -\left\{\frac{J_2-\Delta}{2}\left(\mathsf{F}^{{\rm NNN}x}_{-{\bm q}_2}-\mathsf{F}^{{\rm NNN}x}_{-{\bm q}_2-{\bm q}_3}\right) + \frac{J_2+\Delta}{2}\left(\mathsf{F}^{{\rm NNN}y}_{-{\bm q}_2}-\mathsf{F}^{{\rm NNN}y}_{-{\bm q}_2-{\bm q}_3}\right)\right\}
    \mathsf{s}_{{\bm q}_1}\mathsf{s}_{{\bm q}_2}\mathsf{c}_{{\bm q}_3}\mathsf{s}_{{\bm q}_4} \right] \\
    &\quad + \left( {\bm q}_1\leftrightarrow{\bm q}_2 \right),
    \label{eq:W4_DysonMaleev}
\end{aligned}
\end{align}
\begin{align}
\begin{aligned}
    W^{(7)}_{{\bm q}_1\leftarrow{\bm q}_2,{\bm q}_3,{\bm q}_4} 
    &= \frac{1}{2} \left[
    -\frac{J_1}{4}    \mathsf{F}^{\rm NN}_{ {\bm q}_4} \mathsf{c}_{{\bm q}_1}\mathsf{c}_{{\bm q}_2}\mathsf{c}_{{\bm q}_3}\mathsf{c}_{{\bm q}_4}
    -\frac{J_1}{4}    \mathsf{F}^{\rm NN}_{-{\bm q}_1} \mathsf{s}_{{\bm q}_1}\mathsf{c}_{{\bm q}_2}\mathsf{c}_{{\bm q}_3}\mathsf{s}_{{\bm q}_4}
    -\frac{J_1}{4}    \mathsf{F}^{\rm NN}_{ {\bm q}_1} \mathsf{c}_{{\bm q}_1}\mathsf{s}_{{\bm q}_2}\mathsf{s}_{{\bm q}_3}\mathsf{c}_{{\bm q}_4}
    -\frac{J_1}{4}    \mathsf{F}^{\rm NN}_{-{\bm q}_4} \mathsf{s}_{{\bm q}_1}\mathsf{s}_{{\bm q}_2}\mathsf{s}_{{\bm q}_3}\mathsf{s}_{{\bm q}_4} \right. \\
    &\quad\quad
    -\frac{J_{1}}{2} \mathsf{F}^{\rm NN}_{ {\bm q}_4+{\bm q}_2} \mathsf{c}_{{\bm q}_1}\mathsf{s}_{{\bm q}_2}\mathsf{c}_{{\bm q}_3}\mathsf{c}_{{\bm q}_4}
    -\frac{J_{1}}{2} \mathsf{F}^{\rm NN}_{-{\bm q}_1+{\bm q}_2} \mathsf{s}_{{\bm q}_1}\mathsf{s}_{{\bm q}_2}\mathsf{c}_{{\bm q}_3}\mathsf{s}_{{\bm q}_4} \\
    &\quad\quad
    -\left\{\frac{J_2+\Delta}{2}\left(\mathsf{F}^{{\rm NNN}x}_{ {\bm q}_1}-\mathsf{F}^{{\rm NNN}x}_{ {\bm q}_3+{\bm q}_4}\right) + \frac{J_2-\Delta}{2}\left(\mathsf{F}^{{\rm NNN}y}_{ {\bm q}_1}-\mathsf{F}^{{\rm NNN}y}_{ {\bm q}_3+{\bm q}_4}\right)\right\}
    \mathsf{c}_{{\bm q}_1}\mathsf{c}_{{\bm q}_2}\mathsf{c}_{{\bm q}_3}\mathsf{s}_{{\bm q}_4} \\
    &\quad\quad
    -\left\{\frac{J_2+\Delta}{2}\left(\mathsf{F}^{{\rm NNN}x}_{-{\bm q}_4}-\mathsf{F}^{{\rm NNN}x}_{ {\bm q}_3-{\bm q}_1}\right) + \frac{J_2-\Delta}{2}\left(\mathsf{F}^{{\rm NNN}y}_{-{\bm q}_4}-\mathsf{F}^{{\rm NNN}y}_{ {\bm q}_3-{\bm q}_1}\right)\right\}
    \mathsf{c}_{{\bm q}_1}\mathsf{c}_{{\bm q}_2}\mathsf{c}_{{\bm q}_3}\mathsf{s}_{{\bm q}_4} \\
    &\quad\quad
    -\left\{\frac{J_2-\Delta}{2}\left(\mathsf{F}^{{\rm NNN}x}_{-{\bm q}_1}-\mathsf{F}^{{\rm NNN}x}_{-{\bm q}_1+{\bm q}_3}\right) + \frac{J_2+\Delta}{2}\left(\mathsf{F}^{{\rm NNN}y}_{-{\bm q}_1}-\mathsf{F}^{{\rm NNN}y}_{-{\bm q}_1+{\bm q}_3}\right)\right\}
    \mathsf{s}_{{\bm q}_1}\mathsf{s}_{{\bm q}_2}\mathsf{s}_{{\bm q}_3}\mathsf{c}_{{\bm q}_4} \\
    &\quad\quad \left.
    -\left\{\frac{J_2-\Delta}{2}\left(\mathsf{F}^{{\rm NNN}x}_{ {\bm q}_4}-\mathsf{F}^{{\rm NNN}x}_{ {\bm q}_4+{\bm q}_3}\right) + \frac{J_2+\Delta}{2}\left(\mathsf{F}^{{\rm NNN}y}_{ {\bm q}_4}-\mathsf{F}^{{\rm NNN}y}_{ {\bm q}_4+{\bm q}_3}\right)\right\}
    \mathsf{s}_{{\bm q}_1}\mathsf{s}_{{\bm q}_2}\mathsf{s}_{{\bm q}_3}\mathsf{c}_{{\bm q}_4} \right] \\
    &\quad + \left( {\bm q}_2\leftrightarrow{\bm q}_3 \right),
    \label{eq:W7_DysonMaleev}
\end{aligned}
\end{align}
where $\mathsf{c}_{\bm q}=\cosh (X_{\bm q}/2)$, $\mathsf{s}_{\bm q}=\sinh (X_{\bm q}/2)$, and the form factors are given by
\begin{equation}
\begin{aligned}
    \mathsf{F}^\text{NN}_{\bm q} &= 2( \cos q_x + \cos q_y ), \\
    \mathsf{F}^{\text{NNN}x}_{\bm q} &= \cos (q_x+q_y), \\
    \mathsf{F}^{\text{NNN}y}_{\bm q} &= \cos (q_x-q_y).
\end{aligned}
\end{equation}
The self-energy of the three-magnon forward bubble diagram with the Dyson-Maleev transformation is given by
\begin{equation}
    \Sigma_{{\bm k},\alpha\alpha}^\text{TMFB}(i\omega_n) = \frac{2}{\hbar} \frac{1}{(2\pi)^4} \int_{-\pi}^\pi\int_{-\pi}^\pi\int_{-\pi}^\pi\int_{-\pi}^\pi dq_{1x}dq_{1y}dq_{2x}dq_{2y} \frac{W^{(4)}_{{\bm q}_1,{\bm q}_2,{\bm k}-{\bm q}_1-{\bm q}_2\leftarrow{\bm k}} W^{(7)}_{{\bm k}\leftarrow{\bm q}_1,{\bm q}_2,{\bm k}-{\bm q}_1-{\bm q}_2} }{i\omega_n-\epsilon_+({\bm q}_1)-\epsilon_+({\bm q}_2)-\epsilon_-({\bm k}-{\bm q}_1-{\bm q}_2)}.
    \label{eq:DM_TMFB}
\end{equation}
The long-wavelength scaling of the vertex part $W^{(4)}_{{\bm q}_1,{\bm q}_2,{\bm k}-{\bm q}_1-{\bm q}_2\leftarrow{\bm k}} W^{(7)}_{{\bm k}\leftarrow{\bm q}_1,{\bm q}_2,{\bm k}-{\bm q}_1-{\bm q}_2}$ depends on the specific decay processes. We evaluate the scaling of $W^{(4)}_{{\bm q}_1,{\bm q}_2,{\bm k}-{\bm q}_1-{\bm q}_2\leftarrow{\bm k}} W^{(7)}_{{\bm k}\leftarrow{\bm q}_1,{\bm q}_2,{\bm k}-{\bm q}_1-{\bm q}_2}$ using a Mathematica built-in function $\mathsf{Series}$. 
\section{Power-law scaling of magnon damping in altermagnets}
\renewcommand{\theequation}{D.\arabic{equation}}
\label{sec:scaling}
\subsection{Decay phase space of the upper branch in 2D planar $d$-wave altermagnets}
In the 2D planar $d$-wave altermagnet, the decay phase space is lying in a 4D space spanned by crystal momenta $\{\vec{q}_{1x},\vec{q}_{1y},\vec{q}_{2x},\vec{q}_{2y}\}$.
We assume that the long-wavelength expansion of the bare magnon dispersion is given by
\begin{equation}
    \epsilon_\pm(k_x,k_y) := \nu\sqrt{k_x^2+k_y^2} \pm \Delta\left(k_x^2-k_y^2\right).
\end{equation}
\par We first consider the following one-to-three scattering process with a restriction that one of the three decay products is the Goldstone mode:
\begin{equation}
    \epsilon_+(k,0) \rightarrow \epsilon_+(0,0) + \epsilon_+(q_x,q_y) + \epsilon_-(k-q_x,-q_y),
\end{equation}
where momentum indices were replaced as $\{k_x,k_y,q_{1x},q_{1y},q_{2x},q_{2y}\}=\{k,0,0,0,q_x,q_y\}$.
Note that additional assumptions $0<q_x<k$, $0<q_y\ll q_x$, and $q_y\ll k-q_x$ are satisfied. A Maclaurin expansion with respect to $q_y$ leads to
\begin{align}
    \epsilon_+(q_x,q_y) &= \nu \sqrt{q_x^2+q_y^2} + \Delta (q_x^2-q_y^2) \approx \nu q_x \left( 1+\frac{1}{2}\frac{q_y^2}{q_x^2} \right) + \Delta (q_x^2-q_y^2), \\
    \epsilon_-(k-q_x,-q_y) &= \nu \sqrt{(k-q_x)^2+q_y^2} - \Delta \left\{(k-q_x)^2-q_y^2\right\} \approx \nu (k-q_x) \left\{ 1+\frac{1}{2}\frac{q_y^2}{(k-q_x)^2} \right\} - \Delta \left\{(k-q_x)^2-q_y^2\right\}.
\end{align}
It follows from energy conservation of the scattering process that
\begin{equation}
    \nu q_y^2 = 4\Delta q_x (k-q_x)^2.
    \label{eq:energy_conservation_simplest}
\end{equation}
We find that $q_y$ is maximized when $q_x=k/3$, which indicates that $q_{1x}, \ q_{2x}\propto|\Delta|^0 k^1$. The maximum of $q_y$ is given by
\begin{equation}
    q_{y,\text{max}} = \frac{4}{3\sqrt{3\nu}} \Delta^{1/2}k^{3/2}.
\end{equation}
This result indicates that $q_{2y}$ scales as $q_{2y}\propto|\Delta|^{1/2}k^{3/2}$. Therefore, within this simplest scattering process, the decay phase space scales as $q_{1x}\propto k$, $q_{1y}\propto |\Delta|^{1/2}k^{3/2}$, $q_{2x}\propto k$, and $q_{2y}\propto |\Delta|^{1/2}k^{3/2}$.
\par Next, we consider a general case in which the restriction that one of the three decay products must be the Goldstone mode is lifted:
\begin{equation}
    \epsilon_+(k,0) \rightarrow \epsilon_+(q_{1x},q_{1y}) + \epsilon_+(q_{2x},q_{2y}) + \epsilon_-(k-q_{1x}-q_{2x},-q_{1y}-q_{2y}).
\end{equation}
Up to the second-order in $q_{1y}$ and $q_{2y}$, energy conservation implies
\begin{equation}
    2\Delta k^2 = \frac{\nu}{2} \left\{ \frac{q_{1y}^2}{q_{1x}} + \frac{q_{2y}^2}{q_{2x}} + \frac{(q_{1y}+q_{2y})^2}{k-q_{1x}-q_{1y}} \right\} + 2\Delta \left\{ k(q_{1x}+q_{2x}) - (q_{1x}q_{2x}-q_{1y}q_{2y}) \right\}. \label{eq:encongeneral}
\end{equation}
We set the scaling of $q_{1x}$ and $q_{2x}$ to $q_{1x}\propto |\Delta|^0k^1$ and $q_{2x}\propto |\Delta|^0k^1$. Under these assumptions, Eq.~\eqref{eq:encongeneral} simplifies to
\begin{equation}
    4f_4 \Delta k^3 = \nu \left\{ f_1q_{1y}^2+f_2q_{2y}^2+\left(f_3-\frac{4\Delta k}{\nu}\right)q_{1y}q_{2y} \right\},
\end{equation}
where $\{f_\mu\}$ $(\mu=1,2,3,4)$ are dimensionless constants, and $f_3>0$ and $f_4>0$ are satisfied. Under the additional assumption that $q_{1y}$ and $q_{2y}$ have the same scaling order (i.e., $q_{1y}+f_5q_{2y}=0, \ f_5\in\mathbb{R}$), we obtain the following solutions:
\begin{equation}
\begin{aligned}
    \{q_{1y},q_{2y}\} &= \left\{ \frac{2f_4^{1/2}f_5 \Delta^{1/2}k^{3/2}}{\sqrt{\nu(f_1f_5^2+f_2-f_3f_5)+4f_5\Delta k}}, \ -\frac{2f_4^{1/2} \Delta^{1/2}k^{3/2}}{\sqrt{\nu(f_1f_5^2+f_2-f_3f_5)+4f_5\Delta k}} \right\}, \\
    &\quad \left\{ -\frac{2f_4^{1/2}f_5 \Delta^{1/2}k^{3/2}}{\sqrt{\nu(f_1f_5^2+f_2-f_3f_5)+4f_5\Delta k}}, \ \frac{2f_4^{1/2} \Delta^{1/2}k^{3/2}}{\sqrt{\nu(f_1f_5^2+f_2-f_3f_5)+4f_5\Delta k}} \right\}.
\end{aligned}
\end{equation}
We obtained the universal relation $q_{1y},q_{2y}\propto\Delta^{1/2}k^{3/2}$.
Therefore, in the limit of $k\rightarrow 0$ and $|\Delta|\rightarrow0$, the decay phase space for the upper branch scales as $q_{1x}\propto k$, $q_{1y}\propto |\Delta|^{1/2}k^{3/2}$, $q_{2x}\propto k$, and $q_{2y}\propto |\Delta|^{1/2}k^{3/2}$. This finding is identical to the aforementioned case in which one of the three decay products is the Goldstone mode.
\par As the final step, we perform a dimensional analysis of the decay phase space $V_\textrm{d-wave}(\mathbf{k})$, which is essentially given by Eq.~(\ref{eq:DM_TMFB}), with the vertex part $W^{(4)}W^{(7)}$ replaced by unity. Plugging Eq.~(\ref{eq:energy_conservation_simplest}) into Eq.~(\ref{eq:DM_TMFB}) yields
\begin{equation}
\begin{aligned}
    V_\textrm{d-wave}(\mathbf{k}) &\propto \int{dq_{x}} \int{dq_{y}} \int{dq'_{x}} \int{dq'_{y}} \ \delta \left( \frac{q_y^2}{q_x}-\Delta k^2 \right) \\
    &\sim \int{dq_{x}} \int{dq_{y}} \int{dq'_{x}} \int{dq'_{y}} \ \underbrace{q_x}_{\propto k} \ \delta \left( q_y^2-\Delta k^2\underbrace{q_x}_{\propto k} \right) \\
    &\sim \underbrace{\int{dq_{x}} \int{dq_{y}} \int{dq'_{x}} \int{dq'_{y}}}_{\propto k \cdot\Delta^{1/2}k^{3/2}\cdot k\cdot\Delta^{1/2}k^{3/2}} \ k \ \delta \left( q_y^2-\Delta k^3 \right) \\
    &\sim \Delta k^6 \underbrace{\frac{1}{q_y^2}}_{\propto (\Delta^{1/2} k^{3/2})^{-2}} \underbrace{\left[ \delta\left(1-\frac{\Delta^{1/2}k^{3/2}}{q_y}\right) + \delta\left(1+\frac{\Delta^{1/2}k^{3/2}}{q_y}\right)\right]}_{\propto \Delta^0 k^0} \\
    &\sim \Delta^0 k^3,
    \label{eq:dimensional_analysis}
\end{aligned}
\end{equation}
where some constant coefficients were omitted for simplicity. Thus, we conclude that the scaling of the decay phase space of the upper branch in the 2D planar $d$-wave altermagnet $V_\textrm{d-wave}(\mathbf{k})$ is universally $\Delta^0k^3$. Note that, since the following pairs of momentum coordinates, $q_{1x}\leftrightarrow q_{2x}$ and $q_{1y}\leftrightarrow q_{2y}$, are respectively identical with each other in the dimensional analysis, we replaced $\{q_{1x},q_{1y},q_{2x},q_{2y}\}$ as $\{q_x,q_y,q'_x,q'_y\}$ in Eq.~(\ref{eq:dimensional_analysis}) to avoid confusion.
Note also that the scaling $\Delta^0$ indicates a threshold behavior for a given momentum $k>0$. As we will discuss in the next subsection, in the case of $d$-wave altermagnets, the threshold value of $\Delta$ is $\Delta^{\star\star}_{\textrm{d-wave}}(k\rightarrow0)=0$, which means that an infinitesimal $\Delta$ is enough to open decay phase space of the long-wavelength upper-branch magnon in $d$-wave altermagnets.
\subsection{Generalization to all 2D planar altermagnets}
The dimensional analysis given by Eq.~(\ref{eq:dimensional_analysis}) can be generalized to altermagnets with a higher even-parity spin splitting than $d$-wave. The generalized scaling analysis for the 2D planar $d$- $(n=2)$, $g$- $(n=4)$, and $i$-wave $(n=6)$ altermagnets is given in the form
\begin{equation}
\begin{aligned}
    V_\textrm{d/g/i-wave}(\mathbf{k}) \propto \int{dq_{x}} \int{dq_{y}} \int{dq'_{x}} \int{dq'_{y}} \ \delta \left( \frac{q_y^2}{q_x}-\Delta k^n \right) \sim q_x^2q_y^2 \frac{q_x}{q_y^2} \delta \left( 1-\frac{\Delta k^n q_x}{q_y^2} \right)
    \sim q_x^3 \sim \Delta^0 k^3.
    \label{eq:dimensional_analysis_g_and_i}
\end{aligned}
\end{equation}
Note that the $q_y$ dependence is canceled out irrespective of $n$, which indicates universality of scaling law $V(\mathbf{k})\propto|\Delta|^0k^3$ for all $d$-, $g$-, and $i$-wave spin splittings.
\par As we briefly mentioned in the previous subsection, the scaling $|\Delta|^0$ indicates a threshold behavior. The threshold behavior means that the upper-branch magnon damping $\Gamma_\textrm{d/g/i-wave}(\mathbf{k})$ exhibits a discontinuous jump at a threshold $\Delta^{\star\star}(\mathbf{k})$, i.e., $\Gamma(\mathbf{k})=0$ ($\Gamma(\mathbf{k})>0$) when $|\Delta(\mathbf{k})|<\Delta^{\star\star}(\mathbf{k})$ $\left(|\Delta(\mathbf{k})|>\Delta^{\star\star}(\mathbf{k})\right)$. While the scaling law $V(\mathbf{k}) \propto k^3$ is universal for all 2D planar altermagnets, the difference among $d$-, $g$-, and $i$-wave spin splittings appear in the functional form of the threshold $\Delta^{\star\star}_\textrm{d/g/i-wave}(\mathbf{k})$, which we derive in the following.
\par To simplify the analysis, as we did in the previous subsection, we consider one-to-three scattering process on a maximally split 1D line with a restriction that one of the three decay products is the Goldstone mode:
\begin{equation}
    \epsilon^\textrm{d/g/i-wave}_+(k,0) \rightarrow 
    \epsilon^\textrm{d/g/i-wave}_+(0,0) +
    \epsilon^\textrm{d/g/i-wave}_+(q,0) +
    \epsilon^\textrm{d/g/i-wave}_-(k-q,0),
    \label{eq:1D_energy_conservation}
\end{equation}
where $k>0$, $0<q<k$, and the bare dispersion relations $\epsilon^\textrm{d/g/i-wave}_+(k_x,k_y)$ are respectively given by
\begin{align}
    \epsilon^\textrm{d-wave}_\pm(k,0) &= \pm \Delta k^2 + \nu|k| + \kappa|k|^3 + \chi|k|^5 + \cdots,
    \label{eq:1Ddisp_dwave} \\
    \epsilon^\textrm{g-wave}_\pm(k,0) &= \pm \Delta k^4 + \nu|k| + \kappa|k|^3 + \chi|k|^5 + \cdots,
    \label{eq:1Ddisp_gwave} \\
    \epsilon^\textrm{i-wave}_\pm(k,0) &= \pm \Delta k^6 + \nu|k| + \kappa|k|^3 + \chi|k|^5 + \cdots.
    \label{eq:1Ddisp_iwave}
\end{align}
Substituting Eqs.~(\ref{eq:1Ddisp_dwave})-(\ref{eq:1Ddisp_iwave}) into Eq.~(\ref{eq:1D_energy_conservation}) leads to the following equations:
\begin{align}
    d\textrm{-wave}: &\quad \left[ 2\Delta  + 3\kappa q + 5\chi q \left\{k(k-q)+q^2\right\} + \cdots \right] k (k-q) = 0,
    \label{eq:ecnsv_d} \\
    g\textrm{-wave}: &\quad \left[ 2\Delta \left\{ k(k-q)+2q^2 \right\} + 3\kappa q + 5\chi q \left\{k(k-q)+q^2\right\} + \cdots \right] k(k-q) = 0, 
    \label{eq:ecnsv_g} \\
    i\textrm{-wave}: &\quad \left[ \Delta (2k^4-4k^3q+11k^2q^2-9kq^3+6q^4) + 3\kappa q + 5\chi q \left\{k(k-q)+q^2\right\} + \cdots \right] k(k-q) = 0.
    \label{eq:ecnsv_i}
\end{align}
\begin{figure}
    \centering
    \includegraphics[width=11cm]{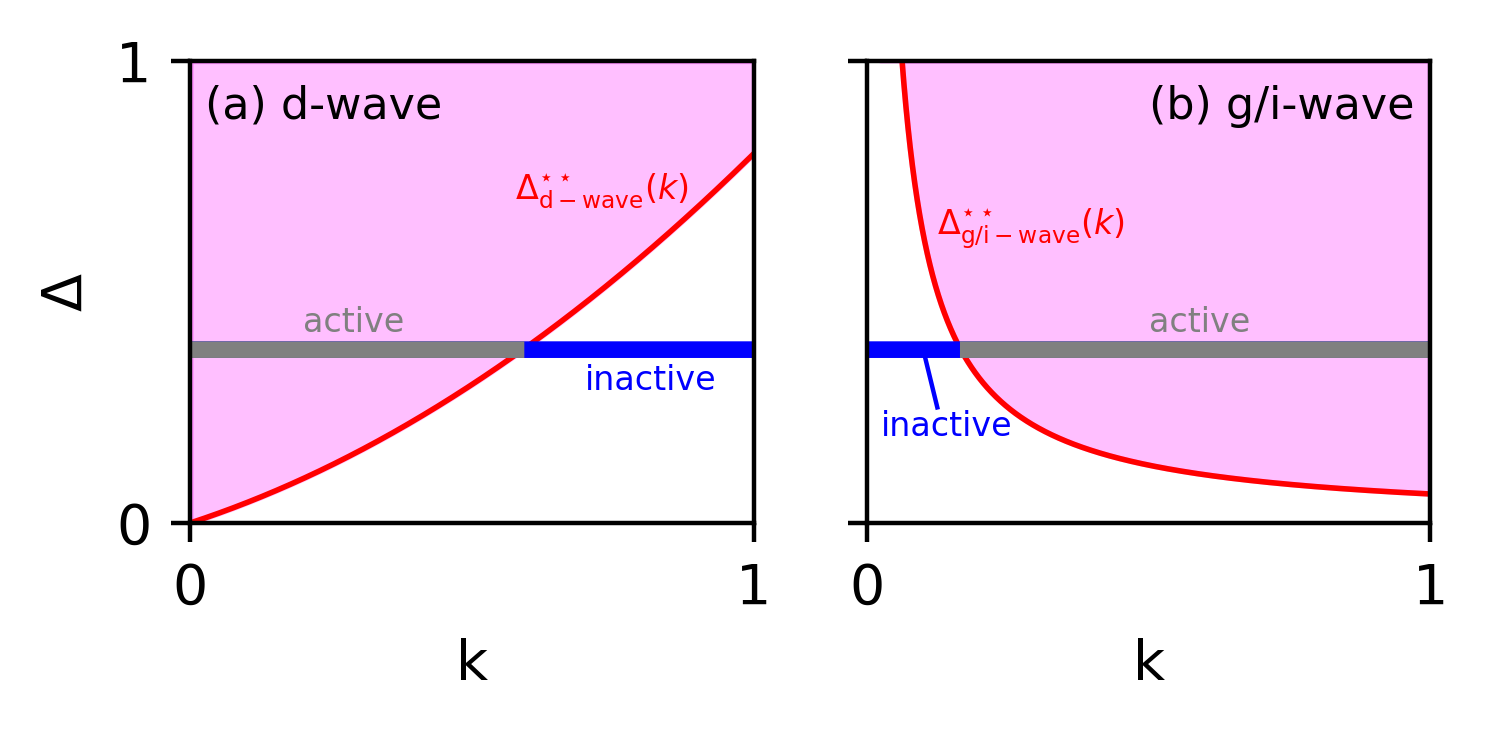}
    \caption{The $k^3$ decay mechanism thresholds (a) $\Delta^{\star\star}_\textrm{d-wave}(k)$ and (b) $\Delta^{\star\star}_\textrm{g/i-wave}(k)$ discussed in the text. This mechanism is active in $d$-wave altermagnets but inactive in $g$- and $i$-wave altermagnets.
    }
    \label{fig:Delta_thresholds}
\end{figure}
Equations~(\ref{eq:ecnsv_d})-(\ref{eq:ecnsv_i}) immediately yield thresholds given by
\begin{align}
    \Delta^{\star\star}_\textrm{d-wave}\left(k\right) &\sim -c^d_1 \kappa k + \mathcal{O}(k^2), \\
    \Delta^{\star\star}_\textrm{g-wave}\left(k\right) 
    &\sim -c^g_{-1} \frac{\kappa}{k} + \mathcal{O}(k^1), \\
    \Delta^{\star\star}_\textrm{i-wave}\left(k\right)
    &\sim - c^i_{-3}\frac{\kappa}{k^3} + \mathcal{O}(k^{-1}),
\end{align}
where $c^d_1$, $c^g_{-1}$, and $c^i_{-3}$ are dimensionless positive constants. These thresholds are visualized in the $k$-$\Delta$ plane in Fig.~\ref{fig:Delta_thresholds}. When the cubic nonlinearity is negative $(\kappa<0)$, $\Delta^{\star\star}_\textrm{g-wave}(k)$ and $\Delta^{\star\star}_\textrm{i-wave}(k)$ diverge into positive infinity in the long-wavelength limit $k\rightarrow0$. This observation signals that the $V(\mathbf{k}) \propto k^3$ decay mechanism is inactive in $g$- and $i$-wave altermagnets. 
\par Since the $V(\mathbf{k}) \propto k^3$ mechanism is not active in the long-wavelength regime in $g$- and $i$-wave altermagnets, we next explore other mechanisms with a higher power of $k$ as candidates for leading-order mechanisms: 

\textbf{g-wave altermagnets:} For $g$-wave altermagnets, Eq.~(\ref{eq:ecnsv_g}) with $\chi=0$ gives the solution $q\propto-\frac{2\Delta}{3\kappa}k^2$. This solution corresponds to a decay mechanism with a decay surface scaling as $q_{1x},q_{2x}\propto|\Delta|^1k^2$ and $q_{1y},q_{2y}\propto|\Delta|^1k^3$. One obtains $V_\textrm{g-wave}(\mathbf{k})\propto|\Delta|^3k^6$. In the limit of $|\kappa|\rightarrow0$, this decay mechanism is inactive, and instead, as seen in Eq.~(\ref{eq:ecnsv_g}), the aforementioned $k^3$ channel becomes active. Therefore, Eq.~(\ref{eq:ecnsv_g}) describes a crossover-like behavior between $V_\textrm{g-wave}(\mathbf{k}) \propto k^6$ and $V_\textrm{g-wave}(\mathbf{k}) \propto k^3$ for varying $\kappa(<0)$. 

\textbf{i-wave altermagnets:} For $i$-wave altermagnets, the previously discussed $k^6$ decay mechanism is inactive because Eq.~(\ref{eq:ecnsv_i}) gives the solution $q\propto\Delta^\gamma$ $(\gamma<0)$, which diverges in the limit of $|\Delta| \rightarrow 0$. Instead, the only possible decay mechanism we have identified is given by the following scattering process on a maximally split 1D line:
\begin{equation}
    \epsilon_+^\textrm{i-wave}(k,0)\rightarrow\epsilon_+^\textrm{i-wave}(q,0)+\epsilon_+^\textrm{i-wave}(-q,0)+\epsilon_-^\textrm{i-wave}(k,0).
\end{equation}
The decay phase space of this decay mechanism scales as $q_{1x},q_{1y},q_{2x},q_{2y}\propto|\Delta|^1k^6$, which gives $V_\textrm{i-wave}(\mathbf{k})\propto|\Delta|^3k^{18}$. Other decay channels with lower powers of $k$ may be found; the comprehensive analysis is left for future study.
\subsection{List of long-wavelength scalings of magnon damping}

Here we summarize the leading-order magnon damping in $d$- and $g$-wave altermagnets in the long-wavelength limit. Scalings of the vertex parts are derived by expanding Eqs.~(\ref{eq:W4_DysonMaleev}) and~(\ref{eq:W7_DysonMaleev}) using a Mathematica built-in function $\mathsf{Series}$.

\par \textbf{d-wave altermagnets:} The decay phase space scales as $q_{1x},q_{2x}\propto|\Delta|^0|\mathbf{k}|^1$ and $q_{1y},q_{2y}\propto|\Delta|^{1/2}|\mathbf{k}|^{3/2}$, which gives $V_\textrm{d-wave}(\mathbf{k}) \propto |\Delta|^0 |\mathbf{k}|^3$. Substituting these scalings into Eqs.~(\ref{eq:W4_DysonMaleev}) and~(\ref{eq:W7_DysonMaleev}) results in $W^{(4)}_{{\bm q}_1,{\bm q}_2,{\bm k}-{\bm q}_1-{\bm q}_2\leftarrow{\bm k}} W^{(7)}_{{\bm k}\leftarrow{\bm q}_1,{\bm q}_2,{\bm k}-{\bm q}_1-{\bm q}_2} \propto |\Delta|^2 |\mathbf{k}|^2$. In addition, when moving away from the maximally split lines, one gets the angular dependence $\propto |\sin (2\varphi_\mathbf{k})|^2$ from the vertex part. Thus, we obtain the following leading-order scaling of the upper-branch decay rate: $\Gamma_\text{d-wave}(\mathbf{k}) \propto |\Delta|^2 |\mathbf{k}|^5 |\sin (2\varphi_\mathbf{k})|^2$.

\par \textbf{g-wave altermagnets:} The decay phase space scales as $q_{1x},q_{2x}\propto|\Delta|^1|\mathbf{k}|^2$ and $q_{1y},q_{2y}\propto|\Delta|^{1}|\mathbf{k}|^{3}$, which gives $V_\textrm{g-wave}(\mathbf{k}) \propto |\Delta|^3 |\mathbf{k}|^6$. Substituting these scalings into Eqs.~(\ref{eq:W4_DysonMaleev}) and~(\ref{eq:W7_DysonMaleev}) results in $W^{(4)}_{{\bm q}_1,{\bm q}_2,{\bm k}-{\bm q}_1-{\bm q}_2\leftarrow{\bm k}} W^{(7)}_{{\bm k}\leftarrow{\bm q}_1,{\bm q}_2,{\bm k}-{\bm q}_1-{\bm q}_2} \propto |\Delta|^0 |\mathbf{k}|^4$. In addition, when moving away from the maximally split lines, one gets the angular dependence $\propto |\sin (4\varphi_\mathbf{k})|^3$ from the decay phase space. Thus, we obtain the following leading-order scaling of the upper-branch decay rate: $\Gamma_\text{g-wave}(\mathbf{k}) \propto |\Delta|^3 |\mathbf{k}|^{10} |\sin (4\varphi_\mathbf{k})|^3$.

\par {\it Note for $i$-wave altermagnets}: In 2D, $i$-wave altermagnetism is realized only in systems which belong to the crystallographic Laue group $6/mmm$. However, we do not have an explicit expression for the vertex part $W^{(4)}W^{(7)}$ for $6/mmm$ systems. In addition, our statement $V_\textrm{i-wave}(\mathbf{k})\propto|\Delta|^3 |\mathbf{k}|^{18}$ holds for the specific decay channel considered. We leave the analysis of the long-wavelength scaling of magnon damping in $i$-wave magnets to future work. 

\subsection{Decay process on the nodal line in $d$-wave altermagnets}

We recall from the main text and Sec.~\ref{sec:decay-d-wave-nodal} that the long-wavelength magnons on the nodal lines in $d$-wave altermagnets decay only if $\Delta>\Delta^\star_\textrm{d-wave}=\sqrt{-3\nu\kappa/2}$. The aforementioned azimuthal-angle-dependent decay process $\Gamma_\text{d-wave}(\mathbf{k}) \propto |\Delta|^2 |\mathbf{k}|^5 |\sin (2\varphi_\mathbf{k})|^2$ is inactive on the nodal lines, as those correspond to momenta $\mathbf{k}$ with $\sin(2\varphi_\mathbf{k})=0$. Instead, we have to explore the long-wavelength scaling of the decay process discussed in Sec.~\ref{sec:decay-d-wave-nodal}, which becomes the leading-order contribution to the magnon decay rate if $\Delta>\Delta^\star_\textrm{d-wave}$.

\par Equations~(\ref{eq:momentumansatz}) and~(\ref{eq:solution_theta}) lead to the following long-wavelength scaling of the corresponding decay phase space: $q_{x}$, $p_{x} \propto k$ and $q_{y}$, $p_{y} \propto k^2$, where $q_{x}$ and $p_{x}$ are momenta along a nodal line. These specific scalings give $V_\textrm{nodes}(\mathbf{k}) \propto k^3$ and $W^{(4)}_{{\bm q},{\bm p},{\bm k}-{\bm q}-{\bm p}\leftarrow{\bm k}} W^{(7)}_{{\bm k}\leftarrow{\bm q},{\bm p},{\bm k}-{\bm q}-{\bm p}} \propto k^4$. Thus, we obtain the scaling law $\Gamma_\textrm{nodes}(\mathbf{k}) \propto \Theta(\Delta-\Delta^\star_\textrm{d-wave}) |\mathbf{k}|^7$. 
\section{Longitudinal spin-structure factor}
\begin{figure}[tbh]
    \centering
    \includegraphics[width=\textwidth]{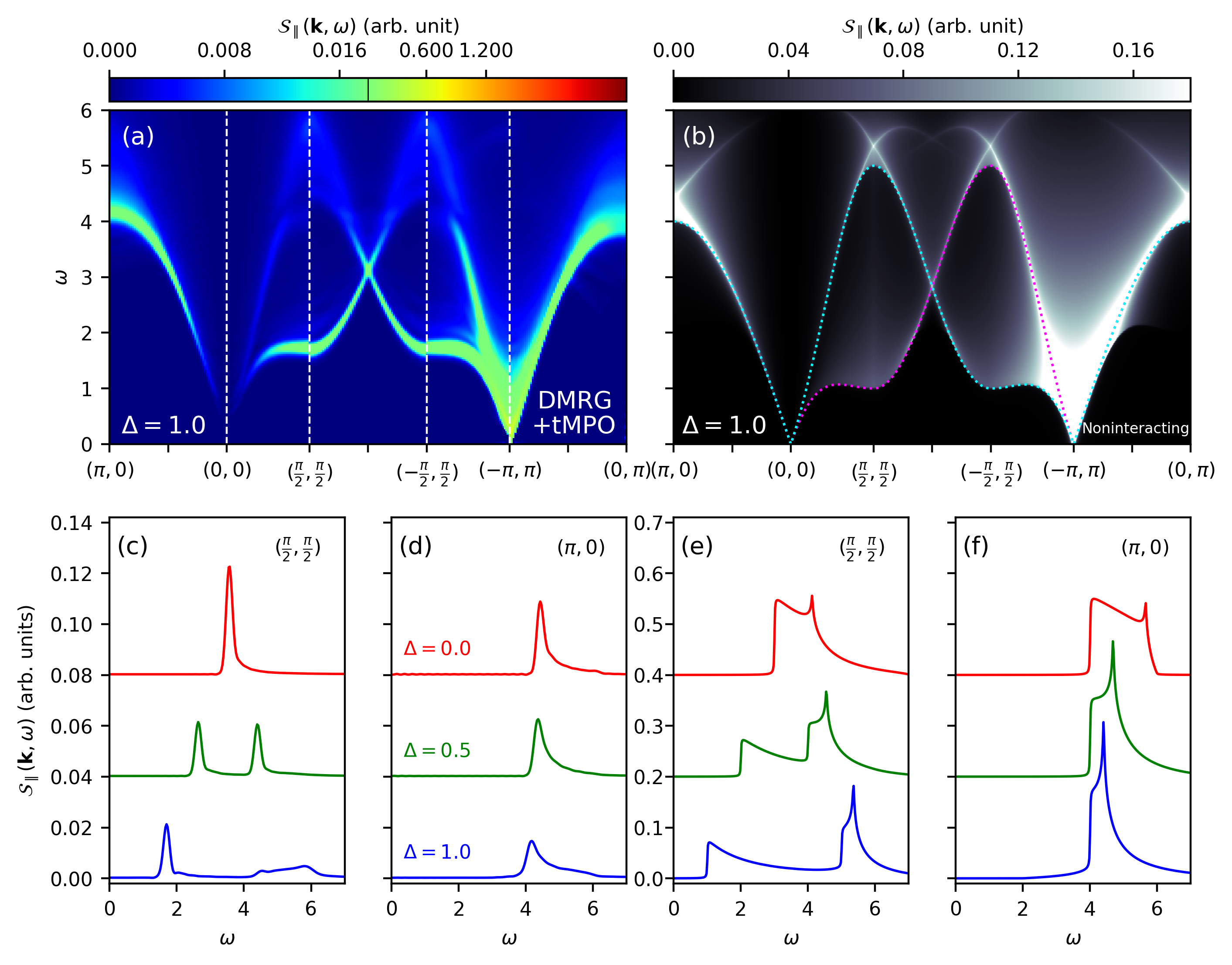}
    \caption{ \textbf{Comparison of nonperturbative and perturbative longitudinal quantum dynamics in two-dimensional d-wave altermagnets.} \textbf{(a)} The longitudinal dynamical spin-structure factor $\mathcal{S}_\parallel({\bm q},\omega)$ obtained with DMRG+tMPO.
    \textbf{(b)} $\mathcal{S}_\parallel({\bm q},\omega)$ obtained with the noninteracting theory given in Sec.~III. Magenta and cyan dotted lines indicate the noninteracting spin-$\uparrow$ and spin-$\downarrow$ magnon band dispersions, respectively. 
    (\textbf{c},\textbf{d}) Line cuts of the nonperturbative $\mathcal{S}_\perp({\bm k},\omega)$ obtained with DMRG+tMPO versus frequency at momenta $(\frac{\pi}{2},\frac{\pi}{2})$ and $(\pi,0)$, respectively.
    (\textbf{e},\textbf{f}) Line cuts of the perturbative $\mathcal{S}_\parallel({\bm k},\omega)$ obtained with noninteracting theory versus frequency at momenta $(\frac{\pi}{2},\frac{\pi}{2})$ and $(\pi,0)$, respectively.
    Parameters read $J_1=1$, $J_2=-0.5$, $S=1/2$, and $\Delta$ as indicated.}
    \label{fig:SzSz_DMRG_noninteracting}
\end{figure}

\renewcommand{\theequation}{E.\arabic{equation}}
We point out that notable features of nonrelativistic spin splitting due to altermagnetism are not only observed in transverse excitations (i.e., magnons) but also in longitudinal excitations. Figure~\ref{fig:SzSz_DMRG_noninteracting}(a) shows the nonperturbative longitudinal dynamical spin structure factor $\mathcal{S}_\parallel({\bm k},\omega)=\mathcal{S}_{zz}({\bm k},\omega)$ of the checkerboard altermagnet obtained with DMRG+tMPO (see main text for technical details). For $\Delta=1.0$, a large altermagnetic spin splitting along the paths $(0,0)\textrm{-to-}(\frac{\pi}{2},\frac{\pi}{2})$ and $(-\frac{\pi}{2},\frac{\pi}{2})\textrm{-to-}(-\pi,\pi)$ can be identified. 
The upper branch is significantly damped at $(\pm\frac{\pi}{2},\frac{\pi}{2})$, as also seen from the line cut in Fig.~\ref{fig:SzSz_DMRG_noninteracting}(c). On the other hand, the lower branch is stable at least within the range of numerical broadening $\sigma_\omega\approx 0.0896$. These features---a stable lower branch and an unstable upper branch---are similar to those of $\mathcal{S}_\perp({\bm k},\omega)$ discussed in the main text. 

Let us contrast these nonperturbative results with spin-wave calculations.
First, note that the Fourier transformation of the longitudinal spin component is given by 
\begin{equation}
\begin{aligned}
    S_{\bm q}^z
    &\simeq \frac{1}{\sqrt{N_{\rm muc}}} \sum_{\bm R} \sum_{\alpha=1}^{N_{\rm sub}} \left[{\bm e}^0_\alpha\right]_z \hat{\mathsf{S}}^0_{{\bm R},\alpha} e^{-i{\bm q}\cdot\left({\bm R}+{\bm r}_\alpha\right)} \\
    &= \frac{1}{\sqrt{N_{\rm muc}}} \sum_{\alpha=1}^{N_{\rm sub}} \sum_{\bm R} \left[{\bm e}^0_\alpha\right]_z \left( S-\hat{a}^\dagger_{{\bm R},\alpha}\hat{a}_{{\bm R},\alpha} \right) e^{-i{\bm q}\cdot\left({\bm R}+{\bm r}_\alpha\right)} \\
    &= \frac{1}{\sqrt{N_{\rm muc}}} \sum_{\alpha=1}^{N_{\rm sub}} \left[{\bm e}^0_\alpha\right]_z \sum_{\bm R} \frac{1}{N_{\rm muc}} \sum_{{\bm k},{\bm k}'} \left( S-\hat{a}^\dagger_{{\bm k},\alpha}\hat{a}_{{\bm k}',\alpha} \right) e^{-i\left({\bm k}+{\bm q}\right)\cdot\left({\bm R}+{\bm r}_\alpha\right)} e^{i{\bm k}'\cdot\left({\bm R}+{\bm r}_\alpha\right)} \\
    &= \frac{1}{\sqrt{N_{\rm muc}}} \sum_{\alpha=1}^{N_{\rm sub}} \left[{\bm e}^0_\alpha\right]_z \sum_{\bm k} \left( S-\hat{a}^\dagger_{ {\bm k},\alpha}\hat{a}_{{\bm q}+{\bm k},\alpha} \right) \\
    &= \frac{1}{\sqrt{N_{\rm muc}}} \sum_{\alpha=1}^{N_{\rm sub}} \left[{\bm e}^0_\alpha\right]_z \sum_{\bm k} 
    \left\{ S - \sum_{\nu_1,\nu_2} 
    \left( U^*_{        {\bm k},\alpha\nu_1}\hat{b}^\dagger_{        {\bm k},\nu_1} + V  _{        -{\bm k},\alpha\nu_1}\hat{b}        _{        -{\bm k},\nu_1} \right) 
    \left( U  _{{\bm q}+{\bm k},\alpha\nu_2}\hat{b}        _{{\bm q}+{\bm k},\nu_2} + V^*_{-{\bm q}-{\bm k},\alpha\nu_2}\hat{b}^\dagger_{-{\bm q}-{\bm k},\nu_2} \right) 
    \right\}
\end{aligned}
\end{equation}
in terms of magnon operators.
Thus, within the noninteracting theory, the dynamical spin-structure factor along the $z$ direction, which is parallel to the local magnetic moments, is given by
\begin{equation}
\begin{aligned}
    &\quad \mathcal{S}_\parallel({\bm q},i\omega_n) \\
    &\propto \frac{1}{\beta\hbar} \int_0^{\beta\hbar} d\tau \ev{\mathcal{T}_\tau\left[ S_{\bm q}^z(\tau)S_{-{\bm q}}^z(0) \right]} e^{i\omega_n\tau} \\
    &\simeq \frac{1}{N_{\rm muc}} \sum_{\alpha,\beta=1}^{N_{\rm sub}} \left[{\bm e}^0_\alpha\right]_z\left[{\bm e}^0_\beta\right]_z \sum_{\nu_1,\nu_2,\nu_3,\nu_4} \sum_{{\bm k},{\bm k}'} 
    V_{-{\bm k},\alpha\nu_1} U_{{\bm q}+{\bm k},\alpha\nu_2}
    U^*_{{\bm k}',\beta\nu_3} V^*_{{\bm q}-{\bm k}',\beta\nu_4}
    \int d\tau \ev{\mathcal{T}_\tau \left[ \hat{b}_{-{\bm k},\nu_1}(\tau) \hat{b}_{{\bm q}+{\bm k},\nu_2}(\tau) \hat{b}^\dagger_{{\bm k}',\nu_3}(0) \hat{b}^\dagger_{{\bm q}-{\bm k}',\nu_4}(0) \right]}_{(0)}^{\rm connected} e^{i\omega_n\tau} \\
    &\simeq
    \frac{1}{N_{\rm muc}} \sum_{\alpha,\beta=1}^{N_{\rm sub}} \left[{\bm e}^0_\alpha\right]_z\left[{\bm e}^0_\beta\right]_z \sum_{\nu_1,\nu_2} \sum_{{\bm k}} 
    \left(
    V  _{{\bm k}        ,\alpha\nu_1} U  _{{\bm q}-{\bm k},\alpha\nu_2} +
    V  _{{\bm q}-{\bm k},\alpha\nu_2} U  _{{\bm k}        ,\alpha\nu_1}
    \right)
    U^*_{{\bm k}        ,\beta \nu_1} V^*_{{\bm q}-{\bm k},\beta \nu_2}
    \frac{1}{i\omega_n-\varepsilon_{\nu_1}({\bm k})-\varepsilon_{\nu_2}({\bm q}-{\bm k})}.
    \label{eq:SzSz}
\end{aligned}
\end{equation}
\par Figure~\ref{fig:SzSz_DMRG_noninteracting}(b) shows $\mathcal{S}_\parallel({\bm k},\omega)$ obtained within the noninteracting (linear) spin-wave theory, and its line cuts are also plotted in Figs~\ref{fig:SzSz_DMRG_noninteracting}(e) and~\ref{fig:SzSz_DMRG_noninteracting}(f). Some quasiparticle-like features in Fig.~\ref{fig:SzSz_DMRG_noninteracting}(e) correspond to one-magnon dispersions indicated by dotted lines. As understood from Eq.~\eqref{eq:SzSz}, $\mathcal{S}_\parallel({\bm k},\omega)$ is related to the spin-compensated two-magnon density of states given by
\begin{equation}
    \mathcal{D}^{(2)}_0({\bm k},\omega) 
    = \frac{1}{N_\textrm{muc}} \sum_{{\bm q}\in\textrm{BZ}} 
    \delta \left( \omega - \epsilon_{+}({\bm q}) - \epsilon_{-}({\bm k}-{\bm q}) \right).
\end{equation}

As an example, in~Fig.~\ref{fig:SzSz_DMRG_noninteracting}(e) with $\Delta=1.0$, a cliff-like van-Hove singularity (vHS) at $\omega \approx 1.0$ and a shoulder-like vHS at $\omega \approx 5.0$ highlight the scattering processes $\epsilon_+({\bm k})\rightarrow\epsilon_+({\bm k})+\epsilon_-({\bm 0})$ and $\epsilon_-({\bm k})\rightarrow\epsilon_-({\bm k})+\epsilon_+({\bm 0})$, respectively. Since one of the scattered products is the Goldstone mode and the other is the original magnon itself, these vHSs directly correspond to the quasiparticle-like features reproducing the one-magnon dispersions.
Therefore, within the noninteracting approximation, the large altermagnetic spin splittings observed in Figs.~\ref{fig:SzSz_DMRG_noninteracting}(a) and~\ref{fig:SzSz_DMRG_noninteracting}(c) may be attributed to the spin-split van-Hove singularities of the two-magnon density of states.
\par However, we emphasize that, as seen in Fig.~\ref{fig:SzSz_DMRG_noninteracting}(a), these quasiparticle-like features 
are significantly renormalized by many-body interaction. Interactions also wash out spike-like vHSs, which are observed at $\omega \gtrsim 4$ in Figs.~\ref{fig:SzSz_DMRG_noninteracting}(b),~\ref{fig:SzSz_DMRG_noninteracting}(e), and~\ref{fig:SzSz_DMRG_noninteracting}(f) and are gone in Figs.~\ref{fig:SzSz_DMRG_noninteracting}(a),~\ref{fig:SzSz_DMRG_noninteracting}(c), and~\ref{fig:SzSz_DMRG_noninteracting}(d). A quantitative evaluation of these interaction-induced renormalizations of $\mathcal{S}_\parallel({\bm k},\omega)$ are left for future study. It may require the incorporation of nonperturbative scattering processes, for example, resummed Feynman diagrams, and fractionalized spinon excitations.

\end{document}